\documentclass[10pt,twocolumn,amsmath,amssymb, aps, prb]{revtex4-1}

\usepackage{graphicx}
\usepackage{dcolumn}
\usepackage{bm}
\usepackage{hyperref}
\usepackage[cp1250]{inputenc}
\usepackage[usenames,dvipsnames]{color}
\usepackage{tabularx}

\usepackage{cancel}
\usepackage{color}

\newcommand{\C}{\gamma_{\rm E}}
\newcommand{\F}{C}
\newcommand{\Tr}{\mathop{\rm Tr}}

\newcommand{\rmi}{{\rm i}}
\newcommand{\rmd}{{\rm d}}
\newcommand{\vecS}{{\rm\bf S}}
\newcommand{\vecs}{{\rm\bf s}}

\newcommand{\veclambda}{\underline{\lambda}}

\newcommand{\arccot}{{\rm arccot}}

\newcommand{\Cot}{\mathop{\rm Cot}}
\newcommand{\sgn}{\mathop{\rm sgn}}

\newcommand{\primesum}{\mathop{\sum\nolimits^{\prime}}}
\newcommand{\doubleprimesum}{\mathop{\sum\nolimits^{\prime\prime}}}
\newcommand{\arctanh}{\mathop{\rm arctanh}}

\def\Xint#1{\mathchoice
   {\XXint\displaystyle\textstyle{#1}}%
   {\XXint\textstyle\scriptstyle{#1}}%
   {\XXint\scriptstyle\scriptscriptstyle{#1}}%
   {\XXint\scriptscriptstyle\scriptscriptstyle{#1}}%
   \!\int}
\def\XXint#1#2#3{{\setbox0=\hbox{$#1{#2#3}{\int}$}
     \vcenter{\hbox{$#2#3$}}\kern-.5\wd0}}

\def\dashint{\Xint-}

\renewcommand{\today}{November 19, 2019}

\begin{document}

\title{Symmetric single-impurity Kondo model on a tight-binding chain:\\
a comparison of analytical and numerical ground-state approaches}
  

\author{Gergely Barcza$^1$}
\author{Kevin Bauerbach$^2$}
\author{Fabian Eickhoff$^3$}
\author{Frithjof B.\ Anders$^3$}
\author{Florian Gebhard$^2$}
\email{florian.gebhard@physik.uni-marburg.de}
\author{\"Ors Legeza$^1$}
\affiliation{$^1$Strongly Correlated Systems Lend\"ulet Research Group, 
Institute for Solid State Physics and Optics, MTA Wigner Research Centre for
Physics, P.O.\ Box 49, H-1525 Budapest, Hungary}
\affiliation{$^2$Fachbereich Physik, Philipps-Universit\"at Marburg,
D-35032 Marburg, Germany}
\affiliation{$^3$ Theoretische Physik 2, Technische Universit\"at Dortmund,
  D-44221 Dortmund, Germany}

\date{\today}

\begin{abstract}%
We analyze the ground-state energy, local spin correlation, 
impurity spin polarization, impurity-induced magnetization, and corresponding
zero-field susceptibilities of the symmetric single-impurity Kondo model (SIKM)
on a tight-binding chain with bandwidth~$W=2{\cal D}$
where a spin-1/2 impurity at the chain center
interacts with coupling strength~$J_{\rm K}$
with the local spin of the bath electrons.
We compare perturbative results and variational upper bounds from Yosida,
Gutzwiller, and first-order Lanczos wave functions
to the numerically exact extrapolations obtained from the
Density-Matrix Renormalization Group (DMRG) method
and from the Numerical Renormalization Group (NRG) method
performed with respect the inverse system size and Wilson parameter, respectively.
%
In contrast to the Lanczos and Yosida wave functions,
the Gutzwiller variational approach becomes exact in the strong-coupling limit,
$J_{\rm K}\gg W$,
and reproduces the ground-state properties from DMRG and NRG
for large couplings, $J_{\rm K}\gtrsim W$, with a high accuracy.
For weak coupling, the Gutzwiller wave function describes a symmetry-broken state
with an oriented local moment, in contrast to the exact solution.
We calculate the impurity spin polarization and its 
susceptibility in the presence of magnetic fields
that are applied globally or only locally to the impurity spin.
The Yosida wave function provides qualitatively correct results in the weak-coupling limit.
In DMRG, chains with about $10^3$ sites are large enough to describe the susceptibilities
down to $J_{\rm K}/{\cal D}\approx 0.5$. For smaller Kondo couplings, only the NRG
provides reliable results for a general host-electron density of states $\rho_0(\epsilon)$.
To compare with results from Bethe Ansatz that become exact in the wide-band limit,
we study the impurity-induced magnetization
and zero-field susceptibility.
For small Kondo couplings, the zero-field susceptibilities
at zero temperature approach
$\chi_0(J_{\rm K}\ll {\cal D})/(g\mu_{\rm B})^2\approx
\exp[1/(\rho_0(0)J_{\rm K})]/(2\F{\cal D}\sqrt{\pi e \rho_0(0)J_{\rm K}})$,
where 
$\ln(\F)$ is the regularized first inverse moment of  the density of states. 
Using NRG, we determine the universal sub-leading corrections 
up to second order in~$\rho_0(0)J_{\rm K}$.
\end{abstract}



\maketitle

\section{Introduction}
\label{sec:Intro}


\subsection{Kondo problem and Kondo model}

Magnetic moments that couple antiferromagnetically to electron spins
of a metallic host pose a difficult many-particle problem
because the spin-flip scattering of the host electrons off the impurity spin
couples the bath degrees of freedom in an intricate way.
Experimentally, this leads to surprising phenomena such as the Kondo resistance
minimum around some characteristic low-temperature
energy scale~$T_{\rm K}$,~\cite{DEHAAS19341115}
often referred to as the Kondo temperature.

Using standard high-temperature perturbation theory to third order
in the coupling between the impurity spin and the host electrons,
Kondo was able to explain the resistance minimum.~\cite{Kondosexplanation}
However, within standard perturbation theory
the resistivity and many other physical quantities like the zero-field magnetic
susceptibility diverge logarithmically at zero temperature.
A summation of the leading logarithmically diverging terms in the perturbation expansion
leads to a divergence at~$T_{\rm K}$.~\cite{PhysRev.158.570,Hewson}
Consequently, approaches beyond perturbation theory are required
to describe adequately the ground state of the coupled system of impurity
spin and host electrons.~\cite{Hewson}

This `Kondo problem' inspired the development of scaling concepts
that were eventually formalized in Wilson's
Renormalization Group (RG).~\cite{RevModPhys.47.773}
Since the Wilson RG can be carried out analytically only to a limited degree,
it found its widespread implementation as
Numerical Renormalization Group (NRG) method which is best suited for the study
of impurity problems;~\cite{RevModPhys.47.773,PhysRevB.21.1003,PhysRevB.21.1044}
for a review, see Ref.~[\onlinecite{RevModPhys.80.395}].

At zero temperature, the impurity spin and the electrons in its surrounding
`Kondo cloud' form a `Kondo singlet' as the many-particle ground state;
its elementary excitations describe a Fermi liquid.~\cite{1974JLTP1731N}
The Bethe Ansatz permits the exact solution of the Kondo model
with infinite bandwidth, see the reviews by
Tsvelick and Wiegmann~\cite{TsvelickWiegmann}
and by Andrei, Furuya, and
Lowenstein.~\cite{RevModPhys.55.331} 
The Bethe Ansatz confirms the findings of (N)RG,
and provides analytical formulae, e.g.,
for the impurity magnetization at finite temperatures, and for the Kondo
temperature in terms of the Bethe-Ansatz parameters.
Since NRG provides explicit results also for dynamical quantities
at finite temperatures, the Kondo problem could be declared `solved'.

The Kondo problem poses one of the fundamental challenges
in theoretical many-body physics. Therefore, one might think that the
ground-state properties of the Kondo model have been studied in very much
detail. Surprisingly, this is not the case. For example, the
dependence of the ground-state energy on the Kondo coupling
is largely unknown, apart from a study by Mancini and Mattis
who used the Lanczos approach.~\cite{PhysRevB.31.7440}
To the best of our knowledge, the large-coupling limit of the Kondo model
has not been analyzed extensively yet. Moreover,
more elaborate variational states such as the Gutzwiller wave function
were not applied to the Kondo model thus far.

It was not until recently that Schnack and H\"ock
used the NRG to investigate the magnetization and zero-field susceptibility
for some weak couplings.~\cite{PhysRevB.87.184408}
They emphasized that the impurity spin polarization differs from the impurity-induced
magnetization for the whole system, as derivable from the free energy.
Moreover, they revived the question how the Bethe Ansatz results can be
used for comparison with NRG data because the Kondo couplings
in Bethe Ansatz and for a lattice model are related in a non-trivial way.
For the series expansion of the Bethe Ansatz coupling $J_{\rm K}^{\rm BA}$
in terms of the bare model parameter $J_{\rm K}$,
only the leading order terms are known analytically
from scaling arguments~\cite{Hewson} and
Wilson's RG.~\cite{RevModPhys.47.773}

With our work, we fill some of the gaps in the quantitative analysis
of the symmetric Kondo model at zero temperature.
We study the ground-state energy, the local spin correlation function,
the impurity spin polarization and the impurity-induced magnetization
as a function of a global and a local magnetic field,
and the corresponding zero-field susceptibilities.
In the absence of an external field, we perform weak-coupling
and strong-coupling perturbation theory.
We employ three analytical variational approaches
(first-order Lanczos,~\cite{Lanczos:1950:IMS,EricKoch,PhysRevB.31.7440}
Yosida,~\cite{PhysRev.147.223,Yosidabook}
and Gutzwiller states~\cite{Gutzwiller1964,TIAM}),
and perform numerical
calculations using the Density-Matrix Renormalization Group
(DMRG)~\cite{White-1992a,White-1992b,White-1993}
and the Numerical Renormalization Group
methods.~\cite{RevModPhys.47.773,PhysRevB.21.1003,PhysRevB.21.1044,RevModPhys.80.395,PhysRevB.87.184408}
We compare to Bethe Ansatz~\cite{TsvelickWiegmann,RevModPhys.55.331}
results where possible.

\subsection{Outline}

Our work is organized as follows.
In Sect.~\ref{sec:SIKMdef} we define the Kondo Hamiltonian
on a chain and the ground-state properties that we investigate
in the thermodynamic limit, namely, the
ground-state energy,
local spin correlation function,
impurity spin polarization and impurity-induced magnetization,
and the corresponding zero-field magnetic susceptibilities.

In Sect.~\ref{sec:PTresutls}
we employ perturbation theory as first analytical method
to derive the ground-state energy and the local
spin correlation for weak and strong Kondo couplings. These results
provide a benchmark test for all
approximate analytical and numerical methods.

Next, in Sect.~\ref{sec:Lanczosresults}, we derive a variational bound for the
ground-state energy from the first-order Lanczos state.
As the energy bound is poor, we refrain from calculating
magnetic properties for this state.

As a more suitable variational state, we study the Yosida wave function
in Sect.~\ref{sec:Yosidaresutls}.
When properly generalized to finite magnetic fields, it permits
the analytic calculation of magnetic ground-state properties
in the presence of a local and a global magnetic field.
Although the Yosida state gives a poor estimate for the ground-state energy,
it provides a qualitatively correct description of the 
zero-field magnetic susceptibilities at small Kondo couplings.

As third analytic variational approach,
we study the Gutzwiller wave function
in Sect.~\ref{sec:Gutzwillerresults}.
It can be viewed as a Hartree-Fock ground state for the Kondo model where the
condition of a spin on the impurity is guaranteed. From the Hartree-Fock perspective
it is not too surprising that the Gutzwiller state contains
an artificial transition from a phase
with a broken local symmetry at small Kondo couplings
to a phase with a local spin singlet at large Kondo couplings.
Apart from this flaw, the ground-state energy and the local spin correlation
are in very good agreement with numerically exact data from
NRG and DMRG. The Gutzwiller state becomes exact for strong couplings.

As the last analytic approach, we recall results from the Bethe Ansatz
in Sect.~\ref{sec:BAnsatz}.
The Bethe Ansatz solves a related Kondo model that has a linear dispersion relation
with an infinite bandwidth so that it is a non-trivial task to establish
the link to the parameters in the lattice model. This is accomplished by
Wilson's Renormalization Group, and we use perturbation theory
to calculate  analytically the leading-order terms for the
zero-field impurity-induced susceptibility.

In Sect.~\ref{sec:Numapproaches} we discuss
two numerically exact approaches to the Kondo problem,
namely the Numerical Renormalization Group (NRG)
and the Density-Matrix Renormalization Group (DMRG) methods.
The DMRG treats finite chains with up to $L\approx 1000$ sites
with a very high numerical accuracy. Thereby, DMRG provides
excellent variational upper bounds for the ground-state energy
and the local spin correlation function. Since it has an essentially constant
energy resolution over the whole band, our present version of DMRG cannot access
the exponentially small Kondo scale that develops for small Kondo couplings.
The NRG was developed and designed to treat these Kondo scales
and therefore provides access to
small Kondo couplings as well.

In Sect.~\ref{sec:Comparison}
we compare the results of all methods.
The Gutzwiller approach provides the best analytic variational state
for the ground-state energy and the local spin correlation function.
The Gutzwiller wave function becomes the exact ground state
for large Kondo couplings, and
reliably describes the physics when the Kondo coupling becomes 
larger than the host-electron bandwidth.
The DMRG provides excellent values for the ground-state properties,
and our analysis of the finite-size data only fails to describe
magnetic properties when the Kondo energy scale becomes exponentially small.
The NRG is found to work very well for all cases. In particular, it permits
to determine the different sub-leading terms of the zero-field magnetic susceptibilities
when they become exponentially large as a function of the Kondo coupling.

In Sect.~\ref{sec:Conclusions}, we summarize and briefly discuss our findings.
We defer technical details to
appendix~\ref{mainappendixA} and provide
extensive calculations 
in the supplemental material, as listed in
appendix~\ref{mainappendixB}.

\section{Single-impurity Kondo model on a chain}
\label{sec:SIKMdef}

We start our investigation with the definition of the model Hamiltonian.
Next, we list the ground-state quantities that we study in this work.

\subsection{Hamiltonian of the single-impurity Kondo model}

In the strong-coupling limit, 
a Schrieffer-Wolff transformation maps the symmetric 
single-impurity Anderson model (SIAM) to the 
the $s$-$d$ (or single-impurity Kondo) model
(SIKM),~\cite{PhysRev.149.491,Hewson,Solyom3}
\begin{equation}
\hat{H}_{\rm K}=\hat{T}+\hat{V}_{\rm sd} +\hat{H}_{\text{m}}\; .
\label{eq:defHKondo}
\end{equation}
We consider a chain with an odd number of sites~$L$, 
$n=-(L-1)/2, \ldots, (L-1)/2$, and we choose $L$ such that $(L+3)/2$ 
is even.

The operator for the kinetic energy of the conduction electrons reads
\begin{equation}
\hat{T}= -t\sum_{n=-(L-1)/2,\sigma}^{(L-3)/2} 
\left(\hat{c}_{n,\sigma}^{+}
\hat{c}_{n+1,\sigma}^{\vphantom{+}} +
\hat{c}_{n+1,\sigma}^{+}
\hat{c}_{n,\sigma}^{\vphantom{+}}\right)\; .
\label{eq:defT} 
\end{equation}
In the absence of an external magnetic field, 
we address
a paramagnetic half-filled system, $N_{\uparrow}=N_{\downarrow}=(L+1)/2$.

The impurity couples purely locally
at the center of the chain. For a local hybridization 
in the symmetric SIAM and for strong coupling, the Kondo coupling becomes
\begin{eqnarray}
  \hat{V}_{\rm sd}&=& J_{\rm K}\, \vecs_0 \cdot \vecS \nonumber \\
  \vecs_0 \cdot \vecS &=& 
\frac{1}{2}\Bigl(\hat{c}_{0,\uparrow}^+\hat{c}_{0,\downarrow}^{\vphantom{+}} 
\hat{d}_{\downarrow}^+\hat{d}_{\uparrow}^{\vphantom{+}} 
+\hat{c}_{0,\downarrow}^+\hat{c}_{0,\uparrow}^{\vphantom{+}} 
\hat{d}_{\uparrow}^+\hat{d}_{\downarrow}^{\vphantom{+}} \Bigr)
\nonumber \\
&& 
+\frac{1}{4}
\Bigl(
\hat{d}_{\uparrow}^+\hat{d}_{\uparrow}^{\vphantom{+}} 
- \hat{d}_{\downarrow}^+\hat{d}_{\downarrow}^{\vphantom{+}}
\Bigr)
\Bigl(\hat{c}_{0,\uparrow}^+\hat{c}_{0,\uparrow}^{\vphantom{+}} 
-\hat{c}_{0,\downarrow}^+\hat{c}_{0,\downarrow}^{\vphantom{+}} \Bigr) \; .
\label{eq:simplestKondo} 
\end{eqnarray}
The host electron spin~$\vecs_0$ at site $n=0$ interacts locally with the
impurity spin~$\vecS$ with coupling strength $J_{\rm K}\geq 0$.
Note that in eq.~(\ref{eq:simplestKondo})
it is implicitly understood
that $\hat{H}_{\rm K}$ only acts in the subspace of singly occupied $d$-levels.

To study the magnetization and magnetic susceptibility,
we add an external magnetic field~${\cal H}>0$,
\begin{equation}
  \hat{H}_{\text{m}}
  =-B   \Bigl(
\hat{d}_{\uparrow}^+\hat{d}_{\uparrow}^{\vphantom{+}} 
- \hat{d}_{\downarrow}^+\hat{d}_{\downarrow}^{\vphantom{+}}
+\sum_{n=-(L-1)/2}^{(L-1)/2 }
\hat{c}_{n,\uparrow}^+\hat{c}_{n,\uparrow}^{\vphantom{+}} 
- \hat{c}_{n,\downarrow}^+\hat{c}_{n,\downarrow}^{\vphantom{+}}
\Bigr)\, ,
\label{eq:Hmagbulk}
     \end{equation}
where we denote the magnetic energy by
\begin{equation}
  B=g_e\mu_{\rm B} {\cal H}/2>0 \; ,
  \label{eq:Benergy}
  \end{equation}
$g_e\approx 2$ is the electronic gyromagnetic factor,
and $\mu_{\rm B}$ is the Bohr magneton.
For completeness, we shall also consider the case where the magnetic field
is applied only at the impurity site,
\begin{equation}
  \hat{H}_{\text{m, loc}}
  =-B   \Bigl(
\hat{d}_{\uparrow}^+\hat{d}_{\uparrow}^{\vphantom{+}} 
- \hat{d}_{\downarrow}^+\hat{d}_{\downarrow}^{\vphantom{+}}
\Bigr)\; .
\label{eq:Hmaglocal}
\end{equation}

The kinetic energy of the host electrons is diagonal in momentum space,
see appendix~\ref{app:diagTandhalfchain},
\begin{equation}
\hat{T}= \sum_{k=1,\sigma}^L \epsilon_k
\hat{a}_{k,\sigma}^+ \hat{a}_{k,\sigma}^{\vphantom{+}} 
\end{equation}
with the dispersion relation $\epsilon_k$.
The corresponding density of states is defined by
\begin{equation}
  \rho_0(\omega)=\frac{1}{L}\sum_{k=1}^L\delta(\omega-\epsilon_k)
  =\frac{1}{\pi}\frac{1}{\sqrt{(2t)^2-\omega^2}}
  \label{eq:DOS}
\end{equation}
for $|\omega| < 2t$.
We use half the bandwidth as our unit of energy, $2t\equiv 1$, $W\equiv 2$,
to make a direct contact with the Bethe Ansatz calculations.
For some of our analytic calculations, we shall treat $\rho_0(\omega)$
as a selectable quantity.

In numerical DMRG calculations, the model is mapped onto a half chain
with the impurity at the left chain end. This is done in appendix~\ref{app:halfchain}.

\subsection{Ground-state properties}
\label{subsec:gspropertiesdefanddiscussion}

In this work, we are interested in the excess
ground-state energy
due to the presence of the coupled impurity spin,
the local spin correlation
the impurity spin polarization for a global and a local field,
and the corresponding susceptibilities. Moreover,
for comparison with Bethe Ansatz, we also address the
impurity-induced magnetization and zero-field susceptibility
for global and local fields.

\subsubsection{Ground-state energy and local spin correlation}

We calculate the excess ground-state energy $e_0(J_{\rm K})$
due to the presence of the impurity, i.e.,
the impurity-induced change of the ground-state energy of free electrons,
\begin{equation}
  e_0(J_{\rm K},L)
  = E_0(J_{\rm K},L)-E_{\rm FS}(L) \; .
  \label{eq:excesse0}
\end{equation}
The impurity-induced energy contribution
$e_0(J_{\rm K})$ is of the order unity and $e_0(J_{\rm K}=0)=0$.
Eventually, we extrapolate to the thermodynamic limit,
\begin{equation}
  e_0(J_{\rm K}) =\lim_{L\to\infty} e_0(J_{\rm K},L) \; .
\end{equation}
This is done explicitly using the DMRG.
The NRG discretizes the continuum model in energy space,
see Sect.~\ref{sec:Numapproaches}.
The analytic calculations are directly performed in the thermodynamic limit.

Another quantity of interest is the local spin correlation function in the ground state,
\begin{equation}
  C_0^S(J_{\rm K})=\langle \vecs_0 \cdot \vecS \rangle \; .
  \label{eq:localspinCFdef}
  \end{equation}
It can either be calculated directly, or from the Hellmann-Feynman
theorem, see appendix~\ref{app:misc},~\cite{Hellmann2,Feynman} 
\begin{equation}
  C_0^S(J_{\rm K})=\frac{\partial e_0(J_{\rm K})}{\partial J_{\rm K}} \; .
  \label{eq:CS0frome0}
\end{equation}
In turn, we may calculate the ground-state energy from
the local spin correlation using
\begin{equation}
e_0(J_{\rm K}) = \int_0^{J_{\rm K}}\rmd J C_0^S(J) \; . \label{eq:e0fromCS0}
\end{equation}
Therefore, eq.~(\ref{eq:e0fromCS0}) can be used to check the
consistency of the ground-state calculations because
eqs.~(\ref{eq:CS0frome0}) and~(\ref{eq:e0fromCS0})
hold for the exact ground state.
Note that the Hellmann-Feynman theorem also applies to variational approaches,
see appendix~\ref{app:misc}.

\subsubsection{Ground-state impurity spin polarization,
  impurity-induced magnetization, and zero-field susceptibilities}
\label{subsec:gspropertiesdefanddiscussionpart2}

\paragraph{Global external field.}

In the presence of an external magnetic field~${\cal H}$,
the spin on the impurity orients itself so that the spin-projection
into the direction of the external field becomes finite. We denote
the impurity spin polarization as
\begin{eqnarray}
  m^S(J_{\rm K},B)
  &=&g_{\rm e}\mu_{\rm B} S^z(J_{\rm K},B)
  \; , \nonumber \\
  S^z(J_{\rm K},B)&=& \langle \hat{S}^z\rangle
  =\frac{1}{2}\langle \hat{n}_{d,\uparrow}-\hat{n}_{d,\downarrow}\rangle \; .
  \label{eq:deflocalm}
\end{eqnarray}
Correspondingly, we define the impurity spin susceptibility via
the relation
\begin{equation}
  \chi^S(J_{\rm K},B)
  = \frac{\partial m^S(J_{\rm K},B)}{\partial {\cal H}}
  = \frac{g_e\mu_{\rm B}}{2}
\frac{\partial m^S(J_{\rm K},B)}{\partial B} \; .
\label{eq:chifromlocalm}
\end{equation}
The impurity spin polarization and susceptibility
can straightforwardly be calculated for our
various ground-state approaches.

The impurity spin polarization must not be confused with the
thermodynamic magnetization of the system,
\begin{equation}
  m(J_{\rm K},B,T)=-\frac{\partial {\cal F}(J_{\rm K},B,T)}{\partial {\cal H}}
  =g_e\mu_{\rm B} S^z_{\rm tot}(J_{\rm K},B,T)\; ,
  \label{eq:defmasderivetiveofcalF}
\end{equation}
where $T=1/\beta$ is the temperature, and the total spin projection in
the direction of the external field is
\begin{eqnarray}
  S^z_{\rm tot}(J_{\rm K},B,T)&=&S^z(J_{\rm K},B,T)+s^z(J_{\rm K},B,T) \; ,
\nonumber \\
  S^z(J_{\rm K},B,T)&=& \langle \hat{S}^z\rangle
  =\frac{1}{2}\langle \hat{n}_{d,\uparrow}-\hat{n}_{d,\downarrow}\rangle \; ,
  \label{eq:mimpuritydef}
\\
  s^z(J_{\rm K},B,T) &=& 
  \langle\hat{s}^z\rangle=
  \sum_{n=-(L-1)/2}^{(L-1)/2}
\frac{\langle
\hat{c}_{n,\uparrow}^+\hat{c}_{n,\uparrow}^{\vphantom{+}} 
- \hat{c}_{n,\downarrow}^+\hat{c}_{n,\downarrow}^{\vphantom{+}}
\rangle}{2}\, ,\nonumber 
\end{eqnarray}
where the angular brackets imply the thermal average.

Since $s^z(J_{\rm K},B,T)$ is proportional to the system size,
the thermodynamic magnetization is not a useful quantity
because the impurity spin contribution $S^z(J_{\rm K},B,T)$ is only of order unity.
Therefore, it is more sensible to define impurity-induced changes
to thermodynamic quantities due to the presence of the
impurity.~\cite{RevModPhys.55.331,TsvelickWiegmann,PhysRevB.87.184408}
The impurity-induced free energy
is defined by
\begin{eqnarray}
  {\cal F}^{\rm ii}(J_{\rm K},B,T)&=&
  -T\ln \Tr \left[ \exp(-\beta \hat{H}_{\rm K})\right] \nonumber \\
  &&   +T\ln \Tr \left[ \exp\left(-\beta (\hat{T}-\hat{H}_{\rm m})\right)\right] ,\;
  \label{eq:defFreeenergyii}
\end{eqnarray}
where the chemical potential is $\mu(T)=0$
for the particle-hole symmetric Kondo and free-fermion Hamiltonians
at all temperatures.~\cite{Barczaetal}
The derivative with respect to ${\cal H}$ gives the impurity-induced
magnetization,
\begin{equation}
  m^{\rm ii}(J_{\rm K},B,T)=g_e\mu_{\rm B}\left(S^z_{\rm tot}(J_{\rm K},B,T)-
  s^{z,{\rm free}}(B,T)\right) .
  \label{eq:miidef}
\end{equation}
It is of the order unity.

In eq.~(\ref{eq:defFreeenergyii}) 
we have ${\cal F}^{\rm ii}(J_{\rm K},B,T=0)=e_0(J_{\rm K},B)$
at zero temperature 
so that we can obtain the impurity-induced magnetization
also from the excess ground-state energy
\begin{equation}
  m^{\rm ii}(J_{\rm K},B)=-\frac{g_e\mu_{\rm B}}{2}
  \frac{\partial e_0(J_{\rm K},B)}{\partial B}
  \; .
  \label{eq:miiande0relation}
\end{equation}
The impurity-induced magnetic susceptibility at zero temperature follows as
\begin{eqnarray}
  \chi^{\rm ii}(J_{\rm K},B)
&=& \frac{g_e\mu_{\rm B}}{2}
\frac{\partial m^{\rm ii}(J_{\rm K},B)}{\partial B} \nonumber \\
&=& - \left(\frac{g_e\mu_{\rm B}}{2}\right)^2
\frac{\partial^2 e_0(J_{\rm K},B)}{\partial B^2}
\; .
\label{eq:chifromm}
\end{eqnarray}
We abbreviate the impurity-induced susceptibility at zero field as
$\chi_0^{\rm ii}(J_{\rm K})\equiv \chi^{\rm ii}(J_{\rm K},B=0)$.

\paragraph{Local external field.} 

When the magnetic field is applied only at the impurity site,
we denote the corresponding quantities by an extra lower index `loc', e.g.,
$m^S_{\rm loc}(J_{\rm K},B)$ and $\chi^S_{\rm loc}(J_{\rm K},B)$.
For a local field, the impurity spin polarization and susceptibility
are the proper thermodynamic quantities. 
They can be calculated from the ground-state energy in the presence of a local field,
\begin{eqnarray}
  m_{\rm loc}^S(J_{\rm K},B) &=& -\frac{g_e\mu_{\rm B}}{2}
  \frac{\partial e_{0,{\rm loc}}(J_{\rm K},B)}{\partial B} \; , \nonumber \\
  \chi_{\rm loc}^S(J_{\rm K},B)
  &=& \frac{g_e\mu_{\rm B}}{2}
  \frac{\partial m_{\rm loc}^S(J_{\rm K},B)}{\partial B} \nonumber \\
  &=& - \left(\frac{g_e\mu_{\rm B}}{2}\right)^2
  \frac{\partial^2 e_{0,{\rm loc}}(J_{\rm K},B)}{\partial B^2}\; .
  \label{eq:definelocalmandchiasderivatives}
\end{eqnarray}
For a local field, the free host electron system is unpolarized.
Therefore, the impurity-induced magnetization in the presence of a local field
describes the impurity spin polarization plus the
induced magnetization of the host electrons and thus is of order unity,
\begin{eqnarray}
  m^{\rm ii}_{\rm loc}(J_{\rm K},B)&=&g_e\mu_{\rm B} S_{\rm tot}^z(J_{\rm K},B)
  \nonumber \\
  &=& g_e\mu_{\rm B} \left(S^z(J_{\rm K},B)+s^z(J_{\rm K},B)\right)
  \end{eqnarray}
at zero temperature. In general,
the impurity-induced magnetization
is smaller than the impurity spin polarization because it is reduced by
the contribution of the bath electron screening cloud,
$s^z(J_{\rm K},B)<0$.

\paragraph{Zero-field susceptibilities.}

There are four different susceptibilities at finite fields but only
three different zero-field susceptibilities because
\begin{equation}
  \chi_0^S(J_{\rm K},T)=\chi_{0,{\rm loc}}^{\rm ii}(J_{\rm K},T)
  \label{eq:twochisareequivalent}
\end{equation}
holds for all temperatures. To see this, we recall the definition of the
impurity-induced magnetization at finite local field~$B$,
\begin{eqnarray}
  \frac{ m^{\rm ii}_{\rm loc}(J_{\rm K},B,T)}{g_e\mu_{\rm B}}
  &=& \langle \hat{S}^z+\hat{s}^z\rangle \\
  &=& \frac{1}{\cal Z} \Tr \left[
  e^{-\beta(\hat{T}+\hat{V}_{\rm sd}-2 B \hat{S}^z)}
\left(  \hat{S}^z+\hat{s}^z \right)\right]\nonumber 
\end{eqnarray}
so that from eq.~(\ref{eq:miidef}) we find
\begin{equation}
\frac{\chi_{0,{\rm loc}}^{\rm ii}(J_{\rm K},T)}{(g_e\mu_{\rm B})^2}= 
\frac{1}{T} \langle \hat{S}^z(\hat{S}^z+\hat{s}^z) 
\rangle \; ,
\label{eq:1stforproof}
\end{equation}
where we used that the system is unpolarized for $B=0$.

On the other hand, the impurity spin polarization 
at finite global field~$B$ is defined by
\begin{eqnarray}
  \frac{ m^S(J_{\rm K},B,T)}{g_e\mu_{\rm B}}
  &=& \langle \hat{S}^z\rangle \\
  &=& \frac{1}{\cal Z} \Tr \left[
  e^{-\beta(\hat{T}+\hat{V}_{\rm sd}-2 B \hat{S}^z-2B\hat{s}^z)}\hat{S}^z \right]\nonumber 
\end{eqnarray}
so that from eq.~(\ref{eq:chifromlocalm}) we find
\begin{equation}
\frac{\chi_0^S(J_{\rm K},T)}{(g_e\mu_{\rm B})^2}= 
\frac{1}{T} \langle (\hat{S}^z+\hat{s}^z) \hat{S}^z
\rangle \; ,
\label{eq:2ndforproof}
\end{equation}
where we used that the system is unpolarized for $B=0$.
A comparison of eqs.~(\ref{eq:1stforproof}) and~(\ref{eq:2ndforproof})
proofs eq.~(\ref{eq:twochisareequivalent}).

Note that the equivalence~(\ref{eq:twochisareequivalent}) does not necessarily hold
for approximate approaches. In Sect.~\ref{sec:Yosidaresutls} we shall see
that it is not fulfilled for the Yosida variational approach.
As shown in Sect.~\ref{sec:Gutzwillerresults}, it is obeyed in the
para\-magnetic Gutzwiller wave function.
For the NRG, eq.~(\ref{eq:twochisareequivalent}) provides a convenient tool
to assess the accuracy of the numerical calculations.

\section{Perturbation theory for the ground-state energy}
\label{sec:PTresutls}

In this section, we derive the excess ground-state energy and local
spin correlation function
from weak-coupling and strong-coupling perturbation theory
at zero magnetic field. 

\subsection{Weak-coupling perturbation theory}

When we ignore the coupling between the impurity spin and the bath electrons,
the ground state is doubly degenerate. Since we are interested in the ground state,
we work with the spin singlet state
\begin{eqnarray}
  |\Phi_0\rangle &=& \sqrt{\frac{1}{2}} \left(
  \hat{d}_{\downarrow}^+\hat{a}_{k_{\rm F},\downarrow}^{\vphantom{+}}
  +
  \hat{d}_{\uparrow}^+\hat{a}_{k_{\rm F},\uparrow}^{\vphantom{+}}
  \right) |\hbox{FS}\rangle|\hbox{vac}_d
  \rangle
  \nonumber \\
  &  \equiv &
   \sqrt{\frac{1}{2}} \bigl(
|A\rangle + |B\rangle
  \bigr)
  \; ,\label{eq:Phizero}
  \end{eqnarray}
where $k_{\rm F}=(L+1)/2$ is the Fermi number in the full chain.
The state $|\Phi_0\rangle$ is normalized to unity.
The ground state of the Kondo Hamiltonian
for $J_{\rm K}=0$ and an empty $d$-level is given by the Fermi sea
\begin{equation}
 |\hbox{FS}\rangle =\prod_{\sigma}|\hbox{FS}_{\sigma}\rangle
 \quad , \quad
 |\hbox{FS}_{\sigma}\rangle = \prod_{k=1}^{k_{\rm F}} \hat{a}_{k,\sigma}^+
    |\hbox{vac}\rangle \; .
   \end{equation}
The calculations from standard perturbation theory are carried out in
appendix~\ref{app:weakcoulingPT}.

In the thermodynamic limit, there is no first-order correction, and the
excess ground-state to second order reads
\begin{equation}
  e_0^{(2)}(J_{\rm K})  =- f b_1^2
  \end{equation}
with
\begin{eqnarray}
  f&=&-4 \int_{-1}^0 \rmd \omega_1 \int_{-1}^0 \rmd \omega_2
  \rho_0(\omega_1) \rho_0(\omega_2)
    \frac{1}{\omega_1+\omega_2} \; ,\\
  b_1^2&=&  \langle \Phi_0 |\hat{V}_{\rm sd}^2 | \Phi_0\rangle\; .
  \label{eq:simplematrixelements}
  \end{eqnarray}
As shown in appendix~\ref{app:A}, we have $b_1^2=3J_{\rm K}^2/32$,
independent of the density of states.
In one dimension, $f^{d=1}=1$, see appendix~\ref{app:weakcoulingPT}, so that
our final result to second order is
\begin{eqnarray}
  e_0^{(2)}(J_{\rm K})   =-\frac{3}{32}J_{\rm K}^2 
\label{eq:secondorderanalyt}
\end{eqnarray}
for the one-dimensional density of states~(\ref{eq:DOS}).

For the local spin correlation function we thus find
\begin{equation}
  C_0^S(J_{\rm K}\ll 1) = -\frac{3}{16}J_{\rm K}
  +{\cal O}\left( J_{\rm K}^2  \right)
  \label{eq:slopeCSweak}
  \end{equation}
in the weak-coupling limit.

\subsection{Strong-coupling perturbation theory}

\subsubsection{Leading order}

To leading order in $J_{\rm K}$, the impurity spin and the
electron spin at the origin form a spin singlet. Since
\begin{equation}
\vecs_0 \cdot \vecS =\frac{1}{2} \left( \left(\vecs_0 + \vecS\right)^2
-\vecs_0^2-\vecS^2\right)
=\frac{1}{2} \left( \vecS_{\rm tot}^2
-\vecs_0^2-\vecS^2\right)
\end{equation}
and $S_{\rm tot}=0$, $S=s_0=1/2$, we have
\begin{equation}
  e_0(J_{\rm K})=-\frac{3J_{\rm K}}{4}
\end{equation}
to leading order in $J_{\rm K}$.

\subsubsection{Next-to-leading order}

To obtain the correction to order $(J_{\rm K})^0$,
we realize that the host electrons experience a scattering center at the origin
of infinite strengths.
As shown in appendix~\ref{app:B},
in the presence of a local impurity potential of strength~$V$,
spinless fermions experience the energy shift
\begin{equation}
  e_0^{\rm ps}(V)=
-\frac{1}{\pi} \int_{-1-\eta}^0 \rmd \omega \omega
 \frac{\partial}{\partial \omega}
 \Cot{}^{-1}\left[\frac{1-V\Lambda_0(\omega)}{\pi V \rho_0(\omega)}\right],
 \end{equation}
where $\rho_0(\omega)$ is the density of states of the free host electrons
and $\Lambda_0(\omega)$ is its Hilbert transform,
\begin{equation}
  \Lambda_0(\omega) = \dashint_{-1}^1\rmd \epsilon
  \frac{\rho_0(\epsilon)}{\omega-\epsilon} \; .
\end{equation}
Moreover,
$\Cot^{-1}(x)=\cot^{-1}(x)+\pi \theta_{\rm H}(-x)$
is continuous and differentiable across $x=0$, where $\theta_{\rm H}(x)$
is the Heaviside step function.

In one dimension, we obtain from appendix~\ref{app:B}
\begin{equation}
  e_0^{\rm ps}(V)=\frac{1}{2}\left(1+V-\sqrt{1+V^2}\right)
\end{equation}
for the energy shift per spin species which reduces to
\begin{equation}
  e_0^{\rm ps}(V\to\infty)=\frac{1}{2}
\end{equation}
for $V\to\infty$.
Summing over both spin species we obtain
\begin{equation}
  e_0(J_{\rm K})=-\frac{3J_{\rm K}}{4} +1
  \label{eq:energyexact1dlargeJK}
\end{equation}
for the strong-coupling limit of the Kondo model, 
with corrections of the order $(J_{\rm K})^{-1}$.

For the local spin correlation function we thus find
\begin{equation}
  C_0^S(J_{\rm K}\gg 1) = -\frac{3}{4} +{\cal O}\left( J_{\rm K}^{-2}  \right)
  \label{eq:CSzerostrongcoupling}
  \end{equation}
in the strong-coupling limit.

\section{Lanczos variational approach}
\label{sec:Lanczosresults}

As a first variational approach, we consider the
Lanczos theory and compile the results for the first-order Lanczos state.
The calculations of higher orders quickly become cumbersome and prone to errors.
Since the Yosida and Gutzwiller variational description are superior
to the Lanczos approach, we only consider the Kondo model
without an external magnetic field.

\subsection{Recursive construction}

The Lanczos approach starts from some
initial state $|\Phi_0\rangle$, e.g., the state defined in eq.~(\ref{eq:Phizero}).
The next states are constructed recursively,~\cite{Lanczos:1950:IMS,EricKoch}
\begin{equation}
|\Phi_{n+1}\rangle = \hat{H}|\Phi_n\rangle - a_n |\Phi_n\rangle - b_n^2 |\Phi_{n-1}\rangle 
\quad , \quad n\geq 0 \; ,
\label{eq:Phindefmain}
  \end{equation}
where we set $b_0\equiv 0 $, and
\begin{eqnarray}
    a_n&=&\frac{\langle \Phi_n |\hat{H}|\Phi_n\rangle}{\langle \Phi_n| \Phi_n\rangle}
  \label{eq:andefmain} \nonumber \; , \\
  b_n^2&=&
  \frac{\langle \Phi_n |\Phi_n\rangle}{\langle \Phi_{n-1}| \Phi_{n-1}\rangle} \geq 0\; .
  \label{eq:bndef}
\end{eqnarray}
The states $|\Phi_n\rangle$ are not normalized to unity but they are
orthogonal to each other, see appendix~\ref{app:A}.

The real parameters $a_l$, $b_l>0$ define the elements of
the $(M+1)\times (M+1)$ tridiagonal
Hamilton-Matrix $\underline{\underline{H}}^{(M)}$
with the entries
\begin{equation}
H_{l,m}=\delta_{l,m+1} b_l + \delta_{l,m} a_l +\delta_{l,m-1} b_{l+1}\; ,
  \end{equation}
for $0\leq l,m\leq M$.
Its lowest eigenvalue, $\Xi_0^{(M)}$, provides a variational upper bound
to the ground-state energy,~\cite{Lanczos:1950:IMS,EricKoch}
\begin{equation}
E_0 \leq \Xi_0^{(M)}\leq \Xi_0^{(M-1)}
\end{equation}
for all $M\geq 1$. For completeness, we include
a simple proof in appendix~\ref{app:A}.

\subsection{Results for the first-order Lanczos state}

The variational Lanczos energy to leading order is
\begin{equation}
\Xi_0^{(0)}=a_0=0\; ,
\label{eq:Lanczosleadingorder}
\end{equation}
see eq.~(\ref{appeq:a0iszero}).
The variational Lanczos energy to first order reads
\begin{equation}
\Xi_0^{(1)}=\frac{1}{2} \left[ a_1 -\sqrt{a_1^2+4b_1^2}\,\right]\; .
\label{eq:Lanczosfirstorder}
\end{equation}
The matrix elements in one spatial  dimension
are calculated in appendix~\ref{app:A} with the result
\begin{equation}
\Xi_0^{(1)}(J_{\rm K})=\frac{1}{2} \left[ 
-\frac{J_{\rm K}}{2} + \frac{4}{\pi}  
-\sqrt{\left(-\frac{J_{\rm K}}{2} + \frac{4}{\pi} \right)^2+
  \frac{3J_{\rm K}^2}{8}}\,\right]\; .
\label{eq:Lanczosenergyfinal}
\end{equation}
To second order in $J_{\rm K}$, the first-order Lanczos energy reads
\begin{equation}
  \Xi_0^{(1)}(J_{\rm K}\ll 1) = -\frac{\pi}{4}
  \frac{3J_{\rm K}^2}{32} + {\cal O}\left(J_{\rm K}^3\right)
  \; .
  \label{eq:piover4lanczos}
  \end{equation}
In comparison with second-order perturbation theory, eq.~(\ref{eq:secondorderanalyt}),
the Lanczos state accounts for $\pi/4\approx 78.5\%$ of the exact second-order term.

For strong coupling, the first-order Lanczos state provides the bound
\begin{eqnarray}
  \Xi_0^{(1)}(J_{\rm K}\gg 1) &=& \frac{1}{8} 
\left(-2 - \sqrt{10}\right) J_{\rm K}
  +2\frac{5 + \sqrt{10}}{5 \pi}\nonumber \\
  &=&-0.645 J_{\rm K} + 1.04\;.
  \label{eq:LanczoslargeJenergy}
  \end{eqnarray}
For $J_{\rm K}\gg 1$, the first-order Lanczos energy accounts for  86.0\%
of the exact ground-state energy given in eq.~(\ref{eq:energyexact1dlargeJK}).

\section{Yosida wave function}
\label{sec:Yosidaresutls}

As the next variational theory, we study the Yosida variational state
that we generalize to the case of a finite external field.
The Yosida state gives a poor variational energy but recovers
the exponentially large magnetic susceptibility for small Kondo couplings.
Moreover, 
the calculations can be carried out analytically to a far degree.

\subsection{Yosida variational state}

\subsubsection{Definition}

Yosida~\cite{PhysRev.147.223,Yosidabook} extended
$|\Phi_0\rangle$ in eq.~(\ref{eq:Phizero}) in a generic way,
and proposed the variational wave function
\begin{equation}
  |\Psi_{\rm Y}\rangle=\sqrt{\frac{1}{2L}}\sum_{k,\epsilon_k>0}\alpha_k\left(
  \hat{a}_{k,\downarrow}^+\hat{d}_{\uparrow}^+
  -
  \hat{a}_{k,\uparrow}^+\hat{d}_{\downarrow}^+
  \right) |\hbox{FS}\rangle|\hbox{vac}_d\rangle \; .
    \label{eq:Yosidawf}
\end{equation}
Here, $\alpha_k$ is real and of the order unity.
Note that $|\Psi_{\rm Y}\rangle$ is a spin singlet state.

To include a spin anisotropy at finite external field, $B \geq 0$,
we generalize the Yosida wave function, 
\begin{eqnarray}
  |\Psi_{\rm Y}(B)\rangle&=&\sqrt{\frac{1}{2L}}\biggl[ \primesum_k
\alpha_{k,\downarrow}
  \hat{a}_{k,\downarrow}^+\hat{d}_{\uparrow}^+  |\hbox{FS}
\rangle|\hbox{vac}_d\rangle  \nonumber \\
&&\hphantom{\sqrt{\frac{1}{2L}}}
  -
 \doubleprimesum_k \alpha_{k,\uparrow}
  \hat{a}_{k,\uparrow}^+\hat{d}_{\downarrow}^+
  |\hbox{FS}\rangle|\hbox{vac}_d\rangle  \biggr] .
    \label{eq:Yosidawfmag}
\end{eqnarray}
Since the Fermi sea depends on the magnetic field,
the prime on the sum restricts the $k$-values to
$\epsilon_k> -\epsilon_{\rm F}$,
the double prime indicates
$\epsilon_k> \epsilon_{\rm F}>0$, where
$\epsilon_{\rm F}$ is a function of the magnetic energy scale $B>0$.

\subsection{Variational ground-state energy}

\subsubsection{Energy equation}
  
The calculations are carried out in appendix~\ref{app:Yosidaenergy}.
We abbreviate the principal-value integral
\begin{equation}
F_1(x,B)= \dashint_B^1 \rmd \omega \rho_0(\omega)
\frac{1}{\omega-x} \; ,
\end{equation}
whereby we assume throughout that $0\leq B<1$, i.e., the host electrons
are not fully polarized. Note that eq.~(\ref{appeq:permittoignoreepsF})
permits to set $\epsilon_{\rm F}=B$ in our further considerations.

The Yosida ground-state energy $\lambda=  e_0^{\rm Y}(J_{\rm K},B)$
follows from the solution of the implicit equation
\begin{equation}
  \left(1-\frac{J_{\rm K}F_+}{4}\right)
  \left(1-\frac{J_{\rm K}F_-}{4}\right)
  -\frac{J_{\rm K}^2F_+F_-}{4}=0 \;, 
  \label{eq:YosidaenergywithBfinal}
  \end{equation}
where we abbreviated $F_{+}\equiv F_1(\lambda+J_{\rm K}s_0/2,B)$
and $F_{-}\equiv F_1(\lambda-J_{\rm K}s_0/2,-B)$.
In one spatial dimension we have $s_0(B)=(1/\pi)\arcsin(B)$
from eq.~(\ref{appeq:MandepsFfromBa}) and
\begin{equation}
F_1(x,B) = 
\frac{1}{\pi\sqrt{1 - x^2}}
\ln \biggl[
\frac{1 - B x + \sqrt{(1-B^2) (1- x^2)}}{B - x}
\biggr] .
\label{eq:defF1explicit}
\end{equation}

Eq.~(\ref{eq:YosidaenergywithBfinal}) provides a solution
only for $B\leq B_{\rm c}^{\rm Y}(J_{\rm K})$
above which the Yosida state becomes unstable.
This problem does not occur in the Gutzwiller description
so that we do not extend the Yosida state to the region 
$B> B_{\rm c}^{\rm Y}(J_{\rm K})$.

\subsubsection{Ground-state energy at zero field}

At $B=0$, eq.~(\ref{eq:YosidaenergywithBfinal}) simplifies to
\begin{eqnarray}
  F(\lambda)  &=&\frac{4}{3J_{\rm K}} \; , \nonumber\\
  F(\lambda)&\equiv& F_1(\lambda,0)=
  \frac{1}{\pi\sqrt{1 - \lambda^2}}
\ln \biggl[
\frac{1 + \sqrt{1- \lambda^2}}{- \lambda}
\biggr]
 \label{eq:Yosidafinalenergy}
\end{eqnarray}
for the ground-state energy $\lambda=e_0^{\rm Y}(J_{\rm K})<0$.
In general, the solution of equation~(\ref{eq:Yosidafinalenergy})
must be determined numerically.

\paragraph{Small Kondo couplings.}

For small $|E|$, we can address a general density of states because
\begin{eqnarray}
  F(\lambda) &=& \rho_0(0) \int_0^1 \rmd \omega \frac{1}{\omega-\lambda}
 +   \int_0^1 \rmd \omega \frac{\rho_0(\omega)-\rho_0(0)}{\omega-\lambda}
 \nonumber \\
 &\approx & \rho_0(0)\left(-\ln(-\lambda) +\ln(\F)\right)  +{\cal O}(\lambda)
\end{eqnarray}
for $|\lambda| \ll 1$. Here, we introduced 
the regularized first negative moment of the density of states
\begin{equation}
\ln(\F) = -\int_{-1}^0 \frac{\rmd \omega}{\omega}
\left(\frac{\rho_0(\omega)}{\rho_0(0)}-1\right) 
\label{eq:defFmain}
\end{equation}
For a constant density of states we have $\F^{\rm const}=1$ by definition.
For the one-dimensional density of states~(\ref{eq:DOS}) we find
$\F^{d=1}=2$.

To leading order we must solve
\begin{equation}
  -\rho_0(0)\ln(|E|/\F) = \frac{4}{3J_{\rm K}}
\end{equation}
so that
\begin{equation}
  e_0^{\rm Y}(J_{\rm K}\ll 1) =
  -\F\exp\left(-\frac{4}{3\rho_0(0)J_{\rm K}}\right)
  \label{eq:yosida1dsmall}
\end{equation}
results from the Yosida wave function for small Kondo couplings.
The density of states only enters via the prefactor~$\F$.
A comparison with the exact second-order
expression~(\ref{eq:secondorderanalyt}) shows that
the exponentially small variational bound provided by the Yosida wave function
is rather poor.

\paragraph{Large Kondo couplings.}

For large Kondo couplings, the structure of the density of states matters
and we restrict ourselves to the one-dimensional case.
For large $|E|$ we must solve to leading and next-to-leading order
\begin{equation}
  \frac{1}{2|E|}-\frac{1}{\pi E^2}=\frac{4}{3J_{\rm K}}
\end{equation}
so that
\begin{equation}
  e_0^{\rm Y}(J_{\rm K}\gg 1) = -\frac{3J_{\rm K}}{8}+\frac{2}{\pi}
  \label{eq:yosida1dlarge}
\end{equation}
results from the Yosida wave function for large Kondo couplings.
The comparison with the perturbative
strong-coupling result~(\ref{eq:energyexact1dlargeJK})
shows that the Yosida wave function does not become exact for 
$J_{\rm K}\gg 1$.
This indicates that the Yosida state does not  properly
describe the strong-coupling singlet state.

\subsection{Zero-field susceptibilities}

The calculations are carried out in appendix~\ref{app:Yosidamagsusz}.
Here, we summarize the results for the various zero-field susceptibilities.

\subsubsection{Zero-field impurity spin susceptibility}

To obtain the zero-field impurity spin susceptibilities, we can replace
$e_0^{\rm Y}(J_{\rm K},B)$
and
$e_{0,{\rm loc}}^{\rm Y}(J_{\rm K},B)$
by
$e_0^{\rm Y}(J_{\rm K})$ in eqs.~(\ref{appeq:Yosidaspinpolglobal})
and~(\ref{appeq:Yosidaspinpollocal});
corrections are of the order $B^2$ because the impurity spin polarization
vanishes at $B=0$.
Using {\sc Mathematica}~\cite{Mathematica11}
and eq.~(\ref{eq:Yosidafinalenergy}) in eq.~(\ref{eq:chifromlocalm}) we find
\begin{equation}
\frac{\chi_0^{S,{\rm Y}}(J_{\rm K})}{(g_e\mu_{\rm B})^2}
=(3J_{\rm K})\frac{2\pi\left(3-[e_0^{\rm Y}(J_{\rm K})]^2  \right)-3J_{\rm K}}{
  128\pi^2\left(1-[e_0^{\rm Y}(J_{\rm K})]^2\right)|e_0^{\rm Y}(J_{\rm K})|}
\label{eq:chiSYosidaresult}
  \end{equation}
for the zero-field impurity spin susceptibility in the presence of a global field, and
\begin{equation}
\frac{\chi_{0,{\rm loc}}^{S,{\rm Y}}(J_{\rm K})}{(g_e\mu_{\rm B})^2}
=\frac{3 \left(3 J_{\rm K} - 4\pi [e_0^{\rm Y}(J_{\rm K})]^2\right)}{
  64\pi \left(1-[e_0^{\rm Y}(J_{\rm K})]^2\right)|e_0^{\rm Y}(J_{\rm K})|}
\label{eq:chiSlocalYosidaresult}
  \end{equation}
for the zero-field impurity spin susceptibility in the presence of a local field
for the one-dimensional density of states~(\ref{eq:DOS}).

\paragraph{Small Kondo couplings.}

The Yosida energy is exponentially small,
$e_0^{\rm Y}(J_{\rm K})\approx -\F\exp[-4/(3\rho_0(0)J_{\rm K})]$, so that we obtain
\begin{eqnarray}
  \frac{\chi_0^{S,{\rm Y}}(J_{\rm K}\ll1) }{(g_e\mu_{\rm B})^2}
  &\approx & \frac{\chi_{0,{\rm loc}}^{S,{\rm Y}}(J_{\rm K}\ll 1)}{(g_e\mu_{\rm B})^2}
  \left(1 - \frac{\rho_0(0)J_{\rm K}}{2}\right) \; , \nonumber \\
  \frac{\chi_{0,{\rm loc}}^{S,{\rm Y}}(J_{\rm K}\ll 1)}{(g_e\mu_{\rm B})^2}
  &\approx & \frac{9\rho_0(0)J_{\rm K}e^{4/(3\rho_0(0)J_{\rm K})}}{64\F}\; ,
    \label{eq:Yosidaimpsussmall}
\end{eqnarray}
which are identical up to a correction factor that goes to unity for 
$\rho_0(0)J_{\rm K}\to 0$.
The zero-field impurity spin susceptibilities
display an exponential increase for small Kondo couplings,
as is characteristic for the Kondo model.

\paragraph{Large Kondo couplings.}

For large Kondo couplings, $J_{\rm K}\gg 1$, we have
$e_0^{\rm Y}(J_{\rm K})\approx -3J_{\rm K}/8+2/\pi$ in one dimension so that
we find
\begin{eqnarray}
  \frac{\chi_0^{S,{\rm Y}}(J_{\rm K}\gg 1) }{(g_e\mu_{\rm B})^2}
  &\approx & \frac{1}{8\pi}+\frac{2}{\pi^2 J_{\rm K}}\; , \label{eq:chiSY}\\
  \frac{\chi_{0,{\rm loc}}^{S,{\rm Y}}(J_{\rm K}\gg 1)}{(g_e\mu_{\rm B})^2}
  &\approx & \frac{1}{2J_{\rm K}} \; .
  \label{eq:chiSYloc}
\end{eqnarray}
Since the Yosida state does not become exact for large Kondo couplings,
the zero-field impurity spin susceptibility for a global field does not vanish
for $J_{\rm K}\to \infty$. The corresponding susceptibility
for a local field behaves properly,
and even reproduces the exact result, as derived in Sect.~\ref{sec:Gutzwillerresults}.

\subsubsection{Zero-field impurity-induced susceptibility}

Using {\sc Mathematica}~\cite{Mathematica11}
we find the impurity-induced magnetic susceptibility at zero field
from eq.~(\ref{appeq:getm})
($\lambda\equiv e_0^{\rm Y}(J_{\rm K})$)
\begin{eqnarray}
\frac{\chi_0^{\rm ii,Y}(J_{\rm K})}{(g_e\mu_{\rm B})^2}&=&
\frac{J_{\rm K}C(\lambda,J_{\rm K})}{
  128 \lambda (\lambda^2-1) \pi^3 (4 \lambda^2 \pi-3 J_{\rm K})}\; , \nonumber \\
C(\lambda,J_{\rm K}) &=&
9 J_{\rm K}^3 + 12 J_{\rm K}^2 (9 \lambda^2-5) \pi \nonumber \\
&&+ 
   4 J_{\rm K} (33 - 62 \lambda^2 + 9 \lambda^4) \pi^2 - 
   96 (\lambda^2-1)^2 \pi^3 , \nonumber \\
   \label{eq:chiiiYosidaresult}
   \end{eqnarray}
where we used eq.~(\ref{eq:Yosidafinalenergy}) to simplify the expressions.
In the presence of a local field, 
eq.~(\ref{appeq:getmlocal}) leads to
\begin{equation}
\frac{\chi_{0,{\rm loc}}^{\rm ii,Y}(J_{\rm K})}{(g_e\mu_{\rm B})^2}=
\frac{9 J_{\rm K}^2 + 24 \pi J_{\rm K} (3 \lambda^2-1)
- 16\pi^2 \lambda^2 (2 + 3 \lambda^2)}{
  32 \lambda (\lambda^2-1) \pi (-3 J_{\rm K} + 4 \pi \lambda^2)} .
\label{eq:chiiilocalYosidaresult}
   \end{equation}

\paragraph{Small Kondo couplings.}

The Yosida energy is exponentially small,
$e_0^{\rm Y}(J_{\rm K})\approx -\F\exp[-4\pi/(3J_{\rm K})]$, so that
\begin{eqnarray}
\frac{\chi_0^{\rm ii,Y}(J_{\rm K}\ll 1)}{(g_e\mu_{\rm B})^2}&\approx&
\frac{e^{4/(3\rho_0(0)J_{\rm K})}}{4\F} \label{eq:YosidachismallJK}
\\
&& \times \left(1 -\frac{11}{8}\rho_0(0)J_{\rm K} +\frac{5}{8} (\rho_0(0)J_{\rm K})^2\right)
\nonumber
\end{eqnarray}
with corrections of order $J_{\rm K}^3$, and
\begin{equation}
\frac{\chi_{0,{\rm loc}}^{\rm ii,Y}(J_{\rm K}\ll 1)}{(g_e\mu_{\rm B})^2}\approx
\frac{e^{4/(3\rho_0(0)J_{\rm K})}}{4\F}
\left(1 -\frac{3}{8}\rho_0(0)J_{\rm K} \right)
\label{eq:YosidachismallJKlocal}
\end{equation}
with exponentially small corrections. Thus,
the zero-field impurity-induced susceptibility has the same exponential prefactor as
in the case of a global field but the correction factor is different
already in linear order.

The zero-field susceptibility is exponentially large for small~$J_{\rm K}$,
in qualitative agreement with the exact solution. However, the exponent
is not quite correct, namely, the factor $4/3$ should be replaced by unity.
Moreover, the exact susceptibility contains a correction factor
proportional to $\sqrt{J_{\rm K}/\pi}$, see Sect.~\ref{sec:BAnsatz}.

Note that the form of the density of states only enters through the
prefactor $\F$ that appears in the Yosida ground-state energy.
Thus, the algebraic correction terms in eqs.~(\ref{eq:YosidachismallJK})
and~(\ref{eq:YosidachismallJKlocal}) are universal in the sense that
they do not depend on the form of the host-electron density of states.
This behavior is also seen in the exact zero-field susceptibilities,
see Sect.~\ref{sec:Comparison}.

\paragraph{Large Kondo couplings.}

For large Kondo couplings, $J_{\rm K}\gg 1$, we have
$e_0^{\rm Y}(J_{\rm K})\approx -3J_{\rm K}/8+2/\pi$ in one dimension so that
\begin{equation}
\frac{\chi_0^{\rm ii,Y}(J_{\rm K}\gg 1)}{(g_e\mu_{\rm B})^2}\approx
-\frac{3 J_{\rm K}}{16 \pi^2} + \frac{\pi^2-12}{2 \pi^3} +{\cal O}(1/J_{\rm K}) <0
\; .
\end{equation}
Since the Yosida wave function does not describe the local spin singlet state properly,
the susceptibility becomes negative for large Kondo couplings
which indicates that the Yosida state is unstable
for $J_{\rm K}>J_{\rm K,c}^{\rm Y}\approx 3.7543$. The instability point is
obtained from a
numerical solution of $\chi_0^{\rm ii,Y}(J_{\rm K,c}^{\rm Y})=0$.
At $J_{\rm K}=J_{\rm K,c}^{\rm Y}$, the critical external field vanishes,
$B_{\rm c}^{\rm Y}(J_{\rm K,c}^{\rm Y})=0$.

For large Kondo couplings, $J_{\rm K}\gg 1$, we have
$e_0^{\rm Y}(J_{\rm K})\approx -3J_{\rm K}/8+2/\pi$ so that
\begin{equation}
\frac{\chi_{0,{\rm loc}}^{\rm ii,Y}(J_{\rm K}\gg 1)}{(g_e\mu_{\rm B})^2}\approx
\frac{1}{J_{\rm K}} +{\cal O}\left(J_{\rm K}^{-3}\right)
\; .
\label{eq:chiiYloc}
\end{equation}
This result is qualitatively correct. Note, however, that the Yosida wave function
fails to reproduce the exact equivalence of
the zero-field impurity-induced susceptibility for a local field
and the zero-field impurity spin susceptibility, eq.~(\ref{eq:twochisareequivalent}).

\begin{figure}[t]
\includegraphics[width=8.5cm]{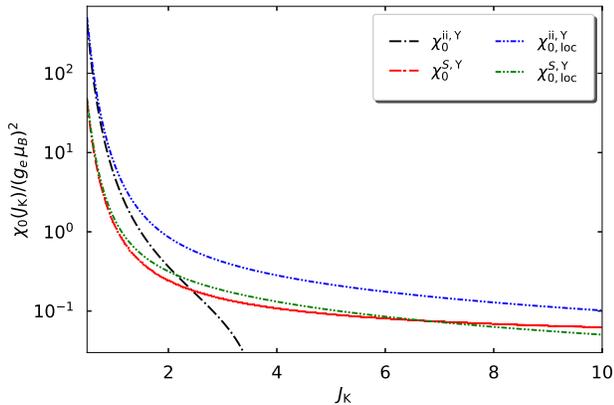}
\caption{(Color online) Zero-field impurity spin susceptibility
  $\chi_0^{S,{\rm Y}}/(g_e\mu_{\rm B})^2$, eqs.~(\ref{eq:chiSYosidaresult})
  and~(\ref{eq:chiSlocalYosidaresult})
  and zero-field impurity-induced susceptibility
  $\chi_0^{\rm ii,Y}/(g_e\mu_{\rm B})^2$, eqs.~(\ref{eq:chiiiYosidaresult})
  and~(\ref{eq:chiiilocalYosidaresult}),
  for global/local magnetic fields as a function of $J_{\rm K}$
  of the one-dimensional symmetric Kondo model from the Yosida
  wave function. The zero-field impurity-induced susceptibility
  becomes negative
  at $J_{\rm K}=J_{\rm K,c}^{\rm Y}\approx 3.7543$.\label{fig:Yosidachis}}
\end{figure}

In Fig.~\ref{fig:Yosidachis} we show the corresponding zero-field
susceptibilities. They all display an exponential increase for small~$J_{\rm K}$,
and only differ in the pre-exponential factor.
For the impurity spin susceptibility in the Yosida wave function, this factor
is proportional to $J_{\rm K}$ and also numerically small,
see eq.~(\ref{eq:Yosidaimpsussmall}). Therefore,
the impurity spin susceptibility is substantially smaller
than the impurity-induced susceptibility; this is an artifact of the Yosida wave function.

For large Kondo couplings,
$J_{\rm K}>1$, the impurity-induced susceptibility becomes negative for
$J_{\rm K}>J_{\rm K,c}^{\rm Y}$, i.e., the Yosida state becomes unstable
against a state with a locally broken symmetry.
The other susceptibilities remain positive for all $J_{\rm K}$.
The impurity spin susceptibility becomes constant for large~$J_{\rm K}$,
see eq.~(\ref{eq:chiSY}) which is at odds with the exact solution.
The local susceptibilities are qualitatively correct for large couplings
inasmuch they decay to zero for strong couplings,
see eqs.~(\ref{eq:chiSYloc}) and~(\ref{eq:chiiYloc}).
In fact, 
the impurity spin susceptibility from the Yosida wave function
in the presence of a local field,
$\chi_{0,{\rm loc}}^{S,{\rm Y}}(J_{\rm K})$,
eq.~(\ref{eq:chiSYloc}),
becomes exact in the limit of strong coupling, see Sect.~\ref{sec:Comparison}.

\section{Gutzwiller wave function}
\label{sec:Gutzwillerresults}

As the third and last analytic variational approach, we study the Gutzwiller wave function.
It becomes exact in  the limit of large Kondo couplings and provides
a very good variational upper bound for the ground-state energy for all Kondo couplings.
However, for weak couplings it describes a symmetry-broken state with an oriented
moment on the impurity, and a transition to the paramagnetic state
at $J_{\rm K,c}^{\rm G}$ that is not contained in the exact solution of the model.

\subsection{Gutzwiller variational state}

We define the Gutzwiller variational state~\cite{Gutzwiller1964,TIAM}
\begin{equation}
  |\Psi_{\rm G}\rangle  = \hat{P}_{\rm G} | \Phi\rangle \; ,
  \label{eq:GWFdef}
\end{equation}
where $|\Phi\rangle$ is a normalized single-particle product state
to be determined variationally.
At half band filling we have
\begin{equation}
  n_{\sigma}^{d,0}=
  \langle \Phi | \hat{n}_{\sigma}^d | \Phi \rangle =\frac{1}{2}+\sigma_n m \; ,
\label{appeq:dsarehalffillednonint}
\end{equation}
where $\sigma_n=1$ for $\sigma=\uparrow$ and
$\sigma_n=-1$ for $\sigma=\downarrow$, and $0\leq m< 1/2$ is the
impurity spin polarization
in the single-particle product state~$|\Phi\rangle$,
\begin{equation}
  m=\frac{1}{2} \langle \Phi | \hat{n}_{\uparrow}^d
  - \hat{n}_{\downarrow}^d
  | \Phi \rangle \; .
    \label{eq:defmimpinPhi}
\end{equation}
For a complete Gutzwiller projection, we choose
\begin{equation}
  \hat{P}_{\rm G} = \lambda_{\uparrow}  \hat{m}_{\uparrow}^d+
  \lambda_{\downarrow}\hat{m}_{\downarrow}^d
\quad, \quad
\hat{m}_{\sigma}^d =\hat{n}_{\sigma}^d(1-\hat{n}_{\bar{\sigma}}^d) \; ,
\label{appeq:Gutzprojector}
\end{equation}
where we use
$\bar{\uparrow}=\downarrow$ and $\bar{\downarrow}=\uparrow$.
Moreover, we demand
\begin{equation}
\hat{P}_{\rm G}^2 
= 1+x(\hat{n}_{\uparrow}^d-n_{\uparrow}^{d,0})
(\hat{n}_{\downarrow}^d-n_{\downarrow}^{d,0})
\; .
\label{appeq:defPGsquared}
\end{equation}
Thus, we have to solve
\begin{eqnarray}
\lambda_{\uparrow}^2  \hat{m}_{\uparrow}^d+
\lambda_{\downarrow}^2\hat{m}_{\downarrow}^d
&=&
\lambda_{\uparrow}^2  \hat{n}_{\uparrow}^d+
\lambda_{\downarrow}^2\hat{n}_{\downarrow}^d
- (\lambda_{\uparrow}^2  +\lambda_{\downarrow}^2)
\hat{n}_{\uparrow}^d\hat{n}_{\downarrow}^d
\nonumber \\
&\stackrel{!}{=}&
1+x(\hat{n}_{\uparrow}^d-n_{\uparrow}^{d,0})
(\hat{n}_{\downarrow}^d-n_{\downarrow}^{d,0}) \; .
\end{eqnarray}
The solution reads
\begin{eqnarray}
  x&=& -\frac{1}{n_{\uparrow}^{d,0}n_{\downarrow}^{d,0}} = -\frac{4}{1-4m^2}
  \; , \nonumber \\
  \lambda_{\sigma}^2&=&\frac{1}{n_{\sigma}^{d,0}}=\frac{2}{1+2\sigma_n m}
  \; .
\end{eqnarray}

Before we proceed, we note the useful relations
\begin{eqnarray}
%
\hat{P}_{\rm G}  
\hat{d}_{\sigma}^+\hat{d}_{\bar{\sigma}}^{\vphantom{+}} 
\hat{P}_{\rm G}   &=&\lambda_{\sigma}\lambda_{\bar{\sigma}}
\hat{d}_{\sigma}^+\hat{d}_{\bar{\sigma}}^{\vphantom{+}}  
\; , \nonumber \\
\hat{P}_{\rm G}  
(\hat{n}_{\uparrow}^d-\hat{n}_{\downarrow}^d)
\hat{P}_{\rm G}   &=& \lambda_{\uparrow}^2\hat{m}_{\uparrow}^d-
\lambda_{\downarrow}^2\hat{m}_{\downarrow}^d
\nonumber \\
&=& \frac{\hat{n}_{\uparrow}^d}{n_{\uparrow}^{d,0}}
- \frac{\hat{n}_{\downarrow}^d}{n_{\downarrow}^{d,0}}
+2m \frac{\hat{n}_{\uparrow}^d\hat{n}_{\downarrow}^d}{
  n_{\uparrow}^{d,0}n_{\downarrow}^{d,0}} \; .
\label{appeq:useful}
\label{eq:useful}
\end{eqnarray}

\subsection{Ground-state energy}
\label{subsec:e0Gutzpara}

The calculation of expectation values and the variational optimization
of the energy functional 
is presented in appendices~\ref{subsec:GWFEvaluation}
and~\ref{subsec:Lagrangian}.
It requires the solution of an effective non-interacting single-impurity Anderson model
that is characterized by a local hybridization parameter~$V$.

\subsubsection{Paramagnetic Gutzwiller state}

For $m=0$,
the ground-state energy for the non-interacting single-impurity Anderson model 
is known explicitly for all relevant cases. For example,
in one dimension it reads~\cite{Barczaetal}
\begin{eqnarray}
e_0^{\rm A}(V)
&=& \frac{1}{\pi} \left[ -\pi +2v_+\arctan\left(\frac{1}{v_-}\right)
+v_-\ln\left(\frac{v_+-1}{v_+ +1}\right)\right] \nonumber \\
&& +2(1-v_+) \;, \label{eq:e0SIAM1d}\\
v_{\pm}&=& \sqrt{\frac{\sqrt{1+4V^4}\pm 1}{2}} \label{eq:defvplusminus}\; .
\end{eqnarray}
The Hellmann-Feynman theorem then gives
\begin{equation}
  \frac{\partial e_0^{\rm A}(V)}{\partial V} =-\frac{8V}{3J_{\rm K}}\; .
\label{eq:selfconstGutzwiller}
\end{equation}
The self-consistency equation~(\ref{eq:selfconstGutzwiller}) 
defines $V(J_{\rm K})$ as a function of~$J_{\rm K}$.

The Gutzwiller variational energy for the Kondo model becomes
\begin{equation}
e_0^{\rm G}(J_{\rm K}) = e_0^{\rm A}(V(J_{\rm K}))
+ \frac{4\left[V(J_{\rm K})\right]^2}{3J_{\rm K}}\; .
\label{eq:Gutzwillerfinalenergy}
\end{equation}
In general, the Gutzwiller variational energy 
for the Kondo model must be determined numerically.

\paragraph{Small Kondo couplings.}

For $J_{\rm K}\ll 1$  and thus $V\ll 1$ we can approximate
\begin{equation}
e_0^{\rm A}(V\ll 1)\approx \frac{2V^2}{\pi} \left( \ln(V^2/2)-1\right)
\end{equation}
in one dimension so that the self-consistency 
equation~(\ref{eq:selfconstGutzwiller}) becomes
\begin{equation}
-\frac{2\pi}{3J_{\rm K}} = \ln(V^2/2) \; , \; 
V^2=2 \exp\left(-\frac{2\pi}{3J_{\rm K}}\right) \; .
\label{eq:VGutzexpsmall}
\end{equation}
Therefore, the Gutzwiller estimate for the ground-state 
energy at small Kondo couplings becomes
\begin{equation}
e_0^{\rm G}(J_{\rm K}\ll 1) 
= -\frac{4}{\pi} \exp\left(-\frac{2}{3\rho_0(0)J_{\rm K}}\right) \;.
\end{equation}
This is much smaller than the Yosida energy eq.~(\ref{eq:yosida1dsmall}),
and even smaller than the value for the Yosida-Yoshimori wave
function,~\cite{Yosidabook,YoshimoriYosida}
\begin{equation}
e_0^{\rm Y,Y}(J_{\rm K}\ll 1)   
\sim -\exp\left(-\frac{1}{\rho_0(0)J_{\rm K}}\right)\;.
\end{equation}
This is not surprising because both variational states miss
the actually quadratic dependence of the ground-state energy 
on $J_{\rm K}$ for small interaction strengths,
$e_0(J_{\rm K})\sim -J_{\rm K}^2$, see eq.~(\ref{eq:secondorderanalyt}).

\paragraph{Large Kondo couplings.}

For $J_{\rm K}\gg 1$ we can approximate 
\begin{equation}
  e_0^{\rm A}(V\gg 1)\approx -2V+1-\frac{1}{2V}+\frac{2}{3\pi V^2}
  -\frac{1}{16V^3}+{\cal O}(1/V^5) 
\end{equation}
in one dimension. The self-consistency 
equation~(\ref{eq:selfconstGutzwiller}) becomes
\begin{eqnarray}
  -\frac{4V}{3J_{\rm K}} &=& -1 +\frac{1}{4V^2}-\frac{2}{3\pi V^3}
+\frac{3}{32V^4}  +{\cal O}(1/V^6) \; , \nonumber \\
V&=&\frac{3J_{\rm K}}{4} -\frac{1}{3J_{\rm K}}+\frac{32}{27\pi J_{\rm K}^2}
-\frac{14}{27 J_{\rm K}^3}+{\cal O}(1/J_{\rm K}^5)\; .\nonumber\\
\label{eq:VlargeGutz}
\end{eqnarray}
Therefore, the Gutzwiller estimate for the ground-state 
energy at large Kondo couplings reads
\begin{equation}
  e_0^{\rm G}(J_{\rm K}\gg 1) = -\frac{3J_{\rm K}}{4} +1 -\frac{2}{3J_{\rm K}}
  +\frac{32}{27\pi J_{\rm K}^2}
-\frac{8}{27 J_{\rm K}^3}\; ,
  \label{eq:GutzenergylargeJK}
\end{equation}
up to and including third order in $1/J_{\rm K}$.
This is much smaller than the Yosida energy~(\ref{eq:yosida1dlarge})
and is actually exact, up to corrections of the order $1/J_{\rm K}$,
see eq.~(\ref{eq:energyexact1dlargeJK}).
Below, we argue that the first-order and second-order corrections
in $1/J_{\rm K}$ are also exact.

The local spin correlation is obtained from the variational Hellmann-Feynman theorem.
For large~$J_{\rm K}$ we find
\begin{equation}
  C_0^{S,{\rm G}}(J_{\rm K}\gg 1)
  = -\frac{3}{4} + \frac{2}{3 J_{\rm K}^2} - \frac{64}{27 \pi J_{\rm K}^3}
  +{\cal O}(1/J_{\rm K}^4)\; .
  \label{eq:C0GutzlargeJ}
\end{equation}

\subsubsection{Magnetic Gutzwiller state for weak coupling}

As shown in appendix~\ref{subsec:Lagrangian}, the numerical optimization
of the variational parameters leads to
  \begin{equation}
  e_0^{\rm G}(J_{\rm K})\approx
  -0.0905 J_{\rm K}^2 -0.051J_{\rm K}^3-0.05 J_{\rm K}^4
  \label{eq:Gutzenergymaganalytmain}
  \end{equation}
for the Gutzwiller variational energy for $J_{\rm K}\lesssim 0.4$.

The quadratic coefficient from the magnetic Gutzwiller wave function
can be compared with the exact result from
perturbation theory, $e_0(J_{\rm K})\approx -3J_{\rm K}^2/32
=-0.09375 J_{\rm K}^2$, see eq.~(\ref{eq:secondorderanalyt}).
The magnetic Gutzwiller states accounts for 96.5\% of the
correlation energy. Hence, the magnetic Gutzwiller provides an excellent energy
estimate but fails to describe the physics properly because 
it breaks the local symmetry, $m>0$ for $B=0^+$.

\subsection{Zero-field susceptibilities}

The calculations are carried out in appendix~\ref{subsec:magGutz}.
Here, we summarize the results for the various zero-field susceptibilities
in the strong-coupling limit.

\subsubsection{Five equations}

The calculation of the zero-field susceptibilities from the Gutzwiller wave
function requires the solution of a $5\times 5$ matrix problem,
 \begin{equation}
   \underline{\underline{M}} \cdot \underline{v} = \underline{g} \; ,
   \label{eq:Matrixequationmain}
 \end{equation}
 where $\underline{g}_{\rm loc}^{\rm T}=(1,0,0,0,0)$ for a local field,
 and $\underline{g}^{\rm T}=(1,0,g_3,g_4,g_5)$ for a global field
 whose non-trivial entries $g_3,g_4,g_5$ are known functions of~$V$, and $V(J_{\rm K})$
 follows from eq.~(\ref{eq:selfconstGutzwiller}), see appendix~\ref{subsec:magGutz}.
 Likewise, the entries of the $5 \times5$ matrix~$\underline{\underline{M}}$
 are known functions of $V$ and $J_{\rm K}$. The vector
 \begin{equation}
   \underline{v} =\left(
   \begin{array}{@{}c@{}}
   \bar{\omega}_p\\
   \bar{E}_d\\
   \bar{K}\\
   \bar{M}_0\\
   \chi
   \end{array}
   \right)
   \label{eq:defvvectormain}
  \end{equation}
 contains the five unknowns that determine the 
 susceptibilities,
 \begin{eqnarray}
   \frac{\chi_0^{S,{\rm G}}(J_{\rm K},B)}{(g_e\mu_{\rm B})^2}&=&   \chi   \; ,
      \label{eq:Gutzchiparamain}\\
   \frac{ \chi_{0}^{\rm ii,G}(J_{\rm K})}{(g_e\mu_{\rm B})^2} 
   &=& \frac{\bar{E}_{d}}{2\pi V^2} \; .
   \label{eq:chizeroiiGutzsmallfieldsmain}
 \end{eqnarray}
 The choice of $\underline{g}$ determines whether the external field is applied
 globally or locally.

Although {\sc Mathematica}~\cite{Mathematica11} provides an analytic solution
of the linear problem,
the expressions are very lengthy and not illuminating.
Eventually, we evaluate them numerically.

\subsubsection{Strong-coupling limit}

 As shown in appendix~\ref{subsec:magGutz}, 
 compact results can be obtained for $J_{\rm K}\gg 1$.
 For the zero-field impurity spin susceptibility we find
 \begin{eqnarray}
  \frac{\chi_0^{S,{\rm G}}(J_{\rm K}\gg 1)}{(g_e\mu_{\rm B})^2}&=&
\frac{20}{9\pi J_{\rm K}^2}   +{\cal O}(1/J_{\rm K}^4)  \; ,
  \label{eq:scformulafullchimain}\\
 \frac{\chi_{0,{\rm loc}}^{S,{\rm G}}(J_{\rm K}\gg 1)}{(g_e\mu_{\rm B})^2}&=&
 =\frac{1}{2 J_{\rm K}}+\frac{28}{27 J_{\rm K}^3}+{\cal O}(1/J_{\rm K}^4) 
 \label{eq:GutzlargeJKlocalfieldmain}
 \end{eqnarray}
 in the presence of a global and a local field, respectively.

 For the zero-field impurity-induced susceptibilities, the Gutzwiller result for
 strong coupling reads
 \begin{eqnarray}
  \frac{\chi_0^{\rm ii,G}(J_{\rm K}\gg 1)}{(g_e\mu_{\rm B})^2}&=&
\frac{8}{9 \pi J_{\rm K}^2} + \frac{416}{81 \pi^2 J_{\rm K}^3} +
  {\cal O}(1/J_{\rm K}^4) \; ,
  \label{eq:chizeroiiGlargeJKmain}\\
\frac{\chi_{0,{\rm loc}}^{\rm ii,G}(J_{\rm K}\gg 1)}{(g_e\mu_{\rm B})^2}&=&
\frac{20}{9 \pi J_{\rm K}^2}+  {\cal O}(1/J_{\rm K}^4) 
  \label{eq:chizeroiilocalGlargeJKmain}
 \end{eqnarray}
 in the presence of a global and a local field, respectively.

 Since the Gutzwiller wave function becomes exact state for strong coupling,
 we argue that these results are correct to the indicated order.
 We shall confirm this assessment from the
 comparison with numerically exact data from NRG and DMRG
 in Sect.~\ref{subsec:NRGDMRGandallstrongcouplingsusc}.

\subsubsection{Critical interaction for the magnetic transition}
\label{subsubsec:JcGutz}

For $J_{\rm K}>J_{\rm K,c}^{\rm G}$, the Gutzwiller state describes a 
spin-isotropic state at the impurity site.
For $J_{\rm K}<J_{\rm K,c}^{\rm G}$, the local spin symmetry 
in the Gutzwiller  state is spontaneously broken, i.e.,
$m> 0$ is optimal even at $B=0^+$.

With the help of the zero-field spin susceptibility, the transition can
accurately be identified because the determinant of the 
matrix~$\underline{\underline{M}}(J_{\rm K})$ changes sign at 
$J=J_{\rm K,c}^{\rm G}$.
Using {\sc Mathematica},~\cite{Mathematica11} the determinant
as a function of~$V$ can be calculated analytically
but the expressions are lengthy.
The solution of 
\begin{equation}
\det\left(\underline{\underline{M}}(V_{\rm c})\right)=0
\end{equation}
is $V_{\rm c}=0.4559222509954975$ 
with $\det\left(\underline{\underline{M}}(V_{\rm c})\right)=
-3.8\cdot 10^{-15}$,
or 
\begin{equation}
J_{\rm K,c}^{\rm G}=0.8392762432533198
\; .
\end{equation}
 
\subsubsection{Comparison of susceptibilities}

The paramagnetic Gutzwiller state is stable only for
$J_{\rm K}>J_{\rm K,c}^{\rm G}\approx 0.839$ so that we focus on  
$J_{\rm K}\geq 1$.
Since the Gutzwiller wave function becomes exact for $J_{\rm K}\to \infty$,
all susceptibilities are positive and well behaved.

\begin{figure}[t]
\includegraphics[width=8.5cm]{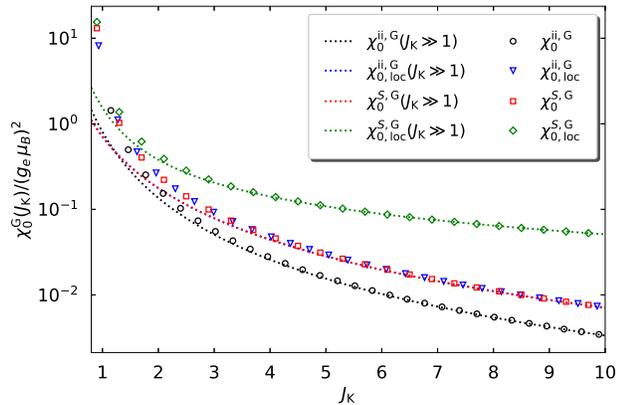}
\caption{(Color online) Zero-field impurity spin susceptibility
  $\chi_0^{S,{\rm G}}/(g_e\mu_{\rm B})^2$, eq.~(\ref{eq:Gutzchiparamain}),
  and zero-field impurity-induced susceptibility
  $\chi_0^{\rm ii,G}/(g_e\mu_{\rm B})^2$, eq.~(\ref{eq:chizeroiiGutzsmallfieldsmain}),
  for global/local magnetic fields as a function of $J_{\rm K}\geq 1$
  of the one-dimensional symmetric Kondo model from the Gutzwiller
  wave function. For comparison, we also include the limiting behavior for
  the susceptibilities,
  eqs.~(\ref{eq:scformulafullchimain}), (\ref{eq:GutzlargeJKlocalfieldmain}),
  (\ref{eq:chizeroiiGlargeJKmain}),
  and~(\ref{eq:chizeroiilocalGlargeJKmain}).\label{fig:Gutzwillerchis}}
\end{figure}

In Fig.~\ref{fig:Gutzwillerchis} we show the global and local zero-field
susceptibilities.
As seen from the figure,
the asymptotic formulae for the impurity spin susceptibilities,
eqs.~(\ref{eq:scformulafullchimain}), (\ref{eq:GutzlargeJKlocalfieldmain}),
  (\ref{eq:chizeroiiGlargeJKmain}),
and~(\ref{eq:chizeroiilocalGlargeJKmain}), are applicable for $J_{\rm K}\gtrsim 4$.

Fig.~\ref{fig:Gutzwillerchis} also shows that
the equivalence
\begin{equation}
  \chi_0^{S,{\rm G}}(J_{\rm K})=\chi_{0,{\rm loc}}^{\rm ii,G}(J_{\rm K})
  \end{equation}
holds for all $J_{\rm K}>1$.
Therefore, the Gutzwiller approach respects the exact
relation~(\ref{eq:twochisareequivalent}) at zero temperature.
Indeed, as seen from the derivation in Sect.~\ref{subsec:Lagrangian},
the Gutzwiller approach solves an effective non-interacting single-impurity
Anderson model for which the exact relation
eq.~(\ref{eq:twochisareequivalent}) is readily shown to hold, too.

\section{Bethe Ansatz results}
\label{sec:BAnsatz}

Using Bethe Ansatz, the Kondo model is solved
for a linear dispersion relation with unit Fermi velocity
in the wide-band limit, i.e., the dispersion relation $\epsilon^{\rm BA}(k)=k$
formally extends from $k_{-}=-\infty$ to $k_+=+\infty$.
Therefore, an appropriate energy
cut-off~$D$ must be introduced in the Bethe Ansatz equations.
This procedure is not unique.
Therefore, there are two Bethe Ansatz solutions for the spin-1/2 Kondo model.
First, the one discussed by Tsvelick and Wiegmann,
referred to as~TW,~\cite{TsvelickWiegmann}
and, second, the one reviewed by Andrei, Furuya, and
Lowenstein, referred to as~AFL.~\cite{RevModPhys.55.331} 
The basic Bethe Ansatz equations agree but the
expressions for the parameters as a function of the Bethe Ansatz
Kondo coupling~$J_{\rm K}^{\rm BA}/D$ differ beyond leading-order.
For a lattice-regularized Bethe-Ansatz solvable impurity model, see
Ref.~[\onlinecite{BortzKluemper}].

In this section we discuss the Bethe Ansatz results
for the zero-field impurity-induced susceptibility and magnetization.
As shown in appendix~\ref{app:GSEBetheAnsatz}, the Bethe Ansatz
solution leads to
\begin{equation}
e_0^{\rm BA}(J_{\rm K}^{\rm BA}) = {\cal O}\left(\left(J_{\rm K}^{\rm BA}\right)^3\right) \; .
\end{equation}
Thus, the Bethe Ansatz results
cannot be used for comparison with the ground-state energy of the
Kondo impurity on a chain.

\subsection{Zero-field impurity-induced magnetic susceptibility}

The Bethe Ansatz leads to equation~(4.30) of AFL~\cite{RevModPhys.55.331}
or equation~(5.1.23) of TW~\cite{TsvelickWiegmann}
for the zero-field magnetic susceptibility
\begin{equation}
  \frac{\chi^{\rm ii}_0(J_{\rm K})}{\left(g_e\mu_{\rm B}\right)^2}
  = \frac{1}{4\pi T_0(J_{\rm K}^{\rm BA})}
  \label{eq:chiBAfinalresultBA}
\end{equation}
in the limit of small $J_{\rm K}^{\rm BA}$ for a half-filled system.

The Bethe Ansatz solves the Kondo model for a Kondo interaction
strength $J_{\rm K}^{\rm BA}$ in the limit of an infinite bandwidth.
To arrive at tangible results, a symmetric bandwidth cutoff, $|\epsilon| < D$,
is imposed on the Bethe Ansatz equations,
and periodic boundary conditions are implemented
so that the electron density remains finite, 
$D_{\rm AFL}\equiv N^e/L=1/2$ at half band-filling, 
see appendix~\ref{app:GSEBetheAnsatz}.
The corresponding bandwidth is $W^{\rm BA}=2D$ with
\begin{equation}
  D\equiv K_{\rm AFL}=\pi D_{\rm AFL}=\frac{\pi}{2}
\end{equation}
so that the density of states is given by
\begin{equation}
\rho_0=\frac{1}{2D}=\frac{1}{\pi}
\end{equation}
at the Fermi energy $E_{\rm F}=0$, and for all $|\epsilon|\leq D$.
Then, the low-temperature magnetic energy scale from Bethe Ansatz
is given by
\begin{equation}
  T_0(J_{\rm K}^{\rm BA})= \frac{D}{\pi}
  \exp\left(-\frac{1}{\rho_0J_{\rm K}^{\rm BA}}\right) \; .
  \label{eq:T0resultBA}
\end{equation}
The remaining problem is to express $T_0(J_{\rm K}^{\rm BA})$
as a function of $J_{\rm K}$, or, equivalently, to find an explicit relation between
$J_{\rm K}^{\rm BA}$ and $J_{\rm K}$. The existence of
such a unique relation is thoroughly discussed in
Sect.~VI of AFL.~\cite{RevModPhys.55.331}

\subsection{Wilson's renormalization group}

Wilson's renormalization group~\cite{RevModPhys.47.773} for the Kondo
model starts from the lattice model in the thermodynamic limit
with its energy cut-off parameter
${\cal D}=W/2=1$ and Kondo coupling $J_{\rm K}$.
By successively integrating out the high-energy 
degrees of freedom, the renormalization group
flows to the Bethe Ansatz model with a linear dispersion relation around
the Fermi energy and the coupling $J_{\rm K}^{\rm BA}$.

The renormalization group (RG) transformation is actually performed
on the Hamiltonian as well as on the matrix representation
of the operators, both influencing the physical quantities such as
the zero-field impurity-induced susceptibility.
In his review, eq.~(IX.91) on p.~835,~\cite{RevModPhys.47.773}
Wilson provides the general series expansion
for the zero-field impurity-induced susceptibility at zero temperature,
\begin{eqnarray}
  \frac{\chi_0^{\rm ii}(J_{\rm K})}{(g_e\mu_{\rm B})^2}
&=&
  \frac{w}{4\widetilde{D}(j)}
  \frac{  \exp\left(1/j\right)}{\sqrt{j}}
  \exp\biggl(-\alpha_1 j-\sum_{n=2}^{\infty}\alpha_n j^n\biggr) \nonumber \; , \\
\frac{\widetilde{D}(j)}{{\cal D}} &=& 
c_0 +\sum_{n=1}^{\infty}c_n j^n \; , \quad j=\rho_0(0)J_{\rm K}\;,
\label{eq:Wilsonsresult}
\end{eqnarray}
where $\rho_0(0)$ is the density of states at the Fermi energy.
In Ref.~[\onlinecite{RevModPhys.47.773}],
$\alpha_1$ and $w$ were determined numerically for a constant 
density of states, $w/4\approx 0.1032$, and $\alpha_1=1.5824$.
AFL calculated the Wilson number~$w$ analytically,~\cite{RevModPhys.55.331}
\begin{equation}
  w=\frac{w_{\rm AFL}}{\pi} = \frac{e^{\C+1/4}}{\pi^{3/2}}
    \approx 0.410705 
\label{eq:defwHewsonfirst}
\end{equation}
with Euler's constant $\C\approx 0.577216$.
We concisely re-derive~$w$ in appendix~\ref{app:freenergyPT}.

The coefficients $c_n$ in eq.~(\ref{eq:Wilsonsresult}) can be
obtained from the high-temperature expansion
of the zero-field impurity-induced susceptibility. To third order it reads,
see eq.~(IX.57) of Ref.~[\onlinecite{RevModPhys.47.773}],
\begin{eqnarray}
\frac{4T\chi_0^{\rm ii}(T,J_{\rm K})}{(g_e\mu_{\rm B})^2} 
&\approx &1-j +j^2\ln(T/\widetilde{D}(j)) \label{eq:givemeDtilde}\\
&&-j^3\left[ \left(\ln(T/\widetilde{D}(j))\right)^2
+\frac{1}{2}\ln(T/\widetilde{D}(j))\right] \; . \nonumber 
\end{eqnarray}
Indeed, to second order the comparison of eq.~(VI-78) of the supplemental material
with eq.~(\ref{eq:givemeDtilde}) gives
\begin{equation}
  \frac{\widetilde{D}(j)}{{\cal D}}\approx \frac{U \F}{2}  \equiv c_0
  \label{eq:c0result}
\end{equation}
with
\begin{equation}
  U = \frac{e^{3/4+\C}}{\pi}= \sqrt{\pi e}w
      \label{eq:defUandFmaintext}
\end{equation}
from eq.~(VI-58) 
and~$\F$ from eq.~(\ref{eq:defFmain}).
Thus, the prefactor of the susceptibility
in eq.~(\ref{eq:Wilsonsresult}) to leading order reads
\begin{equation}
\frac{w}{4\widetilde{D}(j)}
\approx \frac{w}{2 U \F{\cal D} }= \frac{1}{2{\cal D}\F\sqrt{\pi e}}
\; .
\label{eq:nowilsonnumberinhere}
\end{equation}
This result does not contain the Wilson number
but only the prefactor $1/(2{\cal D}\F\sqrt{\pi e})$.
This does not come as a surprise because we consider finite magnetic fields
at zero temperature while $T_{\rm K}$ characterizes
the zero-field susceptibility at finite temperatures.
Note that the prefactor $c_0$ in eq.~(\ref{eq:c0result}) contains information about the
host-electron density of states via the regularized
first negative moment~(\ref{eq:defFmain}).

As shown in AFL,~\cite{RevModPhys.55.331}
and re-derived in Appendix~\ref{app:freenergyPT},
the Kondo temperature $T_{\rm K}$
and the magnetic energy scale $T_{\rm H}=\sqrt{\pi/e}T_0$
are related by
\begin{equation}
  T_{\rm K} = U T_{\rm H}  \; , 
\end{equation}
with corrections of the order $\rho_0(0)J_{\rm K}$.
Using eqs.~(\ref{eq:chiBAfinalresultBA}), (\ref{eq:T0resultBA}), (\ref{eq:Wilsonsresult}),
and~(\ref{eq:nowilsonnumberinhere}), we find
\begin{equation}
T_{\rm K}=  w \sqrt{\pi e}\frac{{\cal D} \F}{2}
\sqrt{\rho_0(0)J_{\rm K}}  \exp\left(-\frac{1}{\rho_0(0)J_{\rm K}}\right) \; ,
\label{eq:thatsmore}
\end{equation}
with corrections of the order $\rho_0(0)J_{\rm K}$.
Thus, in eq.~(\ref{eq:chiBAfinalresultBA})
\begin{equation}
\frac{\chi^{\rm ii}_0(J_{\rm K})}{(g_e\mu_{\rm B})^2}
= \frac{w}{4 T_{\rm K}} 
\; ,
\end{equation}
which is the familiar expression of the zero-temperature susceptibility in terms of the
Kondo temperature~$T_{\rm K}$, see eq.~(4.58) of Hewson's book.~\cite{Hewson}

In general, eq.~(\ref{eq:Wilsonsresult}) can be cast into the form
\begin{eqnarray}
  \frac{\chi_0^{\rm ii}(J_{\rm K})}{(g_e\mu_{\rm B})^2}&=&
  s_0\frac{\bar{\chi}_0(j)}{(g_e\mu_{\rm B})^2}
  \left(1+\sum_{n=1}^{\infty}\frac{s_n}{s_0} j^n\right) \; ,
\;  j=\rho_0(0)J_{\rm K}\;,\nonumber \\
  \frac{\bar{\chi}_0(j)}{(g_e\mu_{\rm B})^2}
  &=&   \frac{  \exp\left(1/j\right)}{\sqrt{j}} 
\; , \quad 
  s_0=\frac{1}{2{\cal D}\F\sqrt{\pi e}}\; .
 \label{eq:mainresultforchi}
\end{eqnarray}
To go beyond the leading order, i.e., to determine the coefficient $s_1$ in
eq.~(\ref{eq:mainresultforchi}) analytically,
requires the cumbersome
calculation of the ground-state energy as a function of magnetic field~$B$
to third order in $\rho_0(0)J_{\rm K}$.
This is beyond the scope of our presentation.

The comparison of the zero-field impurity-induced
susceptibility from the renormalization group
in eq.~(\ref{eq:mainresultforchi})
with the corresponding Bethe-Ansatz expressions~(\ref{eq:chiBAfinalresultBA})
and~(\ref{eq:T0resultBA})
leads to the desired relation between $(D,J_{\rm K}^{\rm BA})$ and 
$({\cal D},J_{\rm K})$,
\begin{eqnarray}
\frac{1}{\rho_0J_{\rm K}^{\rm BA}}&=& 
  \frac{1}{\rho_0(0)J_{\rm K}}-\frac{1}{2} \ln\left(\rho_0(0)J_{\rm K}\right)
  +\ln\left(\sqrt{\frac{1}{\pi e}}\frac{2D}{\F{\cal D}} \right)
  \nonumber \\
  &&+\ln\biggl[1+ \sum_{n=1}^{\infty}
    \frac{s_n}{s_0}\left(\rho_0(0)J_{\rm K}\right)^n\biggr]
    \label{eq:linkJKtoJKBA}
\end{eqnarray}
with ${\cal D}=W/2=1$ for the lattice model and $D=\pi/2$ in the
Bethe-Ansatz solvable model.

\subsection{Impurity-induced magnetization}
\label{subsec:iimagBA}

The Bethe Ansatz provides 
the impurity-induced magnetization of the system $m^{\rm ii}(J_{\rm K}^{\rm BA},B,T)$
at finite temperatures~$T$
and finite external fields~$B$, see eq.~(\ref{eq:miidef}).
For small couplings, $J_{\rm K}\ll 1$, the impurity-induced susceptibility is
exponentially large, see eq.~(\ref{eq:mainresultforchi}),
so that the relevant magnetic fields
that lead to a finite magnetization are exponentially small.
The polarization of the host electrons becomes
negligibly small, and we do not have to distinguish 
between the impurity spin polarization and
the impurity-induced magnetization, i.e.,
$m^{\rm ii}(J_{\rm K}\ll 1,B\to 0)= m^S(J_{\rm K}\ll 1,B\to 0)$,
with exponential accuracy.
Therefore, the zero-field susceptibilities from the impurity-induced
magnetization 
and from the impurity spin polarization become identical.

{}From eqs.~(5.1.34) and (5.1.37) in TW
and (4.29) in AFL, 
the Bethe Ansatz result for the impurity-induced
magnetization reads ($e=\exp(1)$)
\begin{eqnarray}
  \frac{m^{\rm ii}(h\leq 1)}{g_e\mu_{\rm B}}&=&
  \frac{1}{2\sqrt{\pi}} \sum_{n=0}^{\infty} \left(\frac{n+1/2}{e}\right)^{n+1/2}
  \frac{(-1)^nh^{2n+1}}{n!(n+1/2)}   \; ,\nonumber \\
  \label{eq:hsmallerthanunity} \\
  \frac{m^{\rm ii}(h\geq 1)}{g_e\mu_{\rm B}}&=&
  \frac{1}{2}-\frac{1}{2\pi^{3/2}} \int_0^{\infty}
  \rmd \omega
  \frac{\sin(\pi\omega)}{\omega}  \Gamma(1/2+\omega)\nonumber \\
  &&\hphantom{\frac{1}{2}-\frac{1}{2\pi^{3/2}} \int_0^{\infty}}
 \times \left(\frac{\omega}{e}\right)^{-\omega} h^{-2\omega}
  \; .
  \label{eq:hlargerthanunity} 
    \end{eqnarray}
In eqs.~(\ref{eq:hsmallerthanunity}) and~(\ref{eq:hlargerthanunity}),
the external field is scaled by the
universal low-temperature magnetic energy scale $T_1$,
\begin{eqnarray}
  h&=&\frac{B}{T_1} \; , \nonumber \\
  T_1&=&\sqrt{\frac{2\pi}{e}}  T_0 = \sqrt{\frac{2\pi}{e}}
  \left(\frac{4\pi \chi_0^{\rm ii}(J_{\rm K})}{(g_e\mu_{\rm B})^2}  \right)^{-1} \; .
\label{eq:THTWofTKTW} 
\end{eqnarray}
Since $\chi^{\rm ii}_0(J_{\rm K})$
can be calculated analytically in terms of $\rho_0(0)J_{\rm K}$ 
only to leading order, see eq.~(\ref{eq:linkJKtoJKBA}),
we follow the usual approach and determine $T_1$ numerically
from the zero-field susceptibility.~\cite{PhysRevB.87.184408}

\section{Numerical approaches}
\label{sec:Numapproaches}

In this section, we briefly discuss two numerically exact approaches
to the many-body problem.
We begin with the Density-Matrix Renormalization Group (DMRG) method,
and move on to the Numerical Renormalization Group (NRG) technique
that performs the Wilson renormalization scheme numerically.

\subsection{DMRG}
\label{subsec:DMRG!}

\subsubsection{Impurity spin polarization and impurity-induced magnetization}

\paragraph{Impurity spin polarization}
When we apply the magnetic field only at the impurity,
standard DMRG ground-state calculations
provide the results for the impurity spin polarization 
$\langle \hat{S}^z\rangle=m_{\rm loc}^S/(g_e\mu_{\rm B})$.
For a globally applied field ${\cal H}$ the calculation
of $\langle \hat{S}^z\rangle$ is more subtle
because the total spin in $z$-direction $S_{\rm tot}^z=S^z+s^z$ 
is a good quantum number,~\cite{Barczaetal}
see Sect.~\ref{subsec:gspropertiesdefanddiscussionpart2}.
Therefore,
the spin quantum number $S_{\rm tot}^z$ changes from $S_{\rm tot}^z=0$ 
for ${\cal H}=0$ to $S_{\rm tot}^z=1,2,3,\ldots$ 
for increasing external fields
in steps of $g\mu_{\rm B}{\cal H}_n$ whenever 
\begin{equation}
g\mu_{\rm B}{\cal H}_n=E_0(S_{\rm tot}^z=n)-E_0(S_{\rm tot}^z=n-1)
\end{equation}
for $n=1,2,3,\ldots\ $. 
Thus, the impurity spin polarization 
is recorded only at discrete values of
the external field whereby expectation values are calculated with
the ground state for $S_{\rm tot}^z=n$.
For $J_{\rm K}\leq 1$, we use system sizes $L=29,61,125,253,509,637,765$.
Since this approach hampers 
a systematic finite-size extrapolation,
we plot $m^S(B)/(g_e\mu_{\rm B})$ for our largest system sizes.

For $J_{\rm K}\gtrsim 1$ and a global magnetic field, 
the calculation of the impurity spin polarization
faces the problem that the impurity and the electron spin at $n=0$ form
a singlet and tend to separate from the rest of the system.
This reduces the effective length of the half-chain by one site, and
a finite-size gap opens at the Fermi energy. 
To counteract this effect
for $J_{\rm K}\gtrsim 1$,
we subtract two sites form 
the original chain, i.e., we use $L=27,59,123,251,507,635,763$.
Then, the ground state at ${\cal H}=0^+$ 
has total spin $S_{\rm tot}^z=1/2$, and the impurity
magnetization of the ground states at $S_{\rm tot}^z=n+1/2$ ($n=1,2,3,\ldots$)
is recorded. 

\paragraph{Impurity-induced magnetization}
In DMRG, we can calculate the ground-state energy $E_0(J_{\rm K},S^z_{\rm tot},L)$
for given integer $0\leq S^z_{\rm tot}\leq (L+3)/2$.
For very large system sizes, 
$\Delta E(J_{\rm K},S^z_{\rm tot},L)=E_0(J_{\rm K},S^z_{\rm tot},L)
-E_0(J_{\rm K},0,L)$ can be fitted to a positive, continuous function
of $s\equiv S^z_{\rm tot}$.
Then, the global external field is obtained from
\begin{equation}
B= \frac{\partial \Delta E(J_{\rm K},s,L)}{\partial s}\equiv E_0'(s) \;.
\label{eq:Bfroms}
\end{equation}
In turn, we may solve eq.~(\ref{eq:Bfroms}) for 
the total spin~$s_{\rm opt}(B)$ as a function of~$B$,
\begin{equation}
s_{\rm opt}(B)= [E_0']^{-1}(B) \;,
\label{eq:sfromB}
\end{equation}
where $[E_0']^{-1}(x)$ is the inverse function of $E_0'(x)$
for given $J_{\rm K}$ and $L$.
Thus, the impurity-induced magnetization for a global field
is given by
\begin{equation}
\frac{m^{\rm ii,DMRG}(J_{\rm K},B,L)}{g_e\mu_{\rm B}}=
s_{\rm opt}(J_{\rm K},B,L)-s_{\rm opt}^{\rm free}(B,L) \; .
\label{eq:miiDMRG}
\end{equation}
For a local field, one has to also calculate the impurity spin
polarization $\langle \hat{S}^z \rangle$ as a function of $S_{\rm tot}^z$.

In practice, it requires exceedingly large system sizes
to carry out this program because, in the region of small 
Kondo couplings, $J_{\rm K}\lesssim 0.5$,
the susceptibility is very large so that the system is
almost fully polarized for very small fields even for system
sizes $L={\cal O}(10^3)$. For this reason 
the analytic curve $E_0(s)$ is not known with the required accuracy.
Therefore, we do not employ the DMRG to calculate impurity-induced 
quantities.

\subsubsection{Technicalities}

The accuracy of the calculations is controlled using the 
dynamic block-state selection (DBSS) 
scheme.~\cite{PhysRevB.67.125114,PhysRevB.70.205118} 
Setting the control parameter to $\chi=10^{-5}$, the truncation error yields 
around $10^{-7}$ while the number of maximally kept DMRG block-states 
was observed to in the range $M=5000$ for our largest system sizes.

\begin{figure}[t]
  \includegraphics[width=8.5cm]{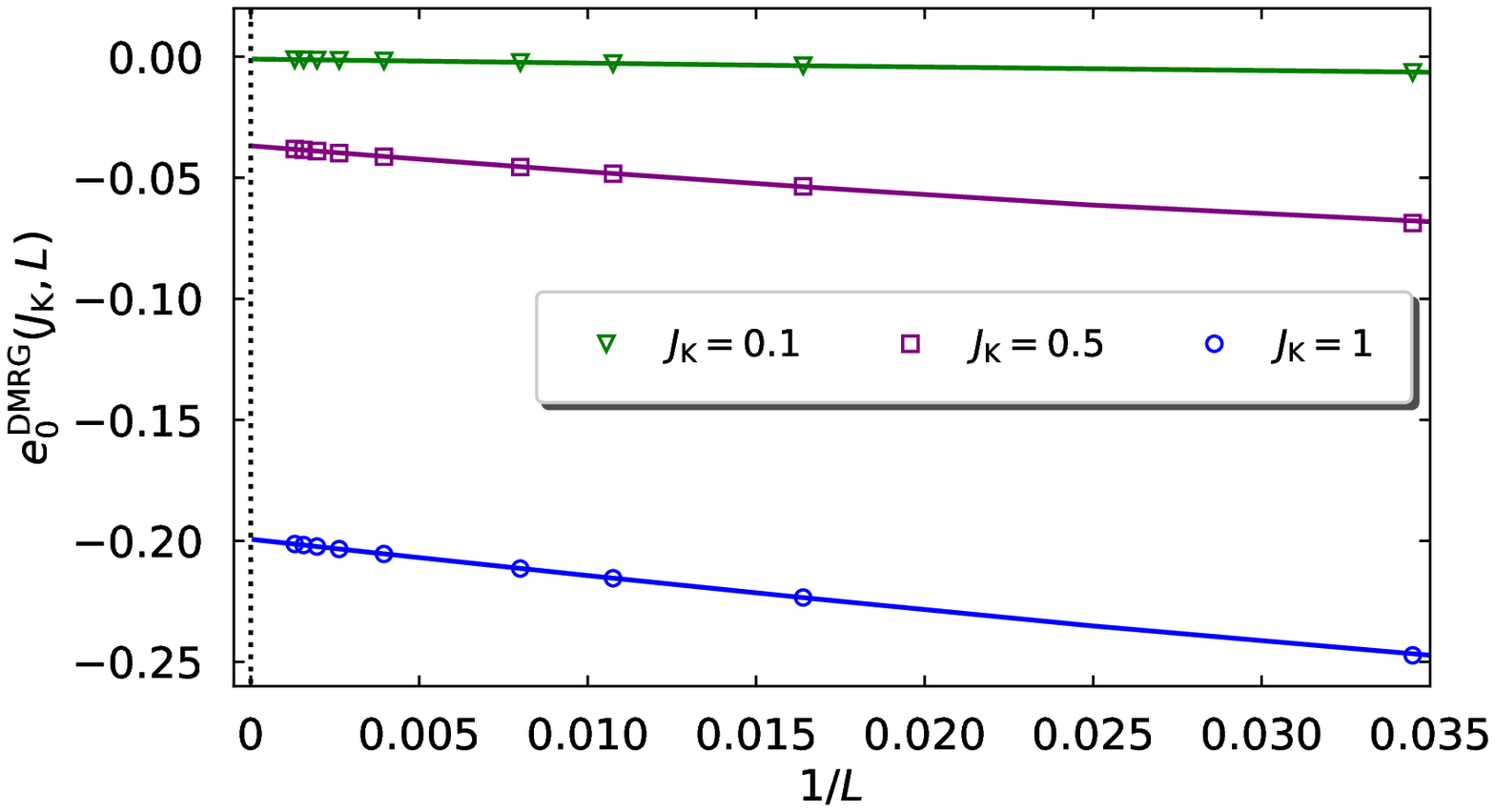}\\[12pt]
  \includegraphics[width=8.5cm]{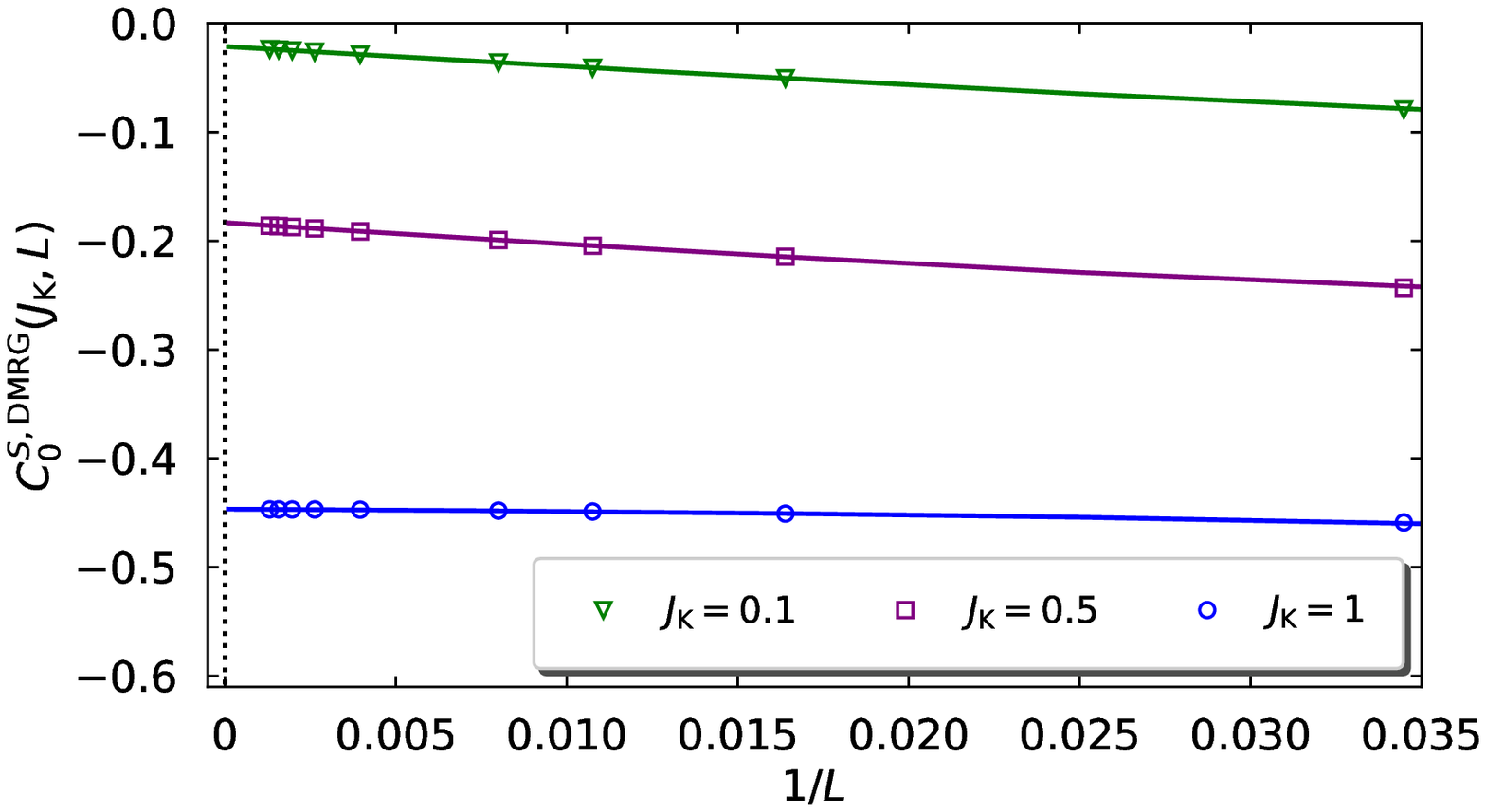}
  \caption{(Color online) Ground-state 
energy $e_0^{\rm DMRG}(J_{\rm K},L)$ from DMRG (upper figure)
    and local spin correlation function 
$C_0^{S,{\rm DMRG}}(J_{\rm K},L)$ (lower figure)
of the one-dimensional symmetric Kondo model
as a function of inverse system size $1/L$ for
$J_{\rm K}=0.1,0.5,1$.
The solid lines represent the second-order polynomial
fit in $1/L$.
\label{fig:finitesize}}
\end{figure}

For the ground-state energy we use DMRG to calculate the excess ground-state energy
$e_0(J_{\rm K},L)$, see eq.~(\ref{eq:excesse0}), 
and extrapolate to the thermodynamic limit,
\begin{equation}
  e_0^{\rm DMRG}(J_{\rm K}) =\lim_{L\to\infty} e_0^{\rm DMRG}(J_{\rm K},L)
  \end{equation}
using a second-order polynomial fit in $1/L$.
As an example, we present the ground-state energy $e_0^{\rm DMRG}(J_{\rm K},L)$
and the local spin-correlation $C_0^{S,{\rm DMRG}}(J_{\rm K},L)$
as a function of inverse system size for $J_{\rm K}=0.1,0.5,1$ 
in Fig.~\ref{fig:finitesize}; the finite-size extrapolation
is unproblematic.

\begin{figure}[t]
  \includegraphics[width=8.5cm]{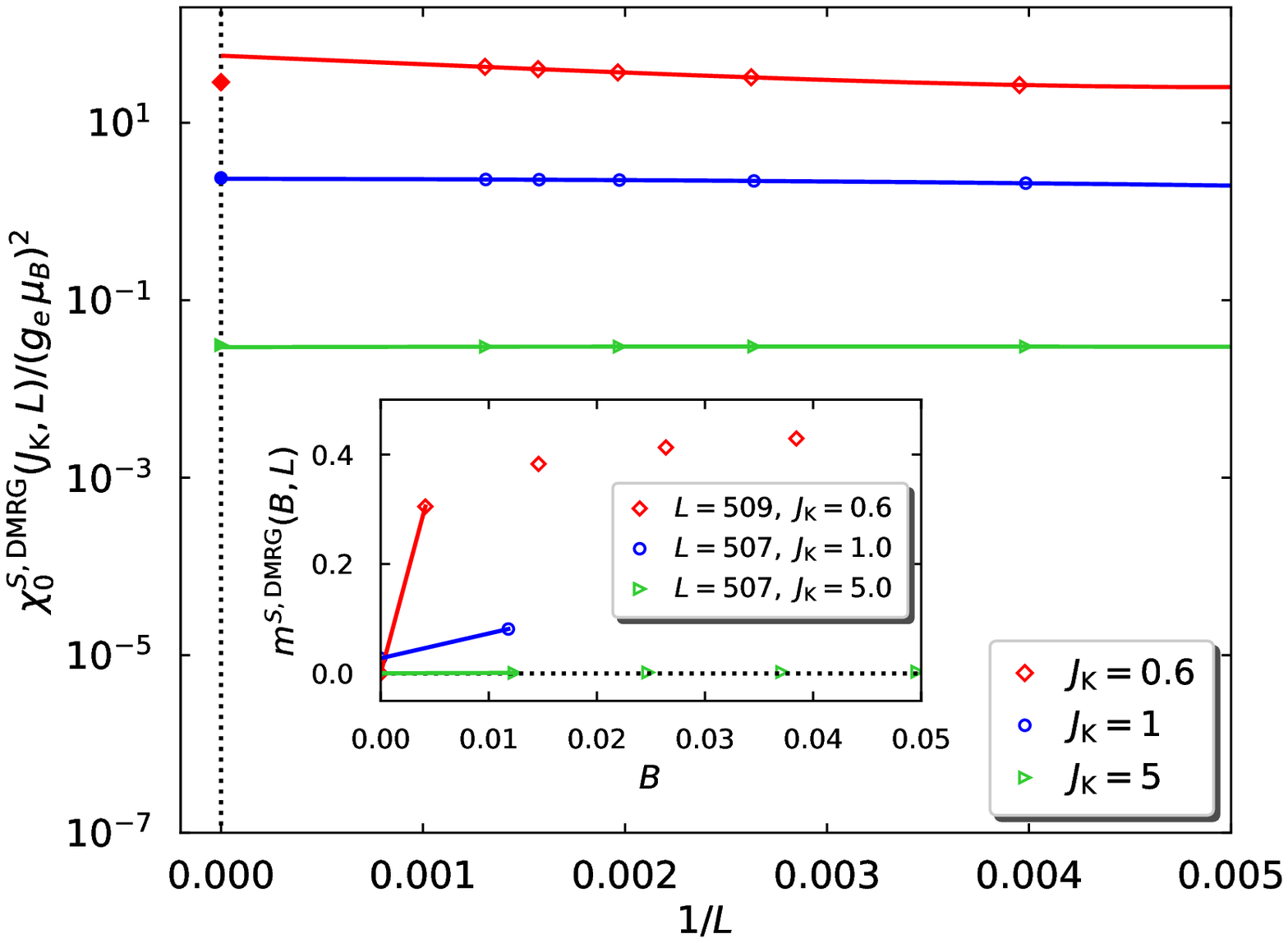}\\[15pt]
   \includegraphics[width=8.5cm]{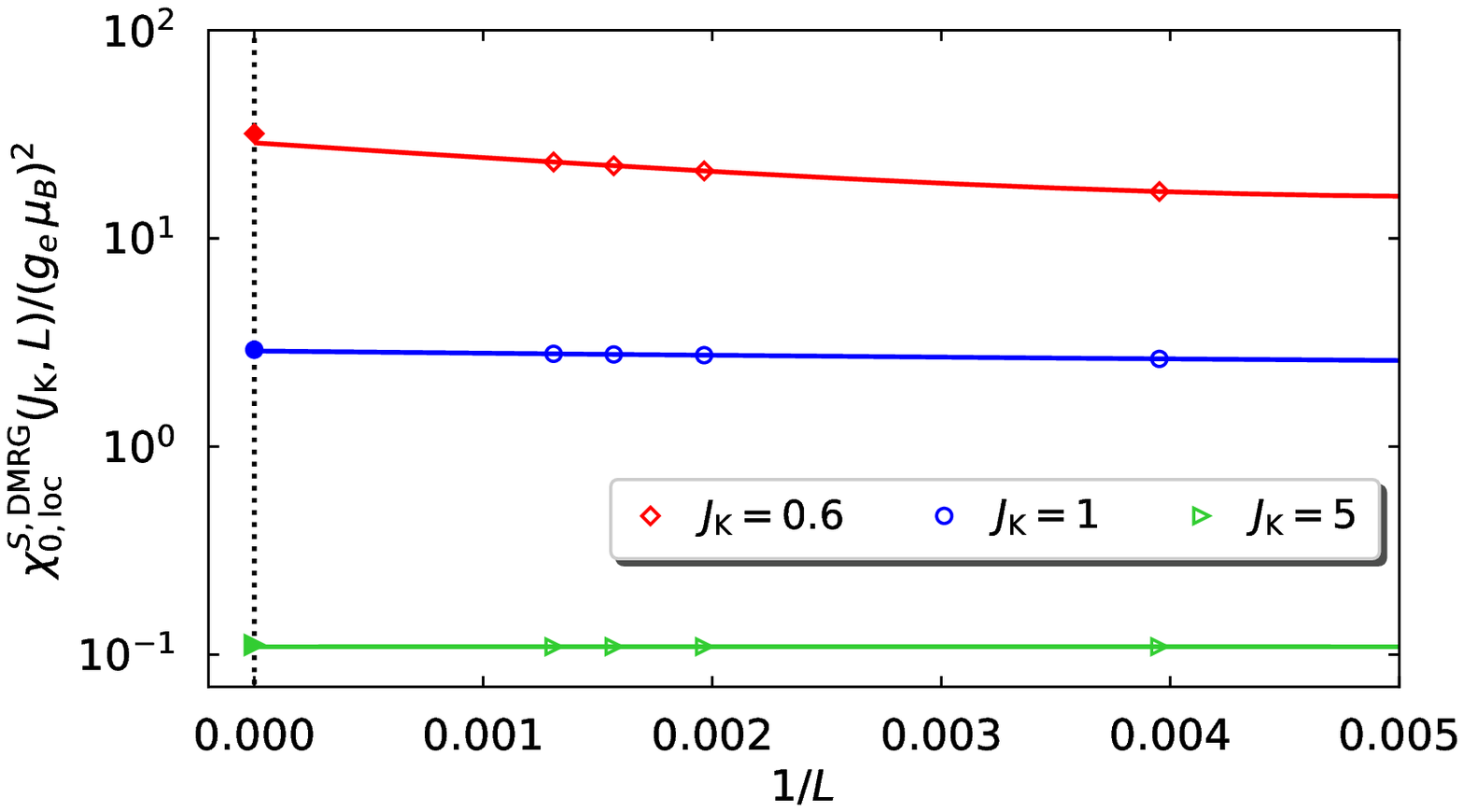}
  \caption{(Color online) Zero-field impurity spin susceptibility
    $\chi_{0}^{S,{\rm DMRG}}(J_{\rm K},L)$ from DMRG
for a global magnetic field (upper figure)
    and $\chi_{0,{\rm loc}}^{S,{\rm DMRG}}(J_{\rm K},L)$ 
for a local magnetic field (lower figure)  on a logarithmic scale
    for the one-dimensional symmetric Kondo model
as a function of inverse system size $1/L$ for 
$J_{\rm K}=0.6,1,5$. For comparison, data from NRG are shown as filled symbols
at $1/L=0$.
The solid lines represent the second-order polynomial fit in $1/L$
on the logarithm of the susceptibility.
The inset shows the magnetization $m^{S,{\rm DMRG}}(J_{\rm K},B,L=507,509)$
as a function of a small applied field
whose slope defines $\chi_0^{S, {\rm DMRG}}(J_{\rm K},L)/(g_e\mu_{\rm B})^2\approx
[m^{S,{\rm DMRG}}(J_{\rm K},B_1,L)-m^{S,{\rm DMRG}}(J_{\rm K},0,L)]/(2B_1)$.
\label{fig:susfinitesizeJKlarge}}
\end{figure}

In Fig.~\ref{fig:susfinitesizeJKlarge} we show 
the zero-field impurity spin susceptibility
$\chi_0^{\rm S, DMRG}(J_{\rm K},L)$
for a global magnetic field 
and $\chi_{0,{\rm loc}}^{\rm S, DMRG}(J_{\rm K},L)$
for a local magnetic field on a logarithmic scale
as a function of inverse system size $1/L$ for $J_{\rm K}=0.6,1,5$.
Apparently, the finite-size extrapolation
can safely be performed for the zero-field impurity spin susceptibility
for moderate to large coupling strengths, $J_{\rm K}\gtrsim 0.6$, because
the NRG data are reasonably well reproduced.

\begin{figure}[t]
  \includegraphics[width=8.5cm]{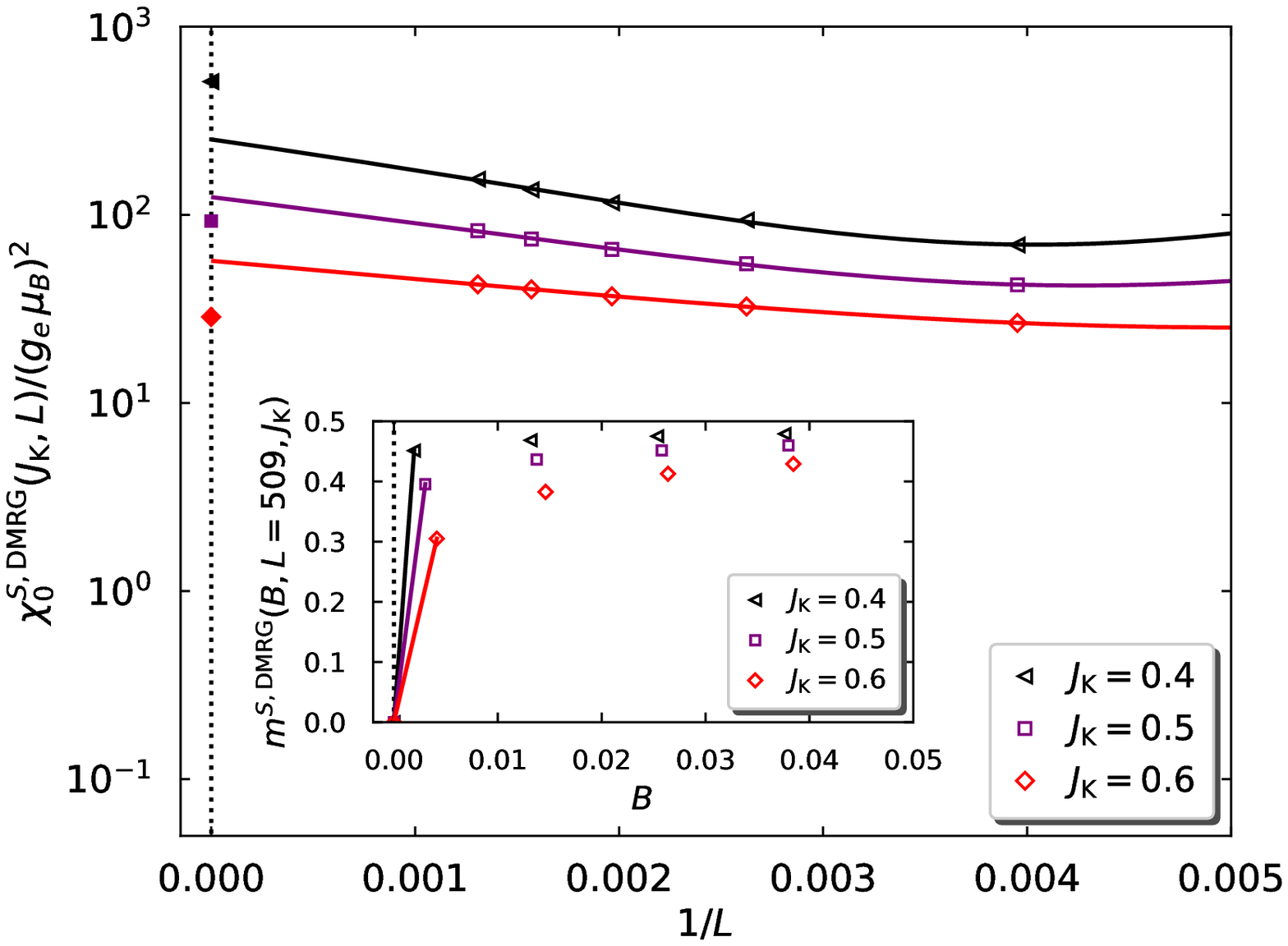}\\[12pt]
    \includegraphics[width=8.5cm]{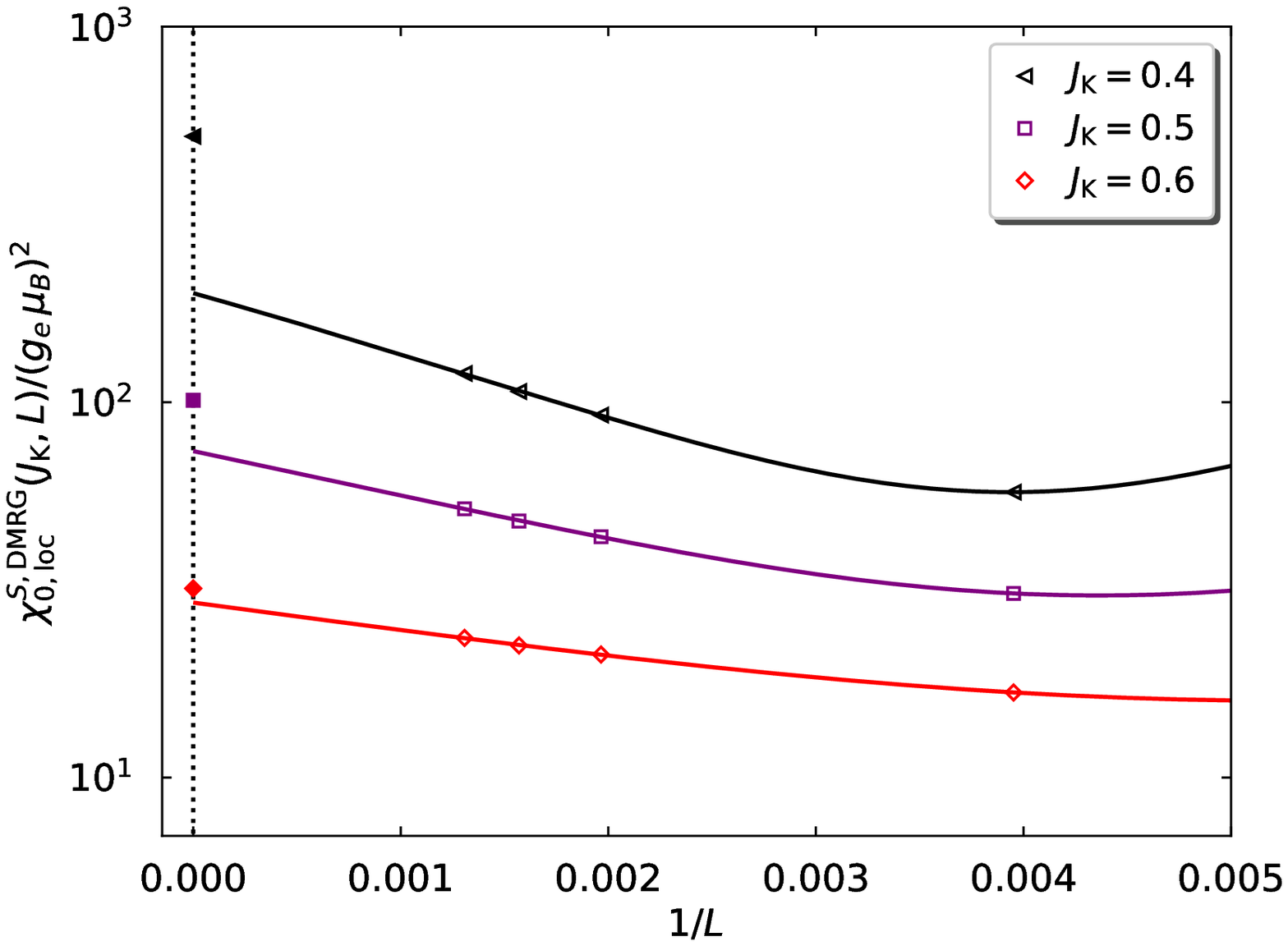}
  \caption{(Color online) Same content as Fig.~\ref{fig:susfinitesizeJKlarge} but for
    $J_{\rm K}=0.4,0.5,0.6$.
    Inset: Magnetization  $m^{S,{\rm DMRG}}(J_{\rm K},B,L=509)$
as a function of a small applied field for $J_{\rm K}=0.4,0.5,0.6$
whose slope defines $\chi_0^{S,{\rm DMRG}}/(g_e\mu_{\rm B})^2$.
\label{fig:susfinitesizeJKsmall}}
\end{figure}

As in the case of the single-impurity Anderson model,~\cite{Barczaetal}
the DMRG calculations for a global magnetic field are troubled 
for small Kondo couplings.
This is shown in Fig.~\ref{fig:susfinitesizeJKsmall} 
for $J_{\rm K}=0.4,0.5,0.6$.
For $J_{\rm K}\ll 1$,
it requires exponentially increasing system size to resolve the exponentially small
energy scale for spin excitations, i.e., for $J_{\rm K}\lesssim 0.5$,
a reliable extrapolation of the susceptibility
to the thermodynamic limit requires
system sizes that already exceed $L=10^3$ by far.
For a local magnetic field, the DMRG can access low fields so that the
zero-field impurity spin susceptibility is much better behaved 
at small interactions.
Nevertheless, the extrapolation is not very stable, as can be seen from the
winding fitting curves, and the NRG values cannot be recovered faithfully.
Again, for $J_{\rm K}\lesssim 0.5$ a reliable extrapolation of the DMRG values
for the zero-field impurity spin susceptibility to the thermodynamic limit requires
exponentially large system sizes. 

\subsection{NRG}
\label{subsec:NRGinfo}

Here, we compile basic information about the NRG algorithm
that we employ in our work; for a review on NRG, 
see Ref.~[\onlinecite{RevModPhys.80.395}].

\subsubsection{Wilson chain}

The NRG starts from the energy representation of the 
Hamiltonian,~\cite{PhysRevB.87.184408}
\begin{equation}
\hat{H}_{\rm K}=\hat{T}^{\rm NRG}+\hat{V}_{\rm sd}^{\rm NRG} 
+ \hat{H}_{{\rm loc},{\rm m}}  \; ,
\end{equation}
with the kinetic energy
\begin{equation}
\hat{T}^{\rm NRG}=  \sum_{\sigma} \int_{-1-\sigma_n B}^{1-\sigma_n B}
\rmd \epsilon\, \epsilon \, \tilde{a}_{\epsilon,\sigma}^+
\tilde{a}_{\epsilon,\sigma}^{\vphantom{+}}
\; , \label{eq:TNRGdef}
\end{equation}
and the local Kondo interaction
\begin{eqnarray}
\hat{V}_{\rm sd}^{\rm NRG}&=& 
\frac{1}{2}\left(\hat{f}_{0,\uparrow}^+\hat{f}_{0,\downarrow}^{\vphantom{+}} 
\hat{d}_{\downarrow}^+\hat{d}_{\uparrow}^{\vphantom{+}} 
+\hat{f}_{0,\downarrow}^+\hat{f}_{0,\uparrow}^{\vphantom{+}} 
\hat{d}_{\uparrow}^+\hat{d}_{\downarrow}^{\vphantom{+}} \right)
\nonumber \\
&& 
+\frac{1}{4}
\left(
\hat{d}_{\uparrow}^+\hat{d}_{\uparrow}^{\vphantom{+}} 
- \hat{d}_{\downarrow}^+\hat{d}_{\downarrow}^{\vphantom{+}}
\right)
\left(\hat{f}_{0,\uparrow}^+\hat{f}_{0,\uparrow}^{\vphantom{+}} 
-\hat{f}_{0,\downarrow}^+\hat{f}_{0,\downarrow}^{\vphantom{+}} \right) \; .
\nonumber
\end{eqnarray}
Here, the electron mode that couples to the impurity is given by
\begin{equation}
\hat{f}_{0,\sigma}=
\int_{-1-\sigma_n B}^{1-\sigma_n B}
\rmd \epsilon \sqrt{\rho_0(\epsilon+\sigma_n B)}
 \tilde{a}_{\epsilon,\sigma}^{\vphantom{+}} \; ,
\label{eq:NRGenergrepresentation}
\end{equation}
where $\sigma_n=1$ for $\sigma=\uparrow$ and
$\sigma_n=-1$ for $\sigma=\downarrow$.
In this step, no approximation is introduced.

The decisive step is the logarithmic discretization of the NRG 
Hamiltonian~(\ref{eq:NRGenergrepresentation}). 
In the presence of a global field, 
the upper and lower band edges differ from each other,
\begin{equation}
W_{\pm,\sigma}=\pm 1 -\sigma_n B \; .
\end{equation}
Thus, we follow Hager~\cite{HagerDiplomarbeit,RevModPhys.80.395}
and define the sampling points
\begin{equation}
x_{n,\sigma,\pm}=W_{\pm,\sigma}\Lambda^{-n} 
\; , \; n=0,1,2, \ldots 
\end{equation}
that depend on the position of the upper ($+$) and lower ($-$) band edges
for spin  $\sigma=\uparrow,\downarrow$.
As usual,~\cite{PhysRevB.87.184408} we approximate the density of
states in each interval $I_{n,\sigma}^+=[x_{n+1,\sigma,+},x_{n,\sigma,+}]$
and $I_{n,\sigma}^-=[x_{n,\sigma,-}, x_{n+1,\sigma,-}]$
by a suitably chosen constant. With the interval width
\begin{equation}
d_{n,\sigma}^{\pm}=
\left|W_{\pm,\sigma}\right|\Lambda^{-n}\left(1-\Lambda^{-1}\right)
\end{equation}
we define
\begin{equation}
\left(\gamma_{n,\sigma}^{\pm}\right)^2=\int_{I_{n,\sigma}^{\pm}}
\rmd \epsilon \rho_0(\epsilon+\sigma_n B) 
\end{equation}
and the expansion operators
\begin{equation}
\tilde{b}_{n,p,\sigma,\pm}^{\vphantom{+}}
=\int_{I_{n,\sigma}^{\pm}}\rmd \epsilon 
\frac{1}{\sqrt{d_{n}^{\pm}}}e^{\mp 2\pi\rmi p\epsilon/d_{n}^{\pm}}
\tilde{a}_{\epsilon,\sigma}^{\vphantom{+}} 
\end{equation}  
such that we can write
\begin{equation}
\hat{f}_{0,\sigma}^{\vphantom{+}}=\sum_{n}\left(\gamma_{n,\sigma}^{+}
\tilde{b}_{n,0,\sigma,+}^{\vphantom{+}}
+\gamma_{n,\sigma}^{-}\tilde{b}_{n,0,\sigma,-}^{\vphantom{+}}\right)
\end{equation}
for the bath state that couples to the impurity.
Note that only the mode $p=0$ appears in the bath-electron operator
$\hat{f}_{0,\sigma}^{\vphantom{+}}$.

The kinetic energy becomes 
\begin{eqnarray}
\hat{T}^{\rm NRG}&=&\sum_{n,p,p^\prime,\sigma}
\zeta_{n,p,p^\prime,\sigma}^{+}
\tilde{b}_{n,p,\sigma,+}^{+}
\tilde{b}^{\vphantom{+}}_{n,p^\prime,\sigma,+} \nonumber \\
&& \hphantom{\sum_{n,p,p^\prime,\sigma}}+
\zeta_{n,p,p^\prime,\sigma}^{-}
\tilde{b}_{n,p,\sigma,-}^{+}
\tilde{b}^{\vphantom{+}}_{n,p^\prime,\sigma,-} \; .
\end{eqnarray}
To construct the Wilson chain we now drop all modes $p\neq 0$ in the kinetic energy,
\begin{equation}
\hat{T}^{\rm NRG}
 \approx \sum_{n,\sigma}
\zeta_{n,\sigma}^{+}
\tilde{b}_{n,\sigma,+}^{+}
\tilde{b}^{\vphantom{+}}_{n,\sigma,+}
+
\zeta_{n,\sigma}^{-}
\tilde{b}_{n,\sigma,-}^{+}
\tilde{b}^{\vphantom{+}}_{n,\sigma,-}
\end{equation}
with $\tilde{b}^{\vphantom{+}}_{n,\sigma,\pm}\equiv 
\tilde{b}^{\vphantom{+}}_{n,0,\sigma,\pm}$ and
\begin{equation}
\zeta_{n,\sigma}^{\pm}\equiv \zeta_{n,0,0,\sigma}^{\pm}
=\frac{\displaystyle \int_{I_{n,\sigma}^{\pm}}\rmd \epsilon\, \epsilon\, 
\rho_0(\epsilon+\sigma_n B)}{\displaystyle
\int_{I_{n,\sigma}^{\pm}}\rmd \epsilon\,\rho_0(\epsilon+\sigma_n B)}\; .
\end{equation}
This approximation becomes exact in the limit $\Lambda\to 1$;
for a thorough discussion, see Ref.~[\onlinecite{RevModPhys.80.395}].

As a final step in the construction of the Wilson chain,
we choose $|0_\sigma\rangle\equiv \hat{f}_{0,\sigma}^+ | \text{vac}\rangle$
as starting vector for the iterative construction
of the Lanczos vectors, see appendix~\ref{app:A}.
The kinetic energy operator is then represented
as a tight-binding Hamiltonian on a chain,
\begin{equation}
\hat{T}^{\rm NRG}
 = \sum_{n=0,\sigma}^{\infty}
\varepsilon_{n,\sigma}
\hat{f}_{n,\sigma}^{+}
\hat{f}^{\vphantom{+}}_{n,\sigma}
+
t_{n,\sigma}\left(\hat{f}_{n,\sigma}^{+}
\hat{f}^{\vphantom{+}}_{n+1,\sigma} +\text{h.c.}\right) \; .
\end{equation}
The matrix elements are calculated according to the equations
(28)-(31) in Ref.~[\onlinecite{RevModPhys.80.395}].
For completeness, we give the details in appendix~\ref{app:NRGchain}.

\begin{figure}[b]
  \includegraphics[width=8.5cm]{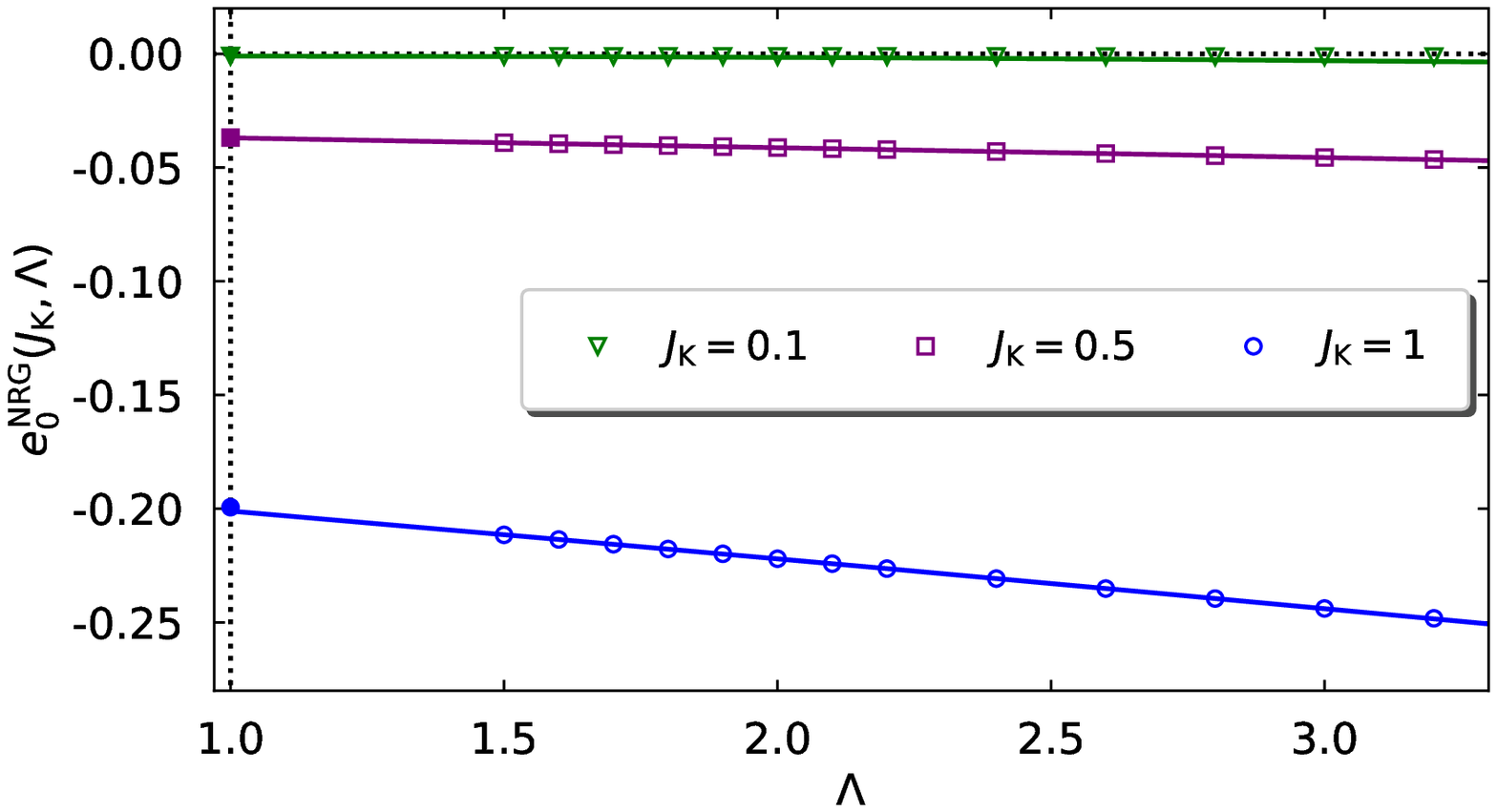}\\[9pt]
    \includegraphics[width=8.5cm]{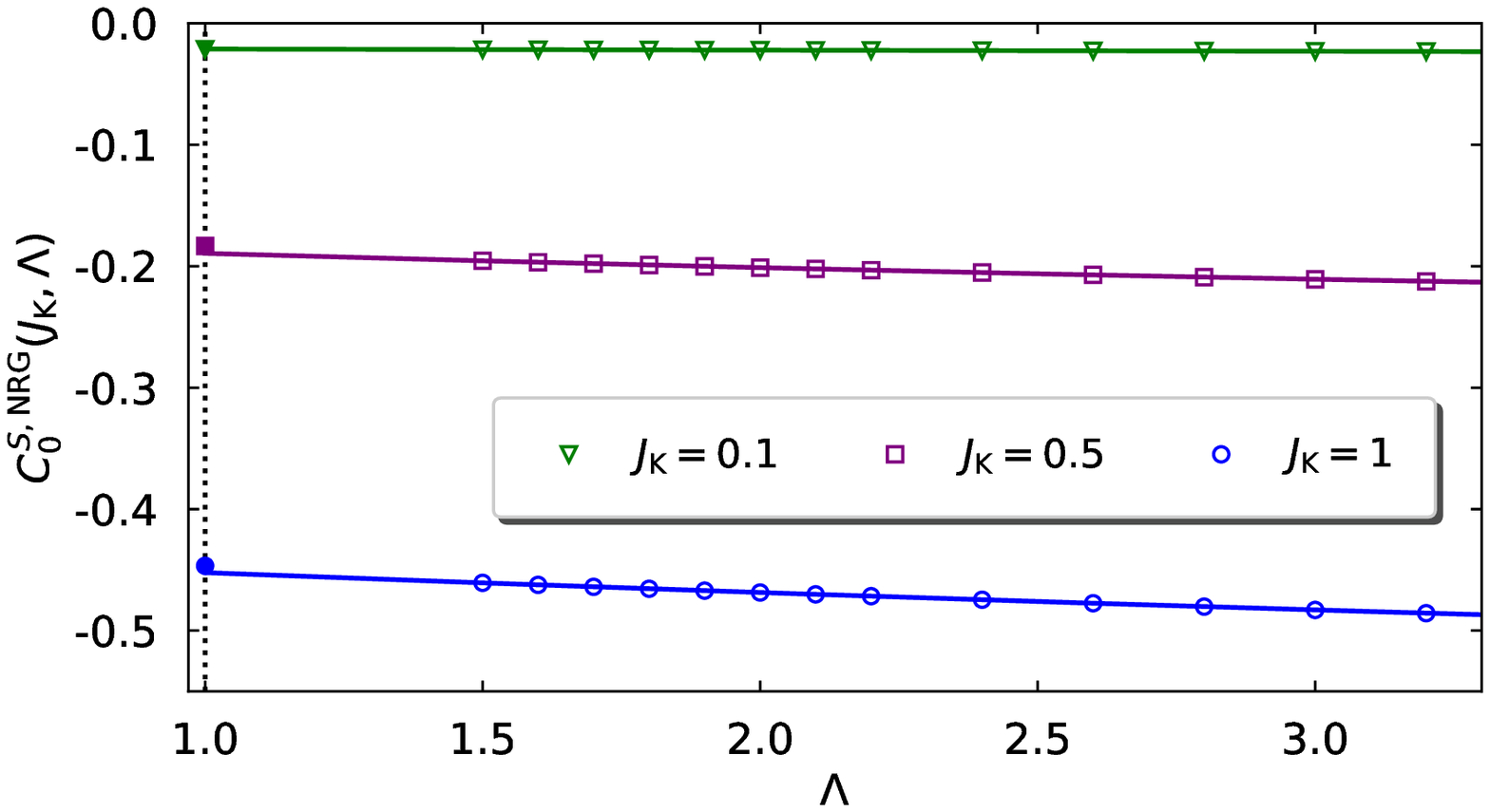}
  \caption{(Color online) 
Ground-state energy $e_0^{\rm NRG}(J_{\rm K},\Lambda)$
 from NRG (upper figure)
    and local spin correlation function $C_0^{S,{\rm NRG}}(J_{\rm K},\Lambda)$
    (lower figure)
of the one-dimensional symmetric Kondo model
as a function of the Wilson parameter $\Lambda$ for 
$J_{\rm K}=0.1,0.5,1$.
The solid lines represent a second-order polynomial fit in $(\Lambda-1)$.
Filled symbols at $\Lambda=1$ represent the DMRG values.
\label{fig:finiteLambda}}
\end{figure} 

\subsubsection{Technicalities}

The Wilson chain is solved iteratively, as described in detail
in Ref.~[\onlinecite{RevModPhys.80.395}]. As maximal chain length
we use $30 \leq n_{\rm max}\leq 100$, depending
on the value of $J_{\rm K}$.
At the end of each diagonalization step
in the renormalization group procedure,
we keep $3000 < N_s < 5000$ lowest-energy eigenstates.

\begin{figure}[t]
  \includegraphics[width=8.5cm]{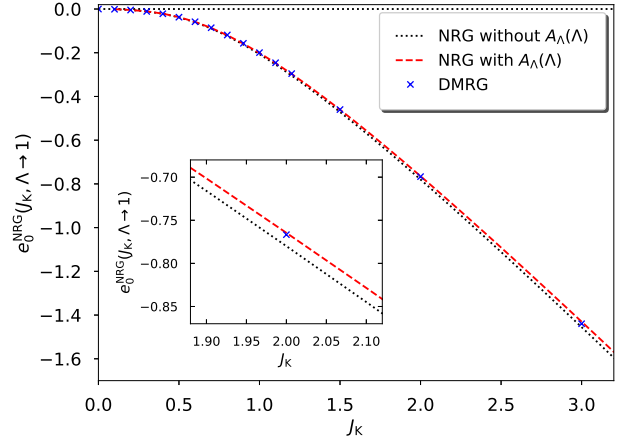}
\caption{(Color online)
  Ground-state energy $e_0^{\rm NRG}(J_{\rm K},\Lambda)$
 from NRG of the one-dimensional symmetric Kondo model
 as a function of the Kondo coupling $J_{\rm K}$.
 NRG data with and without the correction factor $A_{\Lambda}(\Lambda)$
 were extrapolated to $\Lambda\to 1$. DMRG data are shown for comparison.
\label{fig:NRGE0woALambda}}     %
\end{figure}

At the end of the renormalization group calculation,
we thus have $N_s$ states with their global quantum numbers
(energy, particle number, spin component in $z$~direction)
that permit the calculation of thermodynamic quantities
such as the ground-state and free energy, impurity-induced magnetization,
and magnetic susceptibility by taking the derivative
with respect to the external field, see eq.~(\ref{eq:chifromm}).
For large couplings, $J_{\rm K}\gtrsim 5$, 
it is numerically advantageous to calculate
the zero-field impurity-induced susceptibility from the
second-order derivative of the ground-state energy.

For the calculation of local expectation values, e.g., the 
local spin correlation and impurity spin polarization,
the corresponding quantities are expressed in terms
of the Wilson chain operators and are transformed 
in each renormalization group step.

The discretization parameter~$\Lambda$
we choose in the range $1.8\leq \Lambda \leq 3.2$.
To 
include the discretization correction for reconnecting 
  with the original continuum model even for a finite $\Lambda$,
we follow Krishna-murthy, Wilkins, and 
Wilson,~\cite{PhysRevB.21.1003,PhysRevB.21.1044}
and multiply the Kondo coupling with the 
correction factor
\begin{equation}
A_{\Lambda}(\Lambda)=\frac{1}{2}\frac{\Lambda+1}{\Lambda-1}\ln(\Lambda) 
\label{eq:Acorr}
\end{equation}
that becomes unity for $\Lambda\to 1$. 
The factor was derived for a constant density of states but we shall
see that it also works very well for the one-dimensional density of states.

\begin{figure}[b]
  \includegraphics[width=8.5cm]{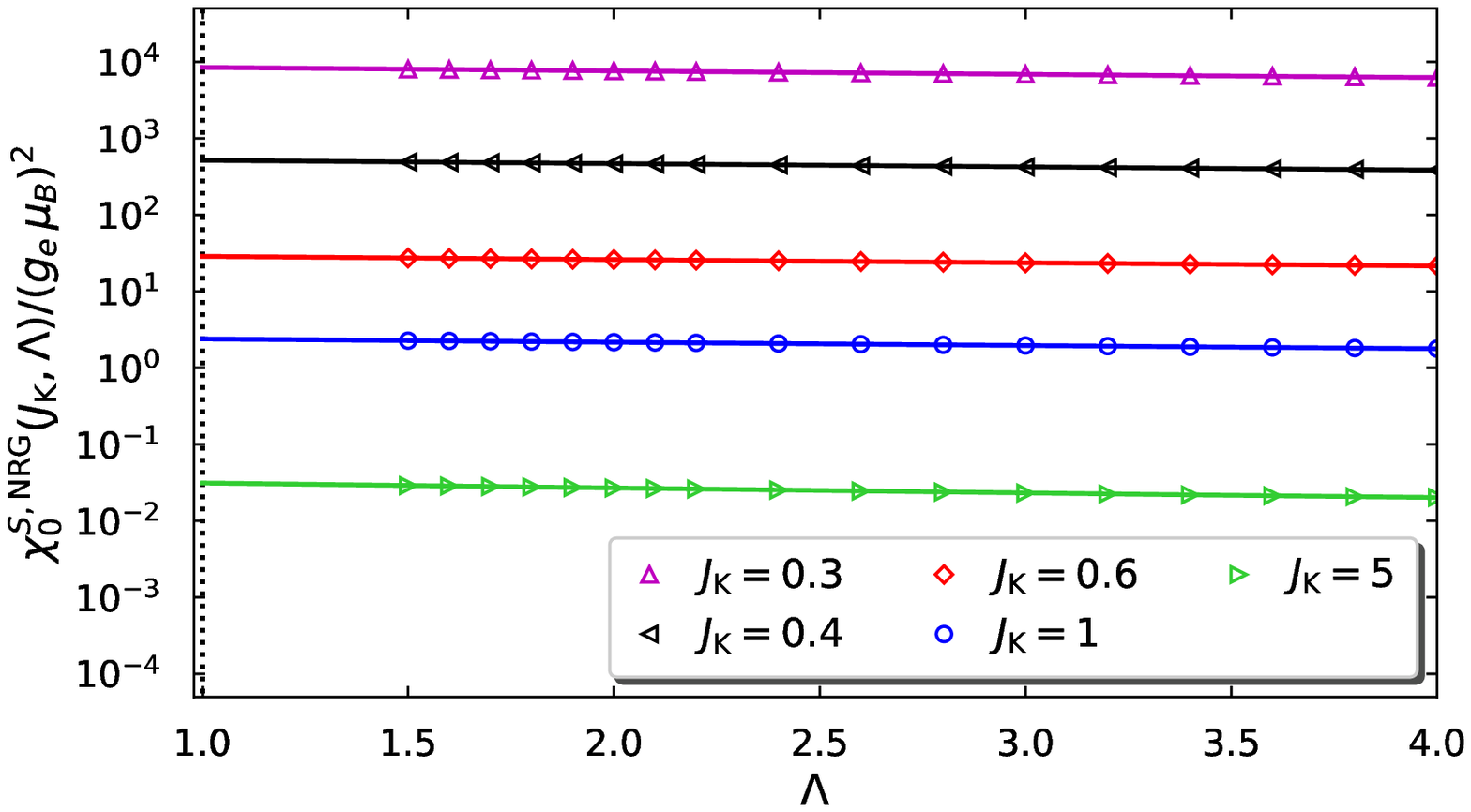}\\
    \includegraphics[width=8.5cm]{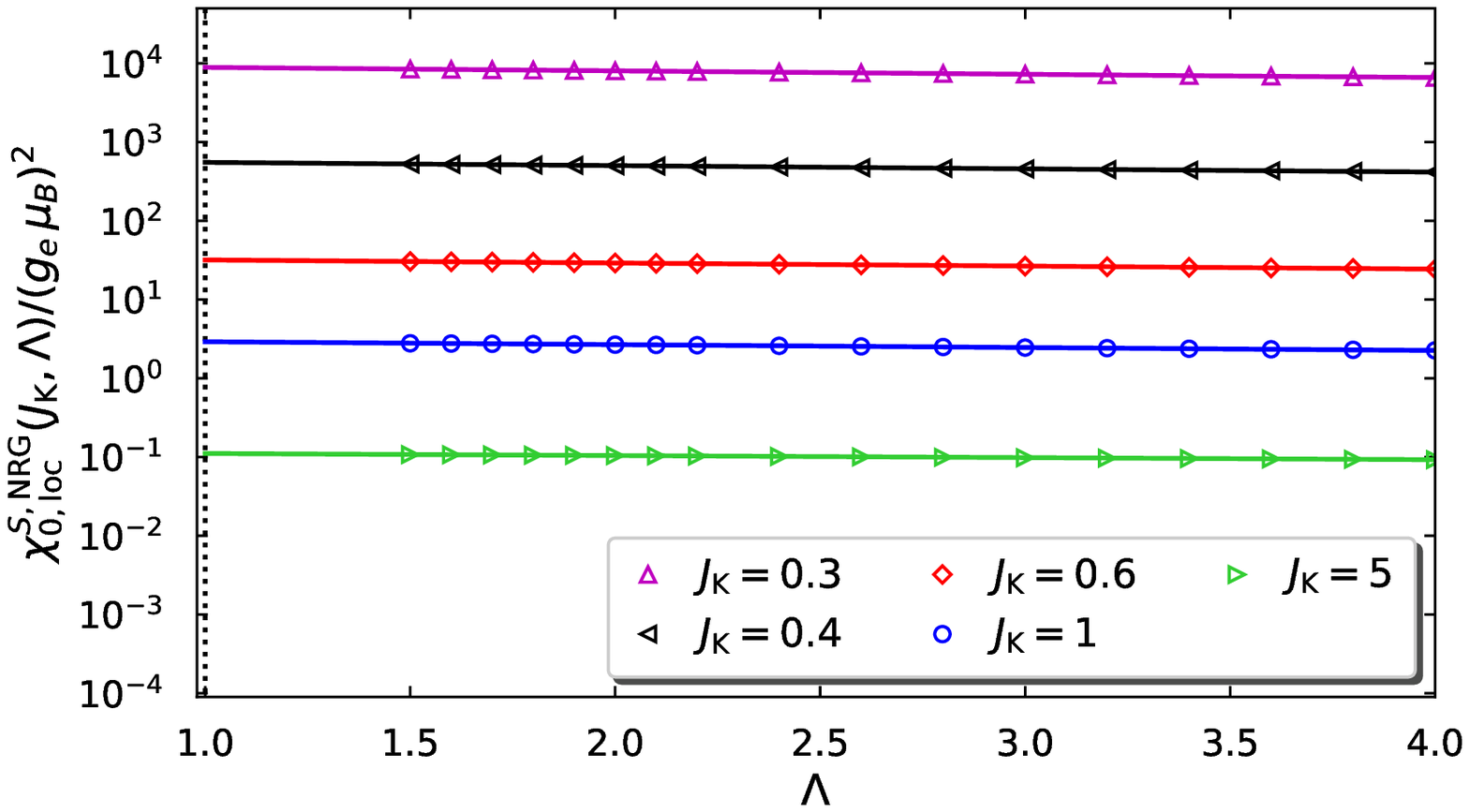}
  \caption{(Color online) Zero-field impurity spin susceptibility
    $\chi_0^{S,{\rm NRG}}(J_{\rm K},\Lambda)$ from NRG
    for a global magnetic field (upper figure)
    and $\chi_{0,{\rm loc}}^{S,{\rm NRG}}(J_{\rm K},\Lambda)$ 
for a local magnetic field (lower figure)
    for the one-dimensional symmetric Kondo model
as a function of the Wilson parameter $\Lambda$ for 
$J_{\rm K}=0.3,0.4,0.6,1,5$.
The solid lines represent the second-order polynomial fit
of $\ln[\chi_0(\Lambda)]$ in $(\Lambda-1)$.
\label{fig:susfinitesizeJKNRG}}
\end{figure}

As an example, we consider the ground-state energy.
We calculate
$ e_0^{\rm NRG}(J_{\rm K},\Lambda)$
and extrapolate to the limit $\Lambda\to 1$,
\begin{equation}
  e_0^{\rm NRG}(J_{\rm K}) =\lim_{\Lambda\to 1} 
e_0^{\rm NRG}(J_{\rm K},\Lambda) \; ,
  \end{equation}
using a second-order polynomial fit in $(\Lambda-1)$.
In Fig.~\ref{fig:finiteLambda}.
we present the ground-state energy
$e_0^{\rm NRG}(J_{\rm K},\Lambda)$
and the local spin-correlation $C_0^{S,{\rm NRG}}(J_{\rm K},\Lambda)$
as a function of the Wilson parameter~$\Lambda$ for $J_{\rm K}=0.1,0.5,1$.

The extrapolation to $\Lambda\to 1$ provides very good results
in comparison with DMRG. Note that it requires $\Lambda$-values as small
as $\Lambda=1.8$ to achieve an agreement of the extrapolated NRG values
and DMRG data within an accuracy of better than one percent.

For an independent assessment of the quality of the $\Lambda$-extrapolation,
we also performed NRG calculations
where we switched off the correction factor $A_{\Lambda}(\Lambda)$.
Recall that the correction factor was derived for a constant density
of states and thus does not necessarily perform perfectly
for the one-dimensional density of states.
As seen from Fig.~\ref{fig:NRGE0woALambda},
the extrapolated values for the ground-state energy differ by less than one percent.
In particular, the resulting energy values are slightly below the DMRG values when 
the correction factor is switched off whereas they remain consistently above
the DMRG energies when the correction factor is employed.
Therefore, we keep the correction factor in all our NRG calculations.

\begin{figure}[b]
\includegraphics[width=8.5cm]{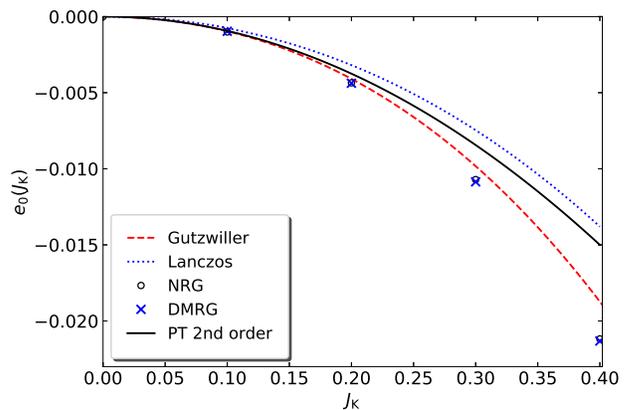}
\caption{(Color online) Ground-state energy of the one-dimensional symmetric
  single-impurity Kondo model as a function of the Kondo coupling
 for small couplings, $0\leq J_{\rm K}\leq 0.4$.
  We compare results from first-order Lanczos
  approximation (blue dotted line), eq.~(\ref{eq:Lanczosenergyfinal}),
magnetic Gutzwiller theory (red dashed line),  eq.~(\ref{eq:Gutzenergymaganalyt}), 
  perturbation theory to second order (black full line),
  eq.~(\ref{eq:secondorderanalyt}), 
  and numerical data from DMRG (blue crosses) and NRG
  (black circles).
  Not shown are the exponentially small Yosida and paramagnetic Gutzwiller
  energies.
  \label{fig:gsenergysmallJ}}
\end{figure}

In Fig.~\ref{fig:susfinitesizeJKNRG} we show 
the zero-field impurity spin susceptibility
$\chi_0^{S,{\rm NRG}}(J_{\rm K},\Lambda)$
for a global magnetic field 
and $\chi_{0,{\rm loc}}^{S,{\rm NRG}}(J_{\rm K},\Lambda)$
for a local magnetic field
as a function of~$\Lambda$ for 
$J_{\rm K}=0.3,0.4,0.6,1,5$.
In contrast to DMRG, the extrapolation
can safely be performed for the zero-field impurity spin susceptibility
for all coupling strengths.

\section{Comparison}
\label{sec:Comparison}

We begin our comparison 
with the ground-state energy and the local spin correlation. Next, we
compare the zero-field susceptibilities, and the impurity spin polarization
and impurity-induced magnetization.

\subsection{Ground-state energy and local spin correlation}

\subsubsection{Ground-state energy at small Kondo couplings}

In Fig.~\ref{fig:gsenergysmallJ} we show the ground-state energy
for small Kondo couplings, $J_{\rm K}\leq 0.4$.
In this parameter region, the Yosida and paramagnetic Gutzwiller energies 
are exponentially small which results 
in a poor variational energy bound for $J_{\rm K}\leq 0.4$.
Therefore, we do not display them.

The Lanczos approach displays the correct quadratic dependence
of the ground-state energy on $J_{\rm K}$. However, the prefactor is
too small by a factor~$\pi/4$, see eqs.~(\ref{eq:secondorderanalyt})
and~(\ref{eq:piover4lanczos}).
The best analytic variational bound is provided 
by the magnetically ordered Gutzwiller state.
As seen from eq.~(\ref{eq:Gutzenergymaganalyt}), it reproduces
96.5\% of the second-order perturbation energy term, 
and gives a very good approximation for
the ground-state energy for weak couplings.
Note, however, that the exact solution has $m=0$ at $B=0^+$, i.e.,
the magnetic Gutzwiller state does not describe the ground-state 
physics correctly.

The NRG and DMRG energies differ by not more than one percent, and thus
provide independent and accurate values for the ground-state energy.
As seen from Fig.~\ref{fig:gsenergysmallJ},
the quadratic approximation to the exact ground-state energy holds
up to $J_{\rm K}\approx 0.1$, beyond which cubic and quartic terms in $J_{\rm K}$
become discernible.

\subsubsection{Ground-state energy at intermediate and large Kondo couplings}

\begin{figure}[t]
\includegraphics[width=8.5cm]{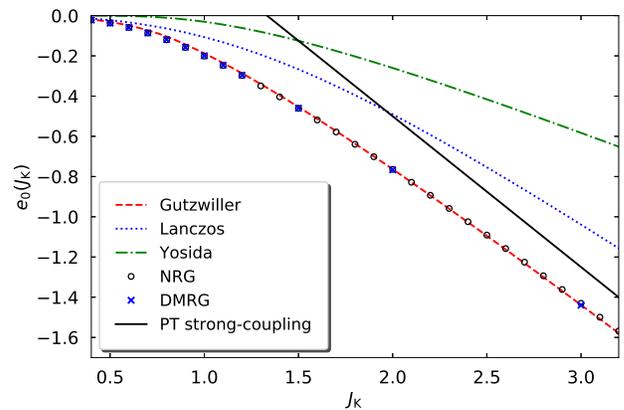}
\caption{(Color online) Ground-state energy of the one-dimensional symmetric
  single-impurity Kondo model as a function of the Kondo coupling
  for intermediate and large couplings, $0.4\leq J_{\rm K}\leq 3.2$.
  We compare results from strong-coupling perturbation theory 
  (black full line),
  eq.~(\ref{eq:energyexact1dlargeJK}), Gutzwiller theory (red dashed line),
 first-order Lanczos
  approximation (blue dotted line), eq.~(\ref{eq:Lanczosenergyfinal}),
  Yosida theory (green dot-dashed line), eq.~(\ref{eq:Yosidafinalenergy}),
  and numerical data from DMRG (blue crosses) and NRG
  (black circles).\label{fig:gsenergylarge}}
\end{figure}

In Fig.~\ref{fig:gsenergylarge} we show the ground-state energy
for intermediate to large Kondo couplings, $0.4 \leq J_{\rm K}\leq 3.2$;
recall that $W=2$ is the bandwidth of the host electrons.
Again, the NRG and DMRG data lie essentially on top of each other and thus
provide independent and accurate values for the ground-state energy.
They converge to the strong-coupling estimate~(\ref{eq:CSzerostrongcoupling})
for the ground-state energy.

Neither the Yosida wave function nor the first-order Lanczos state become
asymptotically exact for strong couplings, 
see eqs.~(\ref{eq:LanczoslargeJenergy})
and~(\ref{eq:yosida1dlarge}). The best analytical variational
upper bound results from the Gutzwiller state that
displays no local symmetry breaking above $J_{\rm K,c}^{\rm G}\approx 0.839$.
In fact, as seen in Fig.~\ref{fig:gsenergylarge},
the Gutzwiller energy for strong coupling~(\ref{eq:GutzenergylargeJK})
is in excellent agreement with the NRG and DMRG data down to $J_{\rm K}
\approx 1$,
with deviations below one percent.
Therefore, we argue that the asymptotic expression~(\ref{eq:GutzenergylargeJK})
is exact up to
and including second order in $1/J_{\rm K}$.

\subsubsection{Local spin correlation}

\begin{figure}[t]       
\includegraphics[width=8.5cm]{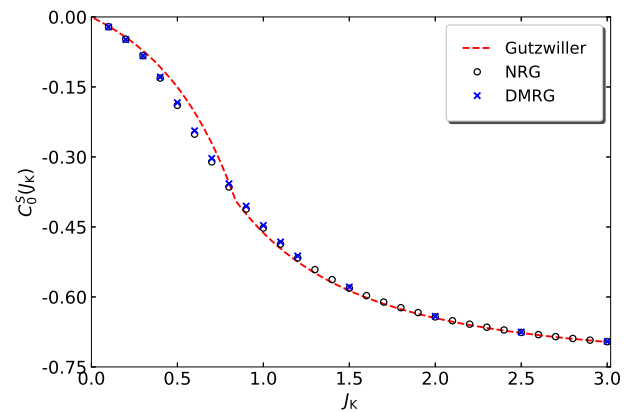}
\caption{(Color online) Local spin correlation function $C_0^S(J_{\rm K})$
  of the one-dimensional symmetric  single-impurity Kondo model 
as a function of the Kondo coupling
  $J_{\rm K}$.
  We compare results from
  Gutzwiller theory (red dashed line),
  eq.~(\ref{eq:Gutzwillerfinalenergy}),
  and numerical data from DMRG (blue crosses) and
  NRG (black circles). The linear behavior for small couplings, 
$C_0^S=-3J_{\rm K}/16$
  is given by eq.~(\ref{eq:slopeCSweak}), the limiting value for
  strong couplings, $C_0^S(J_{\rm K}\to \infty)=-3/4$ is found
  from eq.~(\ref{eq:CSzerostrongcoupling}). \label{fig:CSzero}}
\end{figure}

In Fig.~\ref{fig:CSzero} we show
the local spin correlation function $C_0^S(J_{\rm K})$
as a function of the Kondo coupling $J_{\rm K}$.
It is zero at $J_{\rm K}=0$ and decreases linearly
for small interactions, $C_0^S(J_{\rm K}\ll 1) =-3J_{\rm K}/16$,
see eq.~(\ref{eq:slopeCSweak}). For large interactions, it reaches its limiting value,
$C_0^S(J_{\rm K}\gg 1) =-3/4$, see eq.~(\ref{eq:CSzerostrongcoupling}),
which corresponds to a singlet formed by the impurity spin and
a localized host electron.
The DMRG and NRG data give the local spin correlation for all 
interaction strengths,
and faithfully interpolate between the two limiting cases.
We verified numerically that the Hellmann-Feynman theorem~(\ref{eq:CS0frome0})
is fulfilled both in DMRG and NRG.

While the Yosida and first-order Lanczos states 
are insufficient and thus omitted
from the figure, the Gutzwiller wave function reproduces the numerical data
for all interactions.
The Gutzwiller state with $m>0$ for $J<J_{\rm K,c}^{\rm G}\approx 0.839$
provides a quantitatively
satisfactory value for the local spin correlation function 
but fails qualitatively
because the exact solution does not sustain a locally symmetry-broken state.
For $J>J_{\rm K,c}^{\rm G}$, the Gutzwiller state very well approximates
the local spin correlation function. As seen in Fig.~\ref{fig:CSzero}, %
the Gutzwiller, DMRG, NRG results lie almost on top of each other for
large Kondo couplings. Therefore, we argue that the strong-coupling
expression~(\ref{eq:C0GutzlargeJ}) is actually exact up to and including
third order in $1/J_{\rm K}$.

\subsection{Magnetic susceptibilities for weak coupling}

For the zero-field susceptibilities
only the NRG is capable to examine the weak-coupling limit, $J_{\rm K}\ll 1$,
with the desired high accuracy.
We mostly investigate the case of a constant density of states, $\rho_0^{\rm const}(0)=1/2$,
for which the
correction term $A_{\Lambda}(\Lambda)$ in eq.~(\ref{eq:Acorr})
was originally derived.~\cite{PhysRevB.21.1003,PhysRevB.21.1044}
For a one-dimensional density of states, we  show numerically
that the ratio of the impurity-induced susceptibilities is given by
the regularized first negative moment of the density of states~(\ref{eq:defFmain}).

\subsubsection{Impurity-induced magnetic susceptibility}

\begin{figure}[t]
    \includegraphics[width=8.5cm]{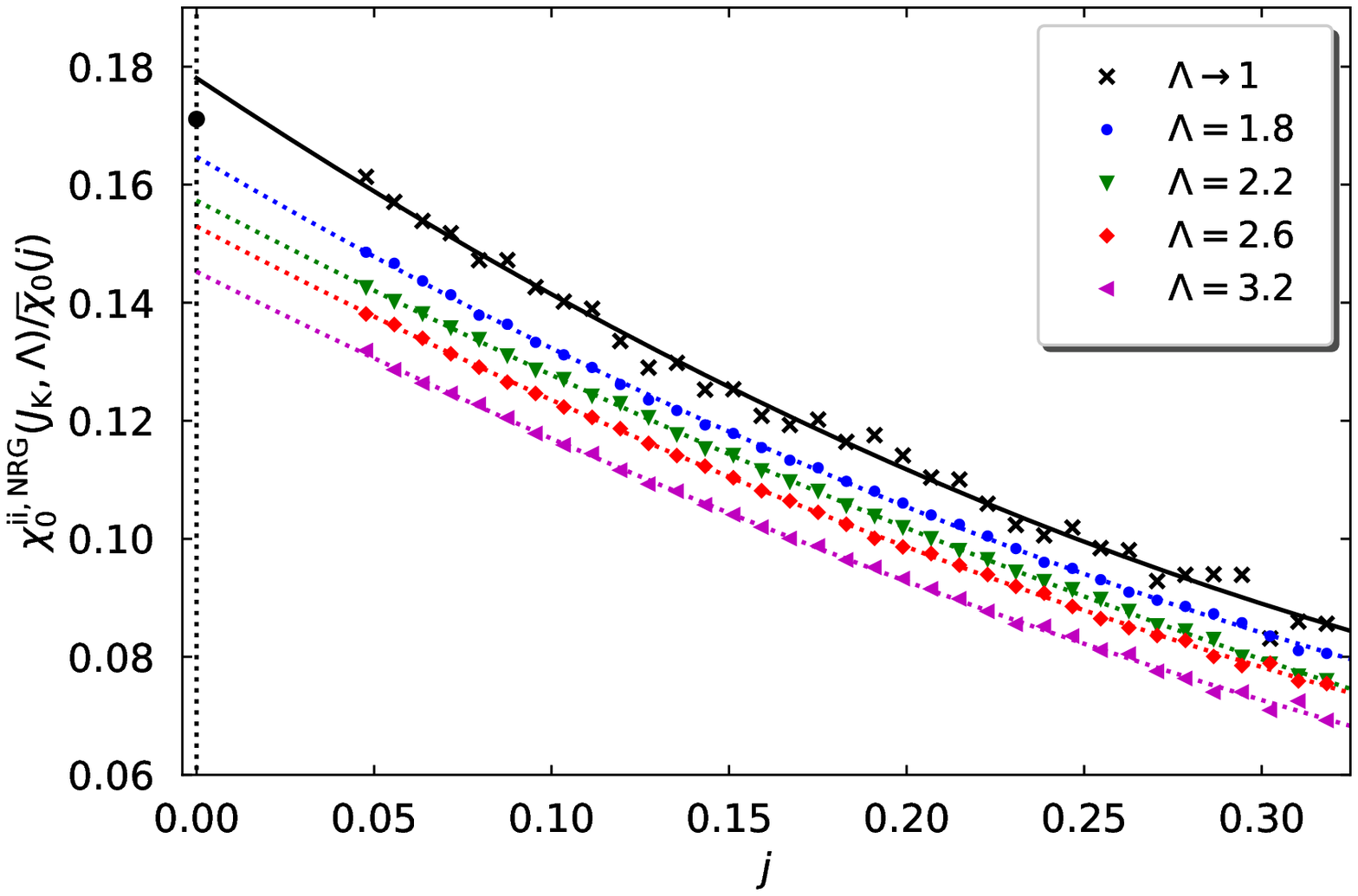}\\[9pt]
  \includegraphics[width=8.5cm]{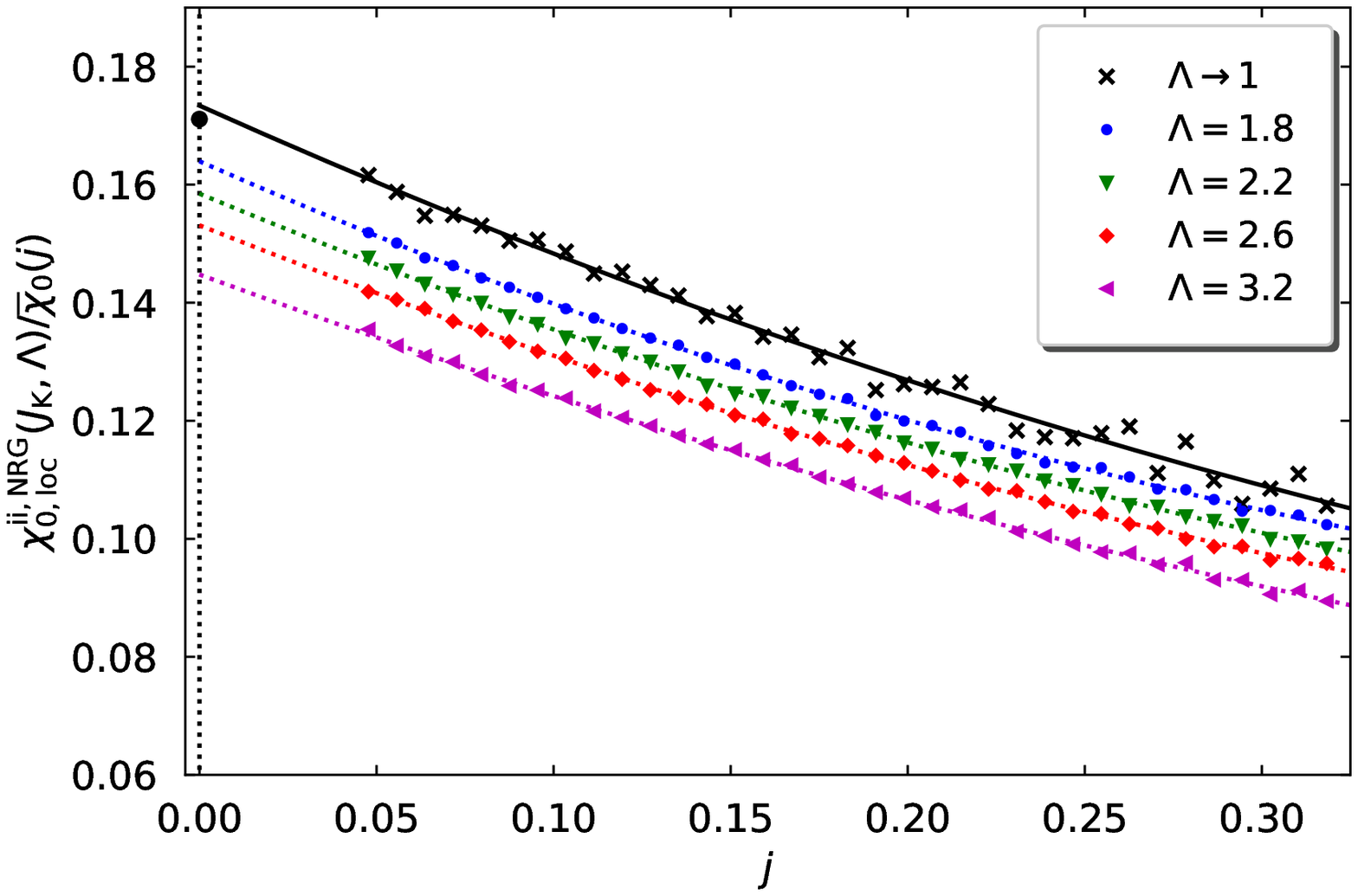}
  \caption{(Color online) Impurity-induced magnetic susceptibility
    $\chi_0^{\rm ii,NRG}(J_{\rm K},\Lambda)$
    for a global magnetic field (upper figure)
   and $\chi_{0,{\rm loc}}^{\rm ii,NRG}(J_{\rm K},\Lambda)$ 
   for a local magnetic field
   (lower figure) for the symmetric Kondo model
  with a constant density of states
  as a function $j=J_{\rm K}\rho_0(0)=J_{\rm K}/2$
  for various values of the Wilson parameter $\Lambda$.
  The lines represent the result of a second-order polynomial fit
  in $j$. To make the sub-leading terms discernible,
  we scale the susceptibilities by the universal part
  $\bar{\chi}_0(j)$, see eq.~(\ref{eq:mainresultforchi}).
  The filled symbols at $j=0$ denote the analytical
  result~(\ref{eq:mainresultforchi}).\label{fig:susJKNRG}}
\end{figure}

In Fig.~\ref{fig:susJKNRG}
we show the zero-field impurity-induced magnetic susceptibility
as a function of $J_{\rm K}$ for various values of the Wilson parameter~$\Lambda$
for a constant density of states,
both for a global magnetic field and a local magnetic field at the impurity.
The data for $\Lambda=1$ are the result
of a quadratic fit in $(\Lambda-1)$ for given $J_{\rm K}$.
We plot the ratio of the susceptibilities and the universal part
$\bar{\chi}_0(j)$, see eq.~(\ref{eq:mainresultforchi}), to focus on
the sub-leading terms. The NRG confirms the quadratic dependence of
these terms on $j=\rho_0(0)J_{\rm K}$ for $J_{\rm K}\to 0$,
\begin{equation}
  \frac{\chi_0^{\rm ii}(J_{\rm K})}{\bar{\chi}_0(j)}
  = s_0 +s_1 j +s_2 j^2 + \ldots \; , \quad j=\rho_0(0)J_{\rm K}\;.
  \label{eq:subleadings}
\end{equation}
We collect the results for $s_0,s_1,s_2$ in table~\ref{tab:one}.

\begin{table}[b]
\begin{tabular}{|l|l|l|l||l|l|}
  \hline
  & $s_0$ & $s_1$& $s_2$ & $s_0$ & $\alpha$ \\
  \hline
  $\hat{H}_{\rm m}$\vphantom{\Large A}
  &0.1781$^{\rm (a)}$  & $-$0.401$^{\rm (a)}$ & 0.348$^{\rm (a)}$
  & 0.1784$^{\rm (a)}$& $-$2.33$^{\rm (a)}$\\
  & 0.1781$^{\rm (b)}$    & $-$0.401$^{\rm (b)}$ & 0.348$^{\rm (b)}$
  & 0.1784$^{\rm (b)}$   & $-$2.35$^{\rm (b)}$\\
  $\hat{H}_{\rm m, loc}$ & 0.1734$^{\rm (a)}$ & $-$0.269$^{\rm (a)}$ 
&0.182$^{\rm (a)}$ & 0.1731$^{\rm (a)}$& $-$1.547$^{\rm (a)}$\\
  &0.1734$^{\rm (b)}$ &$-$0.269$^{\rm (b)}$ &0.182$^{\rm (b)}$ 
& 0.1731$^{\rm (b)}$& $-$1.560$^{\rm (b)}$\\
  exact & 0.171099& & & 0.171099 & \\
  \hline
\end{tabular}
\caption{Coefficients of the sub-leading terms
  in the zero-field impurity-induced magnetic susceptibility for the
  ground state of the symmetric Kondo model with a constant density of states
  in eq.~(\ref{eq:subleadings}) from NRG, see Fig.~\ref{fig:susJKNRG}.
  The two values result from the two sequences of extrapolations~(a) and~(b)
  in $(\Lambda-1)$ and $j$, see text.
  The analytic value for $s_0$ is given
  in eq.~(\ref{eq:mainresultforchi}).
  Also given are the coefficients for the fit 
  in eq.~(\ref{eq:constexpfit}).\label{tab:one}}
\end{table}

We perform two sequences of extrapolations. In extrapolation~(a),
we start with a second-order polynomial fit
in $(\Lambda-1)$ at fixed $J_{\rm K}$ and fit the resulting data
in a second-order polynomial fit in $j=J_{\rm K}\rho_0(0)$,
as shown in Fig.~\ref{fig:susJKNRG}. In extrapolation~(b)
we first extrapolate in $j$ to determine $s_l(\Lambda)$ and 
extrapolate these coefficients in~$\Lambda$ afterwards.
NRG data for $0.15 \leq J_{\rm K}\leq 1.0$ and $1.8 \leq \Lambda\leq 3.2$ are included
in the fit.
As seen from the data in table~\ref{tab:one}, the results agree very well.

The parameter $s_1$ is related to Wilson's coefficients $\alpha_1$ and $c_1$
in eq.~(\ref{eq:Wilsonsresult}) via
\begin{equation}
  s_1=-\frac{c_1/c_0+\alpha_1}{2{\cal D}\sqrt{\pi e}} \; , \quad
  \frac{c_1}{c_0} = -2{\cal D}\sqrt{\pi e} s_1 -\alpha_1 \; , 
\end{equation}
or $c_1\approx 0.45$, where we used $\F=1$ for a constant density of states,
${\cal D}=1$, $s_0\approx -0.40$,
$\alpha_1\approx 1.5824$,~\cite{RevModPhys.47.773} and $c_0\approx 0.6001$
from eq.~(\ref{eq:c0result}).
Since $|c_1/c_0| \lesssim \alpha_1$, we could also have used
\begin{equation}
  \frac{\chi_0^{\rm ii}(J_{\rm K})}{\bar{\chi}_0(j)}
 \approx s_0 e^{-\alpha j} \; , \quad j=\rho_0(0)J_{\rm K}
  \label{eq:constexpfit}
\end{equation}
as our fit function. For completeness, the results for this fitting function
are also included in table~\ref{tab:one}.

The constant term is identical in both cases,
$  s_0^{\rm NRG}\approx  s_{0,{\rm loc}}^{\rm NRG}
\approx s_0= 1/(2\sqrt{\pi e})\approx 0.1709914$,
where we used the analytic result from eq.~(\ref{eq:mainresultforchi})
for comparison. Since the NRG data for a global field show more scatter,
the accuracy of $s_0$ is smaller than for a local field,
the deviations are 4\% for a global field and 1\% for a local field.
In any case, the accuracy is good enough to see that
the prefactors of the first-order and second-order terms are different
for global and local magnetic fields, $s_1\neq s_{1,{\rm loc}}$
and $s_2\neq s_{2,{\rm loc}}$.
Different linear terms are also found in the Yosida wave function,
compare eqs.~(\ref{eq:YosidachismallJK}) and~(\ref{eq:YosidachismallJKlocal}).

\begin{figure}[b]
  \includegraphics[width=8.5cm]{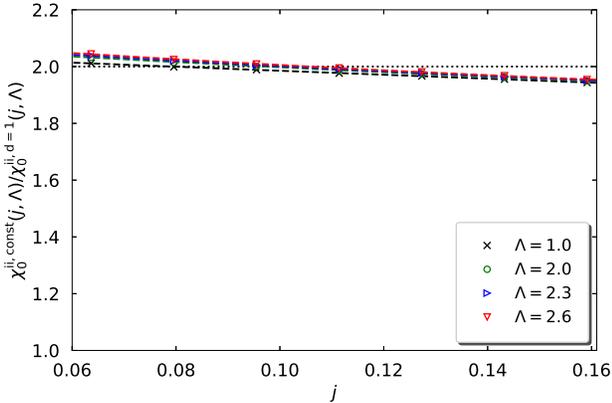}
\caption{(Color online) Ratio between the zero-field impurity-induced
  magnetic susceptibility for a constant density of states,
  $\chi_0^{\rm ii, NRG,const}(J_{\rm K},\Lambda)$,
  and for a one-dimensional density of states,
  $\chi_0^{\rm ii, NRG,d=1}(J_{\rm K},\Lambda)$ 
    for a global magnetic field
  as a function $j=J_{\rm K}\rho_0(0)$
  for various values of the Wilson parameter $\Lambda$.
  The black crosses represent the result of a second-order polynomial fit
  in $\Lambda-1$.
  The line $\F=2$ gives the result for $j=0$.\label{fig:universalratioNRG}}
\end{figure}

\subsubsection{Impurity-induced magnetic susceptibility for a one-dimensional
density of states}

In Fig.~\ref{fig:universalratioNRG} we show the
ratio between the zero-field impurity-induced
  magnetic susceptibility for a constant density of states,
  $\chi_0^{\rm ii, NRG,const}(J_{\rm K},\Lambda)$,
  and for a one-dimensional density of states,
  $\chi_0^{\rm ii, NRG,d=1}(J_{\rm K},\Lambda)$ 
    for a global magnetic field
  as a function $j=J_{\rm K}\rho_0(0)$
  for various values of the Wilson parameter $\Lambda$.
  The extrapolated value for $\Lambda\to 1$ is very close to $\F=2$
  which is the exact result for $j=0$, see eq.~(\ref{eq:mainresultforchi}).

  Apparently, the result holds for {\em all\/}~$j\ll 1$, within the accuracy
  of the NRG calculations. This universality is also seen in the Yosida wave function,
see eq.~(\ref{eq:YosidachismallJK}), where the sub-leading corrections 
are independent of the host-electron density of states.
Therefore, we conjecture that
the algebraic correction terms in eq.~(\ref{eq:mainresultforchi})
are universal in the sense that
$s_1/s_0\approx -2.3$ and $s_2/s_0\approx 2$
do not depend on the form of the host-electron density of states.

\begin{figure}[t]
    \includegraphics[width=8.5cm]{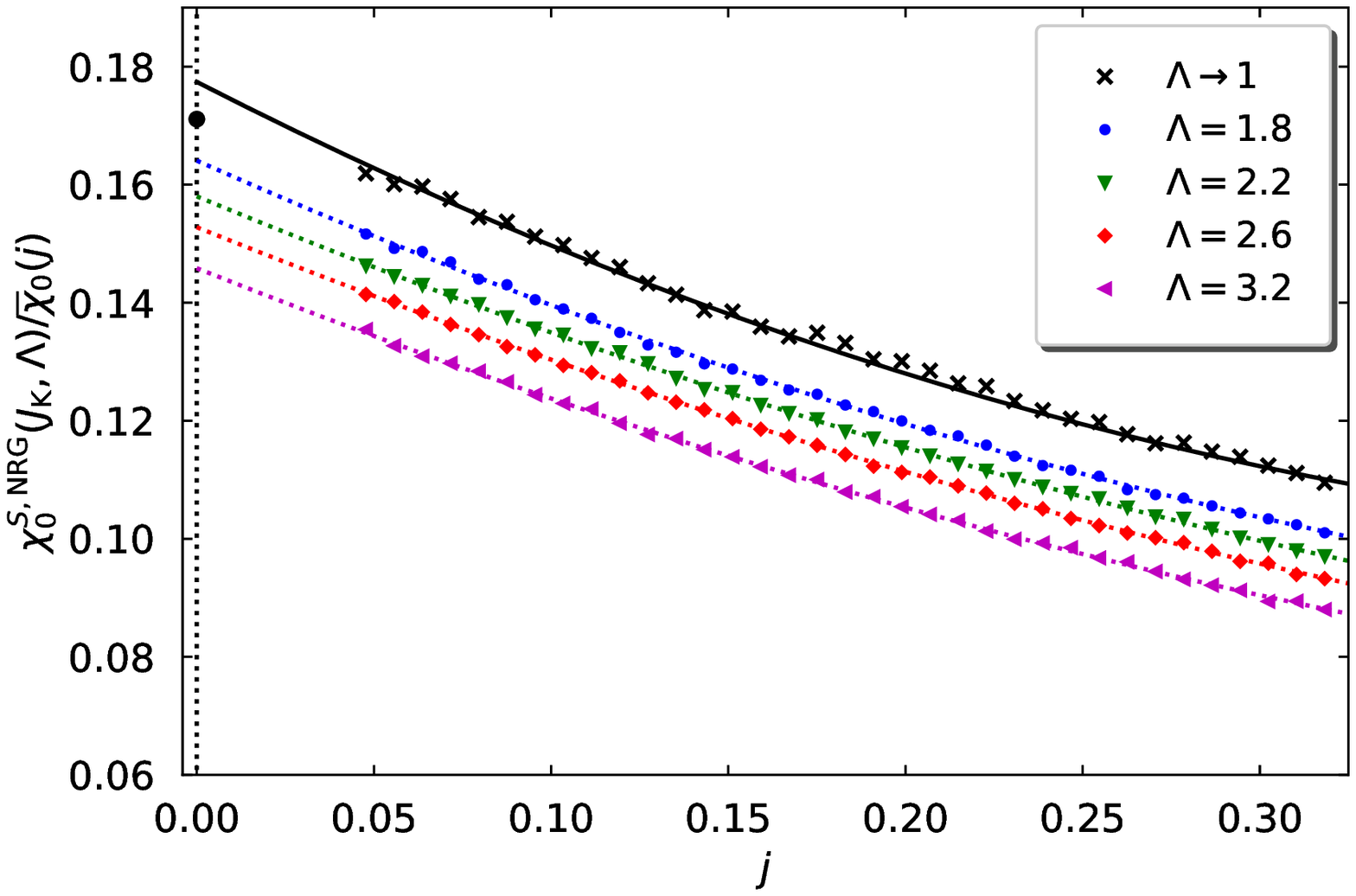}\\[3pt]
  \includegraphics[width=8.5cm]{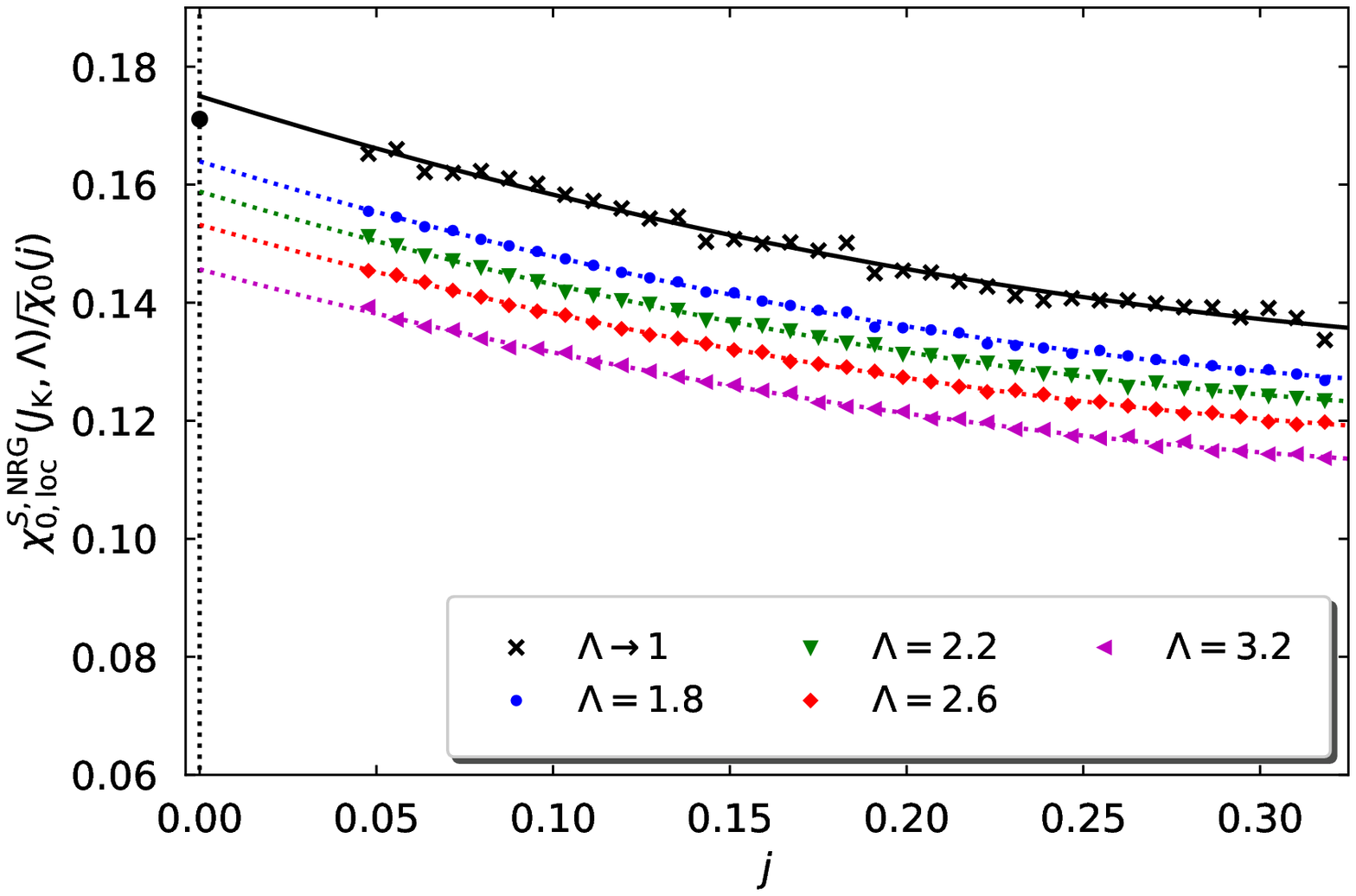}
  \caption{(Color online) Zero-field impurity spin susceptibility
    $\chi_0^{S,{\rm NRG}}(J_{\rm K},\Lambda)$
    for a global magnetic field (upper figure)
   and $\chi_{0,{\rm loc}}^{S,{\rm NRG}}(J_{\rm K},\Lambda)$ 
   for a local magnetic field
   (lower figure) for the symmetric Kondo model
  with a constant density of states
  as a function $j=J_{\rm K}\rho_0(0)=J_{\rm K}/2$
  for various values of the Wilson parameter $\Lambda$.
  The lines represent the result of a second-order polynomial fit
  in $j$. As in Fig.~\ref{fig:susJKNRG}, 
  we scale the susceptibilities by the universal part
  $\bar{\chi}_0(j)$.
The filled symbols at $j=0$ denote the analytical
  result~(\ref{eq:mainresultforchi}).\label{fig:impsusJKNRG}}
\end{figure}

\subsubsection{Impurity spin susceptibility} 

In Fig.~\ref{fig:impsusJKNRG}
we show the zero-field impurity spin susceptibility
as a function of $J_{\rm K}$ for various values of the Wilson parameter~$\Lambda$
for a constant density of states,
both for a global magnetic field and a local magnetic field at the impurity.
The data for $\Lambda=1$ are the result
of a quadratic fit in $(\Lambda-1)$ for given $J_{\rm K}$.
Again, we plot the ratio of the susceptibilities and the universal part
$\bar{\chi}_0(j)$, see eq.~(\ref{eq:mainresultforchi}), to focus on
the sub-leading terms.
The NRG confirms the quadratic dependence of
these terms on $j=\rho_0(0)J_{\rm K}$ for $J_{\rm K}\to 0$,
\begin{equation}
  \frac{\chi_0^S(J_{\rm K})}{\bar{\chi}_0(j)}
  = S_0 +S_1 j +S_2 j^2 + \ldots \; , \quad j=\rho_0(0)J_{\rm K}\;.
  \label{eq:subleadingsimpspin}
\end{equation}
We collect the results for $S_0,S_1,S_2$ in table~\ref{tab:two}.
Note that we use capital letters here to distinguish
the coefficients $s_l$ for the impurity-induced susceptibility 
from the coefficients~$S_l$ for the impurity spin susceptibility.
We also include the results from the extrapolation
analogous to eq.~(\ref{eq:constexpfit}) which provides the 
coefficient $\alpha^S$ from an exponential extrapolation.
NRG data for $0.15 \leq J_{\rm K}\leq 1.0$ and $1.8 \leq \Lambda\leq 3.2$ are included
in the fit.

\begin{table}[t]
\begin{tabular}{|l|l|l|l||l|l|}
  \hline
  & $S_0$ & $S_1$& $S_2$ & $S_0$ & $\alpha^S$\vphantom{\Large A} \\
  \hline
  $\hat{H}_{\rm m}$\vphantom{\Large A}  &0.1774$^{\rm (a)}$  & $-$0.307$^{\rm (a)}$ & 0.300$^{\rm (a)}$
  & 0.1738$^{\rm (a)}$& $-$1.49$^{\rm (a)}$\\
  & 0.1774$^{\rm (b)}$    & $-$0.307$^{\rm (b)}$ & 0.300$^{\rm (b)}$
  & 0.1738$^{\rm (b)}$   & $-$1.50$^{\rm (b)}$\\
  $\hat{H}_{\rm m, loc}$ & 0.1753$^{\rm (a)}$ & $-$0.193$^{\rm (a)}$ 
&0.218$^{\rm (a)}$ & 0.1707$^{\rm (a)}$& $-$0.766$^{\rm (a)}$\\
  &0.1753$^{\rm (b)}$ &$-$0.193$^{\rm (b)}$ &0.218$^{\rm (b)}$ 
& 0.1707$^{\rm (b)}$& $-$0.766$^{\rm (b)}$\\
  exact & 0.171099& & & 0.171099 & \\
  \hline
\end{tabular}
\caption{Coefficients of the sub-leading terms
  in the zero-field impurity spin susceptibility for the
  ground state of the symmetric Kondo model with a constant density of states
  in eq.~(\ref{eq:subleadings}) from NRG, see Fig.~\ref{fig:impsusJKNRG}.
  The two values result from the two sequences of extrapolations~(a) and~(b)
  in $(\Lambda-1)$ and $j$, see text.
  The analytic value for $S_0=s_0$ is given
  in eq.~(\ref{eq:mainresultforchi}).
  Also given are the coefficients for the fit 
  in eq.~(\ref{eq:constexpfit}).\label{tab:two}}
\end{table}

For the impurity spin susceptibility we also find the expected result
$S_0=s_0=1/(2\sqrt{\pi e})\approx 0.1709914$, irrespective of
a global or a local field, with deviations of about 4\%.
Again, the first-order and second-order coefficients $S_1$ and $S_2$ depend on
whether the magnetic field is applied globally or locally.

To assess the accuracy of our extrapolations, we compare the
results for  $\chi_0^S(J_{\rm K})$ and $\chi_{0,{\rm loc}}^{\rm ii}(J_{\rm K})$
which should be equal, see eq.~(\ref{eq:twochisareequivalent}).
We see that $s_{1,{\rm loc}}=-0.269$
agrees reasonably well with $S_1=-0.307$, with a deviation of the order of ten percent.
However, $s_{2,{\rm loc}}=0.182$ and $S_2=0.300$ are off by more than 40~percent.
Since the data for the local susceptibilities are better than those for
the global susceptibilities, we argue that the data for 
$s_{1,{\rm loc}}$ and $s_{2,{\rm loc}}$ are more reliable. Nevertheless, the comparison
indicates that the values for $s_1$ and $S_1$ ($s_2$ and $S_2$) have an uncertainty
of several (ten) percent.

\subsection{Magnetic susceptibilities for strong coupling}
\label{subsec:NRGDMRGandallstrongcouplingsusc}


\subsubsection{Impurity-induced susceptibility}

\begin{figure}[t]
  \includegraphics[width=8.5cm]{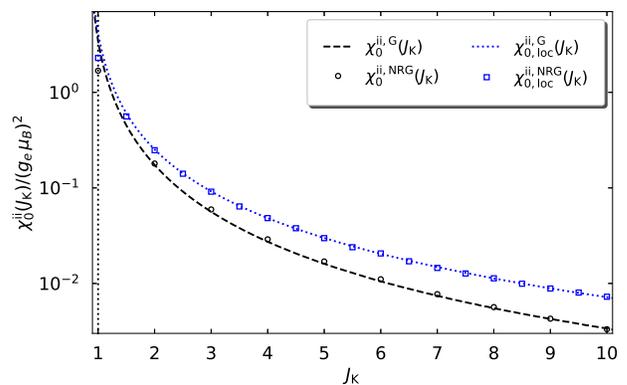}
  \caption{(Color online) Zero-field impurity-induced susceptibilities
  $\chi_0^{\rm ii}(J_{\rm K})$ 
    and $\chi_{0,{\rm loc}}^{\rm ii}(J_{\rm K})$
    on a logarithmic scale
   for a global and a local magnetic field 
   for the one-dimensional
   symmetric Kondo model as a function of $J_{\rm K}$ for $J_{\rm K}\geq 1$
   from NRG and the Gutzwiller wave function.\label{fig:iisusJKlarge}}
\end{figure}

In Fig.~\ref{fig:iisusJKlarge} we show the zero-field impurity-induced susceptibility
with a global and a local field for $J_{\rm K}\geq 1$
from NRG in comparison with the Gutzwiller result.
In general, we calculate the impurity-induced magnetization
for small local fields and determine the susceptibility from the slope.
In the strong-coupling region, this procedure becomes
unstable in NRG so that we determine the susceptibility
from the second-derivative of the ground-state energy with respect
to the global field, see eq.~(\ref{eq:chifromm}).

\begin{figure}[t]
  \includegraphics[width=8.5cm]{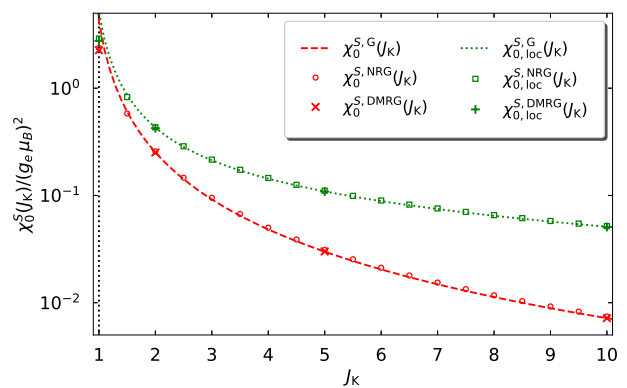}
    \caption{(Color online) Zero-field impurity spin susceptibilities
  $\chi_0^S(J_{\rm K})$ 
    and $\chi_{0,{\rm loc}}^S(J_{\rm K})$
    on a logarithmic scale
   for a global and a local magnetic field 
   for the one-dimensional
   symmetric Kondo model as a function of $J_{\rm K}$ for $J_{\rm K}\geq 1$
   from NRG, DMRG,
   and the Gutzwiller wave function.\label{fig:SzsusJKlarge}}
\end{figure}

As seen from Fig.~\ref{fig:iisusJKlarge},
the Gutzwiller wave function almost perfectly reproduces the NRG data
for $J_{\rm K} \gtrsim 1.5$.
For intermediate to strong couplings,
the Gutzwiller wave function is an excellent trial state for the Kondo model.

The strong-coupling asymptotics is shown in Fig.~\ref{fig:Gutzwillerchis}.
The asymptotic formulae~(\ref{eq:chizeroiiGlargeJKmain})
and~(\ref{eq:chizeroiilocalGlargeJKmain}) are applicable for $J_{\rm K}\gtrsim 4$.

\subsubsection{Impurity spin susceptibility}

In Fig.~\ref{fig:SzsusJKlarge} we show the zero-field
impurity spin susceptibility for $J_{\rm K}\geq 1$.
For intermediate to strong couplings, we find an excellent agreement
between the data from NRG and DMRG both in the presence of
global and local magnetic fields. Again, the Gutzwiller wave function
provides an excellent analytic estimate for the zero-field
susceptibilities for all $J_{\rm K}\gtrsim 1.5$.

The strong-coupling asymptotics is shown in Fig.~\ref{fig:Gutzwillerchis}.
The limiting expressions,
eqs.~(\ref{eq:scformulafullchimain}) and~(\ref{eq:GutzlargeJKlocalfieldmain}),
become applicable for $J_{\rm K}\gtrsim 4$.


\subsection{Impurity-induced magnetization and impurity spin polarization 
  for weak coupling}

Next, we address the impurity-induced magnetization and the impurity spin polarization
as a function of the external field for weak coupling. 
The comparison of Bethe Ansatz results and NRG data was done only
recently.~\cite{PhysRevB.87.184408}

\subsubsection{Impurity-induced magnetization}
\label{subsubsec:iimag}

\begin{figure}[t]
\includegraphics[width=8.5cm]{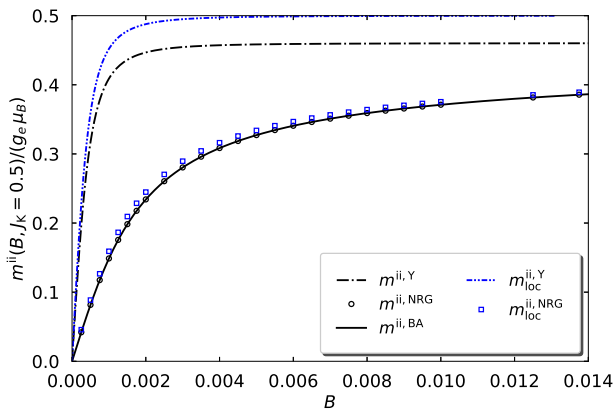}
  \caption{(Color online) Impurity-induced magnetization
    $m^{\rm ii}(B)/(g_e\mu_{\rm B})$  for $J_{\rm K}=0.5$ for the one-dimensional
   symmetric Kondo model as a function of global/local magnetic fields~$B$
   from the Yosida wave function, NRG, and the Bethe Ansatz.\label{fig:impmagJK05}}
\end{figure}

In Fig.~\ref{fig:impmagJK05}
we show the impurity-induced magnetization $m^{\rm ii}(B)/(g_e\mu_{\rm B})$
for $J_{\rm K}=0.5$
as a function of the global field $B\leq 0.014$
as obtained from the Yosida wave function,
in comparison with NRG data and results from the Bethe Ansatz.
We omit the Gutzwiller results because Gutzwiller theory predicts
a finite magnetization even at $B=0^+$
and thus fails to reproduce the paramagnetic phase at $J_{\rm K}=0.5$.
As discussed in Sect.~\ref{subsec:DMRG!}, DMRG cannot faithfully
reproduce the magnetization
for $J_{\rm K}=0.5$ because the tractable system sizes are too small.
Therefore, we do not show DMRG data for the
impurity-induced magnetization in Fig.~\ref{fig:impmagJK05}.

At $J_{\rm K}=0.5$ and in one dimension where $\rho_0(0)=1/\pi$,
we have $j=\rho_0(0)J_{\rm K}=1/(2\pi)\approx 0.159$.
The zero-field susceptibility is quite large already. In units of
$(g_e\mu_{\rm B})^2$ we have
$\chi_0^{\rm ii}(j=1/(2\pi))\approx \bar{\chi}_0(1/(2\pi)) s_0
[1+(s_1/s_0)(2\pi)+(s_2/s_0)(2\pi)^2]\approx 79$,
see eq.~(\ref{eq:mainresultforchi}), where we employ
$s_0=1/(2{\cal D}\F\sqrt{\pi e})\approx0.0856$ for $\F^{d=1}=2$ and ${\cal D}=1$,
and make the assumption that the ratios $s_1/s_0\approx -2.3$
and $s_2/s_0\approx 2$ do not depend on
the host-electron density of states, and are thus obtained from the values in
table~\ref{tab:one}.
A large zero-field susceptibility implies
a sharp increase of the magnetization for small fields
as seen in Fig.~\ref{fig:impmagJK05}. 
The Yosida state overestimates the zero-field susceptibility by more than a factor of five,
$\chi_0^{\rm ii,Y}(j=1/(2\pi))\approx 433$, see eq.~(\ref{eq:YosidachismallJK}).
Thus, the Yosida wave function
also overestimates the magnetization for small and intermediate fields,
see Fig.~\ref{fig:impmagJK05}. 

In the Bethe Ansatz, the sharp increase at small fields
is followed by a very slow convergence to the limiting
value $m^{\rm ii}(B\gg T_{\rm H})=1/2$.
The resulting broad magnetization plateau originates from the logarithmic terms
in the Bethe Ansatz solution, see eq.~(VI-86)  
in the supplemental material. 
The NRG results lie on top of the Bethe Ansatz data which shows that
the Bethe Ansatz expressions~(\ref{eq:hsmallerthanunity})
and~(\ref{eq:hlargerthanunity}) remain valid up to $J_{\rm K}$ of the order
of a quarter of the bandwidth, as long as $T_1$ is determined from
the exact zero-field susceptibility from eq.~(\ref{eq:THTWofTKTW}).

In Fig.~\ref{fig:impmagJK05}
we also show the impurity-induced magnetization $m_{\rm loc}^{\rm ii}(B)$
as a function of a local magnetic field for $J_{\rm K}=0.5$.
For small values of the Kondo coupling, 
the differences between globally and locally applied external fields are fairly small.
The impurity-induced magnetization in the presence of a local field
is a few percent larger than in the presence of a global field.
This can be deduced from eq.~(\ref{eq:subleadings})
which shows that the ratio of the zero-field susceptibilities
is close to unity,  $[1+(s_{1,{\rm loc}}/s_0)j+(s_{2,{\rm loc}}/s_0)j^2]/
[1+(s_1/s_0)j+(s_2/s_0)j^2]\approx 1.13$ at $j=1/(2\pi)$
where the data are taken from table~\ref{tab:one}.
Likewise, the differences between the impurity-induced
magnetization $m^{\rm ii}(B)$ and the impurity spin polarization $m^S(B)$
are small at small~$J_{\rm K}$,
$[1+(S_1/s_0)j+(S_2/s_0)j^2]/[1+(s_1/s_0)j+(s_2/s_0)j^2]\approx 1.12$ at $j=1/(2\pi)$,
where the data are taken from table~\ref{tab:one} and table~\ref{tab:two}.
This has been noted previously in Ref.~[\onlinecite{PhysRevB.87.184408}].

In the Yosida wave function, the differences between the impurity-induced magnetization
for global and local fields are more pronounced.
In the case of a local field, the impurity-induced magnetization in
the Yosida wave function quickly reaches the maximal value
of one half. For a global field, the impurity-induced magnetization
saturates below this value, in contrast to the exact solution.
Altogether, the Yosida wave function correctly describes some gross aspects
of the magnetization curves (large zero-field susceptibility,
monotonous increase to saturation)
but it fails to reproduce them in detail, e.g., the
small difference between global and local fields.

\subsubsection{Impurity spin polarization}

\begin{figure}[t]
  \includegraphics[width=8.5cm]{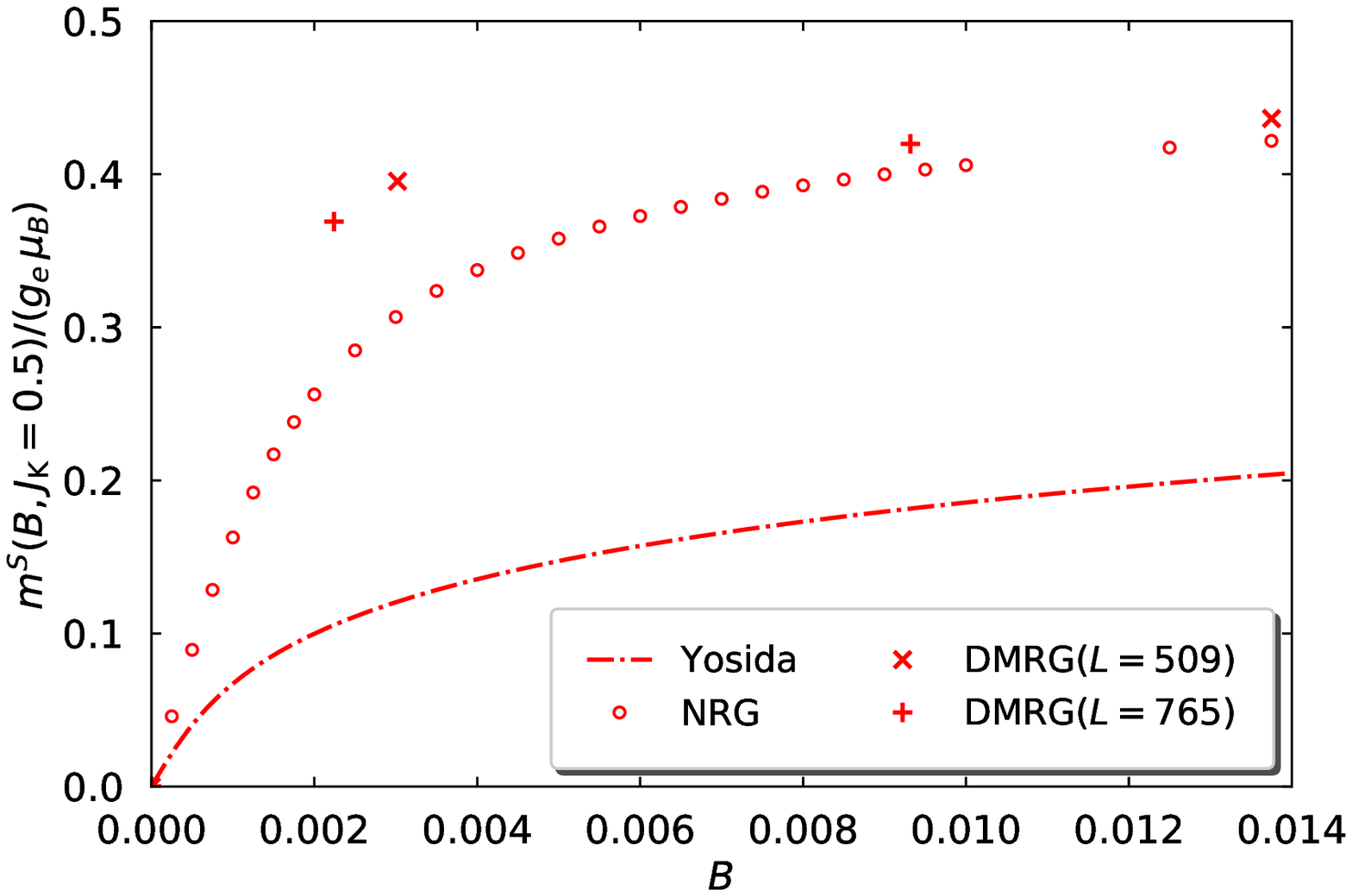}\\[3pt]
  \includegraphics[width=8.5cm]{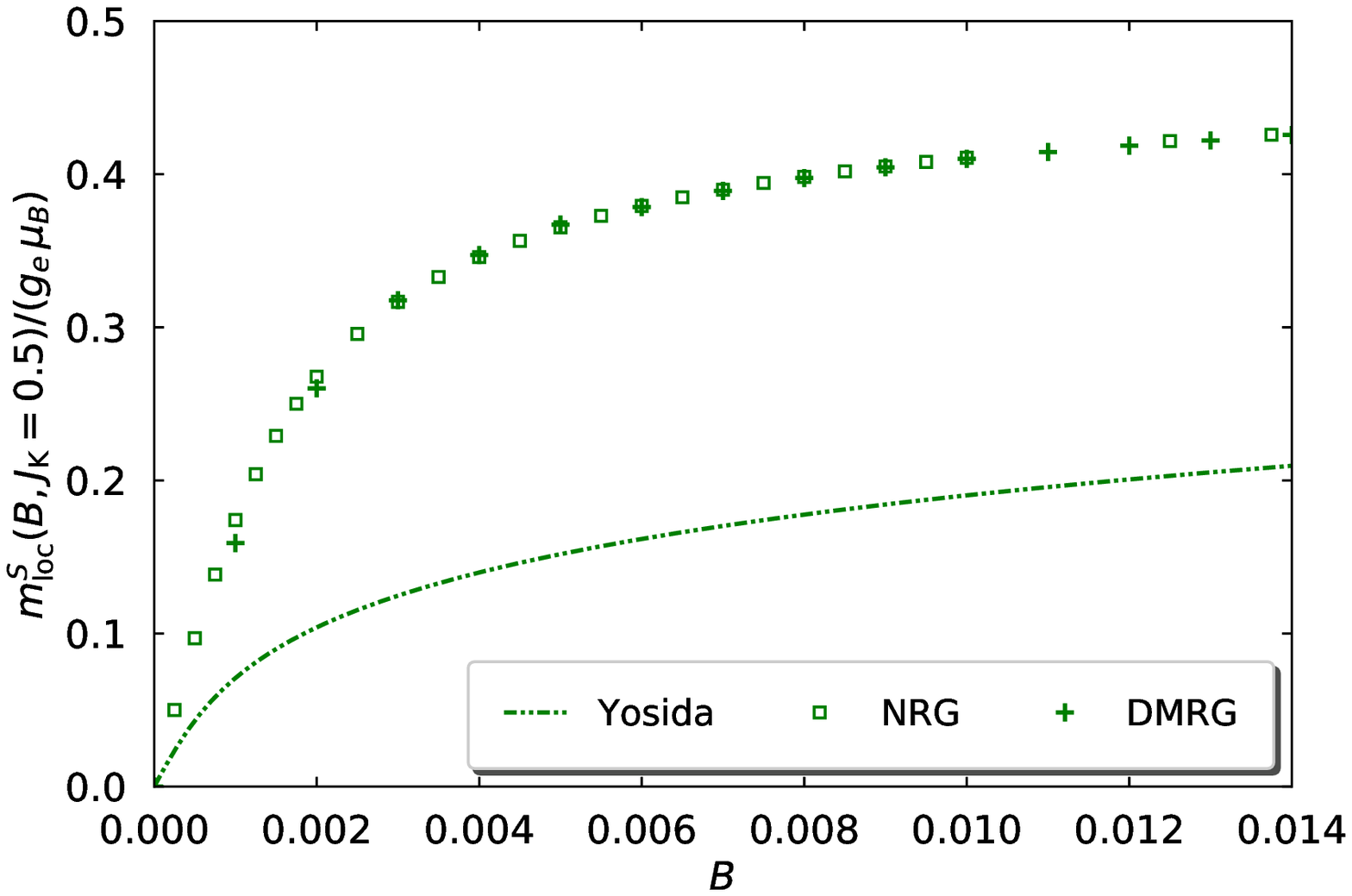}
    \caption{(Color online) Impurity spin polarization $S^z(B)=
      m^S(B)/(g_e\mu_{\rm B})$
      as a function of a global magnetic field (upper figure)
      and
      $S_{\rm loc}^z(B)=m^S_{\rm loc}(B)/(g_e\mu_{\rm B})$
      as a function of a local magnetic field (lower figure)
      for the one-dimensional symmetric Kondo model at $J_{\rm K}=0.5$
   from NRG, DMRG, and the Yosida wave function.\label{fig:SzJK05}}
\end{figure}

In Fig.~\ref{fig:SzJK05} we show the impurity spin polarization in the presence
of a global and a local field, respectively.
As seen from the previous section~\ref{subsubsec:iimag},
the NRG is the best method to study magnetic properties of the Kondo model
at weak coupling. Therefore, its results can be used to assess
the quality of all other methods.

In Fig.~\ref{fig:SzJK05} we
leave out the Gutzwiller results because the spin polarization is finite
at $J_{\rm K}=0.5$, and almost independent of~$B$ for all $0\leq B<0.014$,
in contrast to the NRG data.
The Yosida wave function provides qualitatively correct results
but grossly underestimates the spin polarization in both cases.
Therefore, the Yosida wave function neither provides
an acceptable description of the impurity spin polarization.

As discussed in Sect.~\ref{subsec:DMRG!}, DMRG requires very large system sizes
for small Kondo couplings
to calculate the impurity spin polarization
in the presence of a small global field. Therefore, at $J_{\rm K}=0.5$
the DMRG and NRG data agree only for $B\gtrsim 0.01$ 
For smaller $B$-values, DMRG substantially overestimates the magnetization,
displaying large finite-size effects. Recall that DMRG works
for fixed total spin $S^z_{\rm tot}$ so that only specific values for $B$ are accessible.
This limitation does not apply for a purely local field. It can be tuned freely
also in DMRG so that a finite-size extrapolation
of the DMRG data for the impurity spin polarization is unproblematic.
As seen from Fig.~\ref{fig:SzJK05}, the NRG and DMRG data perfectly agree
for the impurity spin polarization in the presence of a local magnetic field.
Alternatively, since $m^S_{\rm loc}$ is a thermodynamic
quantity, it can also be obtained from
the derivative of the excess ground-state energy with respect to the external field,
see eq.~(\ref{eq:definelocalmandchiasderivatives}).
Both approaches lead to the same results.

\subsection{Impurity-induced magnetization and impurity spin polarization
  for strong coupling}

Lastly, we address the impurity-induced magnetization
and the impurity spin polarization
as a function of the external field for strong coupling.

\begin{figure}[t]
  \includegraphics[width=8.5cm]{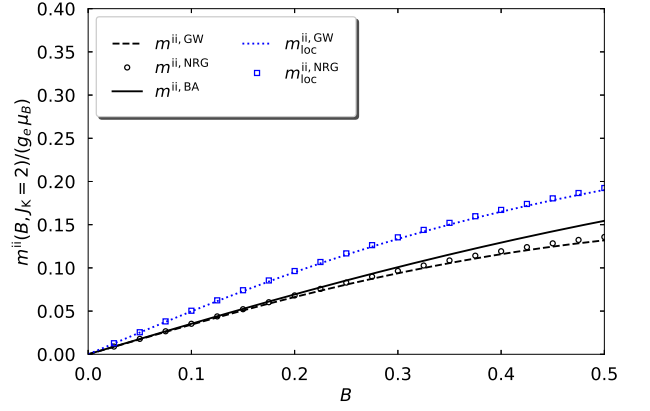}
   \caption{(Color online) Impurity-induced magnetization
    $m^{\rm ii}(B)/(g_e\mu_{\rm B})$  for $J_{\rm K}=2$ for the one-dimensional
     symmetric Kondo model as a function of a global/local magnetic field~$B$
     from Bethe Ansatz, NRG, and the Gutzwiller wave
   function.\label{fig:impmagJK2}}
\end{figure}

\begin{figure}[t]
  \includegraphics[width=8.5cm]{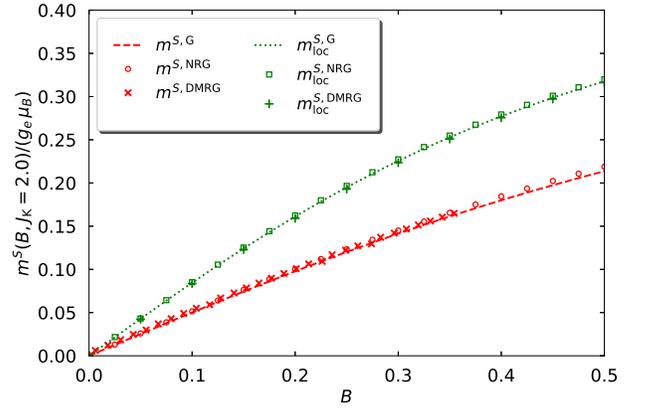}
  \caption{(Color online) Impurity spin polarization
    $S^z(B)=m^S(B)/(g_e\mu_{\rm B})$  for $J_{\rm K}=2$ for the one-dimensional
     symmetric Kondo model as a function of a global/local magnetic field~$B$
     from NRG, DMRG, and the Gutzwiller wave
   function.\label{fig:SzJK2}}
\end{figure}

\subsubsection{Impurity-induced magnetization}

In Fig.~\ref{fig:impmagJK2} we show the impurity-induced magnetization 
as a function of~$B$ at $J_{\rm K}=2$.
When the Kondo coupling reaches the band width,
the singlet state between the impurity spin and the band electron at
the origin is tightly bound so that the susceptibility is small and even a sizable
field can barely polarize the singlet. Therefore, the impurity magnetization
remains small for $B\leq 0.5$.

The NRG data lie on top of the analytic results from the Gutzwiller
wave function. This again shows that the Gutzwiller wave function
is an excellent trial state for the Kondo model at large couplings.
At $J_{\rm K}= 2$, the Bethe Ansatz
is no longer applicable, and sizable differences between NRG/Gutzwiller results and
Bethe Ansatz predictions become discernible at large~$B$,
despite the fact that the linear term is fixed to the exact susceptibility.
Values $J_{\rm K}\gtrsim W$ are beyond the Bethe Ansatz description.

\subsubsection{Impurity spin polarization}

Finally, in Fig.~\ref{fig:SzJK2} we show the impurity spin polarization
as a function of~$B$ at $J_{\rm K}=2$.
For large Kondo couplings, the finite-size restrictions imposed
on DMRG discussed in Sect.~\ref{subsec:DMRG!} are far less severe,
and the results for $L=763$ sites for a global field perfectly reproduce the NRG data
for $0\leq B<0.5$. The same perfect agreement between DMRG and NRG data
is seen for the case of a local field whose value
can be chosen freely also for the DMRG calculations.

The analytic results from the Gutzwiller
wave function lie on top the the NRG/DMRG data for both a global and a local field.
Again, the Gutzwiller wave function
is seen to provide an excellent trial state for the Kondo model at large couplings.

\section{Conclusions}
\label{sec:Conclusions}

In the last section, we summarize our central findings and discuss our main results.

\subsection{Summary}

In our work, we investigated the
symmetric single-impurity Kondo model on a chain at zero temperature.
As a function of the Kondo coupling~$J_{\rm K}$, we studied the 
ground-state energy,
the local spin correlation function,
the impurity spin polarization and impurity-induced magnetization,
and the corresponding zero-field magnetic susceptibilities
for global and local external fields~$B$.
Some of these quantities, e.g.,
the ground-state energy and the local spin correlation function,
are related by the Hellmann-Feynman theorem that also holds
for variational wave functions, see appendix~\ref{app:misc}.

We calculated the ground-state energy and the local spin
correlation function in weak-coupling and strong-coupling perturbation theory
at $B=0$ as benchmark for our approaches. Some of the required
calculations were deferred to appendix~\ref{app:A} for weak coupling
and to appendix~\ref{app:B} for strong coupling.

As the first of three analytical variational methods, we analyzed
the first-order Lanczos state at $B=0$, as done by
Mancini and Mattis~\cite{PhysRevB.31.7440} for a constant density of states.
We recapitulated the Lanczos method and performed
the calculation of the first-order coefficients in appendix~\ref{app:A}.
Second, we extended and evaluated
the Yosida wave function~\cite{PhysRev.147.223,Yosidabook}
to include external magnetic fields.
The Yosida wave function provides a simple description
of the Kondo-singlet ground state with an exponentially small binding energy
at small Kondo couplings
that translates into an exponentially large zero-field magnetic susceptibility.
Third, we introduced and employed the Gutzwiller
variational state~\cite{Gutzwiller1964,TIAM} at finite fields.
The latter provides a Hartree-Fock type
description of the Kondo model that guarantees that the impurity
is singly occupied. The Gutzwiller wave function becomes exact in the limit of
large Kondo couplings, $J_{\rm K}\gg W$.
The evaluation of the Gutzwiller wave function required the solution
of the non-interacting single-impurity Anderson model (SIAM)
in the presence of potential scattering. This was done 
in appendix~\ref{sec:SIAMfull}.

As  numerical techniques, we employed
the Numerical Renormalization Group
(NRG)~\cite{RevModPhys.47.773,PhysRevB.21.1003,PhysRevB.21.1044,%
RevModPhys.80.395} and 
the Density-Matrix Renormalization Group
(DMRG)~\cite{White-1992a,White-1992b,White-1993}
methods.
Using the DMRG method we addressed finite half-chains of length $L/2$,
and performed quadratic fits
in $1/L$ for physical quantities to extrapolate to the thermodynamic limit,
$L\to \infty$.
In DMRG, the system sizes are limited to $L\lesssim 10^3$ so that
we could not address the impurity-induced magnetizations
and the zero-field susceptibilities at weak Kondo couplings.
The NRG permits the accurate calculation of all ground-state quantities
as a function of the Wilson parameter~$\Lambda$.
We performed quadratic fits in $(\Lambda-1)$ to extrapolate our data
to the limit $\Lambda=1$. We employed
the correction factor derived by Krishna-murthy, Wilkins, and 
Wilson~\cite{PhysRevB.21.1003,PhysRevB.21.1044} because
it improved the quality of the extrapolations.

Since the Bethe Ansatz solves a Kondo model with linear dispersion and
infinite bandwidth,
a direct comparison of physical quantities is not easy because there is
a non-trivial relation between the Kondo couplings $J_{\rm K}^{\rm BA}$
and $J_{\rm K}$ 
used in the Bethe Ansatz and in the lattice model, respectively.
As we showed in appendix~\ref{app:GSEBetheAnsatz}, the ground-state energy
in the Bethe Ansatz is zero, up to corrections of the 
order $(J_{\rm K}^{\rm BA})^3$.
Thus, the ground-state energy from Bethe Ansatz
cannot be used for a comparison with the lattice model.
Using known results 
from Andrei, Furuya, and Lowenstein,~\cite{RevModPhys.55.331} 
re-derived in appendix~\ref{app:freenergyPT}
and extended to a general host-electron density of states, we
expressed $J_{\rm K}^{\rm BA}$ as a series expansion in $J_{\rm K}$,
and gave analytic expressions 
for the leading-order terms in eq.~(\ref{eq:linkJKtoJKBA}).

When the zero-field susceptibility from NRG is used,
the impurity-induced magnetization from Bethe Ansatz and from NRG
were seen to agree perfectly. This was observed earlier
in the NRG analysis of Schnack and H\"ock.~\cite{PhysRevB.87.184408}

In our work, we showed that the various zero-field susceptibilities
have a universal small-coupling limit, 
see eq.~(\ref{eq:mainresultforchi}).
For finite $J_{\rm K}$, however, the impurity spin polarization
and the impurity-induced magnetization at global and local fields
differ from each other by a factor that goes to unity for $J_{\rm K}\to 0$.
Using NRG, we calculated the corrections numerically,
with an accuracy of some ten percent.
Since two zero-field susceptibilities agree,
see eq.~(\ref{eq:twochisareequivalent}), it is possible
to assess the accuracy of the NRG calculations for the 
zero-field susceptibilities.

\subsection{Discussion}

The ground state of the symmetric single-impurity Kondo model
describes a Kondo singlet formed by the impurity spin and
its host-electron screening cloud. Unfortunately, it is
by no means easy to formulate a concise, analytically tractable
variational wave function
that adequately describes the ground state for
all couplings. 

In this work we showed that
the Gutzwiller wave function provides an excellent trial state
for large Kondo couplings, $J_{\rm K}\gtrsim W$, where $W=2$
is the bandwidth. It reproduce the ground-state energy, the
local spin correlation, the impurity spin polarization and 
impurity-induced magnetization, and the corresponding 
zero-field susceptibilities from NRG and DMRG with high accuracy. 
Unfortunately, it displays a Hartree-Fock type transition to a state
with an oriented impurity spin below $J_{\rm K,c}^{\rm G}\approx 0.893$
that is not contained in the exact solution of the model.

For weak coupling, the Yosida wave function reproduces the
exponential divergence of the zero-field susceptibilities
known from Bethe Ansatz and NRG but it fails to provide a
good variational bound on the ground-state energy for all couplings.
Eventually, it becomes unstable for large Kondo couplings.
Thus, the Yosida and Gutzwiller variational approaches
provide a complementary view on the Kondo-singlet ground state
of the Kondo model.

The DMRG method numerically determines an optimal
variational ground state for the Kondo model on finite lattices.
Although not specifically designed for impurity models,
the method works very well as long as all energy scales 
lie within the DMRG energy resolution $\Delta \epsilon=W/L$.
However, the Kondo temperature in the Kondo model 
becomes exponentially small for $J_{\rm K}\ll W$ so that
the calculation of magnetic properties using DMRG 
is limited to $J_{\rm K}\gtrsim 0.5$ 
for the one-dimensional density of states.
Other quantities such as the ground-state energy and the local spin correlation
function are unproblematic.
For $J_{\rm K}=2$, the results from NRG and DMRG agree perfectly,
not only for the ground-state energy and local
spin correlation but also for the impurity spin polarization
and zero-field susceptibility. The DMRG can also be applied
to impurity problems as long as all intrinsic energy scales can be resolved
appropriately.

The NRG was specifically designed to treat exponentially small
energy scales in impurity models and thus works exceedingly well for the Kondo model.
In this work we showed that NRG also permits the accurate calculation of the
ground-state energy and local spin correlation function.
In particular, we found that an extrapolation $\Lambda\to 1$ 
is required whereby 
the correction factor derived by Krishna-murthy, Wilkins, and 
Wilson~\cite{PhysRevB.21.1003,PhysRevB.21.1044} is helpful
to improve the extrapolations.

In contrast to the single-impurity Anderson model,~\cite{Barczaetal}
the Bethe Ansatz results for the Kondo model cannot be
directly compared to numerical results because the continuum 
and lattice models differ in their coupling constants.
Therefore, at zero temperature, only the impurity-induced
magnetization for small Kondo couplings can eventually 
be compared to the NRG data. 
For this reason, it was indispensable to generate accurate 
data from NRG for comparison with the analytical
and numerical variational methods (Lanczos, Yosida, Gutzwiller, DMRG)
employed in this work.

\begin{acknowledgments}
This research has been supported in part by
the Hungarian National Research, 
Development and Innovation Office (NKFIH) through Grant No.\ 
K120569 and PD-17-125261 
and the Hungarian Quantum Technology National Excellence Program 
(Project No. 2017-1.2.1-NKP-2017-00001).

Support by the Alexander-von-Humboldt foundation 
(\"O.\ Legeza) and by
the International Collaborative Research Center TRR~160
{\sl Coherent manipulation of interacting spin excitations 
in tailored semiconductors\/} (F.B.\ Anders, F.\ Eickhoff)
is gratefully acknowledged.
\end{acknowledgments}

\appendix

\section{Technical details}
\label{mainappendixA}

\subsection{Diagonalization of the kinetic energy operator}
\label{app:diagTandhalfchain}

We use open boundary conditions and define the operators for standing waves,
\begin{equation}
\hat{a}_{k,\sigma}^{\vphantom{+}} =\sqrt{\frac{2}{L+1}} 
\sum_{n=-(L-1)/2}^{(L-1)/2} 
\sin\Bigl( \frac{\pi kn}{L+1}+\frac{\pi k}{2}\Bigr)
\hat{c}_{n,\sigma}^{\vphantom{+}} \; .
\label{eq:inverseakcn}
\end{equation}
We may formally include the operators 
$\hat{c}_{\pm (L+1)/2,\sigma}$ at sites $n=\pm (L+1)/2$ because they do not enter
the standing-wave operators~$\hat{a}_{k,\sigma}$.
The inverse transformation reads
\begin{equation}
\hat{c}_{n,\sigma}^{\vphantom{+}} =\sqrt{\frac{2}{L+1}} \sum_{k=1}^{L} 
\sin\left[ \frac{\pi k}{L+1}\left(n+\frac{L+1}{2}\right)\right] 
\hat{a}_{k,\sigma}^{\vphantom{+}} \; .
\label{eq:defak}
\end{equation}
The kinetic energy becomes diagonal
\begin{equation}
\hat{T}= \sum_{k=1,\sigma}^L \epsilon_k
\hat{a}_{k,\sigma}^+ \hat{a}_{k,\sigma}^{\vphantom{+}}
\label{eq:defTprime} 
\end{equation}
with the dispersion relation ($1\leq k\leq L$)
\begin{equation}
\epsilon_k=-2t\cos\left(\frac{\pi k}{L+1}\right) \; .
\end{equation}

\subsection{Half-chain geometry}
\label{app:halfchain}

Eq.~(\ref{eq:defak}) shows that 
\begin{equation}
\hat{c}_{0,\sigma}= \sqrt{\frac{2}{L+1}} \sum_{k=1}^{L} 
\sin\left[ \frac{\pi k}{2}\right] 
\hat{a}_{k,\sigma}^{\vphantom{+}} \; .
\end{equation}
Therefore, the standing waves with even $k$ do not couple to the chain center,
and drop out of the problem.

\subsubsection{Canonical transformation to parity eigenstates}

The reason for this decoupling is parity symmetry.
To make it transparent in the real-space representation, 
we introduce the operators for $n=1,2,\ldots,(L-1)/2$
\begin{eqnarray}
\hat{C}_{n,\sigma}^{\vphantom{+}}
 &=& \sqrt{\frac{1}{2}} \left( 
\hat{c}_{n,\sigma}^{\vphantom{+}}+\hat{c}_{-n,\sigma}^{\vphantom{+}}
\right) \; , \nonumber \\
\hat{S}_{n,\sigma}^{\vphantom{+}}
 &=& \sqrt{\frac{1}{2}} \left( 
\hat{c}_{n,\sigma}^{\vphantom{+}}-\hat{c}_{-n,\sigma}^{\vphantom{+}}
\right) \; ,
\end{eqnarray}
with the inverse transformation
\begin{eqnarray}
\hat{c}_{n,\sigma}^{\vphantom{+}}
 &=& \sqrt{\frac{1}{2}} \left( 
\hat{C}_{n,\sigma}^{\vphantom{+}}+\hat{S}_{n,\sigma}^{\vphantom{+}}
\right) \; , \nonumber \\
\hat{c}_{-n,\sigma}^{\vphantom{+}}
 &=& \sqrt{\frac{1}{2}} \left( 
\hat{C}_{n,\sigma}^{\vphantom{+}}-\hat{S}_{n,\sigma}^{\vphantom{+}}
\right) \; .
\end{eqnarray}
Moreover, we set $\hat{C}_{0,\sigma}^{\vphantom{+}}
\equiv \hat{c}_{0,\sigma}^{\vphantom{+}}$
for notational consistency.     %
The transformation is a canonical basis transformation.

Then, the kinetic energy becomes
\begin{eqnarray}
  \hat{T} &=& 
-t\left[\sqrt{2} \hat{C}_{0,\sigma}^{+}\hat{C}_{1,\sigma}^{\vphantom{+}}
+\hbox{h.c.}\right]\nonumber \\
\nonumber \\
&& -t \biggl[
  \frac{1}{2}\sum_{n=1,\sigma}^{(L-3)/2}
\left(
\hat{C}_{n+1,\sigma}^{+}+\hat{S}_{n+1,\sigma}^{+}
\right)
\left(\hat{C}_{n,\sigma}^{\vphantom{+}}+\hat{S}_{n,\sigma}^{\vphantom{+}}
\right) \nonumber \\
&&\hphantom{-t \biggl[} 
+\frac{1}{2}\sum_{n=1,\sigma}^{(L-3)/2}
\left(
\hat{C}_{n+1,\sigma}^{+}-\hat{S}_{n+1,\sigma}^{+}
\right)
\left(\hat{C}_{n,\sigma}^{\vphantom{+}}-\hat{S}_{n,\sigma}^{\vphantom{+}}
\right)\nonumber \\
&&\hphantom{-t \biggl[ } +\hbox{h.c.}\biggr]\nonumber \\
&\equiv& \hat{T}^C +\hat{T}^S \; ,
\end{eqnarray}
where the two commuting parts of the kinetic energy are given by
\begin{eqnarray}
\hat{T}^C &=& 
-\sqrt{2}t \sum_{\sigma}
\left(\hat{C}_{0,\sigma}^{+}\hat{C}_{1,\sigma}^{\vphantom{+}} 
+\hat{C}_{1,\sigma}^{+}\hat{C}_{0,\sigma}^{\vphantom{+}} \right) \nonumber \\
&&+(-t) \sum_{n=1,\sigma}^{(L-3)/2}
\left(\hat{C}_{n+1,\sigma}^{+}
\hat{C}_{n,\sigma}^{\vphantom{+}} +
\hat{C}_{n,\sigma}^{+}
\hat{C}_{n+1,\sigma}^{\vphantom{+}}\right) \nonumber \, , \\[9pt]
\hat{T}^S &=& (-t) 
\sum_{n=1,\sigma}^{(L-3)/2}
\left(\hat{S}_{n+1,\sigma}^{+}
\hat{S}_{n,\sigma}^{\vphantom{+}} +
\hat{S}_{n,\sigma}^{+}
\hat{S}_{n+1,\sigma}^{\vphantom{+}}\right) \; .
\label{eq:TCTS}
\end{eqnarray}
Therefore, the $S$-electrons drop out of the problem, and
the Kondo Hamiltonian reduces to
\begin{equation}
  \hat{H}_{\rm K}^C= \hat{T}^C+ \hat{V}_{\rm sd}+\hat{H}_{\text{m}}
  \label{eq:HKCdef}
\end{equation}
with
\begin{eqnarray}
\hat{V}_{\rm sd}&=& \frac{J_{\rm K}}{2}
\left(\hat{C}_{0,\uparrow}^+\hat{C}_{0,\downarrow}^{\vphantom{+}} 
\hat{d}_{\downarrow}^+\hat{d}_{\uparrow}^{\vphantom{+}} 
+ 
\hat{C}_{0,\downarrow}^+\hat{C}_{0,\uparrow}^{\vphantom{+}} 
\hat{d}_{\uparrow}^+\hat{d}_{\downarrow}^{\vphantom{+}}\right)
\nonumber \\
&& 
+\frac{J_{\rm K}}{4} \left(\hat{d}_{\uparrow}^+\hat{d}_{\uparrow}^{\vphantom{+}} 
- \hat{d}_{\downarrow}^+\hat{d}_{\downarrow}^{\vphantom{+}} \right)
\left(\hat{C}_{0,\uparrow}^+\hat{C}_{0,\uparrow}^{\vphantom{+}} 
-\hat{C}_{0,\downarrow}^+\hat{C}_{0,\downarrow}^{\vphantom{+}} \right) \; ,
\nonumber \\
\hat{H}_{\text{m}} &=& \hat{H}_{\text{m},\text{loc}}
-B\sum_{n=0}^{(L-1)/2 }
\Bigl[\hat{C}_{n,\uparrow}^+\hat{C}_{n,\uparrow}^{\vphantom{+}} 
- \hat{C}_{n,\downarrow}^+\hat{C}_{n,\downarrow}^{\vphantom{+}}\Bigr]
\label{eq:simplestKondoagain} 
\end{eqnarray}
and $\hat{H}_{\text{m},\text{loc}}=
-B(\hat{d}_{\uparrow}^+\hat{d}_{\uparrow}^{\vphantom{+}}-
\hat{d}_{\downarrow}^+\hat{d}_{\downarrow}^{\vphantom{+}})$.
Note that the $C$-parity eigenbasis in this one-dimensional finite-size chain
takes on the role of the $s$-wave scattering in the three-dimensional
continuous model,~\cite{RevModPhys.47.773}
where all other spherical harmonics drop out of the problem.
Now, the impurity is at the left end of the half-chain.
The half-chain has an odd number of sites, $L_{\rm hc}=(L+1)/2$, 
so that the system 
of half-chain plus impurity has a total number of even sites
because we choose $(L+3)/2$ to be even.
In the absence of an external magnetic field, we have an equal number of electrons
with spin $\sigma=\uparrow,\downarrow$ in the half-chain,
$N_{\sigma}=(L+3)/4$, 
so that the ground state has total spin zero.

\subsubsection{Kinetic energy of the half-chain}

We denote for $m=1,\ldots,(L+1)/2$
\begin{eqnarray}
  \hat{b}_{m,\sigma}^{\vphantom{+}}&\equiv&
  (-1)^{m-1} \hat{a}_{2m-1,\sigma}^{\vphantom{+}}
  \; ,\nonumber \\
  \epsilon(m)&=&-2t \cos\left[\frac{\pi(2m-1)}{L+1}\right]
\end{eqnarray}
and find
\begin{equation}
\hat{T}^C = \sum_{m=1,\sigma}^{(L+1)/2}
\epsilon(m) \hat{b}_{m,\sigma}^+\hat{b}_{m,\sigma}^{\vphantom{+}}
\end{equation}
for the kinetic energy of the half-chain in diagonal form.
For half filling of the half-chain, the Fermi energy is at $E_{\rm F}=0$, 
i.e., the last occupied site in Fourier space is $m_{\rm F}=(L+3)/4$
with energy $\epsilon(m_{\rm F})=0$.

We rewrite the dispersion relation as
\begin{equation}
\epsilon(m) =2t \sin\left[\frac{2\pi(m-m_{\rm F})}{L+1}\right] \; ,
\end{equation}
which can be linearized around the Fermi wave number,
\begin{equation}
  \epsilon(m) \approx 2t \frac{2\pi(m-m_{\rm F})}{L+1}
  +{\cal O}\left( (m-m_{\rm F})^3
  \right)
  \label{eq:correctionsthirdorder}
\end{equation}
for $ |m-m_{\rm F}|\ll L $ so that the Fermi velocity becomes
\begin{equation}
v_{\rm F}=\frac{
\epsilon(m_{\rm F}+1)-
\epsilon(m_{\rm F})
}{2\pi/(L+1)}=2t \; .
\end{equation}
In Bethe Ansatz calculations, $v_{\rm F}^{\rm BA}\equiv 1$ is used. Therefore,
it is convenient to set $2t\equiv 1$ so that the bandwidth is $W=2$.

\subsection{Weak-coupling perturbation theory}
\label{app:weakcoulingPT}

\subsubsection{Perturbation theory to leading order}

To leading order, we choose finite system sizes~$L$,
and consider the full-chain geometry.
In Fourier space, the kinetic energy operator and the Kondo coupling for the chain are
given by
\begin{eqnarray}
  \hat{T}&=& \sum_{k=1,\sigma}^L \epsilon_k
  \hat{a}_{k,\sigma}^+
  \hat{a}_{k,\sigma}^{\vphantom{+}}
  \; , \label{appeq:defTbmtilde} \\
  \hat{V}_{\rm sd} &=&\frac{J_{\rm K}}{L+1}
  \sum_{k,p=1}^L \sin(\pi k/2)\sin(\pi p/2)
\label{appeq:TandVsdinkspace} \\
&& \times \biggl(
\hat{d}_{\downarrow}^+\hat{d}_{\uparrow}^{\vphantom{+}} 
\hat{a}_{k,\uparrow}^+\hat{a}_{p,\downarrow}^{\vphantom{+}}
+
\hat{d}_{\uparrow}^+\hat{d}_{\downarrow}^{\vphantom{+}} 
\hat{a}_{k,\downarrow}^+\hat{a}_{p,\uparrow}^{\vphantom{+}}
\nonumber \\ && \quad 
+\frac{1}{2}
\left(
\hat{d}_{\uparrow}^+\hat{d}_{\uparrow}^{\vphantom{+}}
-
\hat{d}_{\downarrow}^+\hat{d}_{\downarrow}^{\vphantom{+}} 
\right)
\left(
\hat{a}_{k,\uparrow}^+\hat{a}_{p,\uparrow}^{\vphantom{+}}
-
\hat{a}_{k,\downarrow}^+\hat{a}_{p,\downarrow}^{\vphantom{+}}
\right)
\biggr)\; . \nonumber 
\end{eqnarray}
Due to spin symmetry we find
\begin{equation}
  \langle \Phi_0 | \hat{T} | \Phi_0\rangle = \langle A | \hat{T} | A\rangle =
  E_{\rm FS}-\epsilon_{k_{\rm F}}=E_{\rm FS}
  \end{equation}
because $\epsilon_{k_{\rm F}}=0$ .
Moreover,
\begin{eqnarray}
  \langle \Phi_0 | \hat{V}_{\rm sd} | \Phi_0\rangle &=&
  -\frac{3J_{\rm K}}{2(L+1)}
  \sum_{k,p=1}^L
  \sin(\pi k/2)\sin(\pi p/2) \nonumber \\
  && \hphantom{-\frac{3J_{\rm K}}{2(L+1)} \sum_{k,p=1}^L}
  \langle A |
\hat{a}_{k,\uparrow}^+\hat{a}_{p,\uparrow}^{\vphantom{+}}
-
\hat{a}_{k,\downarrow}^+\hat{a}_{p,\downarrow}^{\vphantom{+}}
  | A\rangle \nonumber \\
&=&  -\frac{3J_{\rm K}}{2(L+1)}
\end{eqnarray}
because $\sin^2(\pi k_{\rm F}/2)=1$. Thus we obtain
\begin{equation}
  e_0^{(1)}(J_{\rm K})=E_0^{(1)}(J_{\rm K})-E_{\rm FS}
  =-\frac{3J_{\rm K}}{2} \frac{1}{L+1} \; .
  \label{appeq:a0iszero}
  \end{equation}
In first order in $J_{\rm K}$, the energy decrease is only of the order $1/L$.

\subsubsection{Perturbation theory to second order}

To second order, we implicitly work in the thermodynamic limit, $L\to\infty$.
The second-order energy correction reads
\begin{equation}
  e_0^{(2)}(J_{\rm K})=\sum_{|m\rangle\neq |\Phi_0\rangle}
  \frac{\left|
\langle m | \hat{V}_{\rm sd} | \Phi_0\rangle
    \right|^2}{E_0^{(0)}-E_m^{(0)}} \; .
  \end{equation}
All states that can be reached from $|\Phi_0\rangle$ by an application
of $\hat{V}_{\rm sd} $ have an extra particle in
one of the $L/2$ single-particle levels above the Fermi sea at $p>k_{\rm F}$
and a hole in one of the $L/2$ single-particle levels
below $k_{\rm F}$, $k\leq k_{\rm F}$, see eq.~(\ref{appeq:TandVsdinkspace}).
Therefore, $E_0^{(0)}-E_{k,p}^{(0)}=-(\epsilon_p-\epsilon_k)$.
Otherwise, the coupling matrix element is independent
of $k$ and $p$.
Therefore, in the thermodynamic limit, we may sum over all $(L/2)^2$
intermediate states and find
\begin{equation}
  e_0^{(2)}(J_{\rm K})
  =-\left(\frac{2}{L}\right)^2\sum_{k=1}^{k_{\rm F}}\sum_{p=k_{\rm F}}^L
  \frac{ \langle \Phi_0 | \hat{V}_{\rm sd}^2 | \Phi_0\rangle }{
    \epsilon_p-\epsilon_k} \equiv - f b_1^2
  \end{equation}
with
\begin{eqnarray}
  f&=&4 \int_0^1 \rmd \omega_1 \int_{-1}^0 \rmd \omega_2
  \rho_0(\omega_1) \rho_0(\omega_2)
    \frac{1}{\omega_1-\omega_2} \; ,\\
  b_1^2&=&  \langle \Phi_0 |\hat{V}_{\rm sd}^2 | \Phi_0\rangle\; .
  \label{appeq:simplematrixelements}
  \end{eqnarray}
As shown in appendix~\ref{app:A}, we have $b_1^2=3J_{\rm K}^2/32$,
independent of the density of states.

For the one-dimensional density of states~(\ref{eq:DOS}) we find 
\begin{equation}
f= 
  \int_0^{\infty}\rmd \omega e^{-\omega\eta}       J_0(\omega)       \bm{H}_0(\omega) \; ,
  \end{equation}
where $\eta=0^+$, $J_0(\omega)$
is the zeroth order Bessel function and $\bm{H}_0(\omega)$
is the zeroth order Struve function, see eqs.~(9.1.18) and~(12.1.7)
of~Ref.~[\onlinecite{abramowitzstegun}].
Using eq.~(12.1.19) of~Ref.~[\onlinecite{abramowitzstegun}]
this can further be simplified to
\begin{equation}
  f=\frac{8}{\pi} \sum_{k=0}^{\infty} \frac{1}{2k+1}
  \int_0^{\infty}\rmd \omega
  J_0(\omega)J_{2k+1}(\omega)=1\; ,
\end{equation}
where we used eq.~(6.512,2) of Ref.~[\onlinecite{gradshteyn2007}].
Thus, our final result to second order is
\begin{eqnarray}
  e_0^{(2)}(J_{\rm K})   =-\frac{3}{32}J_{\rm K}^2 
\label{appeq:secondorderanalyt}
\end{eqnarray}
for the one-dimensional density of states~(\ref{eq:DOS}).

\subsection{Energy of the magnetic Yosida state}
\label{app:Yosidaenergy}

\subsubsection{Host electron Fermi sea}

At half band-filling and for a symmetric density of states
we have
\begin{equation}
|\hbox{FS} \rangle= \prod_{k; \epsilon_k\leq -\epsilon_{\rm F}}\hat{a}_{k,\downarrow}^+
\prod_{k; \epsilon_k\leq \epsilon_{\rm F}}\hat{a}_{k,\uparrow}^+ 
|\hbox{vac}\rangle \;.
\label{appeq:FSmagnetic}
\end{equation}
The host-electron spin polarization per site 
\begin{equation}
s_0=\frac{1}{2} \frac{1}{L}\sum_k
\langle \hbox{FS} |
\hat{a}_{k,\uparrow}^+\hat{a}_{k,\uparrow}^{\vphantom{+}} 
-\hat{a}_{k,\downarrow}^+\hat{a}_{k,\downarrow}^{\vphantom{+}} 
| \hbox{FS}\rangle
\label{appeq:MfromBingeneral}
\end{equation}
determines $\epsilon_{\rm F}$. 
For the one-dimensional density of states we find
\begin{equation}
s_0=\frac{1}{\pi} \arcsin\epsilon_{\rm F} \quad, \quad \epsilon_{\rm F}
=\sin(\pi s_0) \; .
\end{equation}
In turn, $\epsilon_{\rm F}$ can be viewed as a variational parameter
that optimizes the ground-state energy in the presence of the external
field,
\begin{eqnarray}
  e_{\rm FS}(\epsilon_{\rm F})&= & t_{\rm FS}(\epsilon_{\rm F}) - 2B s_0(\epsilon_{\rm F})
  \nonumber \; , \\
t_{\rm FS}(\epsilon_{\rm F}) &=&  \frac{1}{L}
\langle \hbox{FS} | \hat{T} | \hbox{FS}\rangle \nonumber \\
&=&2 \int_{-1}^0 \rmd \omega \omega\rho_0(\omega)
+ 2 \int_0^{\epsilon_{\rm F}} \rmd \omega \omega \rho_0(\omega) \; .
\label{appeq:eFSofepsF}
\end{eqnarray}
For a symmetric density of states at half band-filling, the variational optimum is at
\begin{equation}
\epsilon_{\rm F}^{(0)}=B
\label{appeq:MandepsFfromBa}
\end{equation}
for $B\leq 1$ and $\epsilon_{\rm F}^{(0)}=1$ for $B\geq 1$.
Thus, in one dimension for $B\leq 1$
\begin{equation}
e_{\rm FS}(B)= -\frac{2}{\pi} \sqrt{1 - B^2} - \frac{2}{\pi}B\arcsin(B) \; .
\label{appeq:E0onedim}
\end{equation}

\subsubsection{Presence of the impurity}

In the presence of the impurity,
we have to determine $\epsilon_{\rm F}$ from the minimization
of
\begin{equation}
  E_0(\epsilon_{\rm F})=Le_{\rm FS}(\epsilon_{\rm F})+e_0(\epsilon_{\rm F})\; .
\end{equation}
This gives the condition
\begin{equation}
  0=Le_{\rm FS}'(\epsilon_{\rm F}^{\rm opt})+e_0'(\epsilon_{\rm F}^{\rm opt}) \; .
  \label{appeq:separateordersinepsF}
  \end{equation}
We separate the terms of order~$L$ and order unity,
\begin{equation}
\epsilon_{\rm F}^{\rm opt} = B +\frac{1}{L} \epsilon_{\rm F}^{(1)} \; ,
\end{equation}
where we use eq.~(\ref{appeq:MandepsFfromBa}) to identify
the leading-order term $\epsilon_{\rm F}^{(0)}=B$, and find
in eq.~(\ref{appeq:separateordersinepsF}) that
\begin{eqnarray}
  0&=&e_{\rm FS}''(B)
  \epsilon_{\rm F}^{(1)}
  +e_0'(B)
  \; , \nonumber \\
  \epsilon_{\rm F}^{(1)}&=& -\frac{e_0'(B)}{e_{\rm FS}''(B) }=
  -\frac{1}{2\rho_0(B)}
  \left.\frac{\partial e_0(\epsilon_{\rm F})}{\partial \epsilon_{\rm F}}
  \right|_{\epsilon_{\rm F}=B}
  \; .
  \label{appeq:epsFone}
  \end{eqnarray}
In the last equation we used eq.~(\ref{appeq:eFSofepsF}) to show that
\begin{eqnarray}
  e_{\rm FS}'(\epsilon_{\rm F})&=& 2(\epsilon_{\rm F}-B)\rho_0(\epsilon_{\rm F})
  \; , \nonumber \\
 e_{\rm FS}''(\epsilon_{\rm F})&=& 2\rho_0(\epsilon_{\rm F}) +
 2(\epsilon_{\rm F}-B) \rho_0'(\epsilon_{\rm F}) \; .
 \label{appeq:e[sFderivatives}
   \end{eqnarray}
Fortunately, we do not need to know $\epsilon_{\rm F}^{(1)}$
when we work with the ground-state energy because
\begin{eqnarray}
  E_0(\epsilon_{\rm F}^{\rm opt})&=&
  L\Bigl(e_{\rm FS}(B)+e_{\rm FS}'(B) \frac{\epsilon_{\rm F}^{(1)}}{L}\Bigr)
  + e_0(B) \nonumber \\
  &=&
  Le_{\rm FS}(B)  + e_0(B)
  \label{appeq:permittoignoreepsF}
  \end{eqnarray}
is independent of $\epsilon_{\rm F}^{(1)}$ due to eq.~(\ref{appeq:eFSofepsF}),
see also eq.~(\ref{appeq:e[sFderivatives}).

\subsubsection{Lagrange functional and minimization}

After calculating all expectation values,
the Lagrange functional for the Yosida state
${\cal L}\equiv {\cal L}\left[\left\{\alpha_{k,\uparrow}\right\},
  \left\{\alpha_{k,\downarrow}\right\},\lambda\right]$ becomes
\begin{eqnarray}
{\cal L}&=&
\frac{1}{2}
\left(\frac{1}{L}\primesum_k \alpha_{k,\downarrow}^2\epsilon_k
+ \frac{1}{L}\doubleprimesum_k \alpha_{k,\uparrow}^2 \epsilon_k\right)
\nonumber \\
&&
- \frac{J_{\rm K}}{2} 
\left(\frac{1}{L}
\primesum_k \alpha_{k,\downarrow}\right)
\left(\frac{1}{L}
\doubleprimesum_k \alpha_{k,\uparrow}\right)
\nonumber \\
&&
+ \frac{J_{\rm K}}{4} s_0 \left(\frac{1}{L}
\primesum_k \alpha_{k,\downarrow}^2-
\frac{1}{L}
\doubleprimesum_k \alpha_{k,\uparrow}^2\right) \nonumber \\
&& - \frac{J_{\rm K}}{8} 
\left(\frac{1}{L}\primesum_k \alpha_{k,\downarrow}\right)^2
- \frac{J_{\rm K}}{8} 
\left(\frac{1}{L}\doubleprimesum_k \alpha_{k,\uparrow}\right)^2\nonumber \\
&& +\lambda \left(1- \frac{1}{2L}
\primesum_k \alpha_{k,\downarrow}^2
-\frac{1}{2L} \doubleprimesum_k \alpha_{k,\uparrow}^2 \right)\; ,
\label{appeq:LagrangeYosidafunctional}
\end{eqnarray}
where we took the normalization into account using the 
Lagrange parameter~$\lambda$.

We define
\begin{eqnarray}
C_{\uparrow} &=& \frac{1}{L}
\doubleprimesum_k \alpha_{k,\uparrow} \; ,\nonumber \\
C_{\downarrow} &=& \frac{1}{L}
\primesum_k \alpha_{k,\downarrow} \; .
\label{appeq:DefCupdown}
\end{eqnarray}
The variation of the 
Lagrange functional~(\ref{appeq:LagrangeYosidafunctional}) gives
\begin{eqnarray}
\alpha_{k,\uparrow} &=& \frac{J_{\rm K}}{4}
\frac{C_{\uparrow}+2 C_{\downarrow}}{
\epsilon_k -(\lambda+J_{\rm K}s_0/2)} \; , \nonumber \\
\alpha_{k,\downarrow} &=& \frac{J_{\rm K}}{4}
\frac{C_{\downarrow}+2 C_{\uparrow}}{
\epsilon_k -(\lambda-J_{\rm K}s_0/2)}
\end{eqnarray}
in the respective regions in $k$-space.
We abbreviate the principal-value integral
\begin{equation}
F_1(x,B)= \dashint_B^1 \rmd \omega \rho_0(\omega)
\frac{1}{\omega-x}
\end{equation}
to find in eq.~(\ref{appeq:DefCupdown})
\begin{eqnarray}
C_{\uparrow} &=& 
\frac{J_{\rm K}}{4}
\left(C_{\uparrow}+2 C_{\downarrow}\right)
F_1(\lambda+J_{\rm K}s_0/2,B)\; ,\nonumber \\
C_{\downarrow} &=& 
\frac{J_{\rm K}}{4}
\left(C_{\downarrow}+2 C_{\uparrow}\right)
F_1(\lambda-J_{\rm K}s_0/2,-B)\; ,
\label{appeq:calculateCupdown}
\end{eqnarray}
whereby we assume throughout that $0\leq B<1$, i.e., the host electrons
are not fully polarized. Note that eq.~(\ref{appeq:permittoignoreepsF})
permits to set $\epsilon_{\rm F}=B$
in our further considerations.

The secular determinant that belongs to eq.~(\ref{appeq:calculateCupdown})
must be zero because the normalization condition
$(C_{\uparrow}+C_{\downarrow})/2=1$ must also be fulfilled. Therefore,
we determine $\lambda$ from the equation
\begin{equation}
  \left(1-\frac{J_{\rm K}F_+}{4}\right)
  \left(1-\frac{J_{\rm K}F_-}{4}\right)
  -\frac{J_{\rm K}^2F_+F_-}{4}=0 \;, 
  \label{appeq:YosidaenergywithBfinal}
  \end{equation}
where we abbreviated $F_{+}\equiv F_1(\lambda+J_{\rm K}s_0/2,B)$
and $F_{-}\equiv F_1(\lambda-J_{\rm K}s_0/2,-B)$.
In one spatial dimension we have $s_0(B)=(1/\pi)\arcsin(B)$
from eq.~(\ref{appeq:MandepsFfromBa}) and
\begin{equation}
F_1(x,B) = 
\frac{1}{\pi\sqrt{1 - x^2}}
\ln \biggl[
\frac{1 - B x + \sqrt{(1-B^2) (1- x^2)}}{B - x}
\biggr] .
\label{appeq:defF1explicit}
\end{equation}
Using the variationally optimal parameters,
it is not difficult to show that 
\begin{equation}
  e_0^{\rm Y}(J_{\rm K},B)=\lambda \; .
\label{appeq:defYosidaenergy}
\end{equation}
Therefore, eq.~(\ref{appeq:YosidaenergywithBfinal}) determines the 
variational ground-state energy as a function of the external field $B$.

Eq.~(\ref{appeq:YosidaenergywithBfinal}) provides a solution
only for $B\leq B_{\rm c}^{\rm Y}(J_{\rm K})$
above which the Yosida state becomes unstable.
This problem does not occur in the Gutzwiller description
so that we do not extend the Yosida state to the region 
$B> B_{\rm c}^{\rm Y}(J_{\rm K})$.

\subsection{Impurity spin polarization and magnetization of the Yosida state}
\label{app:Yosidamagsusz}

\subsubsection{Impurity spin polarization}

By definition, see eq.~(\ref{eq:deflocalm}), the impurity spin polarization is given by
\begin{equation}
  \frac{  m^{S,{\rm Y}}(J_{\rm K},B)}{g_{\rm e}\mu_{\rm B}}
  =S^{z,{\rm Y}}(J_{\rm K},B)=\frac{1}{2}
  \langle \Psi_{\rm Y} |  \hat{n}_{d,\uparrow}-\hat{n}_{d,\downarrow}
  | \Psi_{\rm Y}\rangle \; .
\end{equation}
This expectation value is readily calculated to give
\begin{equation}
  S^{z,{\rm Y}}(J_{\rm K},B)
  = \frac{1}{2} \frac{1-3J_{\rm K}F_+/4}{1+J_{\rm K}F_+/4}
  \label{appeq:Yosidaspinpolglobal}
\end{equation}
with $F_+\equiv F_1(e_0^{\rm Y}(J_{\rm K},B)+J_{\rm K}s_0(B)/2,B)$ and $F_1(x,B)$
from eq.~(\ref{eq:defF1explicit}) in one dimension. The ground-state
energy $e_0^{\rm Y}(J_{\rm K},B)$
in the presence of a global field is derived from the solution
of eq.~(\ref{eq:YosidaenergywithBfinal}), and $s_0(B)=\arcsin(B)/\pi$ in one dimension.

When the external field is applied only locally,
the impurity spin polarization is given by
\begin{equation}
  S_{\rm loc}^{z,{\rm Y}}(J_{\rm K},B)
  = \frac{1}{2} \frac{1-3J_{\rm K}F(e_{0,{\rm loc}}^{\rm Y}(J_{\rm K},B)-B)/4}{
    1+J_{\rm K}F(e_{0,{\rm loc}}^{\rm Y}(J_{\rm K},B)-B)/4}
  \label{appeq:Yosidaspinpollocal}
\end{equation}
with $F(x)$ from eq.~(\ref{eq:Yosidafinalenergy}) in one dimension. 
The ground-state
energy $e_{0,{\rm loc}}^{\rm Y}(J_{\rm K},B)$
in the presence of a local field is derived from the solution
of eq.~(\ref{eq:YosidaenergywithBfinal}) when $F_{\pm}$ is replaced
by $F(\lambda\mp B)$.

\subsubsection{Impurity-induced magnetization}

The solution $e_0^{\rm Y}(J_{\rm K},B)$ 
of eq.~(\ref{eq:YosidaenergywithBfinal}) must be determined numerically
for given Kondo coupling and external field.
The impurity-induced magnetization is then obtained
using eq.~(\ref{eq:miiande0relation}) as
\begin{equation}
  \frac{m^{\rm ii,Y}(J_{\rm K},B)}{g_e\mu_{\rm B}}
  =-\frac{1}{2}
\frac{\partial e_0^{\rm Y}(J_{\rm K},B)}{\partial B}
\label{appeq:mfrome0Y}
\end{equation}
by a numerical derivative of the Yosida ground-state energy in the presence of a
magnetic field. Eq.~(\ref{appeq:mfrome0Y}) holds because of the variational
Hellmann-Feynman theorem,~\cite{Hellmann2,Feynman}
see appendix~\ref{app:misc}.

To obtain an analytic expression,
we take the derivative of eq.~(\ref{eq:YosidaenergywithBfinal}) with
respect to the magnetic field~$B$.
With the abbreviations
\begin{eqnarray}
  s_0'(B)&=&\frac{\partial s_0(B)}{\partial B}=\rho_0(B) \; , \nonumber \\
  y(B)&=&\frac{\rho_0(B)}{e_0^{\rm Y}(J_{\rm K},B)+J_{\rm K}s_0(B)/2-B} \; ,
\end{eqnarray}
and
\begin{eqnarray}
  F_2(x,B)&=&  \frac{\partial F_1(x,B)}{\partial x}\; , \nonumber \\
    F_{2,+}&=& F_2(e_0^{\rm Y}(J_{\rm K},B)+J_{\rm K}s_0(B)/2,B) \; , \nonumber \\
  F_{2,-}&=& F_2(e_0^{\rm Y}(J_{\rm K},B)-J_{\rm K}s_0(B)/2,-B)
\end{eqnarray}
we find
\begin{equation}
  F_{1,+}'+F_{1,-}'+\frac{3J_{\rm K}}{4} \left(F_{1,+}'F_{1,-}+F_{1,-}'F_{1,+}  \right) =0 \; ,
  \label{appeq:needstobesolvedform}
\end{equation}
where
\begin{eqnarray}
  F_{1,+}'&=& y(B)+\left(-2m^{\rm ii,Y}(B)+\frac{J_{\rm K}\rho_0(B)}{2}\right)F_{2,+}\; ,
  \nonumber\\
  F_{1,-}'&=& -y(-B)+\left(-2m^{\rm ii,Y}(B)-\frac{J_{\rm K}\rho_0(B)}{2}\right)F_{2,-}\; .
  \nonumber \\
  \end{eqnarray}
Here, we used eq.~(\ref{appeq:mfrome0Y}),
$e_0^{\rm Y}(J_{\rm K},-B)=e_0^{\rm Y}(J_{\rm K},B)$,
and $s_0(-B)=-s_0(B)$.

We solve eq.~(\ref{appeq:needstobesolvedform}) for $m^{\rm ii,Y}(J_{\rm K},B)$,
\begin{equation}
  \frac{m^{\rm ii,Y}(J_{\rm K},B)}{g_e\mu_{\rm B}}= \frac{1}{4}
  \frac{N_a(J_{\rm K},B)}{N_b(J_{\rm K},B)}
  \label{appeq:getm}
\end{equation}
with
\begin{eqnarray}
  N_a(J_{\rm K},B)&=&
  8y(B) - 8y(-B)\nonumber \\
  &&+4 J_{\rm K}\rho_0(B) \left(F_{2,+}- F_{2,-}\right)\nonumber \\
&&+ 
6J_{\rm K}\left(y(B) F_{1,-}-y(-B) F_{1,+}\right) \nonumber \\
&&+  3J_{\rm K}^2\rho_0(B)
\left(F_{1,-}F_{2,+}- F_{1,+}F_{2,-}  \right) \nonumber \; ,
\end{eqnarray}
and
\begin{eqnarray}
  N_b(J_{\rm K},B) &=& 
  4F_{2,+}+ 4F_{2,-}\nonumber\\
  &&+  3 J_{\rm K}  \left(  F_{1,-}F_{2,+}+    F_{1,+}F_{2,-}      \right)\; .
  \end{eqnarray}

When the external field is applied only locally, eq.~(\ref{eq:YosidaenergywithBfinal})
reduces to
\begin{equation}
  \left(1-\frac{J_{\rm K}F_+}{4}\right)
  \left(1-\frac{J_{\rm K}F_-}{4}\right)
  -\frac{J_{\rm K}^2F_+F_-}{4}=0
  \label{appeq:YosidaenergywithBlocalfinal}
  \end{equation}
with $F_{\pm}=F(\lambda\mp B)$ from eq.~(\ref{eq:Yosidafinalenergy}).
Then, eq.~(\ref{appeq:getm}) reduces to
\begin{equation}
  m^{\rm ii,Y}_{\rm loc}(J_{\rm K},B)=
  \frac{4(F_+'-F_{-}')+3J_{\rm K}(F_+'F_{-}-F_{-}'F_+)}{
    8(F_+'+F_{-}')+6J_{\rm K}(F_+'F_{-}+F_{-}'F_+)}
  \label{appeq:getmlocal}
\end{equation}
with
\begin{eqnarray}
  F'_{\pm}&=&F'(\lambda \pm B)\; , \nonumber \\\
  F'(x) &=& -\frac{1}{\pi x(1-x^2)} +\frac{x}{1 - x^2}F(x) \; .
\end{eqnarray}

\begin{figure}[t]
  \includegraphics[width=8.5cm]{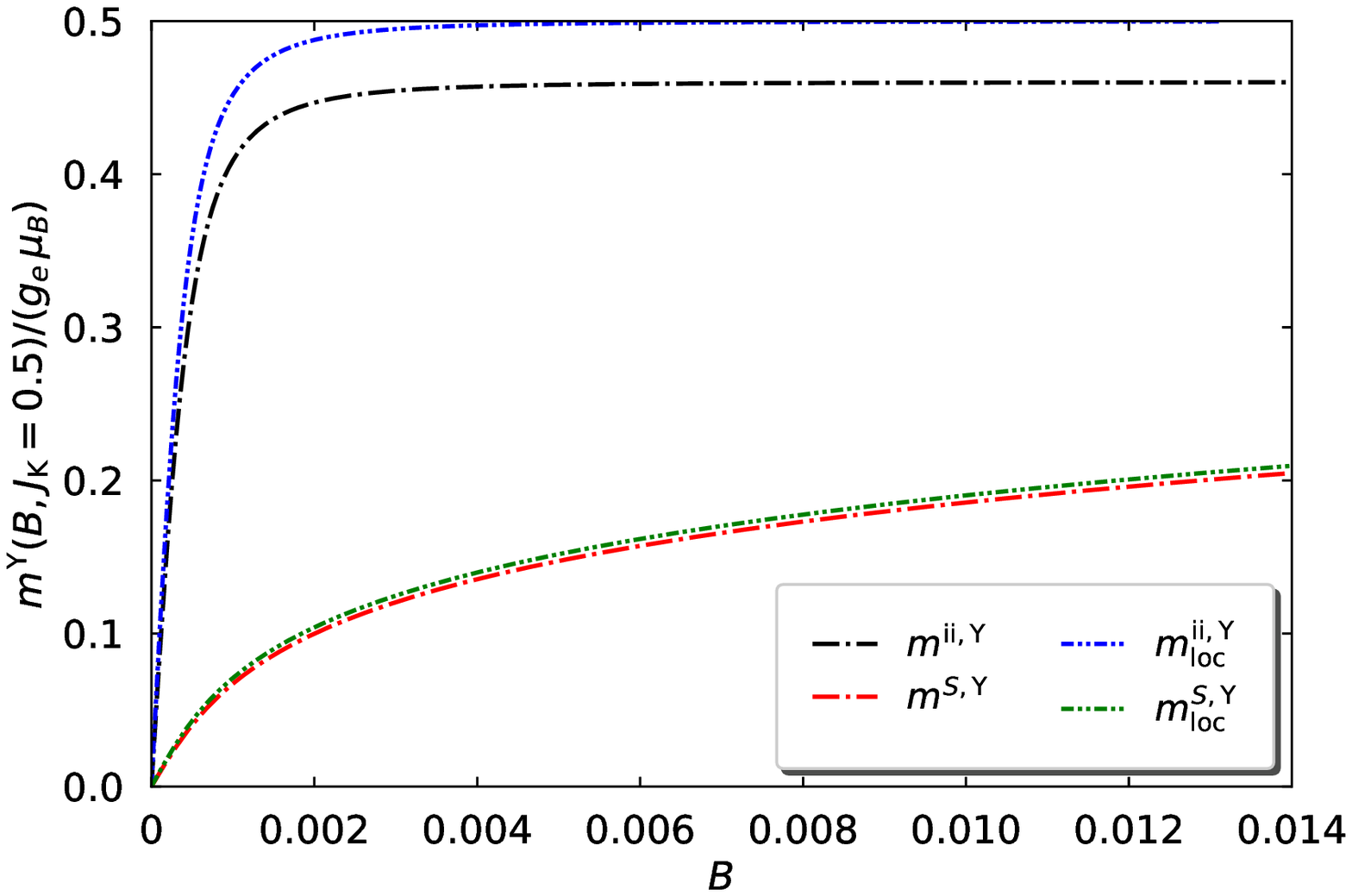}\\
  \includegraphics[width=8.5cm]{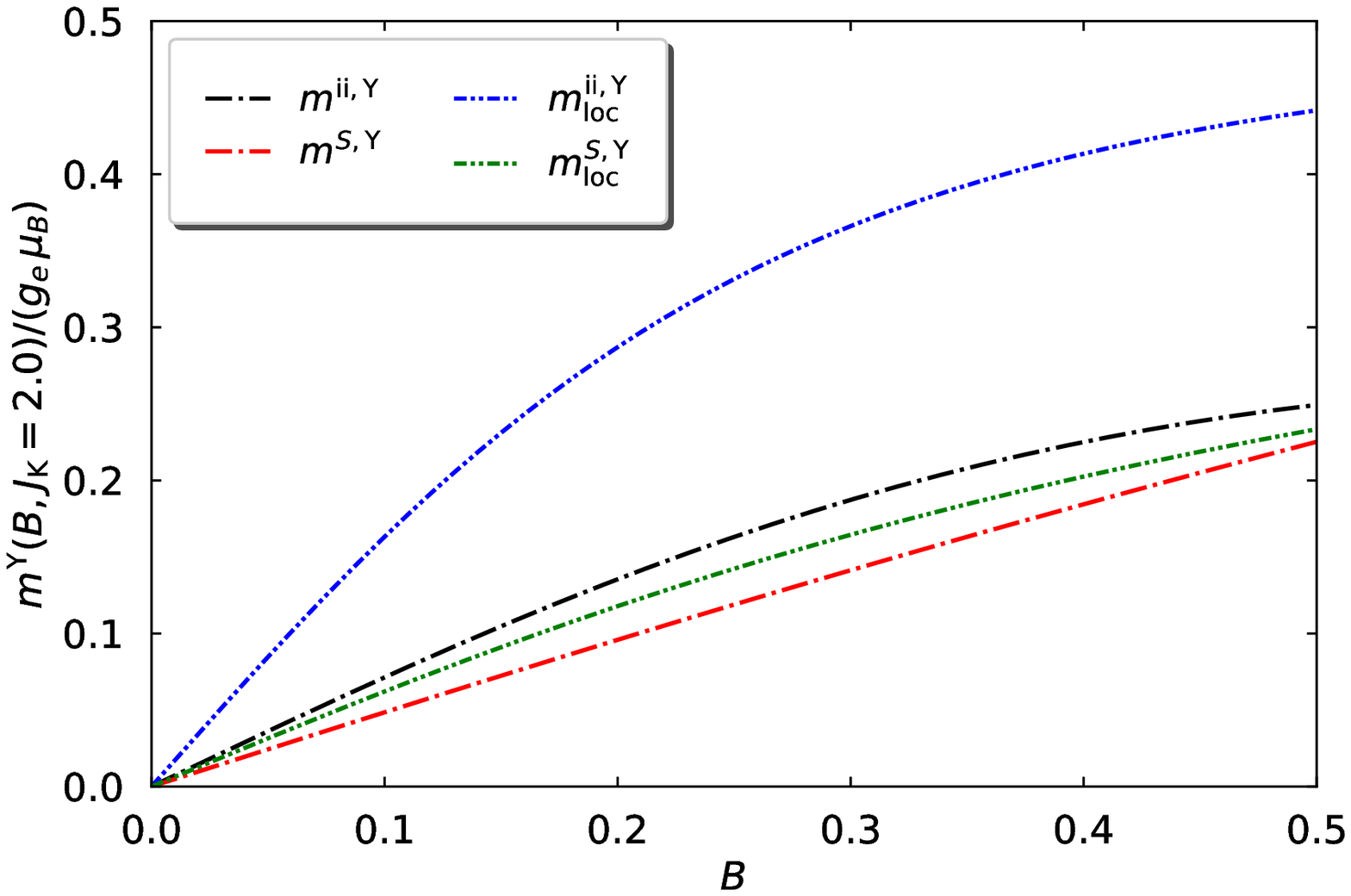}
\caption{(Color online) Impurity spin polarization
  $S^{z,{\rm Y}}=m^{S,{\rm Y}}/(g_e\mu_{\rm B})$, eq.~(\ref{eq:deflocalm}),
  and impurity-induced magnetization $m^{\rm ii,Y}/(g_e\mu_{\rm B})$,
  eq.~(\ref{eq:miidef}), 
  of the one-dimensional symmetric Kondo model
  as a function of global/local fields for $J_{\rm K}=0.5$ (upper figure)
  and $J_{\rm K}=2$ (lower figure)
 from the Yosida wave function.\label{appfig:Yosidams}}
\end{figure}

In Fig.~\ref{appfig:Yosidams} we display
the impurity spin polarization
  $S^{z,{\rm Y}}=m^{S,{\rm Y}}/(g_e\mu_{\rm B})$
and the impurity-induced magnetization $m^{\rm ii,Y}/(g_e\mu_{\rm B})$
from the Yosida wave function as a function of global and local
magnetic fields. The curves noticeably differ
from each other which shows that it is important to distinguish
between the four quantities.

For small interactions, the system has a large impurity-induced
magnetic susceptibility so that
small fields tend to fully polarize the system. Indeed, as seen from the figure,
the impurity-induced magnetization reaches its maximum value for
fields of the order of $10^{-2}{\cal D}$ at $J_{\rm K}=0.5$. For these small fields,
it is not important whether the field is applied globally or only locally
because the magnetic response is mostly determined by the impurity spin and
the electrons in its vicinity.
Note that the impurity spin polarization
$S^{z,{\rm Y}}$ is \emph{smaller\/} than
the impurity-induced magnetization $m^{\rm ii,Y}/(g_e\mu_{\rm B})$;
this is an artifact of the Yosida wave function. 

Large interactions, $J_{\rm K}=2$, require
large fields to polarize the impurity system. Thus, the differences between the curves
are more pronounced.
Recall, however, that the Yosida wave function is not a good variational state
for strong couplings so that the curves in Fig.~\ref{appfig:Yosidams}
are \emph{not} representative for the Kondo model.

\subsection{Evaluation of expectation values for the Gutzwiller wave function}
\label{subsec:GWFEvaluation}

\subsubsection{Norm, kinetic energy, and impurity spin polarization}

The Gutzwiller state is normalized to unity,
\begin{equation}
\langle \Psi_{\rm G} | \Psi_{\rm G}\rangle
=\langle \Phi |
1-x(\hat{n}_{\uparrow}^d-n_{\uparrow}^{d,0})
(\hat{n}_{\downarrow}^d-n_{\downarrow}^{d,0})
|\Phi\rangle=1
\label{appeq:normalizedPsiG}
\end{equation}
because $|\Phi\rangle$ is a normalized single-particle product state
and eq.~(\ref{appeq:dsarehalffillednonint}) holds.
Likewise,~\cite{TIAM} the expectation value of the
kinetic energy in the Gutzwiller wave function gives
\begin{eqnarray}
\langle \hat{T} \rangle_{\rm G} &=& 
\sum_{k,\sigma} \left(\epsilon(k) -\sigma_n B\right)
\langle \Phi | \hat{a}_{k,\sigma}^+\hat{a}_{k,\sigma}^{\vphantom{+}}
\hat{P}_{\rm G}^2| \Phi\rangle \nonumber \\
&=& 
\sum_{k,\sigma} \left(\epsilon(k) -\sigma_n B\right)
\langle \Phi | \hat{a}_{k,\sigma}^+\hat{a}_{k,\sigma}^{\vphantom{+}}
| \Phi\rangle \; ,
\label{eq:EkinGutzexp}
\end{eqnarray}
when we use eqs.~(\ref{appeq:dsarehalffillednonint}) and~(\ref{appeq:defPGsquared}).
Moreover, it is readily seen from eq.~(\ref{eq:useful}) that
\begin{equation}
  S^{z,{\rm G}}=\frac{1}{2}
  \langle \hat{n}_{\uparrow}^d-\hat{n}_{\downarrow}^d  \rangle_{\rm G}
  = m =  \frac{1}{2}
  \langle \Phi |  \hat{n}_{\uparrow}^d-\hat{n}_{\downarrow}^d  |\Phi \rangle 
  \end{equation}
so that we do not have to distinguish between the impurity
spin polarization in  $|\Psi_{\rm G}\rangle$ and in~$|\Phi\rangle$.

\subsubsection{Spin-flip terms in the Kondo coupling}

Using eq.~(\ref{appeq:useful}) we have
\begin{equation}
\langle \hat{V}_{\rm sd}^{\rm sf}\rangle_{\rm G}=
\lambda_{\uparrow}\lambda_{\downarrow}
\frac{J_{\rm K}}{2L}  \sum_{k,k'} 
\langle \Phi | \left[\hat{a}_{k',\uparrow}^+\hat{a}_{k,\downarrow}^{\vphantom{+}} 
\hat{d}_{\downarrow}^+\hat{d}_{\uparrow}^{\vphantom{+}} 
+ \hbox{h.c.}\right]|\Phi\rangle \; .
\end{equation}
We use Wick's theorem for the single-particle product state $|\Phi\rangle$
to find
\begin{eqnarray}
  \langle \hat{V}_{\rm sd}^{\rm sf}\rangle_{\rm G}
&=&
-\frac{2}{\sqrt{1-4m^2}}
\frac{J_{\rm K}}{2L} \nonumber \\
&& \times \biggl(
\sum_{k} 
\langle \Phi | \hat{d}_{\downarrow}^+\hat{a}_{k,\downarrow}^{\vphantom{+}}
|\Phi\rangle
\sum_{k'}
\langle \Phi | \hat{a}_{k',\uparrow}^+ \hat{d}_{\uparrow}^{\vphantom{+}} |\Phi\rangle
\nonumber \\
&&\hphantom{\times\biggl(}
+
\sum_{k} 
\langle \Phi | \hat{d}_{\uparrow}^+\hat{a}_{k,\uparrow}^{\vphantom{+}}
|\Phi\rangle
\sum_{k'}
\langle \Phi | \hat{a}_{k',\downarrow}^+ \hat{d}_{\downarrow}^{\vphantom{+}} |\Phi\rangle
\biggr)\; ,\nonumber\\
\end{eqnarray}

\subsubsection{Ising terms in the Kondo coupling}

Using eq.~(\ref{appeq:useful}) we have
\begin{eqnarray}
  \langle \hat{V}_{\rm sd}^{\rm Is}\rangle_{\rm G}&=&
\frac{J_{\rm K}}{4L}  \sum_{k,k'} 
\langle \Phi |
(\hat{a}_{k',\uparrow}^+\hat{a}_{k,\uparrow}^{\vphantom{+}} 
-\hat{a}_{k',\downarrow}^+\hat{a}_{k,\downarrow}^{\vphantom{+}} )
\nonumber \\
&& \hphantom{\frac{J_{\rm K}}{4L}  \sum_{k,k'} \langle \Phi |}
\biggl(\frac{\hat{d}_{\uparrow}^+\hat{d}_{\uparrow}^{\vphantom{+}} }{n_{\uparrow}^{d,0}}
- \frac{\hat{d}_{\downarrow}^+\hat{d}_{\downarrow}^{\vphantom{+}}}{n_{\downarrow}^{d,0}}
+2m \frac{\hat{n}_{\uparrow}^d\hat{n}_{\downarrow}^d}{
  n_{\uparrow}^{d,0}n_{\downarrow}^{d,0}}
\biggr)|\Phi\rangle
\nonumber \\
    &=&
- \frac{J_{\rm K}}{2L}\biggl(
\sum_{k} 
\langle \Phi | \hat{d}_{\uparrow}^+\hat{a}_{k,\uparrow}^{\vphantom{+}}
|\Phi\rangle
\sum_{k'}
\langle \Phi | \hat{a}_{k',\uparrow}^+ \hat{d}_{\uparrow}^{\vphantom{+}} |\Phi\rangle
\nonumber \\
&&\hphantom{   - \frac{J_{\rm K}}{2L}\biggl( }
+
\sum_{k} 
\langle \Phi | \hat{d}_{\downarrow}^+\hat{a}_{k,\downarrow}^{\vphantom{+}}
|\Phi\rangle
\sum_{k'}
\langle \Phi | \hat{a}_{k',\downarrow}^+ \hat{d}_{\downarrow}^{\vphantom{+}} |\Phi\rangle
\biggr)\nonumber \\
&&+ J_{\rm K} m M_0 \; ,
\end{eqnarray}
where we again used Wick's theorem for the single-particle product state $|\Phi\rangle$,
and defined the host-electron spin polarization on the impurity,
\begin{equation}
  M_0=\frac{1}{2L} \sum_{k,k'}
  \langle \Phi | \hat{a}_{k',\uparrow}^+\hat{a}_{k,\uparrow}^{\vphantom{+}}
  -
  \hat{a}_{k',\downarrow}^+\hat{a}_{k,\downarrow}^{\vphantom{+}}
  | \Phi \rangle \; .   \label{eq:deflocalMhost}
  \end{equation}

\subsection{Lagrange functional and effective non-interacting SIAM
  for the Gutzwiller wave function}
\label{subsec:Lagrangian}

\subsubsection{Optimization of the Lagrange functional}

We must optimize $\langle \hat{H}\rangle_{\rm G}$
with respect to the single-particle product states $|\Phi\rangle$
that are normalized to unity, $\langle \Phi | \Phi \rangle=1$.
Moreover, we must respect the conditions (\ref{eq:defmimpinPhi})
and~(\ref{eq:deflocalMhost}).

We apply Ritz variational principle to the Lagrange
functional ${\cal L}\equiv {\cal L}\left( \left\{|\Phi\rangle\right\},E_{\rm sp},E_d,K\right)$,
see also Ref.~[\onlinecite{TIAM}],
\begin{eqnarray}
{\cal L}&=& \langle \hat{H}\rangle_{\rm G}
+E_{\rm sp}(1-\langle \Phi | \Phi \rangle) \nonumber \\
&&+E_d\left(
2m- \langle \Phi | \hat{n}_{\uparrow}^d  - \hat{n}_{\downarrow}^d  | \Phi \rangle
\right)\nonumber \\
&&-K\Bigl(2M_0-\frac{1}{L} \sum_{k,k'}
  \langle \Phi | \hat{a}_{k',\uparrow}^+\hat{a}_{k,\uparrow}^{\vphantom{+}}
  -
  \hat{a}_{k',\downarrow}^+\hat{a}_{k,\downarrow}^{\vphantom{+}}
  | \Phi \rangle \Bigr)\; ,
  \nonumber \\
  \langle \hat{H}\rangle_{\rm G}&=&\langle \hat{T}\rangle_{\rm G} -2mB
  + \langle \hat{V}_{\rm sd}^{\rm sf}\rangle_{\rm G}+
    \langle \hat{V}_{\rm sd}^{\rm Is}\rangle_{\rm G} \; , 
\label{eq:Lagrangefunctional}
\end{eqnarray}
and find that $|\Phi\rangle$ must obey the Schr\"odinger equation
\begin{equation}
\hat{H}_0^{\rm SIAM}
|\Phi\rangle = E_{\rm sp} |\Phi\rangle 
\label{eq:SeqHeffSIAM}
\end{equation}
with the effective non-interacting single-impurity Anderson model (SIAM)
\begin{eqnarray}
  \hat{H}_0^{\rm SIAM}&=&\widetilde{T} +\widetilde{V}
  -E_d\left(\hat{n}_{\uparrow}^d  - \hat{n}_{\downarrow}^d \right) \nonumber \\
&& +\frac{K}{L} \sum_{k,k'}
  \left( \hat{a}_{k',\uparrow}^+\hat{a}_{k,\uparrow}^{\vphantom{+}}
  -
  \hat{a}_{k',\downarrow}^+\hat{a}_{k,\downarrow}^{\vphantom{+}}\right) \; ,
  \label{eq:HSIAMdefmaintext}
\end{eqnarray}
with the operators for the kinetic energy and the local hybridization 
  \begin{eqnarray}
  \widetilde{T}&=& \sum_{k,\sigma} (\epsilon(k) -\sigma_nB)
 \hat{a}_{k,\sigma}^+\hat{a}_{k,\sigma}^{\vphantom{+}} \; , \nonumber \\
 \widetilde{V} &=&  \frac{1}{\sqrt{L}}
 \sum_{k,\sigma}V_{\sigma}
\left(\hat{a}_{k,\sigma}^+\hat{d}_{\sigma}^{\vphantom{+}}
+\hat{d}_{\sigma}^+\hat{a}_{k,\sigma}^{\vphantom{+}}
\right)\; ,
\label{eq:diagnonalizethis}
\end{eqnarray}
where
\begin{eqnarray}
  V_{\sigma} &=& -\frac{J_{\rm K}}{2} \left(
  \gamma_{\sigma}+\frac{2}{\sqrt{1-4m^2}}\gamma_{\bar{\sigma}}
  \right)\; , \nonumber \\
\gamma_{\sigma}&=& \sqrt{\frac{1}{L}} \sum_{k}  
\langle\Phi | \hat{a}_{k,\sigma}^+\hat{d}_{\sigma}^{\vphantom{+}}
|\Phi\rangle 
\label{eq:selfconsistency}
\end{eqnarray}
have to be determined self-consistently. Hereby, we assumed that 
$\gamma_{\sigma}$ and $V_{\sigma}$ are real; recall $\bar{\uparrow}=\downarrow$,
$\bar{\downarrow}=\uparrow$.

In the following we choose $|\Phi\rangle$ as the ground-state of 
$\hat{H}_0^{\rm SIAM}$ in eq.~(\ref{eq:SeqHeffSIAM}) 
and denote the optimal single-particle product
state by $|\Phi_0\rangle$. Recall that we have to also fulfill
the conditions~(\ref{eq:defmimpinPhi})
and~(\ref{eq:deflocalMhost}),
\begin{eqnarray}
 m&=&\frac{1}{2} \langle \Phi_0 | \hat{n}_{\uparrow}^d
  - \hat{n}_{\downarrow}^d
  | \Phi_0 \rangle \; , \nonumber \\
    M_0&=&\frac{1}{2L} \sum_{k,k'}
  \langle \Phi_0 | \hat{a}_{k',\uparrow}^+\hat{a}_{k,\uparrow}^{\vphantom{+}}
  -
  \hat{a}_{k',\downarrow}^+\hat{a}_{k,\downarrow}^{\vphantom{+}}
  | \Phi_0 \rangle \; ,
  \label{eq:mandM0}
  \end{eqnarray}
which we regain from the minimization of~${\cal L}$ with respect to $E_d$
and $K$. The minimization of~${\cal L}$ with respect to~$m$ and $M_0$
give
\begin{eqnarray}
  K&=&\frac{1}{2} J_{\rm K} m \; , \nonumber \\
  E_d&=&B-\frac{1}{2} J_{\rm K}M_0
  +\frac{4m}{(1-4m^2)^{3/2}}J_{\rm K}\gamma_{\uparrow}\gamma_{\downarrow}\; .
  \label{eq:resetKEd}
  \end{eqnarray}
Note that a finite impurity magnetization~$m$ generates a potential scattering
in the effective single-impurity Anderson model, $K\neq 0$.

\subsubsection{Self-consistency procedure}
\label{sec:selfconsistency}

The remaining task is the calculation of the single-particle density of states
for the effective non-interacting single-impurity Anderson
Hamiltonian~(\ref{eq:SeqHeffSIAM}).
Using the single-particle density of states, we can
calculate  the single-particle energy~$E_{\rm sp}(B,E_d,K,V_{\sigma})$
from which we obtain $m$, $M_0$, and $\gamma_{\sigma}$ 
\begin{eqnarray}
  2m&=&-\frac{\partial E_{\rm sp}(B,E_d,K,V_{\sigma})}{\partial E_d} \; , \nonumber \\
  2M_0&=&\frac{\partial E_{\rm sp}(B,E_d,K,V_{\sigma})}{\partial K} \; , \nonumber \\
  2\gamma_{\sigma}&=&\frac{\partial E_{\rm sp}(B,E_d,K,V_{\sigma})}{\partial V_{\sigma}}
  \; ,
  \label{eq:resetmMOgamma}
\end{eqnarray}
when we use the Hellmann-Feynman theorem that also holds
for variational approaches. The simple proof relies on the
fact that the optimized variational state is stationary with respect to small
wave-function variations, see appendix~\ref{app:misc}.

Therefore, the parameters of the single-impurity Anderson model
are determined self-consistently using the following procedure.
\begin{itemize}
\item[S~1] The self-consistency procedure is initialized by choosing
  the values of the
  paramagnetic solution,  $m=0$, $M_0=0$, $K=0$, and $V_{\sigma}=V$,
  see Sect.~\ref{subsec:e0Gutzpara}.
  This guarantees that the algorithm works for $B=0$.
  To lift the degeneracy, we start at $E_d=B$ for given $B>0,J_{\rm K}>0$.
\item[S~2] The analytic expressions for $E_{\rm sp}(B,E_d,K,V_{\sigma})$
  give new values for $m,M_0,\gamma_{\sigma}$ from eq.~(\ref{eq:resetmMOgamma}),
  and thus new values for $K$ and $E_d$ from eq.~(\ref{eq:resetKEd}),
  and new values for $V_{\sigma}$ from eq.~(\ref{eq:selfconsistency}).
\item[S~3] Check whether or not $K,E_d,V_{\sigma}$ deviate from
  their previous values by more than some small value~$\eta=10^{-12}$. If so, return
  to S~2, otherwise, the algorithm terminates and gives the self-consistent values
  for $K,E_d,V_{\sigma}$ and $m,M_0,\gamma_{\sigma}$.
\end{itemize}
\subsubsection{Ground-state energy}

Note that $E_{\rm sp}$ is not identical to $E_0^{\rm G}=\langle \hat{H}_{\rm G}\rangle$.
Instead, we have from eq.~(\ref{eq:diagnonalizethis})
\begin{eqnarray}
  E_{\rm sp}&=& \langle \widetilde{T}\rangle_0+\langle \widetilde{V}\rangle_0
  -2mE_d+2KM_0 \nonumber \\
  &=& \sum_{k,\sigma} (\epsilon(k) -\sigma_nB)
  \langle \Phi | \hat{a}_{k,\sigma}^+\hat{a}_{k,\sigma}^{\vphantom{+}}  | \Phi\rangle
  \nonumber \\
  &&+ \sum_{\sigma} 2V_{\sigma}\gamma_{\sigma}
  -2mE_d+2KM_0 \nonumber \\
  &=&
   \sum_{k,\sigma} (\epsilon(k) -\sigma_nB)
   \langle \Phi | \hat{a}_{k,\sigma}^+\hat{a}_{k,\sigma}^{\vphantom{+}}  | \Phi\rangle
     -2mE_d+2KM_0 \nonumber \\
     && -J_{\rm K} \left(\gamma_{\uparrow}^2 +\gamma_{\downarrow}^2
     +\frac{4}{\sqrt{1-4m^2}}\gamma_{\uparrow}\gamma_{\downarrow}
     \right) \; ,
     \label{eq:EspGutzwarning}
\end{eqnarray}
where we used eq.~(\ref{eq:selfconsistency}) to replace $V_{\sigma}$ by
$\gamma_{\sigma}$. In contrast, from eq.~(\ref{eq:Lagrangefunctional}) we have
\begin{eqnarray}
  E_0^{\rm G}&=& \sum_{k,\sigma} (\epsilon(k) -\sigma_nB)
  \langle \Phi | \hat{a}_{k,\sigma}^+\hat{a}_{k,\sigma}^{\vphantom{+}}  | \Phi\rangle
  -m(2B-J_{\rm K}M_0) \nonumber \\
  && -\frac{J_{\rm K}}{2}\left(
  \frac{4}{\sqrt{1-4m^2}}\gamma_{\downarrow}\gamma_{\uparrow}
  +\gamma_{\uparrow}^2+\gamma_{\downarrow}^2\right) \; .
  \label{eq:EzeroGutzfull}
\end{eqnarray}
Comparing both equations results in
\begin{eqnarray}
  E_0^{\rm G}&=& E_{\rm sp} +\frac{J_{\rm K}}{2}\left(
  \frac{4}{\sqrt{1-4m^2}}\gamma_{\downarrow}\gamma_{\uparrow}
  +\gamma_{\uparrow}^2+\gamma_{\downarrow}^2\right)\nonumber \\
&&  -m(2B-J_{\rm K}M_0)+2mE_d-2KM_0 \nonumber \\
  &=& E_{\rm sp} +\frac{J_{\rm K}}{2}\left(
  \frac{4}{\sqrt{1-4m^2}}\gamma_{\downarrow}\gamma_{\uparrow}
  +\gamma_{\uparrow}^2+\gamma_{\downarrow}^2\right)\nonumber \\
&&  +\frac{8m^2}{(1-4m^2)^{3/2}}J_{\rm K}\gamma_{\uparrow}\gamma_{\downarrow}
  -J_{\rm K}mM_0  \label{eq:E0Gutzwitheverything}
\end{eqnarray}
for the Gutzwiller variational ground-state energy of the Kondo model.
The excess ground-state energy for the Gutzwiller variational state is given by
\begin{eqnarray}
  e_0^{\rm G}(J_{\rm K},B)&=&E_0^{\rm G}(J_{\rm K},B) \label{eq:e0finalfinal}
  \\
&&-  \sum_{k,\sigma} (\epsilon(k) -\sigma_nB)
  \langle \hbox{FS} | \hat{a}_{k,\sigma}^+\hat{a}_{k,\sigma}^{\vphantom{+}}
  | \hbox{FS}\rangle \; ,\nonumber 
\end{eqnarray}
where $|\hbox{FS}\rangle$ is the Fermi sea of non-interacting electrons.

The variational optimization for $m> 0$ is outlined in appendix~\ref{sec:SIAMfull}.
Here, we summarize the main results.
\begin{itemize}
  \item[--] The Gutzwiller ground state displays a finite local magnetization, $m>0$,
  at $B=0^+$  for all $0<J_{\rm K}<J_{\rm K,c}^{\rm G}\approx 0.839$.
  The precise value is determined in Sect.~\ref{subsubsec:Gutzwilelrstrongcouling}.
\item[--] For small interactions, $J_{\rm K}\to 0$, the values for $V_{\sigma}$,
  $\gamma_{\sigma}$ $E_d$, $K$, $m$, $M_0$, and $\omega_{p,\downarrow}$
  can be determined analytically.
\item[--] The ground-state energy for small interactions in one dimension
  can be approximated by
  \begin{equation}
  e_0^{\rm G}(J_{\rm K})\approx
  -0.0905 J_{\rm K}^2 -0.051J_{\rm K}^3-0.05 J_{\rm K}^4
  \label{eq:Gutzenergymaganalyt}
\end{equation}
for the Gutzwiller variational energy for $J_{\rm K}\lesssim 0.4$.
The quadratic coefficient can be compared with the exact result from
perturbation theory, $e_0(J_{\rm K})\approx -3J_{\rm K}^2/32
=-0.09375 J_{\rm K}^2$, see eq.~(\ref{eq:secondorderanalyt}).
The magnetic Gutzwiller states accounts for 96.5\% of the
correlation energy.
\end{itemize}

\subsection{Zero-field impurity spin susceptibility for the paramagnetic Gutzwiller state}
\label{subsec:magGutz}

{}From the numerical solution of the self-consistency equations,
 we see that $\gamma_{\uparrow}=\gamma_{\downarrow}$
 and $V_{\uparrow}=V_{\downarrow}$ at self-consistency.
 In the following, we use this assumption.

\subsubsection{Impurity spin polarization}

The optimization procedure of Sect.~\ref{subsec:Lagrangian}
directly gives the impurity spin polarization,
\begin{equation}
  \frac{m^{S,{\rm G}}(J_{\rm K},B)}{g_e\mu_{\rm B}}
  =m(J_{\rm K},B) \; .
  \label{eq:mSG}
\end{equation}
When the external field is applied only at the impurity, we simply replace
the expression $(\epsilon(k)-\sigma_nB)$ by $\epsilon(k)$
in eqs.~(\ref{eq:EkinGutzexp}), (\ref{eq:diagnonalizethis}),
(\ref{eq:EspGutzwarning}), (\ref{eq:EzeroGutzfull}), and~(\ref{eq:e0finalfinal})
to arrive at the corresponding `local' expressions for the impurity
spin polarization and impurity-induced magnetization.
Invoking the variational Hellmann-Feynman theorem,~\cite{Hellmann2,Feynman}
see appendix~\ref{app:misc},
we may alternatively use
\begin{equation}
  \frac{m^{{\rm S,G}}_{\rm loc}(J_{\rm K},B)}{g_e\mu_{\rm B}}
   =-\frac{1}{2}
   \frac{\partial e_{0,{\rm loc}}^{\rm G}(J_{\rm K},B)}{\partial B} \; ,
     \label{eq:mSlocG}
\end{equation}
see also eq.~(\ref{eq:definelocalmandchiasderivatives}).

\subsubsection{Impurity-induced magnetization}

Following the steps in Sect.~\ref{subsec:GWFEvaluation}
it is readily shown that
\begin{equation}
  \frac{m^{\rm ii, G}(J_{\rm K},B)}{g_e\mu_{\rm B}}
  = \langle \Phi_0 | \hat{S}^z+\hat{s}^z  | \Phi_0 \rangle
  - \langle \text{FS} |\hat{s}^z  | \text{FS} \rangle \; .
\end{equation}
Here, $|\Phi_0\rangle$ is the optimized ground state of the effective
non-interacting single-impurity Anderson model Hamiltonian
defined in eq.~(\ref{eq:HSIAMdefmaintext}) and $|\text{FS}\rangle$
is the Fermi-sea ground state of non-interacting electrons in the presence
of a magnetic field.
When we use the single-particle density of states
of the non-interacting SIAM, see 
appendix~\ref{sec:SIAMfull}, we find
\begin{eqnarray}
  \frac{m^{\rm ii, G}(J_{\rm K},B)}{g_e\mu_{\rm B}}
  &=&\frac{1}{2} \int_{-\infty}^0 \rmd \omega 
  \left(D_{\uparrow}(\omega)-D_{\downarrow}(\omega)\right) \nonumber \\
  &&   -
  \frac{1}{2} \int_{-\infty}^0 \rmd \omega 
  \left(D^{\rm FS}_{\uparrow}(\omega)-D^{\rm FS}_{\downarrow}(\omega)\right)
  \nonumber \\
  &=&
  \frac{1}{2} \int_{-\infty} \rmd \omega 
  \left(D_{{\rm imp},\uparrow}(\omega)-D_{{\rm imp},\downarrow}(\omega)\right)
  \; ,\nonumber \\
  \label{eq:miiGfromDOS}
\end{eqnarray}
where the impurity density of states is given by
the phase-shift function
\begin{eqnarray}
    D_{{\rm imp},\sigma}(\omega)&=&
  -\frac{1}{\pi} \frac{\partial \varphi_{\sigma}(\omega,B,E_d,K,V)}{\partial \omega}
  \nonumber \;, \\
\cot[  \varphi_{\sigma}(\omega,B,E_d,K,V)] &=&
\frac{{\rm R}_{\sigma}(\omega)}{{\rm I}_{\sigma}(\omega)}
\; ,
\end{eqnarray}
with the real and imaginary parts of the hybridization function
\begin{eqnarray}
{\rm R}_{\sigma}(\omega)
  &=&(\omega+\sigma_n E_d)(1-\sigma_nK\Lambda_0(\omega+\sigma_n B))\nonumber \\
&&-V^2\Lambda_0(\omega+\sigma_n B) \; , \nonumber \\
{\rm I}_{\sigma}(\omega)
&=&\eta\left(1-\sigma_n K\Lambda_0(\omega+\sigma_n B)\right) \nonumber \\
&&+\left[(\omega+\sigma_nE_d)\sigma_nK+V^2\right]\pi \rho_0(\omega+\sigma_nB) \; ,
\nonumber \\
\end{eqnarray}
see eqs.~(IV-43) and~(IV-44) of the supplemental material,
and $\eta=0^+$.

Since the impurity contribution to the density of states is
given by a frequency derivative,
the frequency integration in eq.~(\ref{eq:miiGfromDOS}) is readily carried out.
The density of states vanishes for $\omega \to -\infty$ so that
the density of states at the Fermi energy alone determines
the impurity-induced magnetization.
We focus on the paramagnetic region for the Gutzwiller wave function,
$J_{\rm K}>J_{\rm K,c}^{\rm G}$, so that the band part of the impurity density of states
at the Fermi energy gives, see eq.~(IV-50) of the supplemental material,
\begin{eqnarray}
  \frac{m^{\rm ii,G}(J_{\rm K},B)}{g_e\mu_{\rm B}}
  &=&-\frac{1}{2\pi}\left(X_{\uparrow}(0)-X_{\downarrow}(0)\right) \; , \nonumber \\
    X_{\sigma}(\omega) &\equiv &  X_{\sigma}(\omega,B,E_d,K,V)  \label{eq:Xdefinition}
  \\
&=& \pi \theta_{\rm H}(-\omega-\sigma_nE_d)\nonumber \\
&&+
\arccot\biggl[ 
\frac{(\omega+\sigma_n E_d)\sqrt{1-(\omega+\sigma_n B)^2}}{
  (\omega+\sigma_n E_d)\sigma_nK+V^2}
\biggr]
\nonumber 
\end{eqnarray}
in one dimension. Thus, we obtain the final result
\begin{eqnarray}
  \frac{m^{\rm ii,G}(J_{\rm K},B)}{g_e\mu_{\rm B}}
  &=& \frac{1}{2} -\frac{1}{\pi} \arccot\left[ \frac{E_d\sqrt{1-B^2}}{E_dK+V^2}\right]
  \nonumber \\
  &=& \frac{1}{\pi}\arctan\left[ \frac{E_d\sqrt{1-B^2}}{E_dK+V^2}\right]\; ,
  \label{eq:miiG}
\end{eqnarray}
where $E_d(B)$, $K(B)$, and $V(B)$ are determined from the solution
of the self-consistency cycle in Sect.~\ref{subsec:Lagrangian}.

\begin{figure}[t]
\includegraphics[width=8.5cm]{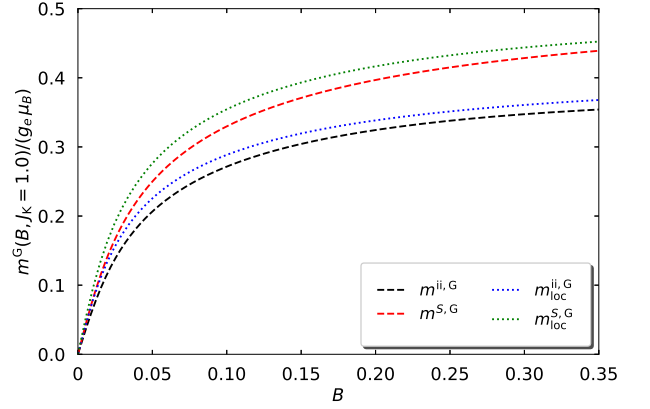}
\caption{(Color online) Impurity spin polarization
  $S^{z,{\rm G}}=m^{S,{\rm G}}/(g_e\mu_{\rm B})$, eqs.~(\ref{eq:mSG})
  and~(\ref{eq:mSlocG}),
  and impurity-induced magnetization $m^{\rm ii,G}/(g_e\mu_{\rm B})$,
  eqs.~(\ref{eq:miiG}) and~(\ref{eq:miilocalG}), 
  of the one-dimensional symmetric Kondo model
  as a function of global/local fields for $J_{\rm K}=1$
   from the Gutzwiller wave function.\label{fig:Gutzwillerms}}
\end{figure}

When the field is only applied locally, the same considerations lead to
\begin{equation}
  \frac{m^{\rm ii,G}_{\rm loc}(J_{\rm K},B)}{g_e\mu_{\rm B}}
  = \frac{1}{\pi} \arctan\left[
    \frac{E_{d,{\rm loc}}}{E_{d,{\rm loc}}K_{\rm loc}+V_{\rm loc}^2}\right]
  \; ,
  \label{eq:miilocalG}
\end{equation}
where the self-consistency problem has to be solved for a local field only.
We show the impurity spin polarization and the impurity-induced
magnetization as a function of an applied global/local magnetic field
in Fig.~\ref{fig:Gutzwillerms}
for $J_{\rm K}=1$, where the Gutzwiller wave function describes a local
spin singlet. 

In contrast to the Yosida wave function, the Gutzwiller wave function
correctly shows that the impurity spin polarization is larger than
the impurity-induced magnetization because the impurity spin is surrounded
by a cloud of conduction electrons that screens the impurity spin.
As in the Yosida wave function, the impurity spin polarization
does not depend much on whether the magnetic field is applied globally
of locally.

\subsubsection{Small fields}

In the paramagnetic phase, $J_{\rm K}>J_{\rm K,c}^{\rm G}$,
and for small fields, $B\to 0$, we can derive explicit results for the
zero-field impurity spin susceptibility because it is sufficient to solve
the self-consistency equations to linear order in the external field.

Keeping all terms up to linear order in~$B$, we make the Ansatz
 \begin{eqnarray}
   \omega_{p,\uparrow}&=&\omega_p +\bar{\omega}_pB  \; , \nonumber \\
   \omega_{p,\downarrow}&=&\omega_p -\bar{\omega}_pB  \; , \nonumber \\
   K&=& \bar{K} B\; , \nonumber \\
   E_d&=& \bar{E}_d B\; , \nonumber \\
   M_0&=& \bar{M}_0B \; , \nonumber \\
   m&=& 2\chi B\; ,
 \end{eqnarray}
 where $\chi$ is the desired zero-field
 impurity-spin susceptibility in units of $(g_e\mu_{\rm B})^2$,
 \begin{equation}
 \frac{\chi_0^{S,{\rm G}}(J_{\rm K},B)}{(g_e\mu_{\rm B})^2}=   \chi
   \; .
   \label{eq:Gutzchipara}
 \end{equation}
 In one dimension at $B=0$,
 the pole is at $  \omega_p=-v_+$, see eq.~(\ref{eq:defvplusminus}).
 Moreover, from eq.~(\ref{eq:chifromm}) and~eq.~(\ref{eq:miiG}) we find
 \begin{equation}
   \frac{ \chi_{0}^{\rm ii,G}(J_{\rm K})}{(g_e\mu_{\rm B})^2} 
   = \frac{\bar{E}_{d}}{2\pi V^2} \; ,
   \label{eq:chizeroiiGutzsmallfields}
  \end{equation}
 where $V_{\uparrow}=V_{\downarrow}=V$ and
 $\gamma_{\uparrow}=\gamma_{\downarrow}=\gamma=-2V/(3J_{\rm K})$,
 with corrections of the order $B^2$, and with
 \begin{equation}
J_{\rm K}(V)=-\frac{8V}{3} \left(\frac{\partial e_0(V)}{\partial V}\right)^{-1} \; ,
 \end{equation}
 where $V$ instead of~$J_{\rm K}$ parameterizes the strength of the
 Kondo interaction. For $e_0(V)$, see eq.~(\ref{eq:e0SIAM1d}).
 
 Apparently, we have five unknowns,
 namely
  \begin{equation}
   \underline{v}^{\rm T} =\left(
        \bar{\omega}_p,     \bar{E}_d, \bar{K}, \bar{M}_0, \chi
   \right) \;,
   \label{eq:defvvector}
 \end{equation}
 and we need five independent linear equations that connect these quantities.
 
\subsubsection{Useful integrals}

For later use we define the following set of integrals,
\begin{equation}
  J_n(V)= \int_{-1}^0 \frac{\rmd \omega}{\pi}
  \frac{\omega^n\sqrt{1-\omega^2}}{(\omega^2-\omega^4+V^4)^2}
  \; .
  \end{equation}
  Using {\sc Mathematica}~\cite{Mathematica11}
  the required integrals read
  \begin{eqnarray}
    J_1(V)&=& -\frac{1}{2\pi V^4 (1 + 4 V^4)} \nonumber \\
    &&
    +    \frac{\left(-2 + \sqrt{1 + 4 V^4}\right)\arctan(1/v_{-})}{2\pi v_{-}(1 + 4 V^4)^{3/2}}
    \nonumber \\
&&   +       \frac{\left(2 + \sqrt{1 + 4 V^4}\right)}{4\pi v_+(1 + 4 V^4)^{3/2}}
        \ln\left(\frac{v_+ - 1}{v_+ + 1}\right)
          \label{eq:J1explicit}
  \end{eqnarray}
  and
  \begin{eqnarray}
    J_3(V)&=& \frac{1}{\pi(1 + 4 V^4)}\nonumber \\
&& + \frac{(-3 - 4 V^4 + \sqrt{1 + 4 V^4}) \arctan(1/v_{-})}{
  4\pi v_{-}(1 + 4V^4)^{3/2}} \nonumber \\
&& +
\frac{\left(3 + 4 V^4 +      \sqrt{1 + 4V^4}\right)}{  8\pi   v_+(1 + 4V^4)^{3/2}}
\ln\left(\frac{v_+ - 1}{v_++ 1}\right)
\; .\nonumber \\
  \label{eq:J3explicit}
\end{eqnarray}
For $v_{\pm}$, see eq.~(\ref{eq:defvplusminus}).
  \subsubsection{Five equations}
As shown in appendix~\ref{sec:SIAMfull},
\begin{equation}
  \bar{E}_d=1-\frac{J_{\rm K}}{2}\bar{M}_0+8J_{\rm K}\chi \gamma^2
    \label{eq:desired1}
\end{equation}
with $\gamma=-2V/(3J_{\rm K})$,
\begin{equation}
  \bar{K}=J_{\rm K}\chi \; ,
    \label{eq:desired2}
\end{equation}
 \begin{equation}
   \bar{E}_d( \omega_p^2-1)+ (2\omega_p^2 -1)\bar{\omega}_p +\omega_p^2
   -    \bar{K} V^2=0
   \; ,   \label{eq:ompbar}
 \end{equation}
 \begin{eqnarray}
   \bar{M}_0 &=&   \bar{M}_0^{\rm b} + \bar{M}_0^{\rm band} \; ,   \label{eq:desired4}
\\
      \bar{M}_0^{\rm b} &=&
   \frac{  \bar{K} (-\omega_p^2 + 2 \omega_p^4 + V^4)
     -\omega_p^2 (3 + \bar{E}_d + 4 \bar{\omega}_p) V^2 }{
     \omega_p (1 - 2 \omega_p^2)^2}
 \nonumber\\
\bar{M}_0^{\rm band} &=&
-2\bar{K}V^2J_3(V)+2\left(\bar{E}_d-1\right)V^4J_1(V)
\; , \nonumber
 \end{eqnarray}
 \begin{eqnarray}
   \chi &=&\chi^{\rm b}+\chi^{\rm band}\; ,   \label{eq:desired5} \\
   \chi^{\rm b} &=&
   \frac{\omega_p (1 + \bar{E}_d+ \omega_p^2 - \bar{E}_d \omega_p^2
     + 2 \bar{\omega}_p - 
   \bar{K} V^2)}{2 (1 - 2 \omega_p^2)^2}
   \; , \nonumber \\
  \chi^{\rm band} &=& \frac{1}{2\pi V^2}\nonumber \\
  &&-V^2\left[\left(\bar{E}_d-1\right)(J_1(V)-J_3(V))+\bar{K}V^2J_1(V)\right]\,.\nonumber
\end{eqnarray}
 Eqs.~(\ref{eq:desired1})--(\ref{eq:desired5})
are the required five equations for the five unknowns in eq.~(\ref{eq:defvvector}).
 
\subsubsection{Matrix problem}

The resulting matrix problem reads with $\omega_p=-v_+$,
see eq.~(\ref{eq:defvplusminus}), and
with $\underline{v}$ from eq.~(\ref{eq:defvvector})
 \begin{equation}
   \underline{\underline{M}} \cdot \underline{v} = \underline{g} \; .
   \label{eq:Matrixequation}
 \end{equation}
 Here, the matrix has the form
 \begin{equation}
    \underline{\underline{M}} =\left(
  \begin{array}{@{}ccccc@{}}
    0 & 1 & 0 & J_{\rm K}/2 & -8 J_{\rm K}\gamma^2 \\
    0 & 0 & 1 & 0 & -J_{\rm K} \\
    2\omega_p^2-1 & \omega_p^2-1 & -V^2 & 0 & 0 \\
    M_{41} & M_{42}  & M_{43} & -1 & 0\\
    M_{51} & M_{52} & M_{53} & 0 & -1 
  \end{array}
   \right) 
   \label{eq:defMmatrix}
  \end{equation}
  with the matrix elements
  \begin{eqnarray}
    M_{41}&=& -\frac{4\omega_pV^2}{(1-2\omega_p^2)^2}
    \nonumber \; ,\\
    M_{42}&=& -\frac{\omega_{p}V^2}{(1-2\omega_p^2)^2}
    +2V^4J_1(V) \nonumber \; ,\\
    M_{43}&=&  \frac{(-\omega_p^2+2\omega_p^4+V^4)}{\omega_p(1-2\omega_p^2)^2}
    -2 V^2J_3(V)\nonumber \; ,\\
M_{51} &=&  \frac{\omega_p}{(1-2\omega_p^2)^2}
\nonumber \; ,\\
M_{52} &=& \frac{\omega_p(1-\omega_p^2)}{2(1-2\omega_p^2)^2}
-V^2\left(J_1(V)-J_3(V)\right)\nonumber \; ,\\
M_{53} &=&  -\frac{V^2\omega_p}{2(1-2\omega_p^2)^2}
-V^4J_1(V)\; .
  \end{eqnarray}
  For a global external field, the inhomogeneity reads
\begin{equation}
   \underline{g}^{\rm T} =\left(
     1, 0,    -\omega_p^2,  g_4,  g_5   \right) \;,
   \label{eq:defgvector}
 \end{equation}
where 
 \begin{eqnarray}
   g_4&=& \frac{3\omega_pV^2}{(1-2\omega_p^2)^2}   +2V^4J_1(V)\; , \nonumber \\
g_5 &=& -\frac{\omega_p(1+\omega_p^2)}{2(1-2\omega_p^2)^2}
\nonumber \\
&& -V^2\left(J_1(V)-J_3(V)\right)-\frac{1}{2\pi V^2}
\; .
 \end{eqnarray}
When the external field is applied only locally,
the matrix $\underline{\underline{M}}$
and the vector~$\underline{v}$ in eq.~(\ref{eq:Matrixequation})
remain unchanged but we have for the inhomogeneity
\begin{equation}
   \underline{g}_{\rm loc}^{\rm T} =\left(
1,0,0,0,0   \right) \; .
   \label{eq:defgvectorlocalfield}
\end{equation}
The matrix problem~(\ref{eq:Matrixequation})
can be solved analytically, generating large expressions.
Eventually, we solve it numerically.

\subsubsection{Strong-coupling limit}
\label{subsubsec:Gutzwilelrstrongcouling}

For the non-trivial entries in the matrix $\underline{\underline{M}}$ we have
\begin{eqnarray}
  M_{14}&=& \frac{2 V}{3} + \frac{1}{6 V} - \frac{4}{9\pi V^2}+\frac{5}{48V^3}
  \nonumber \;,\\
  M_{15} &=& -\frac{8 V}{3} + \frac{2}{3 V} - \frac{16}{9 \pi V^2}+\frac{1}{4V^3}\;,
  \nonumber \\
  M_{25} &=& -\frac{4 V}{3} - \frac{1}{3 V} + \frac{8}{9 \pi V^2}-\frac{5}{24V^3}
  \; ,
  \nonumber \\
M_{31} &=& 2V^2 + \frac{1}{4V^2}\;,\nonumber \\
M_{32} &=& V^2 -\frac{1}{2}+ \frac{1}{8V^2}\; , \nonumber \\
 M_{33} &=& -V^2 \; , \nonumber \\
 M_{41} &=& \frac{1}{V}+\frac{1}{4V^3}\;,\nonumber \\
 M_{42} &=& \frac{1}{4 V}+\frac{1}{16V^3} \; , 
\end{eqnarray}
and
\begin{eqnarray}
 M_{43} &=& -\frac{3}{4V}-\frac{1}{16V^3}\;,\nonumber\\
M_{51} &=& -\frac{1}{4V^3} \; , \nonumber \\
M_{52} &=& \frac{1}{8V}-\frac{1}{32V^3}\nonumber \; , \\
M_{53} &=& \frac{1}{8 V}+\frac{1}{32V^3}\; ,
\label{eq:Mvalsstrong}
\end{eqnarray}
up to and including order $1/V^3$.
To the same order,
    \begin{eqnarray}
     \underline{g}^{\rm T} &  =&
   \Bigl(
   1,0, -V^2-1/2-1/(8V^2),-3/(4V)-3/(16V^3), \nonumber \\
   && \hphantom{\Bigl(}
   1/(8V) - 1/(2\pi V^2)+7/(32V^3)   
   \Bigr)^{\rm T} \;.
   \end{eqnarray}
Lastly, for $J_{\rm K}\gg 1$ we have from eq.~(\ref{eq:VlargeGutz})
 \begin{equation}
   V(J_{\rm K})=\frac{3}{4} J_{\rm K} -
   \frac{1}{3 J_{\rm K}} + \frac{32}{27 \pi J_{\rm K}^2}
   - \frac{14}{27 J_{\rm K}^3} +{\cal O}(1/J_{\rm K}^4) \; .
 \end{equation}
Then, {\sc Mathematica}~\cite{Mathematica11}
gives the vector~$\underline{v}$,
    \begin{eqnarray}
      \underline{v}^{\rm T}   &=&
         \left(     \bar{\omega}_p, \bar{E}_d,\bar{K}, \bar{M}_0, \chi   \right) \nonumber \\
&=&
   \Bigl(
     -1- \frac{4}{3 \pi V}+\frac{29}{18\pi V^3},
     1 + \frac{13}{3 \pi V}-\frac{7}{36\pi V^3},\nonumber \\
&& \hphantom{\Bigl(}     \frac{5}{3 \pi V}+\frac{31}{36\pi V^3},
     -\frac{3}{2 \pi V^2},
     \frac{5}{4 \pi V^2}
   \Bigr)^{\rm T} \;,
   \end{eqnarray}
up to and including order $1/V^3$.
Thus, the strong-coupling limit of the impurity spin
susceptibility in the Gutzwiller wave function is given by
\begin{equation}
  \frac{\chi_0^{S,{\rm G}}(J_{\rm K}\gg 1)}{(g_e\mu_{\rm B})^2}=
  \frac{5}{4 \pi V^2}+{\cal O}(1/V^4) 
  =\frac{20}{9\pi J_{\rm K}^2}   +{\cal O}(1/J_{\rm K}^4)  \; .
  \label{eq:scformulafullchi}
\end{equation}

For a local external field we obtain in the strong-coupling limit
   \begin{eqnarray}
      \underline{v}_{\rm loc}^{\rm T}   &=&
   \Bigl(
     -1 + \frac{7}{12 V^2} -\frac{1}{3 \pi V^3},
     \frac{5}{2} + \frac{2}{3 V^2}+ \frac{1}{3 \pi V^3},\nonumber \\
&&\hphantom{\Bigl(}     \frac{1}{2} + \frac{7}{12 V^2}- \frac{1}{3 \pi V^3},
     - \frac{3}{4 V} + \frac{3}{16 V^3},\nonumber \\
&&\hphantom{\Bigl(}
     \frac{3}{8 V}+\frac{11}{32 V^3} 
     \Bigr)^{\rm T} \;,
   \end{eqnarray}
up to and including order $1/V^3$.
Thus, for a local magnetic field,
the strong-coupling limit of the impurity spin
susceptibility in the Gutzwiller wave function is given by
\begin{equation}
  \frac{\chi_{0,{\rm loc}}^{S,{\rm G}}(J_{\rm K}\gg 1)}{(g_e\mu_{\rm B})^2}=
 \frac{3}{8 V}+\frac{11}{32 V^3} 
 =\frac{1}{2 J_{\rm K}}+\frac{28}{27 J_{\rm K}^3}+{\cal O}(1/J_{\rm K}^4)  \; .
 \label{eq:GutzlargeJKlocalfield}
\end{equation}

For the zero-field impurity-induced susceptibilities
in eq.~(\ref{eq:chizeroiiGutzsmallfields})
we find in the strong-coupling limit
\begin{eqnarray}
  \frac{\chi_0^{\rm ii,G}(J_{\rm K}\gg 1)}{(g_e\mu_{\rm B})^2}&=&
  \frac{1}{2\pi V^2}\biggl[1 + \frac{13}{3\pi V} - \frac{7}{36\pi V^3}
    +{\cal O}(V^{-4})\biggr]   \nonumber \\
&=& \frac{8}{9 \pi J_{\rm K}^2} + \frac{416}{81 \pi^2 J_{\rm K}^3} +
  {\cal O}(1/J_{\rm K}^4) 
  \label{eq:chizeroiiGlargeJK}
\end{eqnarray}
and
\begin{eqnarray}
\frac{\chi_{0,{\rm loc}}^{\rm ii,G}(J_{\rm K}\gg 1)}{(g_e\mu_{\rm B})^2}&=&
\frac{1}{2\pi V^2}\biggl[\frac{5}{2} + \frac{2}{3 V^2} + \frac{1}{3\pi V^3}
  +{\cal O}(V^{-4})\biggr]   \nonumber \\
\nonumber \\
&=&\frac{20}{9 \pi J_{\rm K}^2}+  {\cal O}(1/J_{\rm K}^4) \;.
  \label{eq:chizeroiilocalGlargeJK}
\end{eqnarray}
The impurity spin susceptibility for a local field goes to zero proportional
to $1/J_{\rm K}$, all other susceptibilities vanish proportional to $J_{\rm K}^{-2}$. From
eqs.~(\ref{eq:scformulafullchi}) and~(\ref{eq:chizeroiilocalGlargeJK})
we see that $\chi_0^{S,{\rm G}}(J_{\rm K})$ and
$\chi_{0,{\rm loc}}^{\rm ii,G}(J_{\rm K})$ agree to order $J_{\rm K}^{-2}$.
A closed inspection shows that the expressions indeed agree to order
$J_{\rm K}^{-4}$.

\subsection{Hamiltonian of the Wilson chain}
\label{app:NRGchain}

The matrix elements are calculated recursively from the starting values
\begin{eqnarray}
  \varepsilon_{0,\sigma} &=& \int_{-1-\sigma_n B}^{1-\sigma_n B}\rmd \epsilon\,
  \epsilon \, \rho_0(\epsilon+\sigma_n B)
  \; , \nonumber \\
  t_{0,\sigma}^2 &=& \sum_{m}\Bigl[
    (\zeta_{m,\sigma}^+-\varepsilon_{0,\sigma})^2(\gamma^+_{m,\sigma})^2
    \nonumber \\
    &&\hphantom{\sum_{m}\Bigl[}
    +(\zeta_{m,\sigma}^--\varepsilon_{0,\sigma})^2(\gamma^-_{m,\sigma})^2\Bigr]
  \; , \nonumber \\
u_{0,m,\sigma}&=& \gamma_{m,\sigma}^+ \; , \nonumber \\
 v_{0,m,\sigma}&=& \gamma_{m,\sigma}^- \;, \nonumber \\
 u_{1,m,\sigma}&=&  \frac{(\zeta^+_{m,\sigma}-\varepsilon_{0,\sigma})u_{0,m,\sigma}}{
   t_{0,\sigma}}
  \; , \nonumber \\
  v_{1,m,\sigma}&=& \frac{(\zeta^-_{m,\sigma}-\varepsilon_{0,\sigma})v_{0,m,\sigma}}{
    t_{0,\sigma}}    \;.
\end{eqnarray}
For $n\geq 1$ one has to calculate iteratively
\begin{eqnarray}
  \varepsilon_{n,\sigma} &=& \sum_{m}
  \zeta_{m,\sigma}^+u_{n,m,\sigma}^2+\zeta_{m,\sigma}^-u_{n,m,\sigma}^2
  \; , \nonumber \\
  t_{n,\sigma}^2 &=& \sum_{m}\left[
    (\zeta^+_{m,\sigma})^2u^2_{n,m,\sigma}+(\zeta^-_{m,\sigma})^2v^2_{n,m,\sigma}\right]
\nonumber\\
&&  -t_{n-1,\sigma}^2-\varepsilon_{n,\sigma}^2 \; ,
\end{eqnarray}
and
\begin{eqnarray}
  u_{n+1,m,\sigma}&=& \frac{(\zeta_{m,\sigma}^+
    -\varepsilon_{n,\sigma})u_{n,m,\sigma}-t_{n-1,\sigma}u_{n-1,m,\sigma}}{t_{n,\sigma}}
    \; , \nonumber \\
    v_{n+1,m,\sigma}&=& \frac{(\zeta_{m,\sigma}^-
      -\varepsilon_{n,\sigma})v_{n,m,\sigma}-t_{n-1,\sigma}v_{n-1,m,\sigma}}{t_{n,\sigma}}
        \;.\nonumber \\
\end{eqnarray}
This concludes the derivation of the Wilson chain Hamiltonian.

\section{Outline of the supplemental material}
\label{mainappendixB}

In this appendix, we summarize the content of the supplemental material.

\mbox{}

\subsection{Hellmann-Feynman theorem}
\label{app:misc}

We formulate and prove
the Hellmann-Feynman theorem and its variational counterpart.

\subsection{Lanczos approach}
\label{app:A}

We recapitulate the Lanczos approach and show its variational property.
Moreover, we calculate the first-order Lanczos coefficients for a constant
and a one-dimensional density of states.

\subsection{Scattering off a local impurity}
\label{app:B}

We calculate the single-particle Green function for electrons that scatter off
a local potential. The results permit the calculation of the ground-state
energy of the Kondo model for large Kondo couplings.

\subsection{Non-interacting SIAM
  in the presence of potential scattering and a magnetic field}
\label{sec:SIAMfull}

We calculate the single-particle Green function for electrons in the
non-interacting single-impurity Anderson model in the presence
of a magnetic field and a local scattering potential.
We use it to calculate the ground-state energy, the magnetization,
and the zero-field susceptibility in the SIKM for the Gutzwiller wave function.

\subsection{Ground-state energy from Bethe Ansatz}
\label{app:GSEBetheAnsatz}

We collect the Bethe Ansatz equations and use them to
derive the ground-state energy.
Unfortunately, the Bethe Ansatz does not provide tangible results for this quantity
because the Bethe Ansatz energy is of third order in the Kondo coupling.

\mbox{}

\subsection{Free energy in second-order weak-coupling perturbation theory}
\label{app:freenergyPT}

We re-derive the expressions for the free energy to second order in the Kondo coupling
at finite temperature and external magnetic field for a general density of states
of the host electrons. These results are used
to calculate the Wilson number and the dominant term in the
zero-field susceptibilities whereby we generalize previous results in the literature.


%

\clearpage

\setcounter{section}{0}
\setcounter{subsection}{0}
\setcounter{subsubsection}{0}
\renewcommand{\thesection}{\Roman{section}}
\renewcommand{\theequation}{\Roman{section}-\arabic{equation}}
\renewcommand{\appendixname}{Supplement}
  
\onecolumngrid
\begin{center}
  {\Large Supplemental material}\\[6pt]
  {\large\bf Symmetric single-impurity Kondo model on a tight-binding chain:\\
a comparison of analytical and numerical ground-state approaches}\\[6pt]
  Gergely Barcza, Kevin Bauerbach, Fabian Eickhoff, Frithjof B.\ Anders,
  Florian Gebhard, \"Ors Legeza
\end{center}

{\small
The supplemental material consists of six parts. In supplement~I,
we formulate and prove
the Hellmann-Feynman theorem and its variational counterpart.
In supplement~II, 
we recapitulate the Lanczos approach and show its variational property.
Moreover, we calculate the first-order Lanczos coefficients for a constant
and a one-dimensional density of states.
In supplement~III,
we calculate the single-particle Green function for electrons that scatter off
a local potential. The results permit the calculation of the ground-state
energy of the Kondo model for large Kondo couplings.
In supplement~IV,
we calculate the single-particle Green function for electrons in the
non-interacting single-impurity Anderson model in the presence
of a magnetic field and a local scattering potential.
We use it to calculate the magnetization and the zero-field susceptibility
in the SIKM for the Gutzwiller wave function.
We also discuss the magnetic transition in the Gutzwiller wave function.
In supplement~V,
we collect the Bethe Ansatz equations and use them to
derive the ground-state energy.
Unfortunately, the Bethe Ansatz does not provide tangible results for 
this quantity.
In supplement~VI,
we re-derive the expressions for the free energy to second order 
in the Kondo coupling
at finite temperature and external magnetic field 
for a general density of states
of the host electrons. These results are used
to calculate the Wilson number and the dominant term in the
zero-field susceptibilities whereby we generalize previous results 
in the literature.
\vspace*{12pt}}

\twocolumngrid

\section{Hellmann-Feynman theorem}
\label{supp:misc}

In preparation for the variational Hellmann-Feynman theorem, 
we first recapitulate the proof of the theorem
for the exact ground state, and then move on to its variational counterpart.

\subsection{Exact ground state}
\label{appsub:HFexaact}

Let $|\Psi_0(\lambda)\rangle$
be the exact ground state of the Hamiltonian~$\hat{H}(\lambda)$,
\begin{equation}
  \hat{H}(\lambda)  | \Psi_0(\lambda)\rangle = E_0(\lambda) | \Psi_0(\lambda)\rangle\;,
  \label{eq:Schroedingerequation}
\end{equation}
where $\lambda$ is some real parameter, e.g., the
interaction strength.
Taking the derivative with respect to $\lambda$ results in
\begin{eqnarray}
  \frac{\partial \hat{H}(\lambda)}{\partial \lambda}
  | \Psi_0(\lambda)\rangle
  &=& \left(E_0(\lambda) - \hat{H}(\lambda)\right)  | \dot{\Psi}_0(\lambda)\rangle
  \nonumber \\
  &&  + \frac{\partial E_0(\lambda)}{\partial \lambda}| \Psi_0(\lambda)\rangle
  \;, 
\end{eqnarray}
where the dot on the wave function implies the derivative with respect to $\lambda$.
We multiply this equation with $\langle \Psi_0(\lambda)|$ and use
the Schr\"odinger equation~(\ref{eq:Schroedingerequation}) to find
\begin{equation}
  \langle \Psi_0(\lambda) |
  \frac{\partial \hat{H}(\lambda)}{\partial \lambda}
  | \Psi_0(\lambda)\rangle
  =
  \langle \Psi_0(\lambda) |
  \frac{\partial E_0(\lambda)}{\partial \lambda}| \Psi_0(\lambda)\rangle
\end{equation}
or
\begin{equation}
    \frac{\partial E_0(\lambda)}{\partial \lambda}
=\frac{  \langle \Psi_0(\lambda) |
  \frac{\displaystyle \partial \hat{H}(\lambda)}{\displaystyle \partial \lambda}
  | \Psi_0(\lambda)\rangle}{
  \langle \Psi_0(\lambda) |\Psi_0(\lambda)\rangle} \; ,
\label{eq:HFtheorem}
\end{equation}
which constitutes the Hellmann-Feynman theorem.~\cite{Hellmann2,Feynman}

\subsection{Variational state}

The variationally optimized state $|\Phi_{\rm G}(\lambda)\rangle$
provides an upper bound to the
ground-state energy of $\hat{H}(\lambda)$,
\begin{equation}
  E_0(\lambda) \leq 
  E_0^{\rm var}(\lambda)=
  \frac{  \langle \Phi_{\rm G}(\lambda) |
  \hat{H}(\lambda)  | \Phi_{\rm G}(\lambda)\rangle}{
    \langle \Phi_{\rm G}(\lambda) |\Phi_{\rm G}(\lambda)\rangle} \; .
  \label{eq:Evardef}
\end{equation}
The variational optimization implies that a linear deviation from
$|\Phi_{\rm G}(\lambda)\rangle$ results in a quadratic change in the variational
energy,
\begin{equation}
  \frac{  \langle \Phi_{\rm G}(\lambda) +\delta \Phi |
  \hat{H}(\lambda)  | \Phi_{\rm G}(\lambda)+\delta \Phi \rangle}{
    \langle \Phi_{\rm G}(\lambda)+\delta \Phi |\Phi_{\rm G}(\lambda)+\delta \Phi\rangle}
  =  E_0^{\rm var}(\lambda) +{\cal O}\left( (\delta\Phi)^2\right)\, .
  \label{eq:PhiGstationary}
\end{equation}
We shift $\lambda$ by $(\rmd \lambda)$ in eq.~(\ref{eq:Evardef})
to find up to linear order in $(\rmd\lambda)$
\begin{eqnarray}
  \hbox{lhs}&=&\hbox{rhs} \; , \nonumber \\
\hbox{lhs}&=& E_0^{\rm var}(\lambda)+ 
\frac{\partial E_0^{\rm var}(\lambda)}{\partial \lambda} (\rmd \lambda) \; ,\nonumber \\
\hbox{rhs} &=& 
  \frac{  \langle \Phi_{\rm G}(\lambda+\rmd \lambda) |
  \hat{H}(\lambda)  | \Phi_{\rm G}(\lambda+\rmd \lambda)\rangle}{
    \langle \Phi_{\rm G}(\lambda+\rmd \lambda)
    |\Phi_{\rm G}(\lambda+\rmd \lambda)\rangle} \nonumber \\
  && +
\frac{  \langle \Phi_{\rm G}(\lambda) |
  \frac{\displaystyle \partial \hat{H}(\lambda)}{\displaystyle\partial \lambda}
| \Phi_{\rm G}(\lambda)\rangle}{\langle \Phi_{\rm G}(\lambda)|\Phi_{\rm G}(\lambda)\rangle}
(\rmd \lambda)
\nonumber \\
&=& E_0^{\rm var}(\lambda) +{\cal O}\left((\rmd\lambda)^2\right) \nonumber \\
&&+
\frac{  \langle \Phi_{\rm G}(\lambda) |
  \frac{\displaystyle \partial \hat{H}(\lambda)}{\displaystyle\partial \lambda}
| \Phi_{\rm G}(\lambda)\rangle}{\langle \Phi_{\rm G}(\lambda)|\Phi_{\rm G}(\lambda)\rangle}
(\rmd \lambda) \; ,
\end{eqnarray}
where we used eq.~(\ref{eq:PhiGstationary}) in the last step; note that
$|\delta\Phi\rangle =|\Phi_{\rm G}(\lambda+\rmd\lambda)\rangle -
|\Phi_{\rm G}(\lambda)\rangle$ is proportional to $(\rmd \lambda)$.
Therefore, we obtain the variational Hellmann-Feynman theorem,
\begin{equation}
  \frac{\partial E_0^{\rm var}(\lambda)}{\partial \lambda}
  =
  \frac{  \langle \Phi_{\rm G}(\lambda) |
  \frac{\displaystyle \partial \hat{H}(\lambda)}{\displaystyle\partial \lambda}
| \Phi_{\rm G}(\lambda)\rangle}{\langle \Phi_{\rm G}(\lambda)|\Phi_{\rm G}(\lambda)\rangle}
  \; .
\label{eq:VariationalHFtheorem}
\end{equation}

\section{Lanczos approach}
\label{supp:A}

\subsection{Lanczos method}
\label{sec:Lanczosmethod}

Let $\hat{H}$ be the Hamiltonian of the system
and let $|\Psi_0\rangle$ be its (non-degenerate) ground state,
\begin{equation}
\hat{H} |\Psi_0\rangle = E_0 |\Psi_0\rangle \; ,
  \end{equation}
where $E_0$ is the exact ground-state energy.
Moreover, we assume that the Hilbert space of dimension $d=\hbox{dim}(\hat{H})$
does not split into orthogonal subspaces.

\subsubsection{Lanczos vectors}

To approximate the ground state and its energy,
we start from some properly chosen state $|\Phi_0\rangle$
that is not orthogonal to the ground state,
$\langle \Phi_0 | \Psi_0\rangle\neq 0$.
The next states are constructed recursively,~\cite{Lanczos:1950:IMS,EricKoch}
\begin{equation}
|\Phi_{n+1}\rangle = \hat{H}|\Phi_n\rangle - a_n |\Phi_n\rangle - b_n^2 |\Phi_{n-1}\rangle 
\quad , \quad n\geq 0 \; ,
\label{eq:Phindef}
  \end{equation}
where we set $b_0\equiv 0 $, and we assume that $a_n$ and $b_n>0$ are real.
The states $|\Phi_n\rangle$ are not normalized to unity but they are
orthogonal to each other when $a_n$ and $b_n^2$ are chosen properly.

The proof is accomplished by induction. Let as assume for some $n\geq 2$ that
\begin{equation}
  \langle \Phi_i| \Phi_n\rangle =0 \quad , \quad \hbox{for} 
\quad i=0,1,2,\ldots,n-1
  \label{eq:orthoindassumption}
\end{equation}
holds true. Then, we calculate for $0\leq i\leq n-2$,
\begin{eqnarray}
  \langle \Phi_i| \Phi_{n+1}\rangle &=&\langle \Phi_i| \hat{H}| \Phi_n\rangle
  -a_n \langle \Phi_i| \Phi_n\rangle -b_n^2\langle \Phi_i| \Phi_{n-1}\rangle \nonumber \\
    &=& \left(\langle \Phi_n| \hat{H}| \Phi_i\rangle \right)^* \nonumber \\
  &=& \left(\langle \Phi_n |\Phi_{i+1}\rangle+a_i\langle\Phi_n|\Phi_i\rangle
  +b_i^2\langle \Phi_n|\Phi_{i-1}\rangle  \right)^* \nonumber \\
  &=& 0 \; , 
  \end{eqnarray}
where we used eq.~(\ref{eq:Phindef})
in the first and third step and eq.~(\ref{eq:orthoindassumption}) in the second
and fourth step. For $i=n-1$ we find along the same lines that
\begin{eqnarray}
  \langle \Phi_{n-1}| \Phi_{n+1}\rangle  &=&  -b_n^2\langle \Phi_{n-1}| \Phi_{n-1}\rangle
+  \left(\langle \Phi_n| \hat{H}| \Phi_{n-1}\rangle \right)^* \nonumber \\
&=& -b_n^2\langle \Phi_{n-1}| \Phi_{n-1}\rangle + \langle \Phi_n |\Phi_n\rangle
\stackrel{!}{=} 0 \; , 
  \end{eqnarray}
which is fulfilled if we choose
\begin{equation}
  b_n^2=\frac{\langle \Phi_n |\Phi_n\rangle}{\langle \Phi_{n-1}| \Phi_{n-1}\rangle} \geq 0\; .
  \label{eq:bnresult}
\end{equation}
Furthermore, for $i=n$ we find
\begin{eqnarray}
  \langle \Phi_n| \Phi_{n+1}\rangle &=&\langle \Phi_n| \hat{H}| \Phi_n\rangle
  -a_n \langle \Phi_n| \Phi_n\rangle -b_n^2\langle \Phi_n| \Phi_{n-1}\rangle
  \nonumber \\
  &=&    \langle \Phi_n| \hat{H}| \Phi_n\rangle -a_n\langle \Phi_n| \Phi_n\rangle
  \stackrel{!}{=} 0 \; , 
  \end{eqnarray}
which is fulfilled if we choose 
\begin{equation}
  a_n=\frac{\langle \Phi_n |\hat{H}|\Phi_n\rangle}{\langle \Phi_n| \Phi_n\rangle} \; .
  \label{eq:andef}
\end{equation}
This closes the induction.

\subsubsection{Hamilton matrix}

Using the orthonormal set of basis states
\begin{equation}
|\varphi_n\rangle = \frac{|\Phi_n\rangle}{\sqrt{\langle \Phi_n | \Phi_n\rangle}}
\quad , \quad 0\leq n\leq d-1\; ,
\end{equation}
we can write the Hamiltonian in the form
\begin{equation}
  \hat{H}= \sum_{l,m} |\varphi_l\rangle H_{l,m} \langle \varphi_m|
  \quad, \quad
  H_{l,m} =    \langle \varphi_l |\hat{H}| \varphi_m\rangle \; .
\end{equation}
We have from the definition~(\ref{eq:Phindef})
\begin{eqnarray}
  H_{l,m}&=& \frac{
      \langle \Phi_l |\Phi_{m+1}\rangle + a_m \langle \Phi_l|\Phi_m\rangle
      +b_m^2\langle \Phi_l|\Phi_{m-1}\rangle
    }{
    \sqrt{\langle \Phi_l | \Phi_l\rangle}
    \sqrt{\langle \Phi_m | \Phi_m\rangle}
    } \nonumber \\
  &=& \delta_{l,m+1} \frac{ \sqrt{\langle \Phi_l | \Phi_l\rangle}}{
    \sqrt{\langle \Phi_{l-1} | \Phi_{l-1}\rangle}}
  + \delta_{l,m} a_l \nonumber \\
  && +\delta_{l,m-1} b_{l+1}^2
     \frac{ \sqrt{\langle \Phi_l | \Phi_l\rangle}}{
       \sqrt{\langle \Phi_{l+1} | \Phi_{l+1}\rangle}} \nonumber \\
     &=& \delta_{l,m+1} b_l + \delta_{l,m} a_l +\delta_{l,m-1} b_{l+1}\; ,
     \label{eq:Hlmelements}
\end{eqnarray}
where we used eq.~(\ref{eq:bnresult}).
This shows that the Hamilton matrix $\underline{\underline{H}}$ with the
entries $H_{l,m}$ is tridiagonal with real entries in the diagonal and
in its neighboring sub-diagonals.~\cite{Lanczos:1950:IMS,EricKoch}

\subsubsection{Variational Lanczos Ansatz}

We choose the variational Lanczos Ansatz
\begin{equation}
  |\Psi\rangle = \sum_{n=0}^M \lambda_n |\varphi_n\rangle \quad, \quad
    \lambda_n\in \mathbb{C} 
\end{equation}
with $M\geq 0$. Its norm is given by
\begin{equation}
  \langle \Psi| \Psi\rangle
  =  \sum_{n=0}^M \lambda_n^*\lambda_n \; .
  \label{eq:norm}
  \end{equation}
The expectation value of the Hamiltonian is given by
\begin{eqnarray}
  \langle \Psi |\hat{H}|\Psi\rangle
  &=& \sum_{n,m=0}^M \lambda_m^* \lambda_n
  \langle \varphi_m|\hat{H}| \varphi_n \rangle \nonumber \\
  &=& \sum_{n=0}^M\left(\lambda_n^* \lambda_na_n+
  b_n \left(  \lambda_{n-1}^*\lambda_n +\lambda_n^*\lambda_{n-1}\right)\right) \;,
  \nonumber \\
    \end{eqnarray}
where we used eq.~(\ref{eq:Hlmelements}) and $b_0=0$.

We employ Ritz' variational principle to find an approximation to the ground-state
energy by minimizing the energy functional
\begin{eqnarray}
  L\left[\left\{\lambda_n\right\},\Xi\right]
  &=&
  \langle \Psi |\hat{H}|\Psi\rangle -\Xi \left(\langle \Psi| \Psi\rangle -1\right)\nonumber \\
  &=&
\sum_{n=0}^M\left(\lambda_n^* \lambda_na_n+
b_n \left(  \lambda_{n-1}^*\lambda_n +\lambda_n^*\lambda_{n-1}\right)\right)
\nonumber \\
&& -\Xi\biggl(\sum_{n=0}^M \lambda_n^*\lambda_n -1\biggr)
  \end{eqnarray}
with respect to $\lambda_n$ and the Lagrange parameter $\Xi$
that takes into account the normalization condition~(\ref{eq:norm}).

{}From
\begin{equation}
\frac{\partial L\left[\lambda_n,\Xi\right]}{\partial \lambda_l^*}=0
\end{equation}
we obtain
\begin{equation}
  \lambda_l a_l +b_{l+1}\lambda_{l+1}+b_l\lambda_{l-1} = \Xi \lambda_l
  \label{eq:Minieq}
\end{equation}
for $l=0,1,\ldots,M$. Equation~(\ref{eq:Minieq})
shows that the vector $\veclambda^{(M)}$ with the entries
$\veclambda_{\,l}^{(M)}=\lambda_l$ for $l=0,1,\ldots,M$ is an eigenvector of the 
truncated $(M+1)\times (M+1)$
Hamilton-Matrix $\underline{\underline{H}}^{(M)}$ with the
entries $H_{l,m}^{(M)}=H_{l,m}$ for $0\leq l,m\leq M$,
\begin{equation}
  \underline{\underline{H}}^{(M)} \cdot \veclambda^{(M)}= \Xi^{(M)} \veclambda^{(M)} \;.
  \label{eq:Minieqvectorform}
\end{equation}
Consequently, the lowest eigenvalue of $\underline{\underline{H}}^{(M)}$
gives an upper bound to the ground-state energy,
\begin{equation}
  E_0 \leq \Xi_0^{(M)} \; .
  \label{eq:EXibound}
\end{equation}
More importantly, the estimate improves systematically with~$M$
because we extend our variational space as we increase~$M$.
Therefore, we have for all $M\geq 1$,
\begin{equation}
  E_0 \leq \Xi_0^{(M)}\leq \Xi_0^{(M-1)} \; .
  \label{eq:EXiXibound}
\end{equation}

\subsection{Matrix elements for the Lanczos calculations}
\label{app:A2}

For $B=0$, we consider the two Lanczos states $|\Phi_0\rangle$
from eq.~(\ref{eq:Phizero})
and
\begin{equation}
  |\Phi_1\rangle =\left(\hat{H}_{\rm K}-E_{\rm FS}\right)
  |\Phi_0\rangle \; ,
  \end{equation}
where we used that $a_0=\langle \Phi_0 | \hat{H}_{\rm K}-E_{\rm FS}| \Phi_0\rangle
=0$ in the thermodynamic limit, see eq.~(\ref{appeq:a0iszero}). More importantly,
\begin{equation}
  \left(\hat{T}-E_{\rm FS}\right) |\Phi_0\rangle = 0\; .
  \label{eq:TonPhi0iszero}
\end{equation}
We employ this relation repeatedly in this section.

\subsubsection{Calculation of $b_1$}

Using eq.~(\ref{eq:TonPhi0iszero}) we find from eq.~(\ref{eq:bnresult})
\begin{equation}
b_1^2= \langle \Phi_0 |\hat{V}_{\rm sd}^2| \Phi_0\rangle \; ,
\end{equation}
where we used that $|\Phi_0\rangle$ is normalized, $\langle \Phi_0| \Phi_0\rangle=1$.

We consider local spins for the calculation of the square of $V_{\rm sd}$,
\begin{eqnarray}
\frac{\hat{V}_{\rm sd}}{J_{\rm K}}&=& 
\frac{1}{2}\left(\hat{s}_0^+\hat{S}^-+\hat{s}_0^-\hat{S}^+\right) 
+\hat{s}_0^z\hat{S}^z 
\nonumber \; , \\
\left(\frac{\hat{V}_{\rm sd}}{J_{\rm K}}\right)^2&=& 
\frac{1}{4}\left[ \hat{s}_0^+\hat{s}_0^-\hat{S}^-\hat{S}^+
+
\hat{s}_0^-\hat{s}_0^+\hat{S}^+\hat{S}^-\right] +(\hat{s}_0^z)^2(\hat{S}^z)^2
\nonumber \\
&& + \frac{1}{2} \left[ 
\hat{s}_0^z\hat{S}^z \left(\hat{s}_0^+\hat{S}^-+\hat{s}_0^-\hat{S}^+\right)
+\hbox{h.c.}\right]\; ,
\label{eq:alltermsinVsquare}
\end{eqnarray}
where we used $(\hat{S}^+)^2=(\hat{S}^-)^2=0$ for the spin-1/2 impurity.
Moreover, for the same reason we have 
\begin{eqnarray}
\hat{S}^+\hat{S}^- &=& \hat{d}_{\uparrow}^+\hat{d}_{\uparrow}^{\vphantom{+}}
=\hat{n}_{\uparrow}^d \;, \nonumber \\
\hat{S}^-\hat{S}^+ &=&\hat{d}_{\downarrow}^+\hat{d}_{\downarrow}^{\vphantom{+}}
=\hat{n}_{\downarrow}^d \;, \nonumber \\
(\hat{S}^z)^2&=& \frac{1}{4} \; , \nonumber \\
\hat{S}^z\hat{S}^+
&=& \frac{1}{2}\hat{S}^+ = - \hat{S}^+\hat{S}_z\; , \nonumber \\
\hat{S}^z\hat{S}^-&=& -\frac{1}{2}\hat{S}^-
= - \hat{S}^-\hat{S}^z\; , \nonumber \\
\hat{s}_0^+\hat{s}_0^- &=& \hat{m}_{0,\uparrow}^c\equiv
\hat{n}_{0,\uparrow}^c (1-\hat{n}_{0,\downarrow}^c)\;, \nonumber \\
\hat{s}_0^-\hat{s}_0^+ &=& \hat{m}_{0,\downarrow}^c\equiv
\hat{n}_{0,\downarrow}^c (1-\hat{n}_{0,\uparrow}^c)
\; , \nonumber \\
(\hat{s}_0^z)^2&=&
\frac{1}{4} \left(\hat{m}_{0,\uparrow}^c+\hat{m}_{0,\downarrow}^c\right)
\; , \nonumber \\
\hat{s}_0^z\hat{s}_0^+&=& \frac{1}{2}\hat{s}_0^+
= - \hat{s}_0^+\hat{s}_0^z\; , \nonumber \\
\hat{s}_0^z\hat{s}_0^-&=& -\frac{1}{2}\hat{s}_0^-= - \hat{s}_0^-\hat{s}_0^z
\end{eqnarray}
with the projection operators $m_{0,\sigma}^c$.
Therefore, 
\begin{eqnarray}
\left(2\frac{\hat{V}_{\rm sd}}{J_{\rm K}}\right)^2&=&
\hat{n}_{\uparrow}^d\hat{m}_{0,\downarrow}^c
+
\hat{n}_{\downarrow}^d\hat{m}_{0,\uparrow}^c
+
\frac{1}{4}\left( \hat{m}_{0,\uparrow}^c+\hat{m}_{0,\downarrow}^c\right)
\nonumber \\
&& -\left(\hat{s}_0^+\hat{S}^-+\hat{s}_0^-\hat{S}^+\right) \nonumber \\
&=& 
\frac{3}{4}\left( \hat{m}_{0,\uparrow}^c+\hat{m}_{0,\downarrow}^c\right) 
- \left(\frac{2\hat{V}_{\rm sd}}{J_{\rm K}}\right) \; .
\label{eq:Vsdsquarecompact}
\end{eqnarray}
Since $a_0=0$, see eq.~(\ref{appeq:a0iszero}),
the first term in eq.~(\ref{eq:Vsdsquarecompact})
leads to 
\begin{eqnarray}
b_1^2&=&\frac{J_{\rm K}^2}{4} \frac{3}{4} \langle \hbox{FS} | 
\left(\hat{n}_{0,\uparrow}^c+
\hat{n}_{0,\downarrow}^c-2\hat{n}_{0,\uparrow}^c\hat{n}_{0,\uparrow}^c\right)
|\hbox{FS}\rangle \nonumber \\
&=& \frac{3J_{\rm K}^2}{16}
\left(\frac{1}{2} +\frac{1}{2} -2 \frac{1}{2} \frac{1}{2} \right)
= \frac{3}{32} J_{\rm K}^2 \; .
\end{eqnarray}
For comparison with Mancini and Mattis we use $J_{\rm K}=-2J$ to find
$b_1^2=3J^2/8$, as given in table~I of Ref.~[\onlinecite{PhysRevB.31.7440}].

\subsubsection{Calculation of $a_1$}

{}From eq.~(\ref{eq:andef}) we have 
\begin{eqnarray}
a_1 b_1^2 &= & \langle \Phi_0 |\hat{H}^3| \Phi_0\rangle \nonumber \\
&=& 
\langle \Phi_0 |\hat{V}_{\rm sd} \left(\hat{T}-E_{\rm FS}\right) \hat{V}_{\rm sd}
| \Phi_0\rangle +
\langle \Phi_0 |\hat{V}_{\rm sd}^3| \Phi_0\rangle \; ,\nonumber \\
\label{eq:aonecalc}
\end{eqnarray}
where we used eq.~(\ref{eq:TonPhi0iszero}) and the fact
that $|\Phi_0\rangle$ is normalized to unity,
$\langle \Phi_0| \Phi_0\rangle=1$.

\paragraph{First term in eq.~(\ref{eq:aonecalc}):}

Again, the spin-flip terms give only a contribution of the order $1/L$ 
so that we are left with
\begin{eqnarray}
1^{\rm st}&=& \left(\frac{J_{\rm K}}{2}\right)^2
2 \frac{1}{2}\biggl[
  \langle A | \hat{s}_0^-\left(\hat{T}-E_{\rm FS}\right)
  \hat{s}_0^+
  |A\rangle \nonumber \\
  && \hphantom{\left(\frac{J_{\rm K}}{2}\right)^2
2 \frac{1}{2}\biggl[}
+\langle A | \hat{s}_0^z \left(\hat{T}-E_{\rm FS}\right)\hat{s}_0^z |A\rangle
\biggr]\nonumber \\
&=& \frac{3J_{\rm K}^2}{16}
\langle \hbox{FS} | 
\left(\hat{n}_{0,\uparrow}^c-\hat{n}_{0,\downarrow}^c\right)
\bigl(\hat{T}-E_{\rm FS}\bigr)
\left(\hat{n}_{0,\uparrow}^c-\hat{n}_{0,\downarrow}^c\right)
| \hbox{FS}\rangle \nonumber \\
&=&  \frac{3J_{\rm K}^2}{16} 2 \frac{1}{L^2} \sum_{p_1,p_2,k_1,k_2}
\left(\epsilon_{k_1}-\epsilon_{k_2}\right)
\nonumber \\
&& \hphantom{\frac{3J_{\rm K}^2}{16} 2 \frac{1}{L^2} \sum_{p_1,p_2,k_1,k_2}}
\times 
\langle \hbox{FS} | 
\hat{a}_{p_1,\uparrow}^+\hat{a}_{p_2,\uparrow}^{\vphantom{+}}
\hat{a}_{k_1,\uparrow}^+\hat{a}_{k_2,\uparrow}^{\vphantom{+}}
| \hbox{FS}\rangle \nonumber \\
&=& 
 \frac{3J_{\rm K}^2}{8}  \frac{1}{L^2} \sum_{\epsilon_{k_1}>E_{\rm F},\epsilon_{k_2}<E_{\rm F}}
\left(\epsilon_{k_1}-\epsilon_{k_2}\right) \nonumber \\
&=& \frac{3J_{\rm K}^2}{8}  \frac{1}{L} \sum_{\epsilon_k>E_{\rm F}}
\epsilon_k = \frac{3J_{\rm K}^2}{8} \int_0^1\rmd \omega \omega
\rho_0(\omega)\; , 
\label{eq:final1st}
\end{eqnarray}
where we used spin symmetry in the second and third step,
$\sum_k\epsilon_k=0$, and $(1/L)\sum_{\epsilon_k>E_{\rm F}}=
(1/L)\sum_{\epsilon_k<E_{\rm F}}=1/2$ at half band-filling in the last step.

For a constant density of states, $\rho_0^{\rm const}(\omega)=1$
for $|\omega|\leq 1$,
and $J_{\rm K}=-2J$ we find $(3J^2/2)(1/8)=3J^2/16$ for the first term,
in agreement with table~I of Ref.~[\onlinecite{PhysRevB.31.7440}].
For the one-dimensional density of states~(\ref{eq:DOS1d}),
we find $(3J_{\rm K}^2/8)(1/\pi)$ for the first term.

\paragraph{Second term in eq.~(\ref{eq:aonecalc}):}

We use eq.~(\ref{eq:Vsdsquarecompact}) to find
\begin{eqnarray}
\left(\frac{2\hat{V}_{\rm sd}}{J_{\rm K}}\right)^3&=& 
\left[-  \left(\frac{2\hat{V}_{\rm sd}}{J_{\rm K}}\right) +
\frac{3}{4}\left( \hat{m}_{0,\uparrow}^c+\hat{m}_{0,\downarrow}^c\right)
\right]\left(\frac{2\hat{V}_{\rm sd}}{J_{\rm K}}\right)
\nonumber \\
&=&
\frac{7}{4}\left(\frac{2\hat{V}_{\rm sd}}{J_{\rm K}}\right) 
-\frac{3}{4}\left( \hat{m}_{0,\uparrow}^c+\hat{m}_{0,\downarrow}^c\right)  
\; .
\end{eqnarray}
Using $a_0=0$, see eq.~(\ref{appeq:a0iszero}), we find
\begin{equation}
2^{\rm nd}= \frac{J_{\rm K}^3}{8} \left(-\frac{3}{4}\right)
\left(\frac{1}{2} +\frac{1}{2} -2 \frac{1}{2} \frac{1}{2} \right)
=-\frac{3J_{\rm K}^3}{64} \; .
\label{eq:final2nd}
\end{equation}
Using $J_{\rm K}=-2J$ we find $(3J^2)/8$ for the second term, 
as given in table~I of Ref.~[\onlinecite{PhysRevB.31.7440}].

Thus, we find from eq.~(\ref{eq:final1st}) and~eq.~(\ref{eq:final2nd})
that
\begin{eqnarray}
a_1&=& \frac{-(3J_{\rm K}^3/64)+
(3J_{\rm K}^2/8)\int_0^1 \rmd \omega \omega\rho_0(\omega)}{
3J_{\rm K}^2/32} \nonumber \\
&=&-\frac{J_{\rm K}}{2} + 4 \int_0^1\rmd \omega \omega\rho_0(\omega)
\; .
\end{eqnarray}
For a constant density of states with unit bandwidth and $J_{\rm K}=-2J$, 
this reduces to $a_1^{\rm const}=J+1/2$, 
as given in table~I of Ref.~[\onlinecite{PhysRevB.31.7440}].
For the one-dimensional density of states~(\ref{eq:DOS1d})
we find 
\begin{equation}
a_1^{\rm 1d}=-\frac{J_{\rm K}}{2} + \frac{4}{\pi} 
\label{eq:a1for1d}\; .
\end{equation}

The second-order Lanczos matrix $\underline{\underline{H}}^{(2)}$
requires the calculation of expectation values $\langle \Phi_0 | \hat{H}_{\rm K}^5
|\Phi_0\rangle$ which is cumbersome and prone to errors.
For example, the matrix $\underline{\underline{H}}^{(2)}(J)$
given in Ref.~[\onlinecite{PhysRevB.31.7440}] has a negative
eigenvalue at $J=0$ which violates the variational property~(\ref{eq:EXibound}).

\section{Scattering off a local impurity}
\label{supp:B}

We investigate the potential scattering problem
\begin{equation}
  \hat{H}_{\rm ps}=\sum_k\tilde{\epsilon}(k) \hat{a}_k^{+}\hat{a}_k^{\vphantom{+}}
+\frac{V}{L} \sum_{k,p} \hat{a}_k^{+}\hat{a}_p^{\vphantom{+}}
\label{eq:defHpotential}
\end{equation}
for a system with $L$ sites, and periodic boundary conditions apply.

\subsection{Calculation of the Green function}

We need to calculate the retarded Green function
\begin{equation}
  G_{k,p}^{\rm ret}(t)
  = (-\rmi) \theta_{\rm H}(t) \langle
  \left[ \hat{a}_k^{\vphantom{+}}(t),\hat{a}_p^+\right]_+\rangle
  \; ,
  \end{equation}
where $\hat{A}(t)=\exp(\rmi\hat{H}_{\rm ps}t)\hat{A}\exp(-\rmi\hat{H}_{\rm ps}t)$
is the Heisenberg operator assigned to the Schr\"odinger operator~$\hat{A}$.

\subsubsection{Equation-of-motion method}

The time derivative of the retarded Green function obeys
\begin{equation}
  \rmi \dot{G}_{k,p}^{\rm ret}(t) =\delta_{k,p}\delta(t)
    +\tilde{\epsilon}(k) G_{k,p}^{\rm ret}(t)
    +\frac{V}{L} \sum_{p'} G_{p',p}^{\rm ret}(t)  \; .
  \end{equation}
A Fourier transformation leads to the result ($\eta=0^+$)
\begin{equation}
  \tilde{G}_{k,p}^{\rm ret}(\omega)
  =\frac{\delta_{k,p}+V H_p(\omega)}{
    \omega-\tilde{\epsilon}(k)+\rmi \eta}
  \label{eq:halfwayGF}
  \end{equation}
with the abbreviation
\begin{equation}
  H_p(\omega) =\frac{1}{L} \sum_{p'} \tilde{G}_{p',p}^{\rm ret}(\omega) \;.
  \label{eq:Hpomegadef}
  \end{equation}
We insert eq.~(\ref{eq:halfwayGF}) into eq.~(\ref{eq:Hpomegadef})
to find
\begin{eqnarray}
  H_p(\omega) &=&\frac{1}{L} \sum_{p'}
  \frac{\delta_{p',p}+V H_p(\omega)}{    \omega-\tilde{\epsilon}(p')+\rmi \eta} \nonumber \\
  &=& \frac{1}{L} \frac{1}{\omega-\tilde{\epsilon}(p)+\rmi \eta}+
  Vg_0(\omega) H_p(\omega) 
  \nonumber \; ,\\
  H_p(\omega)&=& \frac{1}{L} \frac{1}{1-Vg_0(\omega)}
    \frac{1}{\omega-\tilde{\epsilon}(p)+\rmi \eta} \; ,
    \end{eqnarray}
where
\begin{equation}
  g_0(\omega) = \frac{1}{L}\sum_p \frac{1}{\omega-\tilde{\epsilon}(p)+\rmi \eta}
  = \Lambda_0(\omega) -\rmi \pi \rho_0(\omega)
  \end{equation}
is the local Green function of the non-interacting electrons.
Therefore, eq.~(\ref{eq:halfwayGF}) has the solution
\begin{eqnarray}
  \tilde{G}_{k,p}^{\rm ret}(\omega) &=&
  \frac{\delta_{k,p}}{\omega-\tilde{\epsilon}(k)+\rmi \eta}
  \nonumber \\
  && +\frac{1}{L} \frac{V}{1-Vg_0(\omega)}
  \frac{1}{\omega-\tilde{\epsilon}(k)+\rmi \eta}
  \frac{1}{\omega-\tilde{\epsilon}(p)+\rmi \eta}
  \; .\nonumber \\
  \label{eq:GFomegafinal}
  \end{eqnarray}
For our further considerations, only the diagonal part, $k=p$, is required.

\subsubsection{Density of states}

The density of states is given by
\begin{eqnarray}
  D(\omega)&=&-\frac{1}{\pi} {\rm Im}\left( \tilde{G}_{k,k}^{\rm ret}(\omega)  \right)
  \nonumber \\
  &=& L\rho_0(\omega) \nonumber \\
&&  -\frac{1}{\pi} {\rm Im}\left[
  \frac{1}{L}\sum_k \frac{V}{1-Vg_0(\omega)}
  \left(\frac{1}{\omega-\tilde{\epsilon}(k)+\rmi \eta}
  \right)^2 \right] \; . \nonumber \\
  \end{eqnarray}
The impurity-induced contribution to the density of states becomes
\begin{equation}
  D_0(\omega)\equiv D(\omega)-L\rho_0(\omega)
  = -\frac{1}{\pi} \frac{\partial}{\partial \omega}
  {\rm Im}\left[\ln\left(1-Vg_0(\omega)\right)
         \right]  .
  \end{equation}

\subsubsection{One spatial dimension}

In the following we use
\begin{equation}
\rho_0(\omega)=\frac{1}{\pi\sqrt{1-\omega^2}} \quad \hbox{for}\;
|\omega|< 1 \; ,
\label{eq:DOS1d}
\end{equation}
$\Lambda_0(\omega)=0$ for $|\omega|< 1$,
and
\begin{equation}
\Lambda_0(\omega)=\frac{\sgn(\omega)}{\sqrt{\omega^2-1}}
\end{equation}
for $|\omega|> 1$ where $\sgn(x)=x/|x|$ is the sign function.

Let $|\omega|>1$. We obtain the (anti-)bound state from
\begin{equation}
1-V\Lambda_0(\omega_{\rm b,ab})=0 \; .
  \end{equation}
For the one-dimensional case we thus find
\begin{equation}
\omega_{\rm b,ab}=\pm  \sqrt{1+V^2} \; .
  \end{equation}
There is a bound state at $\omega_{\rm b}=-\sqrt{1+V^2}$ for $V<0$
and an anti-bound state at $\omega_{\rm ab}=\sqrt{1+V^2}$ for $V>0$.
To calculate the contribution to the density of states from
the bound-state contribution outside the band
where we have $\rho_0(\omega)=\eta\equiv 0^+$, we expand
\begin{equation}
  {\rm R}(\omega) \equiv  1-V\Lambda_0(\omega)
  \approx {\rm R}'(\omega_p)(\omega-\omega_p)
\end{equation}
in the vicinity of $\omega_p\equiv\omega_{\rm b,ab}$. Then,
\begin{eqnarray}
D_{\rm imp}^{\rm b,ab}(\omega)&=&
 - \frac{1}{\pi} 
\frac{\partial }{\partial \omega}
\left[
\cot^{-1}\left( 
\frac{{\rm R}'(\omega_p)  (\omega-\omega_p)}{\pi V \eta}
\right)
\right] \nonumber \\
&=& \frac{1}{\pi}
\frac{\tilde{\eta}}{\tilde{\eta}^2+(\omega-\omega_p)^2} \nonumber \\
  &=& \delta(\omega-\omega_p) 
\end{eqnarray}
with $\tilde{\eta}=\pi V \eta/{\rm R}'(\omega_p)\to 0^+$.
Thus, the bound and anti-bound states contribute
\begin{equation}
  D_0^{\rm b,ab}(\omega) = \delta(\omega-\omega_{\rm b})\theta_{\rm H}(-V)
  +\delta(\omega-\omega_{\rm ab})\theta_{\rm H}(V) 
  \end{equation}
to the impurity part of the density of states.

For the band contribution we consider the region that includes the band edges,
$|\omega|\leq 1^+$.
In general, we obtain 
\begin{eqnarray}
  D_0^{\rm band}(\omega)&=& -\frac{1}{\pi}\sgn(V)
  \frac{\partial}{\partial \omega}
  \Cot\nolimits^{-1}\left[ \varphi(\omega)  \right]\; , \nonumber \\
\varphi(\omega) &=& 
  \frac{1-V\Lambda_0(\omega)}{\pi |V| \rho_0(\omega)}\; ,
  \end{eqnarray}
where $\Cot^{-1}(x)=\pi\theta_{\rm H}(-x)+\cot^{-1}(x)$ is continuous and differentiable
across $x=0$; $\theta_{\rm H}(x)$ is the Heaviside step function.
For $V>0$, the phase $\phi(\omega)$ jumps by $\pi/2$ when going from
$\omega=(-1)^-$ to $\omega=(-1)^+$. The same jump appears at $\omega=1$.
For $V<0$, we obtain the same discontinuities.
Inside the band we have $\Lambda_0(|\omega|<1)=0$ so that we find altogether
\begin{eqnarray}
  D_0(\omega)&=&
\delta(\omega-\omega_{\rm b})\theta_{\rm H}(-V)
  +\delta(\omega-\omega_{\rm ab})\theta_{\rm H}(V)   \nonumber \\
&&  -\frac{1}{2} \delta(\omega+1)-\frac{1}{2} \delta(\omega-1)
  \nonumber \\
&&  -\theta_{\rm H}(1-|\omega|)\frac{1}{\pi}  \frac{\partial}{\partial \omega}
 \arctan\left[ \pi V\rho_0(\omega) \right]  .
\end{eqnarray}

\subsection{Ground-state energy in one dimension}

The Hamiltonian~(\ref{eq:defHpotential}) is not particle-hole symmetric.
Therefore, the chemical potential $\epsilon_{\rm F}(V)$ depends on~$V$.
Since the scattering only appears at a single site, we have
\begin{equation}
\epsilon_{\rm F}=\epsilon_{\rm F}^{(0)} + \frac{\epsilon_{\rm F}^{(1)}}{L}
\end{equation}
to leading order in $1/L$. We can calculate the $\epsilon_{\rm F}^{(1)}$ from
\begin{equation}
  0 = L \int_{\epsilon_{\rm F}^{(0)}}^{\epsilon_{\rm F}^{(0)}+\epsilon_{\rm F}^{(1)}/L} \rmd \omega
\rho_0(\omega) + \int_{-\infty}^{\epsilon_{\rm F}^{(0)}} \rmd \omega D_0(\omega)
\; .
\end{equation}
At half band-filling, we do not need to know the correction to calculate the ground-state
energy because $\epsilon_{\rm F}^{(0)}=0$ and the bulk contribution
to the energy is
\begin{eqnarray}
  E_0^{\rm bulk}(V)&=& L \int_{-\infty}^{\epsilon_{\rm F}^{(1)}/L} \rmd \omega
  \omega   \rho_0(\omega) \nonumber \\
  &=& E_0^{\rm bulk}(V=0) + L\rho_0(0) \frac{1}{2} \left(
  \frac{\epsilon_{\rm F}^{(1)}}{L}  \right)^2 \nonumber \\
  &=& E_0^{\rm bulk}(V=0) + {\cal O}(1/L) \; .
\end{eqnarray}
Thus, we can calculate the scattering contribution
to the ground-state energy from the single-particle
density of states as
\begin{equation}
e_0(V) = E_0(V)-E_0^{\rm bulk}(V=0) 
= \int_{-\infty}^0\rmd \omega \omega D_0(\omega) \; .
\end{equation}

\subsubsection{Repulsive interaction}
For $V>0$ there is no bound state and the ground-state energy can be calculated from
the band contribution alone,
\begin{eqnarray}
  e_0(V>0)&=& \frac{1}{2}
    -\frac{1}{\pi} \biggl[
\omega \arctan\left(\pi V \rho_0(\omega)\right)
\biggr]_{-1^+}^0 \nonumber \\
  && + \frac{1}{\pi} \int_{-1}^0
  \rmd \omega  \arctan \left[\pi V \rho_0(\omega)\right] \nonumber \\
  &=&  \frac{1}{\pi} \int_{-1}^0
  \rmd \omega  \arctan\left[\pi V \rho_0(\omega)\right]  \nonumber \\
  &=& \frac{1}{2}\left(1+V-\sqrt{1+V^2}\right)\;.
    \label{eq:Vpositive}
\end{eqnarray}
For the last step we rely on {\sc Mathematica}.~\cite{Mathematica11}

\subsubsection{Attractive interaction}

For attractive interactions, we can investigate the particle-hole transformed
Hamiltonian,
\begin{equation}
  \tau_{\rm ph}^+ \hat{H}_{\rm ps}(V)\hat{\tau}_{\rm ph} 
= \hat{H}_{\rm ps}(-V)+V
  \; .
\end{equation}
At half filling, this implies 
for the scattering contribution to the ground-state energy
\begin{equation}
e_0(V)=V+e_0(-V) \; .
\end{equation}
Thus, we find $(V<0)$
\begin{eqnarray}
  e_0(V) &=&
  V+\frac{1}{2}\left(1-V-\sqrt{1+V^2}\right)\nonumber \\
  &=&
  \frac{1}{2}\left(1+V-\sqrt{1+V^2}\right)\;.
    \label{eq:Vnegative}
\end{eqnarray}
Eq.~(\ref{eq:Vnegative}) is formally identical to eq.~(\ref{eq:Vpositive}).

Alternatively, we can calculate $e_0(V)$ for $V<0$ from the density of states.
We include the bound state and find
\begin{eqnarray}
  e_0(V<0)&=&
-\sqrt{1+V^2}+  \frac{1}{2}  
  \nonumber \\
&&+\frac{1}{\pi} \biggl[
\omega \arctan\left(\pi |V|\rho_0(\omega)\right)\biggr]_{-1^+}^0 \nonumber \\
  && - \frac{1}{\pi} \int_{-1^+}^0
  \rmd \omega  \arctan\left[\pi |V| \rho_0(\omega)\right]
  \nonumber \\
  &=& \frac{1}{2}\left(1+V-\sqrt{1+V^2}\right)\;,
  \label{eq:Vnegativeagain}
\end{eqnarray}
which is identical to eq.~(\ref{eq:Vnegative}), and
\begin{equation}
  e_0(V)=\frac{1}{2}\left(1+V-\sqrt{1+V^2}\right)
  \label{eq:potentialproblemenergy}
\end{equation}
holds for all~$V$.

\section{Non-interacting SIAM
  in the presence of potential scattering and a magnetic field}
\label{supp:SIAMfull}

\subsection{Retarded Green functions}

For the equation-of-motion approach, it is convenient
to study the retarded Green function,
\begin{equation}
G_{A,B}^{\rm ret}(t)
= (-\rmi) \theta_{\rm H}(t) \langle 
\left[\hat{A}^{\vphantom{+}}(t), \hat{B}^+\right]_+\rangle \; . 
\end{equation}
Its Fourier transformation is defined by
\begin{equation}
\tilde{G}_{A,B}^{\rm ret}(\omega)= 
\int_{0}^{\infty} \!\!\rmd t e^{(\rmi \omega-\eta) t}G_{A,B}^{\rm ret}(t)
= \int_{-\infty}^{\infty} \!\!\rmd \omega' \frac{D_{A,B}(\omega')}{
\omega-\omega'+\rmi \eta} \, .
\end{equation}
Here, the spectral function is defined by 
\begin{eqnarray}
D_{A,B}(\omega)
&=& \sum_m \Bigl[
\langle 0 | \hat{B}^+ | m\rangle 
\langle m | \hat{A}^{\vphantom{+}} | 0\rangle
\delta(\omega-E_0+E_m) \nonumber \\
&&  \hphantom{\sum_m\Bigl[}
+ 
\langle 0 | \hat{A}^{\vphantom{+}} | m\rangle
\langle m | \hat{B}^+ | 0\rangle 
\delta(\omega+E_0-E_m) \Bigr] \; ,\nonumber \\
\label{eq:defDOSAB}
\end{eqnarray}
where $|m\rangle$ denotes the eigenstates of $\hat{H}_0$ with
energy $E_m$ (Lehmann representation).
Using the Lehmann representation it is readily shown that
\begin{equation}
\tilde{G}_{A,B}^{\rm c}(\omega)
= \int_{-\infty}^{\infty} \rmd \omega' \frac{D_{A,B}(\omega')}{
\omega-\omega'+\rmi \eta \sgn(\omega)}
\label{eq:lehmann}
\end{equation}
with the sign function $\sgn(x)=|x|/x$.
Therefore, the causal Green function is obtained from the
retarded Green function by replacing $\omega+\rmi \eta $
by $\omega+\rmi \eta \sgn(\omega)$.

When $\hat{A}\neq \hat{B}$, the spectral function $D_{A,B}(\omega)$
is not necessarily real. We separate the real and imaginary part,
\begin{equation}
D_{A,B}(\omega)
= \frac{D_{A,B}(\omega) +D_{A,B}^*(\omega) }{2} 
+ \rmi 
\frac{D_{A,B}(\omega) -D_{A,B}^*(\omega) }{2\rmi} 
\end{equation}
and use $D_{A,B}^*(\omega) =D_{B,A}(\omega)$ to find
\begin{eqnarray}
D_{A,B}(\omega) &=& -\frac{1}{\pi}
{\rm Im}\left[ \frac{G_{A,B}^{\rm ret}(\omega)+G_{B,A}^{\rm ret}(\omega)}{2}\right] 
\nonumber \\
&& -\frac{\rmi }{\pi}
{\rm Im}\left[ \frac{G_{A,B}^{\rm ret}(\omega)-G_{B,A}^{\rm ret}(\omega)}{2\rmi }\right] 
\; .
\label{eq:DOSAB}
\end{eqnarray}
For $\hat{A}=\hat{B}$ we recover the standard expression
\begin{equation}
D_{A,A}(\omega) = -\frac{1}{\pi}
{\rm Im}\left[ G_{A,A}^{\rm ret}(\omega)\right] \; .
\label{eq:DOSAA}
\end{equation}

\subsubsection{Green functions in the time domain}

For $\sigma=\uparrow$, we study the Hamiltonian
\begin{eqnarray}
\hat{H}_0&=&\sum_{k} (\epsilon(k) -B)
\hat{a}_{k}^+\hat{a}_{k}^{\vphantom{+}}
+\frac{V}{\sqrt{L}}
 \sum_{k}\left(\hat{a}_{k}^+\hat{d}^{\vphantom{+}}
 +\hat{d}^+\hat{a}_{k}^{\vphantom{+}} \right)
 \nonumber \\
 &&-E_d\hat{d}^+\hat{d}^{\vphantom{+}}
+\frac{K}{L} \sum_{k,k'}
\hat{a}_{k'}^+\hat{a}_{k}^{\vphantom{+}} \nonumber \\
&\equiv & \hat{T} +\hat{V}+\hat{P}+\hat{K}
 \; .
  \end{eqnarray}
The expressions for $\sigma=\downarrow$ are obtained by
changing the sign in $(B,E_d,K)$.
We study the four retarded Green functions for spin $\sigma=\uparrow$
\begin{eqnarray}
G_{k,p}^{\rm ret}(t) &=& (-{\rm i}) \theta_{\rm H}(t) \langle 
\left[ \hat{a}_{k}^{\vphantom{+}}(t),\hat{a}_{p}^+\right]_+
\rangle \; , \label{eq:defGFretarded1}\\
G_{d,p}^{\rm ret}(t) &=& (-{\rm i}) \theta_{\rm H}(t) \langle 
\left[\hat{d}^{\vphantom{+}}(t),\hat{a}_{p}^+\right]_+
\rangle \; , \label{eq:defGFretarded2}\\
G_{k,d}^{\rm ret}(t) &=& (-{\rm i}) \theta_{\rm H}(t) \langle 
\left[ \hat{a}_{k}^{\vphantom{+}}(t),\hat{d}^+\right]_+
\rangle \; , \label{eq:defGFretarded3}\\
G_{d,d}^{\rm ret}(t) &=& (-{\rm i}) \theta_{\rm H}(t) \langle 
\left[ \hat{d}^{\vphantom{+}}(t),\hat{d}^+\right]_+
\rangle \; . 
\label{eq:defGFretarded4}
\end{eqnarray}
Taking the time derivative leads to
\begin{eqnarray}
{\rm i} \dot{G}^{\rm ret}_{k,p}(t) &=& \delta(t) \delta_{k,p}
+(-{\rm i}) \theta_{\rm H}(t) 
\langle \Bigl[ \left[ \hat{a}_{k}^{\vphantom{+}}(t),\hat{H}_0\right]_-,
\hat{a}_{p}^+\Bigr]_+\rangle \nonumber \\
&=&
\delta(t) \delta_{k,p}+ (\epsilon(k)-B) G_{k,p}^{\rm ret}(t) 
+\frac{V}{\sqrt{L}} G_{d,p}^{\rm ret}(t) \nonumber \\
&&+\frac{K}{L}\sum_{k'}G_{k',p}^{\rm ret}(t) \; ,
\label{eq:GFdot1}\\
{\rm i} \dot{G}_{d,p}^{\rm ret}(t) &=& 
(-{\rm i}) \theta_{\rm H}(t) 
 \langle  
\Bigl[ \left[\hat{d}^{\vphantom{+}}(t),\hat{H}_0\right]_-,
\hat{a}_{p}^+\Bigr]_+\rangle  \nonumber \\
&=& 
\sum_k\frac{V}{\sqrt{L}} G_{k,p}^{\rm ret}(t) 
-E_{d} G_{d,p}^{\rm ret}(t)\; ,
\label{eq:GFdot2}\\
{\rm i} \dot{G}_{k,d}^{\rm ret}(t) &=& 
(-{\rm i})\theta_{\rm H}(t)  \langle  
\Bigl[\left[ \hat{a}_{k}^{\vphantom{+}}(t),\hat{H}_0\right]_-,
\hat{d}^+\Bigr]_+\rangle \nonumber \\
&=& 
(\epsilon(k)-B) G_{k,d}^{\rm ret}(t) +\frac{V}{\sqrt{L}} G_{d,d}^{\rm ret}(t)
\nonumber \\
&&+\frac{K}{L}\sum_{k'}G_{k',d}^{\rm ret}(t) \; ,
\label{eq:GFdot3}\\
{\rm i} \dot{G}_{d,d}^{\rm ret}(t) &=& 
\delta(t) 
+ (-{\rm i}) \theta_{\rm H}(t) \langle 
\Bigl[\left[ \hat{d}^{\vphantom{+}}(t),\hat{H}_0\right]_-,
\hat{d}_{\sigma}^+ \Bigr]_+\rangle  \nonumber \\
&=& \delta(t) + \sum_k\frac{V}{\sqrt{L}} G_{k,d}^{\rm ret}(t) 
-E_{d} G_{d,d}^{\rm ret}(t)\, .
\label{eq:GFdot4}
\end{eqnarray}
Here, we used the anticommutation relations of the Fermi operators 
and the commutation relations
\begin{eqnarray}
\left[\hat{a}_{k}^{\vphantom{+}},\hat{T}\right]_- 
= (\epsilon(k)-B) \hat{a}_{k}^{\vphantom{+}} &\; , \;  & 
\left[\hat{d}^{\vphantom{+}},\hat{T}\right]_- 
= 0 \; , \\
\left[\hat{a}_{k}^{\vphantom{+}},\hat{V}\right]_- 
= \frac{V}{\sqrt{L}} \hat{d}^{\vphantom{+}} &\; , \;  & 
\left[\hat{d}^{\vphantom{+}},\hat{V}\right]_- 
= \frac{V}{\sqrt{L}} \sum_k\hat{a}_{k}^{\vphantom{+}}
\; , \nonumber 
\end{eqnarray}
and
\begin{eqnarray}
\left[\hat{a}_k^{\vphantom{+}},\hat{P}\right]_- = 0 &\; , \;  & 
\left[\hat{d}^{\vphantom{+}},\hat{P}\right]_- =
-E_d \hat{d}^{\vphantom{+}} \; , \nonumber \\
\left[\hat{a}_k^{\vphantom{+}},\hat{K}\right]_- =
\frac{K}{L} \sum_{k'}\hat{a}_{k'}^{\vphantom{+}}
&\; , \;  & 
\left[\hat{d}^{\vphantom{+}},\hat{K}\right]_- =0\; .
\end{eqnarray}
For non-interacting electrons, the equations of motion
lead to a closed 
set of differential equations~(\ref{eq:GFdot1})--(\ref{eq:GFdot4}).

\subsubsection{Green functions in the frequency domain}

To solve the equations~(\ref{eq:GFdot1})--(\ref{eq:GFdot4})
we transform them into frequency space.~\cite{TIAM}
We find
\begin{eqnarray}
\left(\omega+E_d+{\rm i}\eta\right)\tilde{G}_{d,d}^{\rm ret}(\omega) &=&
1+VR_d(\omega) \;,
\label{eq:GFdot4om}\\
\left(\omega-(\epsilon(k)-B)+{\rm i}\eta\right)\tilde{G}_{k,d}^{\rm ret}(\omega)&=& 
\frac{V}{\sqrt{L}} \tilde{G}_{d,d}^{\rm ret}(\omega) \nonumber \\
&&+\frac{K}{\sqrt{L}}R_d(\omega)\; , \label{eq:GFdot3om}\\
\left(\omega+E_d+{\rm i}\eta\right)
\tilde{G}_{d,p}^{\rm ret}(\omega)&=& 
VQ_p(\omega)\; ,
\label{eq:GFdot2om}\\
\left(\omega-(\epsilon(k)-B) +{\rm i}\eta\right)
 \tilde{G}_{k,p}^{\rm ret}(\omega)
&=& \delta_{k,p}
 +\frac{V}{\sqrt{L}} \tilde{G}_{d,p}^{\rm ret}(\omega)\nonumber \\
&& +\frac{K}{\sqrt{L}}Q_p(\omega)\;,
\label{eq:GFdot1om}
\end{eqnarray}
where we introduced the abbreviations
\begin{equation}
  Q_p(\omega)=\frac{1}{\sqrt{L}}\sum_{k'}G_{k',p}^{\rm ret}(\omega)
  \; , \;
  R_d(\omega)=\frac{1}{\sqrt{L}}\sum_{k'}G_{k',d}^{\rm ret}(\omega) \; .
\end{equation}
The resulting set of algebraic equations is readily solved.

From eq.~(\ref{eq:GFdot3om}) we find
\begin{equation}
  \tilde{G}_{k,d}^{\rm ret}(\omega)= \frac{1}{\sqrt{L}}
  \frac{V\tilde{G}_{d,d}^{\rm ret}(\omega)  +KR_d(\omega)
  }{\omega-(\epsilon(k)-B) +{\rm i}\eta}
\end{equation}
so that
\begin{equation}
  R_d(\omega)= \left(V\tilde{G}_{d,d}^{\rm ret}(\omega)  +KR_d(\omega)\right)g_0(\omega)
  \label{eq:Rdequation}
\end{equation}
with the retarded local non-interacting Green function
\begin{eqnarray}
  g_0(\omega+B)&=&\frac{1}{L}\sum_k \frac{1}{\omega-(\epsilon(k)-B) +{\rm i}\eta}
  \nonumber \\
  &=&\Lambda_0(\omega+B)-\rmi \pi \rho_0(\omega+B)
  \; .
  \label{appeq:retardedg0}
\end{eqnarray}
In one dimension,
\begin{equation}
\Lambda_0(|\omega|>1) =\frac{\sgn(\omega)}{\sqrt{\omega^2-1}}
\end{equation}
and $\Lambda_0(|\omega|<1)=0$
with $\rho_0(\omega)$ from eq.~(\ref{eq:DOS}).

Eq.~(\ref{eq:Rdequation}) has the solution
\begin{equation}
  R_d(\omega)=
  \frac{V\tilde{G}_{d,d}^{\rm ret}(\omega)g_0(\omega+B)}{1-Kg_0(\omega+B)} \; .
  \label{eq:Rdequationsolved}
\end{equation}
Thus, the $k$-$d$ Green function becomes
\begin{equation}
  \tilde{G}_{k,d}^{\rm ret}(\omega)
  =
  \frac{1}{\sqrt{L}}
  \frac{V\tilde{G}_{d,d}(\omega)}{\omega-(\epsilon(k)-B) +{\rm i}\eta}
  \frac{1}{1-Kg_0(\omega+B)}
  \label{appeq:GKdret}
  \end{equation}
in terms of $\tilde{G}_{d,d}(\omega)$ which is obtained from eq.~(\ref{eq:GFdot4om}), %
\begin{eqnarray}
    \tilde{G}_{d,d}^{\rm ret}(\omega) &=&
    \frac{1+VR_d(\omega)}{\omega+E_d+{\rm i}\eta}
    \nonumber \\
    &=& \frac{1-Kg_0(\omega+B)+V^2g_0(\omega+B)\tilde{G}_{d,d}^{\rm ret}(\omega)
    }{(\omega+E_d+{\rm i}\eta)(1-Kg_0(\omega+B))}\nonumber \\
    \label{eq:GFdot4omfinal}
\end{eqnarray}
and
\begin{equation}
\tilde{G}_{d,d}^{\rm ret}(\omega) =
\frac{1}{\omega+E_{d}-\Delta^{\rm ret}(\omega+B)}
\; ,
\label{eq:EOMimpurityGF4}
\end{equation}
where we defined the retarded hybridization function
in the presence of potential scattering
\begin{equation}
\Delta^{\rm ret} (\omega)= 
\frac{V^2g_0(\omega)}{1-Kg_0(\omega)} \; .
\label{eq:Deltacandret}
\end{equation}
Thus, we can rewrite~(\ref{appeq:GKdret}) in the form
\begin{eqnarray}
\tilde{G}_{k,d}^{\rm ret}(\omega) &=&
\sqrt{\frac{1}{L}}
\frac{V}{(\omega-\epsilon(k)+{\rm i}\eta)
  (\omega+E_{d}-\Delta^{\rm ret}(\omega+B))}
\nonumber \\
&& \times \frac{1}{1-Kg_0(\omega+B)}
\; . \label{eq:EOMimpurityGF3}
\end{eqnarray}
For the other two Green functions we consider
\begin{eqnarray}
 \tilde{G}_{k,p}^{\rm ret}(\omega) &=& \frac{1}{\omega-(\epsilon(k)-B) +{\rm i}\eta}
 \nonumber \\
 &&\times \left[\delta_{k,p}+ \frac{Q_p(\omega)}{\sqrt{L}}
\left( \frac{V^2}{\omega+E_d+{\rm i}\eta}+K \right)\right]\; , 
 \nonumber \\
 Q_p(\omega)&=& \frac{1}{\sqrt{L}} \frac{1}{\omega-(\epsilon(p)-B) +{\rm i}\eta}
\nonumber \\
&& +g_0(\omega+B) Q_p(\omega)
 \left( \frac{V^2}{\omega+E_d+{\rm i}\eta}+K \right)
 \nonumber \\
 &=&
 \frac{1}{\sqrt{L}} \frac{1}{\omega-(\epsilon(p)-B) +{\rm i}\eta}
 \frac{\omega+E_d+\rmi \eta}{1-Kg_0(\omega+B)}\nonumber \\
&&\times  \frac{1}{\omega+E_d-\Delta^{\rm ret}(\omega+B)}
\end{eqnarray}
so that
\begin{eqnarray}
  \tilde{G}_{d,p}^{\rm ret}(\omega) &=&
   \frac{1}{\sqrt{L}} \frac{1}{\omega-(\epsilon(p)-B) +{\rm i}\eta}
 \frac{V}{1-Kg_0(\omega+B)} \nonumber \\
&&\times  \frac{1}{\omega+E_d-\Delta^{\rm ret}(\omega+B)}
  \end{eqnarray}
and
\begin{eqnarray}
 \tilde{G}_{k,p}^{\rm ret}(\omega) &=& 
\frac{\delta_{k,p}}{\omega-(\epsilon(k)-B)+{\rm i}\eta}
\nonumber\\
&& +\frac{1}{L}\frac{1}{\omega-(\epsilon(k)-B)+{\rm i}\eta}
\frac{1}{\omega-(\epsilon(p)-B)+{\rm i}\eta}\nonumber \\
&&
\times \frac{V^2+K(\omega+E_d+\rmi \eta)}{[1-Kg_0(\omega+B)]
[\omega+E_{d}-\Delta^{\rm ret}(\omega+B)]} \nonumber \\
\label{eq:EOMimpurityGF1}
\end{eqnarray}
for the other two retarded Green functions.

\subsection{Density of states}
\label{sec:DOSprops}

The single-particle density of states is defined by
\begin{equation}
D(\omega) = \sum_m \delta(\omega-E_m) \; .
\end{equation}
To make contact with the retarded Green functions,
we write the single-particle density of states in the form
\begin{eqnarray}
D(\omega)&=& - \frac{1}{\pi} {\rm Im}\biggl(\sum_m 
\langle \hat{A}_{m}^+ \frac{1}{\omega-(\hat{H}_0-E_0)+{\rm i}\eta}
\hat{A}_{m}^{\vphantom{+}}\rangle \nonumber \\
&&\hphantom{- \frac{1}{\pi} {\rm Im}\biggl(\sum_m }
+\langle \hat{A}_{m}^{\vphantom{+}}
\frac{1}{\omega-(\hat{H}_0-E_0)+{\rm i}\eta}
\hat{A}_{m}^+ 
\rangle
\biggr) \, , \nonumber \\
\end{eqnarray}
where we used the fact that $\hat{A}_{m}^+$
($\hat{A}_{m}^{\vphantom{+}}$)
creates (annihilates) an electron with exact single-particle energy $E_m$ 
in the ground state.
The sum over $m$ runs over all single-particle
excitations of the ground state and thus represents
the trace over all single-particle eigenstates,
\begin{equation}
D(\omega)= - \frac{1}{\pi} {\rm Im}{\rm Tr}_1
\Bigl(\frac{1}{\omega-(\hat{H}_0-E_0)+{\rm i}\eta}\Bigr) \; .
\end{equation}
We can equally use the excitations
$\hat{a}_{k}^+|\Phi_0\rangle$, 
$\hat{a}_{k}^{\vphantom{+}}|\Phi_0\rangle$,
and 
$\hat{d}^+|\Phi_0\rangle$, 
$\hat{d}^{\vphantom{+}}|\Phi_0\rangle$, respectively,
to perform the
trace over the single-particle excitations of the ground state.
Therefore, we may write
\begin{eqnarray}
D(\omega)&=& - \frac{1}{\pi} {\rm Im}
\biggl[\sum_k \biggl(
\langle \hat{a}_{k}^+ \frac{1}{\omega-(\hat{H}_0-E_0)+{\rm i}\eta}
\hat{a}_{k}^{\vphantom{+}}\rangle 
\nonumber \\
&& \hphantom{ - \frac{1}{\pi} {\rm Im}\biggl( \sum_k}
+ \langle \hat{a}_{k}^{\vphantom{+}}
\frac{1}{\omega-(\hat{H}_0-E_0)+{\rm i}\eta}
\hat{a}_{k}^+ 
\rangle\biggr)
\nonumber \\
&& 
\hphantom{ - \frac{1}{\pi} {\rm Im}\biggl( }
+ 
\langle \hat{d}^+ \frac{1}{\omega-(\hat{H}_0-E_0)+{\rm i}\eta}
\hat{d}^{\vphantom{+}}\rangle
\nonumber \\
&& \hphantom{ - \frac{1}{\pi} {\rm Im}\biggl( }
+\langle \hat{d}^{\vphantom{+}}
\frac{1}{\omega-(\hat{H}_0-E_0)+{\rm i}\eta}
\hat{d}^+ 
\rangle \biggr] \nonumber \\
&=& - \frac{1}{\pi} {\rm Im}
\biggl[
\sum_k \tilde{G}_{k,k}^{\rm ret}(\omega)+ \tilde{G}_{d,d}^{\rm ret}(\omega) 
\biggr] \; .
\label{eq:totalDOSforHF}
\end{eqnarray}
Equation~(\ref{eq:EOMimpurityGF1}) shows that the band Green function
consists of the undisturbed host Green function for $V\equiv 0$
and a $1/L$ correction due to the hybridization.
Therefore, using eqs.~(\ref{eq:EOMimpurityGF4}) and~(\ref{eq:EOMimpurityGF1}),
the contribution due to a finite hybridization is given by
\begin{eqnarray}
D_{\rm imp}(\omega)&=& -\frac{1}{\pi}{\rm Im}
\biggl[ \tilde{G}_{d,d}^{\rm ret}(\omega) 
  \nonumber \\
&& 
+
\frac{1}{L}\sum_k 
\frac{\tilde{G}_{d,d}^{\rm ret}(\omega)
[V^2+K(\omega+E_d+\rmi \eta)]}{
  (1-Kg_0(\omega+B))(\omega-\epsilon(k)+{\rm i}\eta)^2}
\biggr] \nonumber \\
&=& 
- \frac{1}{\pi} \frac{\partial }{\partial \omega}
{\rm Im}
\left[\ln\left({\rm R}(\omega)+\rmi {\rm I}(\omega \right)\right]
\; ,
\end{eqnarray}
where we use 
\begin{equation}
  g_0'(\omega+B)
  =-\frac{1}{L}\sum_k \frac{1}{(\omega-(\epsilon(k)-B) +{\rm i}\eta)^2}\; ,
  \end{equation}
see eq.~(\ref{appeq:retardedg0}), 
and the real and imaginary parts of the argument of the logarithm read
\begin{eqnarray}
  {\rm R}(\omega)
  &=&(\omega+E_d)(1-K\Lambda_0(\omega+B))
-V^2\Lambda_0(\omega+B) \; , \nonumber \\
{\rm I}(\omega)
&=&\eta\left(1-K\Lambda_0(\omega+B)\right) \nonumber \\
&&+\left[(\omega+E_d)K+V^2\right]\pi \rho_0(\omega+B) \; .
\label{appeq:RandIdefforphi}
\end{eqnarray}
We thus have
\begin{eqnarray}
  D_{\rm imp}(\omega)&=&
  -\frac{1}{\pi} \frac{\partial \varphi(\omega,B,E_d,K,V)}{\partial \omega}
  \nonumber \;, \\
\cot[  \varphi(\omega,B,E_d,K,V)] &=&
  \frac{{\rm R}(\omega,B,E_d,K,V)}{{\rm I}(\omega,B,E_d,K,V)}
  \; .
  \label{appeq:phaseshiftfunction}
\end{eqnarray}
Here, we made apparent the dependence of all quantities also
on $(B,E_d,K,V)$.
The impurity-contribution to the density of states has a band part and
a contribution from the bound states,
\begin{equation}
D_{\rm imp}(\omega)=D_{\rm imp}^{\rm band}(\omega)+D_{\rm imp}^{\rm b}(\omega) \; .
\end{equation}
We discuss them separately.

For the bound-state contribution outside the band, $|\omega_p+B|>1$
we have $\rho_0(\omega)=0$ and we expand
\begin{equation}
  {\rm R}(\omega,B,E_d,K,V)
  \approx {\rm R}'(\omega_p)(\omega-\omega_p)
\end{equation}
in the vicinity of $\omega_p\equiv\omega_p(B,E_d,K,V)$. Then,
\begin{eqnarray}
D_{\rm imp}^{\rm b}(\omega)&=&
 - \frac{1}{\pi} 
\frac{\partial }{\partial \omega}
\left[
\cot^{-1}\left( 
\frac{{\rm R}'(\omega_p)  (\omega-\omega_p)}{
  (1-K\Lambda_0(\omega_p+B))\eta}
\right)
\right] \nonumber \\
&=& \frac{1}{\pi}
\frac{\tilde{\eta}}{\tilde{\eta}^2+(\omega-\omega_p)^2} \nonumber \\
  &=& \delta(\omega-\omega_p) 
\end{eqnarray}
with $\tilde{\eta}=\eta(1-K\Lambda_0(\omega_p+B))/{\rm R}'(\omega_p)\to 0^+$.
For the bound-state contribution, the task therefore is to find $\omega_p$
from the solution of
\begin{eqnarray}
  {\rm R}(\omega_p,B,E_d,K,V)&=&
  (\omega_p+E_d)(1-K\Lambda_0(\omega_p+B))\nonumber \\
  &&-V^2\Lambda_0(\omega_p+B)\stackrel{!}{=} 0   \label{eq:getmeomegap}
  \end{eqnarray}
for $|\omega_p+B|>1$.
If a bound state exists for $\omega_p<-1-B$,
we know from $\Lambda_0(\omega_p+B)<0$
and $\omega_p-E_d<0$ that eq.~(\ref{eq:getmeomegap})
has a solution only if
$[(\omega_p+E_d)K+V^2)][-\Lambda_0(\omega_p+B)]>0$ or
$(\omega_p+E_d)K+V^2>0$. Thus, for a bound state to exist, we have to demand
that $K(-1-B+E_d)+V^2>0$.

Since $\rho_0(\omega+B)$ develops a square-root singularity
at $\omega\uparrow-(1+B)$ and $\Lambda_0(\omega+B)$ diverges
to negative infinity
at $\omega\downarrow -(1+B)$ in one dimension,
it is subtle to determine the band contribution to the density of states.
If there is a bound state, the real part ${\rm R}(\omega,B,E_d,K,V)$
diverges to infinity for $\omega\uparrow-(1+B)$,
and the imaginary part ${\rm I}(\omega,B,E_d,K,V)$ goes to zero.
Therefore, $\varphi(-(1+B)-0^+,B,E_d,K,V)=0$.
For $\omega\downarrow-(1+B)$,
the real part ${\rm R}(\omega,B,E_d,K,V)$ is finite 
but the imaginary part ${\rm I}(\omega,B,E_d,K,V)$ goes to infinity.
Therefore, $\varphi(-(1+B)+0^+,B,E_d,K)=\pi/2$.
If there is no bound state, 
the real part ${\rm R}(\omega,B,E_d,K,V)$
diverges to negative infinity for $\omega\uparrow-(1+B)$,
and the imaginary part ${\rm I}(\omega,B,E_d,K,V)$ goes to zero.
Therefore, $\varphi(-(1+B)-0^+,B,E_d,K,V)=\pi$.
For $\omega\downarrow-(1+B)$,
the real part ${\rm R}(\omega,B,E_d,K,V)$ is finite 
but the imaginary part ${\rm I}(\omega,B,E_d,K,V)$ goes to negative infinity.
Therefore, $\varphi(-(1+B)+0^+,B,E_d,K,V)=3\pi/2$.
The step discontinuity is $\pi/2$ in both cases
and contributes
\begin{eqnarray}
  \delta D_{\rm imp}^{\rm band}(\omega)&=&-\frac{1}{\pi} \frac{\partial}{\partial \omega}
  \left[ \frac{\pi}{2}\theta_{\rm H}(\omega+B+1)    \right]\nonumber \\
  &=&
  -\frac{1}{2} \delta(\omega+B+1)
  \label{eq:Dimpdeltacontribution}
\end{eqnarray}
to the band part of the impurity density of states.

The remaining band contribution results from $|\omega+B|< 1$.
We use $\Lambda_0(\omega+B)=0$ in that region
in one dimension to find the total band contribution 
\begin{eqnarray}
D_{\rm imp}^{\rm band}(\omega)
&=&   -\frac{1}{2} \delta(\omega+B+1) \nonumber \\
&& - \frac{1}{\pi} 
\frac{\partial X(\omega,B,E_d,K,V)}{\partial \omega} \; ,
\label{eq:DOSfromGFagain}
\end{eqnarray}
where
\begin{eqnarray}
  X(\omega) &\equiv &  X(\omega,B,E_d,K,V)  \label{appeq:Xdefinition}
  \\
&=& \pi \theta_{\rm H}(-\omega-E_d)\nonumber \\
&&+
\arccot\left( 
\frac{(\omega+E_{d})\sqrt{1-(\omega+B)^2}}{(\omega+E_d)K+V^2}
\right)\; .
\nonumber 
\end{eqnarray}
Note that $X(\omega,B,E_d,K,V)$
is continuous and differentiable for $|\omega+B|<1$.

\subsection{Ground-state energy}
\label{appsubsec:gsenergy}

The ground-state energy for the SIAM becomes
\begin{eqnarray}
  E_{\rm sp}^{\rm tot}(B)&=&
  E_{\rm sp}^{\rm bulk}(B) +E_{\rm sp}(B)
  \nonumber \; , \\
  E_{\rm sp}^{\rm bulk}(B)&=&L \int_{-1}^{-B}\rmd\omega \omega \rho_0(\omega)
 -2BL\int_0^B\rmd \omega \rho_0(\omega) \; , \nonumber \\
E_{\rm sp}(B)&=&
E^{\rm b}(B,E_d,K,V_{\uparrow},V_{\downarrow}) \nonumber \\
&&+ E^{\rm band}(B,E_d,K,V_{\uparrow},V_{\downarrow})\; . 
  \label{eq:Esingleparticletotal}
\end{eqnarray}
If existing, the bound states contribute
\begin{eqnarray}
  E^{\rm b}(B,E_d,K,V_{\uparrow},V_{\downarrow})
  &=& \omega_{p,\uparrow}(B,E_d,K,V_{\uparrow}) \\
  &&+\omega_{p,\downarrow}(B,E_d,K,V_{\downarrow})   \; , \nonumber 
\end{eqnarray}
where the poles follow from the solution of
\begin{eqnarray}
   (\omega_{p,\uparrow}+E_d)(1-K\Lambda_0(\omega_{p,\uparrow}+B))-
  V_{\uparrow}^2\Lambda_0(\omega_{p,\uparrow}+B) &=&0 \, , \nonumber \\
     (\omega_{p,\downarrow}-E_d)(1+K\Lambda_0(\omega_{p,\downarrow}-B))
  -V_{\downarrow}^2\Lambda_0(\omega_{p,\downarrow}-B) &=&0 \nonumber \\
  \label{eq:betterpolerepresentation}
\end{eqnarray}
with $\Lambda_0(x)=-1/\sqrt{x^2-1}$ for $x<-1$.
The bound states can also be obtained from 
\begin{eqnarray}
P(\omega_{p,\uparrow},B,E_d,K,V_{\uparrow})&=& 0\; ,\nonumber \\
P(\omega_{p,\downarrow},-B,-E_d,-K,V_{\downarrow})&=& 0\; ,
\label{eq:getmeomegapupanddown}\\
  P(x,B,E_d,K,V)&=&    (x+E_d)^2\left((x+B)^2-1\right)\nonumber \\
&&-  \left(
V^2+K(x+E_d)\right)^2  \nonumber
    \end{eqnarray}
for $\omega_{p,\uparrow}<-(1+B)$ and $\omega_{p,\downarrow}<-(1-B)$.
Recall that the root $\omega_{p,\uparrow}$ exists only if $K(-1-B+E_d)+V_{\uparrow}^2>0$.
As a starting point for the root search we can use
\begin{equation}
  \omega_{p,\uparrow}^{\rm s}=-1-B-\frac{\delta^2}{2} \quad ,
  \quad
  \delta= \frac{K(-1-B+E_d)+V_{\uparrow}^2}{1+B-E_d} \; .
\end{equation}
The root $\omega_{p,\downarrow}$ exists only if $K(1-B+E_d)+V_{\downarrow}^2>0$.
As a starting point for the root search we can use
\begin{equation}
  \omega_{p,\downarrow}^{\rm s}=-1+B-\frac{\delta^2}{2} \quad ,
  \quad
  \delta= \frac{K(1-B+E_d)+V_{\downarrow}^2}{1-B+E_d} \; .
\end{equation}
Note that, in general, we have $B,E_d,K\ll 1$.

The band energy $E^{\rm band}\equiv E^{\rm band}(B,E_d,K,V_{\uparrow},V_{\downarrow})$
is given by
\begin{eqnarray}
  E^{\rm band}  &=&
-\frac{1}{2}(-1-B) -\frac{1}{2}(-1+B)\nonumber \\
&&  -\int_{-(1+B)}^0\frac{\rmd \omega}{\pi}   \omega 
\frac{\partial X(\omega,B,E_d,K,V_{\uparrow})}{\partial \omega}
\nonumber \\
&&
-\int_{-(1-B)}^0\frac{\rmd \omega}{\pi}   \omega 
 \frac{\partial X(\omega,-B,-E_d,-K,V_{\downarrow})}{\partial \omega} \nonumber \\
 &=&
 1- \left[ \vphantom{\biggl(}
   \frac{\omega}{\pi} X(\omega,B,E_d,K,V_{\uparrow})\right]_{-1-B}^{0}
 \nonumber \\
&& + \int_{-1-B}^0 \frac{\rmd\omega}{\pi} X(\omega,B,E_d,K,V_{\uparrow}) 
\nonumber \\
&& -\left[ \vphantom{\biggl(}
  \frac{\omega}{\pi} X(\omega,-B,-E_d,-K,V_{\downarrow})\right]_{-1+B}^{0}
\nonumber \\
&& + \int_{-1+B}^0 \frac{\rmd\omega}{\pi} X(\omega,-B,-E_d,-K,V_{\downarrow}) 
 \nonumber \\
 &=&1-(1+B)\Bigl(\frac{3}{2}-\theta_{\rm H}\left[K(-1-B+E_d)+V_{\uparrow}^2\right]\Bigr)
 \nonumber \\
&& -(1-B)\Bigl(\frac{3}{2}-\theta_{\rm H}\left[-K(-1+B-E_d)+V_{\downarrow}^2\right]\Bigr)
  \nonumber \\
  &&+ \int_{-1-B}^0 \frac{\rmd\omega}{\pi} X(\omega,B,E_d,K,V_{\uparrow})
  \nonumber \\
&&  + \int_{-1+B}^0 \frac{\rmd\omega}{\pi} X(\omega,-B,-E_d,-K,V_{\downarrow})
  \nonumber\\
  &=&-2 +(1+B)\theta_{\rm H}\left[V_{\uparrow}^2-K(1+B-E_d) \right] \nonumber \\
&&  +(1-B)\theta_{\rm H}\left[V_{\downarrow}^2+K(1-B+E_d) \right]\nonumber \\
  &&+ \int_{-1-B}^0 \frac{\rmd\omega}{\pi} X(\omega,B,E_d,K,V_{\uparrow})
  \nonumber \\
&&  + \int_{-1+B}^0 \frac{\rmd\omega}{\pi} X(\omega,-B,-E_d,-K,V_{\downarrow}) \; .
  \label{eq:defAofkrams} 
\end{eqnarray}
Here, we used that $X(\omega=0,B,E_d,K,V)$ is finite, and that
$\arccot(x)$ jumps by $\pi$ at $x=0$.
Note that the Heaviside step functions become unity if the corresponding bound state
exists.
Equations~(\ref{eq:getmeomegapupanddown}) and~(\ref{eq:defAofkrams})
are suitable for a numerical calculation of the single-particle
energy for known parameters $(B,E_d,K,V_{\uparrow},V_{\downarrow})$.

\subsection{Self-consistency cycle}
\label{appsubsec:sccycle}

Here, we collect the equations for the self-consistency procedure
of Sect.~\ref{sec:selfconsistency}. The external parameters,
$B$ and $J_{\rm K}$, are given and fixed during the iteration.

The value after the $n$-th iteration are denoted by
$(E_d,K,m,M_0,V_{\uparrow},V_{\downarrow})_{(n)}$.
As the first of values we may use
$(E_d,K,m,M_0,V_{\uparrow},V_{\downarrow})_{(0)}=(B,0,0,0,V,V)$
with $V$ from the paramagnetic solution, see Sect.~\ref{subsec:e0Gutzpara}.
At $B=0$ and for $m=0$, the self-consistency is guaranteed.

For $B>0$ or $m\neq 0$, we evaluate eqs.~(\ref{eq:resetmMOgamma})
\begin{eqnarray}
  2m_{(n+1)}
  &=&-\left.
  \frac{\partial E_{\rm sp}(B,E_d,K,V_{\uparrow},V_{\downarrow})}{\partial E_d}
  \right|_{(n)}
  \; , \nonumber \\
  2M_{0,(n+1)}&=&\left.
  \frac{\partial E_{\rm sp}(B,E_d,K,V_{\uparrow},V_{\downarrow})}{\partial K}
  \right|_{(n)}
  \; , \nonumber \\
  2\gamma_{\sigma,(n+1)}&=&\left.
  \frac{\partial E_{\rm sp}(B,E_d,K,V_{\uparrow},V_{\downarrow})}{\partial V_{\sigma}}
  \right|_{(n)}
  \; ,
  \label{eq:resetmMOgammaagain}
\end{eqnarray}
where the right-hand-side is evaluated with the values of the $n$th iteration, namely
$(E_d,K,m,M_0,V_{\uparrow},V_{\downarrow})_{(n)}$. Using eq.~(\ref{eq:resetKEd}),
we then find
\begin{eqnarray}
  K_{(n+1)}&=&\frac{1}{2} J_{\rm K} m_{(n+1)} \; , \nonumber \\
  E_{d,(n+1)}&=&B-\frac{1}{2} J_{\rm K}M_{0,(n+1)}   \label{eq:resetKEdagain}
\\
&&  +\frac{4m_{(n+1)}}{(1-4m_{(n+1)}^2)^{3/2}}J_{\rm K}
  \gamma_{\uparrow,(n+1)}\gamma_{\downarrow,(n+1)}\nonumber
  \end{eqnarray}
for the values of $(E_d,K)_{(n+1)}$ for the next iteration.
Recall the dependence of $V_{\sigma}$ on $\gamma_{\sigma}$ in
eq.~(\ref{eq:selfconsistency}),
\begin{equation}
  V_{\sigma,(n+1)} = -\frac{J_{\rm K}}{2} \left(
  \gamma_{\sigma,(n)}+\frac{2}{\sqrt{1-4m_{(n)}^2}}\gamma_{\bar{\sigma},(n)}
  \right) \; .
\label{eq:selfconsistencyVagain}
\end{equation}
The derivatives in eq.~(\ref{eq:resetmMOgammaagain}) can be calculate numerically
from the ground-state energy. For a further analytic treatment
for large and small Kondo couplings and to avoid numerical inaccuracies,
we perform the derivatives analytically. All formulae apply for the one-dimensional
density of states.

In the following we employ
\begin{eqnarray}
  E_{\rm sp}(B,E_d,K,V_{\uparrow},V_{\downarrow})
  &=& A+ \omega_{p,\uparrow}(B,E_d,K,V_{\uparrow}) \nonumber \\
&&  + \omega_{p,\downarrow}(B,E_d,K,V_{\downarrow}) \nonumber \\
  &&  + E_{\uparrow}(B,E_d,K,V_{\uparrow})\\
  &&+E_{\uparrow}(-B,-E_d,-K,V_{\downarrow})\nonumber 
\end{eqnarray}
with the constant
\begin{eqnarray}
A&=&(1+B)\theta_{\rm H}\left[V_{\uparrow}^2-K(1+B-E_d) \right] \\
&&  +(1-B)\theta_{\rm H}\left[V_{\downarrow}^2+K(1-B+E_d) \right]-2 \; .
\nonumber
\end{eqnarray}
We always work in parameter regimes where the bound states either exist
or do not exist. Therefore, $A$ actually does not depend
on $(E_d,K,V_{\uparrow},V_{\downarrow})$.
Moreover,
\begin{equation}
  E_{\uparrow}^{\rm band}(B,E_d,K,V)=
  \int_{-1-B}^0 \frac{\rmd\omega}{\pi} X(\omega,B,E_d,K,V)
\end{equation}
with $X(\omega,B,E_d,K,V)$ from eq.~(\ref{appeq:Xdefinition}).
We determine the bound states from eq.~(\ref{eq:getmeomegapupanddown})
for $\omega_{p,\uparrow}<-(1+B)$ and $\omega_{p,\downarrow}<-(1-B)$.
Recall that the root $\omega_{p,\uparrow}$ exists only if $K(-1-B+E_d)+V_{\uparrow}^2>0$.
In all practical cases, the bound state $\omega_{p,\downarrow}$ exists.

\subsubsection{Derivatives with respect to $V_{\uparrow}$, $V_{\downarrow}$}

The coefficient $\gamma_{\uparrow}$ obeys
\begin{eqnarray}
  2\gamma_{\uparrow,(n+1)}&=&\frac{\partial \omega_{p,\uparrow}}{\partial V_{\uparrow}}+
  \frac{\partial E_{\uparrow}}{\partial V_{\uparrow}}
  \nonumber \\
  \frac{\partial \omega_{p,\uparrow}}{\partial V_{\uparrow}}&=&
  \frac{2V_{\uparrow}\left(V_{\uparrow}^2+K(\omega_{p,\uparrow}+E_d)\right)}{
    N_1(\omega_{p,\uparrow})} \nonumber \; ,\\
N_1(\omega_{p,\uparrow}) &=& 
    (\omega_{p,\uparrow}+E_d)(\omega_{p,\uparrow}+B)(2\omega_{p,\uparrow}+B+E_d)
\nonumber \\
&&-KV_{\uparrow}^2   -(K^2+1)(\omega_{p,\uparrow}+E_d) 
\nonumber  \; , \\
  \frac{\partial E_{\uparrow}^{\rm band}}{\partial V_{\uparrow}}&=&
  \int_{-(1+B)}^0\frac{\rmd \omega}{\pi} 
  \frac{2 V_{\uparrow} (E_d + \omega) \sqrt{1 - (\omega+B)^2}}{N_2(\omega)}
  \nonumber \; , \\
N_2(\omega)&=& 
  (K (E_d + \omega) +  V_{\uparrow}^2)^2\nonumber \\
&&    + (E_d + \omega)^2(1 - (\omega+B)^2) \; .
  \label{eq:neededforgammaup}
  \end{eqnarray}
The right-hand-side of these equations is evaluated at $V_{\uparrow}\equiv V_{\uparrow,(n)}$,
$E_d\equiv E_{d,(n)}$, and $K\equiv K_{(n)}$.
For the derivative with respect to $V_{\downarrow}$ we note that
we simply have to reverse
$(B,E_d,K)$. This follows from eq.~(\ref{eq:Esingleparticletotal}),
\begin{eqnarray}
  2\gamma_{\downarrow,(n+1)}&=&\frac{\partial \omega_{p,\downarrow}}{
    \partial V_{\downarrow}}+
  \frac{\partial E_{\uparrow}}{\partial V_{\downarrow}} \; ,
  \nonumber \\
  \frac{\partial \omega_{p,\downarrow}}{\partial V_{\downarrow}}&=&
  \frac{2V_{\downarrow}\left(V_{\downarrow}^2-K(\omega_{p,\downarrow}-E_d)\right)}{
    N_3(\omega_{p,\downarrow})} \nonumber \; ,\\
N_3(\omega_{p,\downarrow}) &=& 
    (\omega_{p,\downarrow}-E_d)(\omega_{p,\downarrow}-B)(2\omega_{p,\downarrow}-B-E_d)
\nonumber \\
&&+KV_{\downarrow}^2   -(K^2+1)(\omega_{p,\downarrow}-E_d) 
\nonumber  \; , \\
  \frac{\partial E_{\uparrow}^{\rm band}}{\partial V_{\downarrow}}&=&
  \int_{-(1-B)}^0\frac{\rmd \omega}{\pi} 
  \frac{2 V_{\downarrow} (-E_d + \omega) \sqrt{1 - (\omega-B)^2}}{N_4(\omega)}
  \nonumber \; , \\
N_4(\omega)&=& 
\left(  K (E_d - \omega) +  V_{\downarrow}^2\right)^2\nonumber \\
&&    + (-E_d + \omega)^2(1 - (\omega-B)^2) \; .
  \label{eq:neededforgammadown}
  \end{eqnarray}

\subsubsection{Derivative with respect to $K$}

We determine the bound states from eq.~(\ref{eq:getmeomegap})
for $\omega_{p,\uparrow}<-(1+B)$ and $\omega_{p,\downarrow}<-(1-B)$ and
using $E_d\equiv E_{d,(n)}$ and $K\equiv K_{(n)}$.
Then, the contribution from the bound states is
\begin{equation}
  2M_{0,(n+1)}^{\rm b}= \frac{\partial \omega_{p,\uparrow}}{\partial K}
  + \frac{\partial \omega_{p,\downarrow}}{\partial K}
\end{equation}
with
\begin{eqnarray}
  \frac{\partial \omega_{p,\uparrow}}{\partial K} &=&
  \frac{(\omega_{p,\uparrow}+E_d)\left(V_{\uparrow}^2+K(\omega_{p,\uparrow}+E_d)\right)
  }{N_1(\omega_{p,\uparrow})} \; , \nonumber 
  \\
  \frac{\partial \omega_{p,\downarrow}}{\partial K} &=&
  \frac{(E_d-\omega_{p,\downarrow})\left(V_{\downarrow}^2
    -K(\omega_{p,\downarrow}-E_d)\right)}{N_3(\omega_{p,\downarrow})} \; . 
\nonumber\\
   \label{eq:Kderivatives}
\end{eqnarray}
The band contribution reads
\begin{equation}
  2M_{0,(n+1)}^{\rm band}=
  \frac{\partial E_{\uparrow}^{\rm band}(V_{\uparrow})}{\partial K}
  + \frac{\partial E_{\uparrow}^{\rm band}(V_{\downarrow})}{\partial K}
\end{equation}
with
\begin{eqnarray}
  \frac{\partial E_{\uparrow}^{\rm band}(V_{\uparrow})}{\partial K}
  &=& \int_{-(1+B)}^0\frac{\rmd \omega}{\pi}
  \frac{(\omega+E_d)^2\sqrt{1-(\omega+B)^2}}{N_2(\omega)} \; , \nonumber \\
    \frac{\partial E_{\uparrow}^{\rm band}(V_{\downarrow})}{\partial K}&=&
  -\int_{-(1-B)}^0\frac{\rmd \omega}{\pi}
  \frac{(\omega-E_d)^2\sqrt{1-(\omega-B)^2}}{N_4(\omega)}\; , \nonumber \\
  \label{eq:Kderivativeband}
  \end{eqnarray}
to be evaluated
at $V_{\sigma}\equiv V_{\sigma,(n)}$,
$E_d\equiv E_{d,(n)}$, and $K\equiv K_{(n)}$.
In total,
 \begin{equation}
 2M_{0,(n+1)}= 2M_{0,(n+1)}^{\rm band }+ 2M_{0,(n+1)}^{\rm b} \; .
 \end{equation}

\subsubsection{Derivative with respect to $E_d$}

 We determine the bound states from eq.~(\ref{eq:getmeomegap})
 for $\omega_{p,\uparrow}<-(1+B)$ and $\omega_{p,\downarrow}<-(1-B)$ and
using $E_d\equiv E_{d,(n)}$ and $K\equiv K_{(n)}$.
Then, the contribution from the bound states is
\begin{equation}
  2m_{(n+1)}^{\rm b}= -\frac{\partial \omega_{p,\uparrow}}{\partial E_d}
  - \frac{\partial \omega_{p,\downarrow}}{\partial E_d}
\end{equation}
with
\begin{eqnarray}
  \frac{\partial \omega_{p,\uparrow}}{\partial E_d} &=&
  \frac{Y_1(\omega_{p,\uparrow})}{    N_1(\omega_{p,\uparrow})}
  \; ,  \nonumber \\
  Y_1(\omega_{p,\uparrow})&=&
  K\left(V_{\uparrow}^2+K(\omega_{p,\uparrow}+E_d)\right)\nonumber \\
&&    -(\omega_{p,\uparrow}+E_d)\left((\omega_{p,\uparrow}+B)^2-1\right)\; , \nonumber \\
  \frac{\partial \omega_{p,\downarrow}}{\partial E_d} &=&
\frac{Y_2(\omega_{p,\downarrow})}{    N_3(\omega_{p,\downarrow})}
  \; ,   \label{eq:Edderivatives} \\
  Y_2(\omega_{p,\downarrow})&=&
  K\left(V_{\downarrow}^2-K(\omega_{p,\downarrow}-E_d)\right) \nonumber \\
&&  +(\omega_{p,\downarrow}-E_d)\left((\omega_{p,\downarrow}-B)^2-1\right)
  \; .\nonumber 
\end{eqnarray}
The band contribution reads
\begin{equation}
  2m_{(n+1)}^{\rm band}=
  -\frac{\partial E_{\uparrow}^{\rm band}(V_{\uparrow})}{\partial E_d}
  - \frac{\partial E_{\uparrow}^{\rm band}(V_{\downarrow})}{\partial E_d}
\end{equation}
with
\begin{eqnarray}
  \frac{\partial E_{\uparrow}^{\rm band}(V_{\uparrow})}{\partial E_d}
  &=& -\int_{-(1+B)}^0\frac{\rmd \omega}{\pi}
  \frac{V_{\uparrow}^2\sqrt{1-(\omega+B)^2}}{N_2(\omega)}  \; ,
  \nonumber \\
    \frac{\partial E_{\uparrow}^{\rm band}(V_{\downarrow})}{\partial E_d}&=&
\int_{-(1-B)}^0\frac{\rmd \omega}{\pi}
    \frac{V_{\downarrow}^2\sqrt{1-(\omega-B)^2}}{N_4(\omega)}
    \label{eq:EupbandderivativeEd}
  \end{eqnarray}
to be evaluated
at $V_{\sigma}\equiv V_{\sigma,(n)}$,
$E_d\equiv E_{d,(n)}$, and $K\equiv K_{(n)}$.
In total,
 \begin{equation}
 2m_{(n+1)}= 2m_{(n+1)}^{\rm band }+ 2m_{(n+1)}^{\rm b} \; .
\label{eq:mvaluesfrombandandbound}
 \end{equation}
 When the magnetic field is applied only locally,
all explicit $B$-dependencies
in the integration boundaries and the density of states must be dropped
in the formulae in subsections~\ref{sec:DOSprops}, \ref{appsubsec:gsenergy},
and~\ref{appsubsec:sccycle}. In this case, only $E_d$ depends on~$B$.
 
 \subsection{Magnetic susceptibility}
 \label{appsubsec:magGutz}

{}From the numerical solution of the self-consistency equations,
 we see that $\gamma_{\uparrow}=\gamma_{\downarrow}=\gamma$
 and $V_{\uparrow}=V_{\downarrow}=V$ at self-consistency.
 
 \subsubsection{Small fields}

We keep all terms up to linear order in~$B$. Thus, we make the Ansatz
 \begin{eqnarray}
   \omega_{p,\uparrow}&=&\omega_p +\bar{\omega}_pB  \; , \nonumber \\
   \omega_{p,\downarrow}&=&\omega_p -\bar{\omega}_pB  \; , \nonumber \\
   K&=& \bar{K} B\; , \nonumber \\
   E_d&=& \bar{E}_d B\; , \nonumber \\
   M_0&=& \bar{M}_0B \; , \nonumber \\
   m&=& 2\chi B\; ,
 \end{eqnarray}
 where $\chi$ is the desired susceptibility in units of $(g_e\mu_{\rm B})^2$,
 \begin{equation}
\frac{\chi_0^{S,{\rm G}}}{(g_e\mu_{\rm B})^2}=   \chi    \; .
 \end{equation}
 Moreover, $\gamma=-2V/(3J_{\rm K})$ with
 \begin{equation}
J_{\rm K}(V)=-\frac{8V}{3} \left(\frac{\partial e_0(V)}{\partial V}\right)^{-1} \; ,
 \end{equation}
 where $V$ instead of~$J_{\rm K}$ parameterizes the strength of the
 Kondo interaction. For $e_0(V)$, see eq.~(\ref{eq:e0SIAM1d}).

 As a first step, we turn to eq.~(\ref{eq:getmeomegap}).
 To lowest order in~$B$ we have 
 \begin{equation}
   0= \omega_p-V^2\Lambda_0(\omega_p) \; .
      \label{appeq:ompequation}
 \end{equation}
 In one dimension, equation~(\ref{appeq:ompequation}) has the  solution
 \begin{equation}
   \omega_p=-\sqrt{\frac{\sqrt{1+4V^4}+1}{2}}=-v_+
   \label{appeq:ompofV}
 \end{equation}
 as a function of~$V$.

 Apparently, we have five unknowns,
 namely
 \begin{equation}
   \underline{v} =\left(
   \begin{array}{@{}c@{}}
     \bar{\omega}_p\\
     \bar{E}_d\\
     \bar{K}\\
     \bar{M}_0\\
     \chi
     \end{array}
   \right) \;,
   \label{appeq:defvvector}
 \end{equation}
and we need five independent linear equations that connect these quantities.

\subsubsection{Derivation of the five equations}

First, we use eq.~(\ref{eq:resetKEdagain}) to find from its leading order in~$B$
\begin{equation}
\bar{E}_d=1-\frac{J_{\rm K}}{2}\bar{M}_0+8J_{\rm K}\chi \gamma^2
\end{equation}
with $\gamma=-2V/(3J_{\rm K})$.

Second, from eq.~(\ref{eq:resetKEdagain}) we also find to leading order in~$B$
\begin{equation}
\bar{K}=J_{\rm K}\chi \; .
\end{equation}


Third, to first order in~$B$ we have from eq.~(\ref{eq:betterpolerepresentation})
 \begin{equation}
   \bar{E}_d( \omega_p^2-1)+ (2\omega_p^2 -1)\bar{\omega}_p +\omega_p^2
   -    \bar{K} V^2=0
   \; .   \label{appeq:ompbar}
 \end{equation}
 The equation for $\omega_{p,\downarrow}$ does not give any new information
  because of the symmetry in $B\to -B$.
 
  Forth, we have from eqs.~(\ref{eq:Kderivatives})
 and~(\ref{eq:Kderivativeband})
 \begin{eqnarray}
   2M_0
   &\approx& 2 (\bar{M}_0^{\rm b} +\bar{M}_0^{\rm band})B  \; , \nonumber \\
   \bar{M}_0^{\rm b} &=&
   \frac{  \bar{K} (-\omega_p^2 + 2 \omega_p^4 + V^4)
     -\omega_p^2 (3 + \bar{E}_d + 4 \bar{\omega}_p) V^2 }{
     \omega_p (1 - 2 \omega_p^2)^2} \; ,
 \nonumber\\
\bar{M}_0^{\rm band} &=&
-2\int_{-1}^0\frac{\rmd \omega}{\pi}
\frac{\sqrt{1 -   \omega^2}
  \left(\bar{K}\omega^3 V^2+\left(1-\bar{E}_d\right)\omega V^4\right) }{
  (\omega^2-\omega^4+V^4)^2}
\nonumber \\
&=& 
-2\bar{K}V^2J_3(V)+2\left(\bar{E}_d-1\right)V^4J_1(V)
\; ,
 \end{eqnarray}
 where $J_{1,3}(V)$ are defined in eqs.~(\ref{eq:J1explicit})
 and~(\ref{eq:J3explicit}).
 
Fifth, we have from eq.~(\ref{eq:Edderivatives})
 and eq.~(\ref{eq:EupbandderivativeEd})
 \begin{eqnarray}
   2m&\approx&
   4 (\chi^{\rm b}+\chi^{\rm band}) B  \; , \nonumber \\
   \chi^{\rm b} &=&
   \frac{\omega_p (1 + \bar{E}_d+ \omega_p^2 - \bar{E}_d \omega_p^2
     + 2 \bar{\omega}_p - 
   \bar{K} V^2)}{2 (1 - 2 \omega_p^2)^2}
   \; , \nonumber \\
  \chi^{\rm band} &=& \frac{1}{2\pi V^2}
  - \int_{-1}^0\frac{\rmd \omega}{\pi} Y_3(\omega)\; , \nonumber \\
Y_3(\omega)  &=&
\frac{V^2\sqrt{1 -   \omega^2}      \left(    (\omega-\omega^3)\left(\bar{E}_d-1\right)
    + \bar{K} \omega V^2 \right)}{(\omega^2-\omega^4+V^4)^2}
\nonumber   \; ,\\
  \chi^{\rm band} &=& \frac{1}{2\pi V^2}\\
&&-V^2\left[\left(\bar{E}_d-1\right)(J_1(V)-J_3(V))+\bar{K}V^2J_1(V)\right]\,,\nonumber
\end{eqnarray}
 where the first term results from the integral
 \begin{equation}
   \int_0^B 
  \frac{V_{\uparrow}^2\sqrt{1-\omega^2}\rmd \omega/\pi}{
  (\omega+E_d-B)^2(1-\omega^2)+((\omega+E_d-B)K+V_{\uparrow}^2)^2   }
   \end{equation}
to leading order in~$B$.

\subsection{Magnetic state}
\label{subsec:magstateGutzwiller}

Lastly, we address some properties of the magnetic state
for $J<J_{\rm K,c}^{\rm G}\approx 0.839$.

\subsubsection{Magnetic transition}

\begin{figure}[t]
  \includegraphics[width=8.2cm]{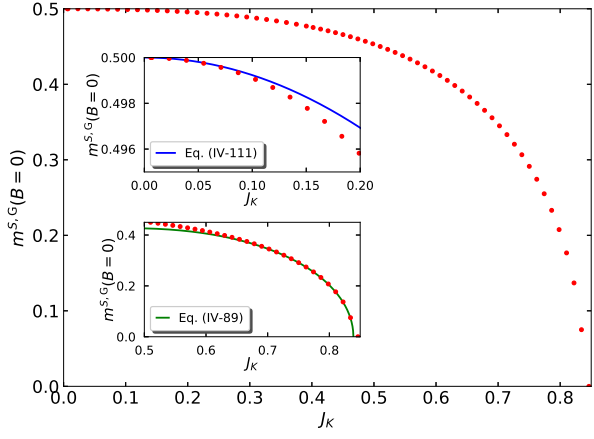}
  \caption{(Color online) Impurity spin polarization $m^{S,{\rm G}}(J_{\rm K})$
    in the Gutzwiller wave function for the one-dimensional symmetric
    Kondo model as a function of $J_{\rm K}$
    with the transition at $J_{\rm K,c}^{\rm G}\approx 0.839$.
    The mean-field behavior~(\ref{eq:Gutzmmeanfield}) close to the transition
    and the expansion for small couplings~(\ref{eq:allvaluesanalytical}) are
    shown by straight lines.\label{fig:Gutzmtransition}}
\end{figure}

The numerical solution of the self-consistency equations shows that
$V_{\uparrow}=V_{\downarrow}$ and
$\gamma_{\uparrow}=\gamma_{\downarrow}$ hold at self-consistency.
At the transition $J_{\rm K}=J_{\rm K,c}^{\rm G}$, 
the ground-state impurity magnetization $m(J_{\rm K})$ 
goes to zero continuously as a function of $J_{\rm K}$.
Moreover, the transition actually is of mean-field character,
i.e., we have
\begin{equation}
m(J_{\rm K}\lesssim J_{\rm K,c}^{\rm G}) \approx r_1 (J_{\rm K,c}^{\rm G}-J_{\rm K})^{1/2}
+r_3 (J_{\rm K,c}^{\rm G}-J_{\rm K})^{3/2} +\ldots
\label{eq:Gutzmmeanfield}
\end{equation}
with 
$r_1 \approx 1.0645$ and $r_3\approx -0.981$.
The Gutzwiller impurity magnetization is shown
in Fig.~\ref{fig:Gutzmtransition}.

\subsubsection{Expansion for small couplings}

The numerical solution of the self-consistency equations for small couplings
suggests the following behavior, using  $V_{\uparrow} = V_{\downarrow} \equiv V$
and $\gamma_{\uparrow} = \gamma_{\downarrow} \equiv \gamma$
\begin{eqnarray}
V &\approx& V_1 J_{\rm K} 
+{\cal O}(J_{\rm K}^2) \; , \nonumber \\
\gamma &\approx& -\gamma_1 J_{\rm K} 
+{\cal O}(J_{\rm K}^2) \; , \nonumber \\
E_d & \approx & E_{d,0} -E_{d,1} J_{\rm K}
+{\cal O}(J_{\rm K}^2) \; , \nonumber \\
K&\approx & K_1 J_{\rm K} 
+{\cal O}(J_{\rm K}^2) \; , \nonumber \\
m&\approx &  \frac{1}{2} - m_2 J_{\rm K}^2
+{\cal O}(J_{\rm K}^3) \; , \nonumber \\
M_0&\approx & -M_{0,1} J_{\rm K} 
+{\cal O}(J_{\rm K}^2) \; , \nonumber \\
\omega_{p,\uparrow} &:& \hbox{no bound state} 
\; , \nonumber \\
\omega_{p,\downarrow} &\approx & -1 -\omega_2 J_{\rm K}^2
+{\cal O}(J_{\rm K}^4)\; .
\label{eq:allvaluesnumerical}
\end{eqnarray}
Knowing that $E_d$ is of order unity, and $m\to 1/2$ for $J_{\rm K}\to 0$,
we find from equation~(\ref{eq:resetKEdagain}) that
\begin{equation}
K= J_{\rm K}/4  \;, 
\end{equation}
so that $K_1=1/4$ in eq.~(\ref{eq:allvaluesnumerical}).
Next, from eq.~(\ref{eq:Kderivativeband}) we readily see
that the band contribution
to $M_0$ is proportional to $J_{\rm K}^3$ so that we are left with
\begin{equation}
  \frac{\partial \omega_{p,\downarrow}}{\partial K} \approx
\frac{(E_d+1)
\left(-K(-1-E_d)\right)}{
 (-1-E_d)(-1)(-2-E_d)
  -(-1-E_d)     }=-K
\end{equation}
to leading order in $J_{\rm K}$.
Therefore, eq.~(\ref{eq:Kderivatives}) gives
\begin{equation}
M_0 = -\frac{K}{2} = -\frac{J_{\rm K}}{8}\; ,
\end{equation}
so that $M_{0,1}=1/8$ in eq.~(\ref{eq:allvaluesnumerical}).
As a consequence, $K(-1+E_d)+V^2<0$ so that $\omega_{p,\uparrow}$ does not 
exist and 
\begin{equation}
\omega_{p,\downarrow} = -1-\frac{K^2}{2} = -1 -\frac{J_{\rm K}^2}{32} \; ,
\label{eq:ompdownsecondorder}
\end{equation}
so that $\omega_2=1/32$ in  eq.~(\ref{eq:allvaluesnumerical}).

Eq.~(\ref{eq:neededforgammaup}) we then find
\begin{eqnarray}
\gamma &\approx & V \int_{-1}^0\frac{\rmd \omega}{\pi} 
\frac{(E_d + \omega)
    \sqrt{1 - \omega^2}}{(E_d + \omega)^2(1 - \omega^2)} \equiv V  I_1(E_d) \; , 
\nonumber \\
I_1(E_d) &=& \int_{-1}^0\frac{\rmd \omega}{\pi} 
\frac{1}{(E_d + \omega)\sqrt{1 - \omega^2}}\nonumber \\
&=& \frac{1}{\pi \sqrt{1-E_d^2}}
\ln \left(\frac{E_d}{1+\sqrt{1-E_d^2}}\right)
\end{eqnarray}
for $0<E_d<1$. Moreover, eq.~(\ref{eq:selfconsistencyVagain}) links
$V$ and $\gamma$,
\begin{equation}
  V = -\frac{J_{\rm K}}{\sqrt{1-4m^2}}\gamma \; ,
\label{eq:selfconsistencyVagain222}
\end{equation}
where we used that $m\to 1/2$ for $J_{\rm K}\to 0$.
Therefore, we have obtained a first equation that relates $m$ and $E_d$.
We have
\begin{equation}
\sqrt{1-4m^2}\approx  \sqrt{4m_2} J_{\rm K} = -J_{\rm K}I_1(E_d) \; ,
\end{equation}
which implies
\begin{equation}
\sqrt{4m_2}=-I_1(E_d) \quad , \quad m_2= \frac{I_1(E_d)^2}{4}\; .
\label{eq:firstalpha}
\end{equation}
In addition, from eq.~(\ref{eq:resetKEdagain}) we find
\begin{equation}
E_d\approx \frac{2}{(1-4m^2)^{3/2}}J_{\rm K}\gamma^2
= -2 \left(\frac{\gamma}{J_{\rm K}}\right)^2
\frac{1}{I_1(E_d)^3}
\end{equation}
so that
\begin{eqnarray}
\gamma&=& -J_{\rm K} \sqrt{\frac{E_d [-I_1(E_d)]^3}{2}} 
\; , \nonumber \\
V&=& J_{\rm K} \sqrt{\frac{-E_d I_1(E_d)}{2}}
\label{eq:weknowVandgammaofJK}
\end{eqnarray}
are known as a function of $E_d$. 

It remains to derive a second equation for $m_2$ as a function of $E_d$
from eq.~(\ref{eq:mvaluesfrombandandbound}). There is no contribution
from the bound states up to second order in $J_{\rm K}$, as seen from 
eq.~(\ref{eq:ompdownsecondorder}). The first non-negative contribution
originates from
\begin{equation}
    \frac{\partial E_{\uparrow}^{\rm band}(V_{\downarrow})}{\partial E_d}
  \approx  \int_{-1}^0\frac{\rmd \omega}{\pi}
  \frac{V^2\sqrt{1-\omega^2}}{(\omega-E_d)^2(1-\omega^2) }
  =V^2 I_2(E_d) 
\end{equation}
with
\begin{eqnarray}
I_2(E_d) &=& \int_{-1}^0\frac{\rmd \omega}{\pi}
  \frac{1}{(\omega-E_d)^2\sqrt{1-\omega^2} }\nonumber \\
  &=& \frac{1}{\pi E_d(1 - E_d^2)}\\
  &&+ 
  \frac{E_d}{\pi(1 - E_d^2)^{3/2}}\ln\left(\frac{1-\sqrt{1-E_d^2}}{E_d}\right)
  \nonumber
\end{eqnarray}
using {\sc Mathematica}.~\cite{Mathematica11}
In the remaining term we isolate the singularity for $V\to 0$,
\begin{equation}
    \frac{\partial E_{\uparrow}^{\rm band}(V_{\uparrow})}{\partial E_d}
    \approx  T_a(E_d) +T_b(E_d)   \; ,
\end{equation}
where
\begin{eqnarray}
T_a(E_d) &=& -\int_{-1}^0\frac{\rmd \omega}{\pi}
\sqrt{1-\omega^2}
\frac{V^2}{(\omega+E_d)^2(1-E_d^2) +V^4}
\; , \nonumber \\
T_b(E_d) &=& \int_{-1}^0\frac{\rmd \omega}{\pi}\sqrt{1-\omega^2} \biggl[  
\frac{V^2}{(\omega+E_d)^2(1-E_d^2) +V^4} \nonumber \\
&& 
-\frac{V^2}{(\omega+E_d)^2(1-\omega^2) +(V^2+K(\omega+E_d))^2}
\biggr] \; . \nonumber \\
\end{eqnarray}
Using {\sc Mathematica}~\cite{Mathematica11}
we find
\begin{eqnarray}
T_a(x) &=& -1 + V^2 I_3(x) \; , \nonumber \\
I_3(x) &=& \frac{2 + \pi x}{2\pi x (1 - x^2)} \nonumber \\
&&+
  \frac{x}{\pi(1 - x^2)^{3/2}(\sqrt{1 - x^2} + 1)}
\ln\biggl(\frac{1 + \sqrt{1 - x^2}}{x}\biggr)\nonumber \\
&&+ \frac{x}{2\pi(1 - x^2)(\sqrt{1 - x^2} + 1)}\nonumber \\
&&\hphantom{+\; }\times
\ln\biggl(\frac{2 - x^2 + 2 \sqrt{1 - x^2}}{x^2}\biggr)\;.
\end{eqnarray}
In the term $T_b(E_d)$ we are in the position to let $K,V\to 0$ 
in the denominator such that
\begin{eqnarray}
T_b(x) &\approx & V^2
\int_{-1}^0\frac{\rmd \omega}{\pi}\frac{\sqrt{1-\omega^2}}{(\omega+x)^2} 
\left[\frac{1}{(1-x^2)} -\frac{1}{(1-\omega^2)}\right] \nonumber \\
&=& \frac{V^2 }{1-x^2}
\int_{-1}^0\frac{\rmd \omega}{\pi}\frac{1}{\sqrt{1-\omega^2}}
\frac{x-\omega}{x+\omega}
\equiv V^2 I_4(x)\; , \nonumber \\
I_4(x)&=& -\frac{1}{2 (1 - x^2)} -
 \frac{2x}{\pi(1 - x^2)^{3/2}}
\ln\biggl(\frac{1 + \sqrt{1 -x^2}}{x}\biggr) \, .\nonumber \\
\end{eqnarray}
In total we find to second order in $V$ and thus in $J_{\rm K}$
\begin{eqnarray}
m&\approx &\frac{1}{2} -\frac{V^2}{2}\left(I_2(E_d)+I_3(E_d)+I_4(E_d)\right)\\
&\approx & \frac{1}{2} -J_{\rm K}^2 \frac{E_d(-I_1(E_d))}{4} 
\left(I_2(E_d)+I_3(E_d)+I_4(E_d)\right)\nonumber
\end{eqnarray}
using eq.~(\ref{eq:weknowVandgammaofJK}) so that
\begin{eqnarray}
m_2&=&\frac{E_d(-I_1(E_d))}{4} 
\left(I_2(E_d)+I_3(E_d)+I_4(E_d)\right)\nonumber \\
&\stackrel{!}{=} &\frac{I_1(E_d)^2}{4}
\end{eqnarray}
from eq.~(\ref{eq:firstalpha}).
The numerical solution of this equation gives
\begin{equation}
E_d=0.3857965059132358 \; .
\end{equation}
Therefore, to leading order we have the analytic results,
\begin{eqnarray}
V &=& \sqrt{\frac{-E_d I_1(E_d)}{2}} J_{\rm K} 
+{\cal O}(J_{\rm K}^2) \; , \nonumber \\
&= & 0.32694495035854915 J_{\rm K} 
+{\cal O}(J_{\rm K}^2) \; , \nonumber \\
\gamma &=& -\sqrt{\frac{E_d [-I_1(E_d)]^3}{2}} J_{\rm K} 
+{\cal O}(J_{\rm K}^2) \; , \nonumber \\
&=&  -0.18117388948693602 J_{\rm K} 
+{\cal O}(J_{\rm K}^2) \; ,\nonumber \\
\end{eqnarray}
and
\begin{eqnarray}
E_d & = & 0.3857965059132358
+{\cal O}(J_{\rm K}) \; , \nonumber \\
K&= & \frac{J_{\rm K}}{4} 
+{\cal O}(J_{\rm K}^2) \; , \nonumber \\
m&= &  \frac{1}{2} - m_2 J_{\rm K}^2
+{\cal O}(J_{\rm K}^3) \; , \nonumber \\
m_2&=& \frac{I_1(E_d)^2}{4} = 0.07676830582531644 \; , \nonumber \\
M_0&= & -\frac{J_{\rm K}}{8} 
+{\cal O}(J_{\rm K}^2) \; , \nonumber \\
\omega_{p,\uparrow} &:& \hbox{no bound state} 
\; , \nonumber \\
\omega_{p,\downarrow} &= & -1 -\frac{J_{\rm K}^2}{32}
+{\cal O}(J_{\rm K}^4)\; .
\label{eq:allvaluesanalytical}
\end{eqnarray}

\subsubsection{Ground-state energy}

The ground-state energy for small values of $J_{\rm K}$ must be determined numerically
because the two integrals in eq.~(\ref{eq:defAofkrams}) are too cumbersome
to analyze in the weak-coupling limit.
However, it can be shown that the Gutzwiller energy in
eq.~(\ref{eq:E0Gutzwitheverything}) is quadratic in the coupling for small~$J_{\rm K}$.
A quartic fit of the numerical data gives
\begin{equation}
  E_0^{\rm G}(J_{\rm K})\approx
-0.0908979J_{\rm K}^2  - 0.047363 J_{\rm K}^3  - 0.0450817 J_{\rm K}^4\; .
\end{equation}
The approximation is excellent up to $J_{\rm K}=0.4$.
We may safely state that
\begin{equation}
  E_0^{\rm G}(J_{\rm K})\approx
 -0.0905 J_{\rm K}^2 -0.051J_{\rm K}^3-0.05 J_{\rm K}^4
\end{equation}
for the Gutzwiller variational energy for $J_{\rm K}\lesssim 0.4$.
The quadratic coefficient can now be compared with the exact result from
perturbation theory,
\begin{equation}
  E_0(J_{\rm K})\approx -\frac{3}{32} J_{\rm K}^2 =-0.09375 J_{\rm K}^2\; .
\end{equation}
The magnetic Gutzwiller states accounts for 96.5\% of the
correlation energy.

\section{Ground-state energy from Bethe Ansatz}
\label{supp:GSEBetheAnsatz}

Using Bethe Ansatz, the Kondo model is solved
for a linear dispersion relation with unit Fermi velocity
in the wide-band limit, i.e., the dispersion relation $\epsilon^{\rm BA}(k)=k$
formally extends from $-\infty$ to $\infty$. Therefore, an appropriate energy
cut-off~$D$ must be introduced. This procedure is not unique.
Therefore, there are two Bethe Ansatz solutions for the spin-1/2 Kondo model.
First, the one discussed in Tsvelick and Wiegmann,~\cite{TsvelickWiegmann}
referred to as~TW, and the one reviewed by Andrei, Furuya, and
Lowenstein,~\cite{RevModPhys.55.331} referred to as~AFL.
The basic Bethe Ansatz equations agree but the
expressions for the parameters as a function of the Kondo coupling
differ beyond leading-order in $J_{\rm K}^{\rm BA}/D$.
For a lattice-regularized Bethe-Ansatz solvable impurity model, see
Ref.~[\onlinecite{BortzKluemper}].

\subsection{Bethe Ansatz equations}

\subsubsection{Basic relations}

According to equation~(4.2.A) of~TW,
the electronic momenta are obtained from
\begin{equation}
  e^{\rmi k_j L}
  = e^{-\rmi \phi}
\prod_{\gamma=1}^{M} 
\frac{\tilde{\lambda}_{\gamma}+\rmi/2}{\tilde{\lambda}_{\gamma}-\rmi/2}
\; ,
\label{eq:kdefeBATW}
\end{equation}
where $L$ is a length scale that helps to quantize the momenta.
$M=(N^e+1)/2$ in the ground state where $S^z=0$.
The spin momenta are found from eq.~(4.2.B) of~TW,
\begin{equation}
\left[ 
\frac{\tilde{\lambda}_{\gamma}+\rmi/2}{\tilde{\lambda}_{\gamma}-\rmi/2}
\right]^{N^e}
\left[ 
\frac{\tilde{\lambda}_{\gamma}+1/g+\rmi/2}{\tilde{\lambda}_{\gamma}+1/g-\rmi/2}
\right]
=-\prod_{\delta=1}^M 
\left[ 
\frac{\tilde{\lambda}_{\gamma}-\tilde{\lambda}_{\delta}+\rmi}{
\tilde{\lambda}_{\gamma}-\tilde{\lambda}_{\delta}-\rmi}
\right]
\label{eq:lambdadefeBATW}
\end{equation}
for $\gamma=1,\ldots,M$. Here, $g$ is the interaction parameter.

To make contact with the work by AFL,~\cite{RevModPhys.55.331} we substitute
\begin{equation}
\tilde{\lambda}_{\gamma}=\frac{\lambda_{\gamma}-1}{g}
\end{equation}
and set $g=c$ to find
\begin{equation}
e^{\rmi k_j L} = e^{-\rmi \phi}
\prod_{\gamma=1}^{M} 
\frac{\rmi (1-\lambda_{\gamma})+c/2}{\rmi(1-\lambda_{\gamma})-c/2}
\label{eq:kdefeBA}
\end{equation}
and $L(\lambda_{\gamma})=R(\lambda_{\gamma})$ with
\begin{eqnarray}
L(\lambda_{\gamma})&=&\left[ 
\frac{\rmi (1-\lambda_{\gamma})+c/2}{\rmi(1-\lambda_{\gamma})-c/2}
\right]^{N^e}
\left[ 
\frac{\rmi (-\lambda_{\gamma})+c/2}{\rmi(-\lambda_{\gamma})-c/2}
\right] \; , \nonumber\\
R(\lambda_{\gamma})&=&-\prod_{\delta=1}^M
\left[ 
\frac{\rmi (\lambda_{\delta}-\lambda_{\gamma})+c}{
\rmi(\lambda_{\delta}-\lambda_{\gamma})-c}
\right]
\label{eq:lambdadefeBA}
\end{eqnarray}
for $\gamma=1,\ldots,M=(N^e+1)/2$.
Eqs.~(\ref{eq:kdefeBA}) and~(\ref{eq:lambdadefeBA})
are eq.~(2.47) and eq.~(2.48') of~AFL.
The energy is obtained from
\begin{equation}
E=\sum_{j=1}^{N^e} k_j \; ,
\label{eq:energyBA}
\end{equation}
see eq.~(2.10') of~AFL and eq.~(4.2.C) of~TW.

Note that TW and AFL give different expressions for $\phi$ and $c$,
namely 
\begin{eqnarray}
  c^{\rm TW}&=& \tan(J_{\rm K}^{\rm BA}) \approx J_{\rm K}^{\rm BA}
  +\frac{(J_{\rm K}^{\rm BA})^3}{3} +\ldots
  \label{eq:cvaluesBA}\; , \\
  c^{\rm AFL} &=& \frac{J_{\rm K}^{\rm BA}}{1-3(J_{\rm K}^{\rm BA})^2/16}
  \approx J_{\rm K}^{\rm BA}
  +\frac{3}{16}(J_{\rm K}^{\rm BA})^3 +\ldots \; ,\nonumber
\end{eqnarray}
and
\begin{eqnarray}
 \phi^{\rm TW}&=&-\frac{J_{\rm K}^{\rm BA}}{2}\; ,  \nonumber \\
\phi^{\rm AFL}&=&\rmi
\ln\left[\frac{1-3(J_{\rm K}^{\rm BA})^2/16+\rmi J_{\rm K}^{\rm BA}}{1+
    3(J_{\rm K}^{\rm BA})^2/16+\rmi J_{\rm K}^{\rm BA}/2}
  \right]\nonumber\\
&=&\rmi \ln\left[\frac{4\rmi- J_{\rm K}^{\rm BA}}{4\rmi +J_{\rm K}^{\rm BA}}\right]
\nonumber \\
&\approx &-\frac{J_{\rm K}^{\rm BA}}{2}+\frac{(J_{\rm K}^{\rm BA})^3}{96}\; ,
  \label{eq:phivaluesBA}
\end{eqnarray}
see eq.~(4.2.A) and (4.2.55') of~TW with $I\equiv 2J_{\rm K}^{\rm BA}$
and  eq.~(2.42) of~AFT, where $J=-J''=J_{\rm K}^{\rm BA}/2$ was used.
To leading order, the interaction parameters agree
but they differ to third order in $J_{\rm K}^{\rm BA}$.
As a consequence, the ground-state energy agrees 
to leading order in $J_{\rm K}^{\rm BA}$ only.

\subsubsection{Non-interacting case}

We can take the limit $c\to 0^+$
in eqs.~(\ref{eq:kdefeBA}) and~(\ref{eq:lambdadefeBA})
because the parameters $\lambda_{\gamma}$ remain finite in the non-interacting limit,
see below. Therefore, we find that eq.~(\ref{eq:lambdadefeBA}) is fulfilled
if we assume that all $\lambda_{\gamma}$ are pairwise different
and different from zero. This is fulfilled for finite system sizes.
Moreover, eq.~(\ref{eq:kdefeBA}) reduces to
\begin{equation}
e^{\rmi k_j^{(0)} L} =1 
\end{equation}
because $\phi(0)=0$ so that
\begin{equation}
k_j^{(0)}=\frac{2\pi}{L} \tilde{n}_j
\end{equation}
with integer $\tilde{n}_j$ and
\begin{equation}
E^{\rm (0), BA}=\frac{2\pi}{L}\sum_{j=1}^{N^e} \tilde{n}_j \; .
\label{eq:BAnonintenergydef}
\end{equation}
In the following, we are interested in the ground state and its energy
in the thermodynamic limit.

We want to make contact with our lattice Hamiltonian.
Its dispersion around the Fermi wave vector is linear, 
\begin{equation}
\epsilon(m)\approx 2t \frac{2\pi(m-m_{\rm F})}{L+1} \; .
\end{equation}
When we work with a half-filled band, $N^e=(L+3)/2\approx L/2$,
the corrections to the linear dispersion relation
are not of second order, but of order $(m-m_{\rm F})^3$,
see eq.~(\ref{eq:correctionsthirdorder}).
Therefore, the Bethe Ansatz results agree with those from 
DMRG for the lattice Hamiltonian in a larger region of $J_{\rm K}$-values.
For $t=1/2$ and $m_{\rm F}=(L+3)/4\approx L/4$ ($L\gg 1$) we have
\begin{equation}
E_0^{\rm (0), latt}=2 (2t) \frac{2\pi}{L} \sum_{m=1}^{m_{\rm F}}(m-m_{\rm F})
=-\frac{\pi L}{8} \; ,
\label{eq:latticeEzero}
\end{equation}
where the additional factor two accounts for the spin degeneracy.

Therefore, in eq.~(\ref{eq:BAnonintenergydef}) we choose the $N^e\approx L/2$ 
integers
\begin{equation}
\tilde{n}_j= -\frac{3N^e}{4},\ldots,\frac{N^e}{4}-1
\label{eq:ntilda}
\end{equation}
for our sequence of Bethe-Ansatz charge quantum numbers.
Then,
\begin{equation}
E_0^{\rm (0), BA}=\frac{2\pi}{L}\sum_{\tilde{n}=-3N^e/4}^{N^e/4-1} \tilde{n} 
=-\frac{\pi L}{8}\; ,
\label{eq:BAnonintenergyresult}
\end{equation}
in agreement with the expression~(\ref{eq:latticeEzero}) 
for the lattice Hamiltonian
$E_0^{\rm (0), BA}=E_0^{\rm (0), latt}$.

\subsubsection{Energy from Bethe quantum numbers}

We use the following identity,
\begin{equation}
\ln\left[\frac{\rmi x+1}{\rmi x-1}\right] = (-2\rmi) \arccot(x) \; .
\label{eq:logidentity}
\end{equation}
Note that $\arccot(x)$ has a jump discontinuity of $-\pi$ at $x=0$.

Therefore, taking the complex logarithm of eqs.~(\ref{eq:kdefeBA}) gives
\begin{equation}
  \rmi k_j L = 2\pi\rmi \tilde{n}_j -\rmi \phi +  (-2\rmi) 
\sum_{\gamma=1}^{(N^e+1)/2} \arccot\left[2(1-\lambda_{\gamma})/c\right] \; .
\label{eq:BAmomentacorrected}
\end{equation}
The integer numbers $\tilde{n}_j$ (charge Bethe quantum numbers)
distinguish between the various electron momenta. 

The energy~(\ref{eq:energyBA}) thus becomes
\begin{eqnarray}
E&=&\sum_{j=1}^{N^e} \frac{2\pi}{L}\tilde{n}_j -\frac{N^e}{L}\phi\nonumber \\
&& +\frac{N^e}{L} 
\sum_{\gamma=1}^{(N^e+1)/2} 
\left(-2\arccot\left[2(1-\lambda_{\gamma})/c\right]\right) \;.\nonumber \\
\label{eq:BA31corrected}
\end{eqnarray}
Eq.~(\ref{eq:BA31corrected}) replaces eq.~(3.1) of~AFL
where the $\phi$-term was dropped as `inessential'.
Of course, it is important for the complete ground-state energy.

Next, taking the complex logarithm of eq.~(\ref{eq:lambdadefeBA}) 
we find
\begin{eqnarray}
L(\lambda_{\gamma})&=& R(\lambda_{\gamma})\nonumber \; ,\\
L(\lambda_{\gamma})&=& 
N^e\left(-2 \arccot\left[2(1-\lambda_{\gamma})/c\right]\right)\nonumber \\
&&+\left(-2\arccot\left[2(-\lambda_{\gamma})/c\right]\right)
\nonumber \; , \\
R(\lambda_{\gamma})&=& 2\pi \tilde{I}_{\gamma} 
+\sum_{\delta=1}^{(N^e+1)/2}
\left(-2 
\arccot\left[(\lambda_{\delta}-\lambda_{\gamma})/c\right]
\right) \; ,\nonumber\\
\label{eq:improved32}
\end{eqnarray}
where the $(N^e+1)/2$ (half-)integer numbers $\tilde{I}_{\gamma}$ 
(spin Bethe quantum numbers)
distinguish the various spin momenta. 

\subsubsection{Contact with energy formulae in AFL}

In our formulae~(\ref{eq:BA31corrected}) and~(\ref{eq:improved32})
we replace
\begin{equation}
-2\arccot(-x)=-2\arctan(x)-\pi +2\pi \theta_{\rm H}(x) \; ,
\end{equation}
where $\theta_{\rm H}(x)$ is the Heaviside step function.
We define 
\begin{equation}
\Theta(x)=-2\arctan(x/c)
\end{equation}
to make close contact with the formulae in~AFL.
Then, eq.~(\ref{eq:BA31corrected}) becomes
\begin{equation}
E=\sum_{j=1}^{N^e} \frac{2\pi}{L}n_j -\frac{N^e}{L}\phi
+\frac{N^e}{L} 
\sum_{\gamma=1}^{(N^e+1)/2} 
\left[\Theta\left(2\lambda_{\gamma}-2\right)-\pi\right] \;,
\label{eq:BA31correctedtransformed}
\end{equation}
where we set
\begin{equation}
n_j=\tilde{n}_j +\sum_{\gamma=1}^{(N^e+1)/2} 
\theta_H\left((2\lambda_{\gamma}-2)/c\right) \; .
\label{eq:njvalues}
\end{equation}
Apart from the $\phi$-term,
eq.~(\ref{eq:BA31correctedtransformed}) is eq.~(3.1) of~AFL.

Furthermore, eq.~(\ref{eq:improved32}) becomes
\begin{equation}
N^e\Theta_{\rm H}(2\lambda_{\gamma}-2)+\Theta(2\lambda_{\gamma})
=-2\pi I_{\gamma} +
\sum_{\delta=1}^{(N^e+1)/2} \Theta(\lambda_{\gamma}-\lambda_{\delta})
\label{eq:BA32transformed}
\end{equation}
with a proper re-definition of the (half-)integers $I_{\gamma}$. 
Eq.~(\ref{eq:BA32transformed}) is eq.~(3.2) of~AFL.

\subsection{Spin distribution function}

\subsubsection{Integral equation for the spin momenta}

We start from eq.~(\ref{eq:BA32transformed}) and define
the spin distribution function
\begin{equation}
  \sigma(\lambda_{\gamma})= \frac{1}{\lambda_{\gamma+1}-\lambda_{\gamma}}
  =\frac{1}{\Delta \lambda_{\gamma}}\; ,
\label{eq:defsigmafirst}
\end{equation}
which is of the order $1/L$ for $L\gg 1$. 
As usual, we take the difference of two 
consecutive values for $\gamma$,
\begin{eqnarray}
-2\pi &=& -2\pi I_{\gamma+1}+2\pi I_{\gamma} 
\nonumber \\
&=& N^e\left[ \Theta(2\lambda_{\gamma+1}-2)-\Theta(2\lambda_{\gamma}-2)\right]
\nonumber \\
&&+\Theta(2\lambda_{\gamma+1})-\Theta(2\lambda_{\gamma})
\nonumber \\
&& -\int\rmd \lambda' \sigma(\lambda')
\left[ \Theta(\lambda_{\gamma+1}-\lambda')
-\Theta(\lambda_{\gamma}-\lambda')\right]
\; , \nonumber \\
\label{eq:intermediate}
\end{eqnarray}
where we used the definition of the Riemann integral 
in the thermodynamic limit,
\begin{equation}
  \sum_{\delta}f(\lambda_{\delta}) =\int \frac{\rmd \lambda'}{\Delta \lambda'}  f(\lambda')
  =\int \rmd \lambda' \sigma(\lambda')   f(\lambda')\; .
\label{eq:defsigma}
\end{equation}
Using eq.~(\ref{eq:defsigmafirst}) again we find
\begin{equation}
\Theta(2\lambda_{\gamma+1}-2)-\Theta(2\lambda_{\gamma}-2)
= \Theta'(2\lambda_{\gamma}-2)\frac{2}{\sigma(\lambda_{\gamma})}
\end{equation}
with
\begin{equation}
\Theta'(x)=-\frac{2c}{x^2+c^2} \; .
\end{equation}
Thus, eq.~(\ref{eq:intermediate}) becomes by setting 
$\lambda_{\gamma}\equiv \lambda$
\begin{eqnarray}
-2\pi\sigma(\lambda) &=& 
2N^e\Theta'(2\lambda-2)+2\Theta'(2\lambda)\nonumber \\
&&-\int\rmd \lambda' \sigma(\lambda')
\Theta'(\lambda-\lambda')
\; . \label{eq:integralequation}
\end{eqnarray}
This is the integral equation~(3.8) of~AFL, 
\begin{equation}
\sigma(\lambda) = f(\lambda)-\int\rmd\lambda'
K(\lambda-\lambda')\sigma(\lambda')
\label{eq:39a}
\end{equation}
with
\begin{eqnarray}
f(\lambda)&=& \frac{2c}{\pi}\left(\frac{N^e}{c^2+4(\lambda-1)^2}
+ \frac{1}{c^2+4\lambda^2}\right)  \; , \nonumber \\
K(\lambda)&=& \frac{c}{\pi} 
\frac{1}{c^2+\lambda^2}\;  \label{eq:39b}\; ,
\end{eqnarray}
in agreement with eq.~(3.9) of~AFL.

\subsubsection{Solution of the integral equation}

Upon Fourier transformation
\begin{eqnarray}
g(\lambda) &=& \int\rmd x g(x)e^{\rmi\lambda x} \nonumber \; ,\\
g(x) &=& \int \frac{\rmd \lambda}{2\pi} g(\lambda)e^{-\rmi \lambda x}\; ,
\label{eq:Fouriertrafo}
\end{eqnarray}
we obtain from~eq.~(\ref{eq:39a})
\begin{equation}
\sigma(x)=\frac{f(x)}{1+K(x)}
\label{eq:thatsit}
\end{equation}
with
\begin{eqnarray}
f(x)&=&\int\rmd\lambda e^{\rmi\lambda x}\frac{2c}{\pi}
\left(\frac{N^e}{c^2+4(\lambda-1)^2}
+ \frac{1}{c^2+4\lambda^2}
\right) \nonumber \\
&=& e^{-c|x|/2}\left(N^e e^{\rmi x}+1\right) ,\\
K(x)&=& \int\rmd\lambda e^{\rmi\lambda x}\frac{c}{\pi}
\frac{1}{c^2+\lambda^2}= e^{-c|x|} \; .
\end{eqnarray}
Therefore, we obtain $\sigma(\lambda)$ from eq.~(\ref{eq:Fouriertrafo})
and eq.~(\ref{eq:thatsit}) as
\begin{eqnarray}
\sigma(\lambda)&=& \int \frac{\rmd x}{2\pi} e^{-\rmi \lambda x-c|x|/2}
\frac{N^e e^{\rmi x}+1}{1+e^{-c|x|}}  \nonumber \\
&=& 
\int \frac{\rmd x}{2\pi} 
\frac{e^{-\rmi \lambda x}}{2\cosh(cx/2)}\left(N^e e^{\rmi x}+1\right)
\nonumber \\
&=& \frac{1}{2c} \left(\frac{N^e}{\cosh(\pi(\lambda-1)/c)}
+ \frac{1}{\cosh(\pi\lambda/c)}
\right)\; ,\nonumber\\
\label{eq:310}
\end{eqnarray}
using {\sc Mathematica}.~\cite{Mathematica11}
Eq.~(\ref{eq:310}) is eq.~(3.10) of~AFL.

\subsection{Ground-state energy}

\subsubsection{Charge quantum numbers in the ground state}

In eq.~(\ref{eq:ntilda}) we determined the values for the
charge Bethe-Ansatz quantum numbers $\tilde{n}_j$ in the ground state.
According to eq.~(\ref{eq:njvalues}), the quantum numbers $n_j$
obey
\begin{eqnarray}
n_j&=&\tilde{n}_j+\int\rmd\lambda \sigma(\lambda)\theta_{\rm H}((2\lambda-2)/c)
\nonumber \\
&=& \tilde{n}_j +\frac{N^e}{2c} 
\int_1^{\infty}\rmd\mu \frac{1}{\cosh(\pi\mu/c)}\nonumber \\
&=& \tilde{n}_j +\frac{N^e}{4}  
\end{eqnarray}
in the thermodynamic limit.
Consequently, we find from eq.~(\ref{eq:ntilda})
that the quantum numbers $n_j$ are symmetrically distributed around zero,
\begin{equation}
n_j= -\frac{N^e}{2},\ldots,\frac{N^e}{2}-1 \; .
\label{eq:njvaluesfinal}
\end{equation}
as discussed in~TW, Sect.~5.1.1.

\subsubsection{Ground-state energy in the thermodynamic limit}

In the thermodynamic limit, the ground-state 
energy in eq.~(\ref{eq:BA31correctedtransformed}) becomes
\begin{eqnarray}
E_0&=& -\frac{N^e}{L}\phi
+ \frac{N^e}{L}
\int \rmd \lambda \sigma(\lambda)
\left[\Theta(2\lambda-2)-\pi\right] \nonumber \\
&\equiv & 
 \frac{N^e}{L}\left(-\phi+{\cal B}(c)+{\cal I}(c)\right) \; ,
\end{eqnarray}
because
the quantum numbers $n_j$ are distributed symmetrically around zero so that
the sum over all occupied $n_j$ gives zero.
Here,
\begin{eqnarray}
  {\cal B}(c) &\equiv& \int\rmd \lambda \frac{N^e}{2c}
  \frac{\Bigl[-\pi -2\arctan\left[2(\lambda-1)/c\right]\Bigr]}{\cosh(\pi(\lambda-1)/c)}
\nonumber \\
&=& -\frac{\pi}{2}N^e \; ,
\end{eqnarray}
and
\begin{eqnarray}
{\cal I}(c)&=&  {\cal I}_1(c)+{\cal I}_2(c)\; , \nonumber \\
 {\cal I}_1(c) &=& -\pi \int_{-\infty}^{\infty}\rmd \lambda
  \frac{1}{2c\cosh[\pi\lambda/c]} = -\frac{\pi}{2} \; , \\
{\cal I}_2(c) &=& 
\int_{-\infty}^{\infty}\rmd \lambda
  \frac{1}{2c\cosh[\pi\lambda/c]}
\Bigl[-2\arctan\left[2(\lambda-1)/c\right]\Bigr]  \; . \nonumber \\
    \label{eq;startIc}
\end{eqnarray}
Altogether, we have 
\begin{equation}
E_0= -\frac{\pi (N^e)^2}{2L}
+\frac{N^e}{L}\left(-\phi-\frac{\pi}{2} + {\cal I}_2(c)\right) \; .
\end{equation}
Apart from the $\phi$-term this is the first line of eq.~(3.12) of~AFL.

When we subtract the energy $E_0^{\rm (0),BA}$ for $c=0$,
see eq.~(\ref{eq:BAnonintenergyresult}), 
we find
\begin{equation}
e_0(J)=E_0-E_0^{(0),{\rm BA}}=\frac{1}{2}
\left(-\phi-\frac{\pi}{2} + {\cal I}_2(c)\right) \; ,
\label{eq:finalBAAFLresult}
\end{equation}
where we used $N^e=L/2$.

\subsubsection{Evaluation of the integral ${\cal I}_2(c)$}
\label{sec:eq312}

The calculation of the ground-state energy requires the evaluation of
the integral ${\cal I}_2(c)$, eq.~(\ref{eq;startIc}).
For its calculation we use the integral representation of the arctan function
\begin{equation}
  \arctan(\lambda) = \int_0^{\infty}\rmd \omega e^{-\omega}
  \frac{\sin(\omega\lambda)}{\omega}
\label{eqapp:arctan2}
  \end{equation}
to find
\begin{equation}
  {\cal I}_2(c)=-\frac{1}{c} \int_0^{\infty}\frac{\rmd \omega}{\omega} e^{-\omega}
  \int_{-\infty}^{\infty}\rmd \lambda\frac{\sin\left(2\omega(\lambda-1)/c\right)
  }{\cosh(\pi\lambda/c)}
    \;.
  \end{equation}
With the help of {\sc Mathematica}~\cite{Mathematica11}
we can write ${\cal I}_2(c)$ as
\begin{equation}
  {\cal I}_2(c)=\int_0^{\infty}\frac{\rmd \omega}{\omega} e^{-\omega}
  \frac{\sin(2\omega/c)}{\cosh\omega}=
  \int_{-\infty}^{\infty}\frac{\rmd \omega}{\rmi \omega}
  \frac{ e^{\rmi \omega}}{e^{|\omega|c}+1} \; .
\end{equation}
The latter integral can be found in eq.~(17.18) in Ref.~[\onlinecite{Essler}]
\begin{equation}
  {\cal I}_2(c)=\rmi \ln\left[
    \frac{\displaystyle
      \Gamma\left(\frac{1}{2}+\frac{\rmi}{2c}\right)
      \Gamma\left(1-\frac{\rmi}{2c}\right)
    }{\displaystyle
      \Gamma\left(\frac{1}{2}-\frac{\rmi}{2c}\right)
      \Gamma\left(1+\frac{\rmi}{2c}\right)
    }
    \right] \; .
  \label{eq:I2fromBA}
\end{equation}
It can be derived using the series expansion of $1/\cosh(\omega)$, performing
the $\omega$-integral, and using eq.~(6.3.13) of~[\onlinecite{abramowitzstegun}].
Eqs.~(\ref{eq:finalBAAFLresult}) and~(\ref{eq:I2fromBA})
correct the typographical errors in equation~(3.12) of~AFL.

\subsubsection{Final result for the ground-state energy}

Lastly, we summarize the result for the ground-state energy from Bethe Ansatz
that originates from the interaction of the impurity spin 
and the bath spin at the origin.

\paragraph{Andrei, Furuya, Lowenstein}

\begin{eqnarray}
e_0^{\rm AFL}(J)&=&\arctan(J/2)+\frac{1}{2}
\left(-\frac{\pi}{2} + {\cal I}_2(c)\right) \; ,\nonumber \\
c^{\rm AFL}&=& \frac{2J}{1-3J^2/4} \; , \nonumber \\
{\cal I}_2(c) &=&
\rmi \ln\left[
    \frac{\displaystyle
      \Gamma\left(\frac{1}{2}+\frac{\rmi}{2c}\right)
      \Gamma\left(1-\frac{\rmi}{2c}\right)
    }{\displaystyle
      \Gamma\left(\frac{1}{2}-\frac{\rmi}{2c}\right)
      \Gamma\left(1+\frac{\rmi}{2c}\right)
    }
    \right]
\end{eqnarray}
with $J=J_{\rm K}^{\rm BA}/2$.
The  expansion for small couplings~$J$ reads
\begin{eqnarray}
e_0^{\rm AFL}(J) &=& \frac{J}{2} -\frac{J^3}{24}+\frac{J^5}{160}
-\frac{1}{2}\left(\frac{c}{2}+\frac{c^3}{12}+\frac{c^5}{10}\right) \nonumber \\
&=& 
\frac{J}{2} -\frac{J^3}{24}+\frac{J^5}{160}
-\left(\frac{J}{2} +\frac{17J^3}{24}+\frac{421J^5}{160}\right)\nonumber \\
&=& -\frac{3J^3}{4}-\frac{21J^5}{8} \nonumber \\
&=& -\frac{3}{32} (J_{\rm K}^{\rm BA})^3 - \frac{21}{256} (J_{\rm K}^{\rm BA})^5\; ,
\label{eq:finale0AFL}
\end{eqnarray}
up to corrections of the order $(J_{\rm K}^{\rm BA})^7$.
The result was obtained using {\sc Mathematica}.~\cite{Mathematica11}

\paragraph{Tsvelick and Wiegmann}

\begin{eqnarray}
e_0^{\rm TW}(I)&=&\frac{I}{8}+\frac{1}{2}
\left(-\frac{\pi}{2} + {\cal I}_2(c)\right) \; ,\nonumber \\
c^{\rm TW}&=& \tan(I/2)\; , \nonumber \\
{\cal I}_2(c) &=&
\rmi \ln\left[
    \frac{\displaystyle
      \Gamma\left(\frac{1}{2}+\frac{\rmi}{2c}\right)
      \Gamma\left(1-\frac{\rmi}{2c}\right)
    }{\displaystyle
      \Gamma\left(\frac{1}{2}-\frac{\rmi}{2c}\right)
      \Gamma\left(1+\frac{\rmi}{2c}\right)
    }
    \right] 
\end{eqnarray}
with $I=2J_{\rm K}^{\rm BA}$.
The  expansion for small couplings~$I$ reads
\begin{eqnarray}
e_0^{\rm TW}(I) &=& \frac{I}{8}
-\frac{1}{2}\left(\frac{c}{2}+\frac{c^3}{12}+\frac{c^5}{10}\right) \nonumber \\
&=& 
\frac{I}{8} -\left(\frac{I}{8} +\frac{I^3}{64}+\frac{I^5}{256}\right)
=-\frac{I^3}{64}-\frac{I^5}{256} \nonumber \\
&=& -\frac{4}{32} (J_{\rm K}^{\rm BA})^3 - \frac{32}{256} (J_{\rm K}^{\rm BA})^5\; ,
\label{eq:finale0TW}
\end{eqnarray}
up to corrections of the order $(J_{\rm K}^{\rm BA})^7$.
The result was obtained using {\sc Mathematica}.~\cite{Mathematica11}

The comparison of eq~(\ref{eq:finale0AFL}) and~(\ref{eq:finale0TW})
show that the two expressions differ to third order.
More importantly, neither of the approaches contains a second-order
term, see eq.~(\ref{eq:secondorderanalyt}).
Consequently, the Bethe Ansatz results for the ground-state energy
only apply to first order in $J_{\rm K}/W$, as is implicit in the wide-band limit,
$W\to \infty$.

\subsection{Two-particle problem}

It is instructive to solve the two-particle problem.
We address a linear dispersion relation.

\subsubsection{Analytic solution}

We focus on the subspace that contains the ground state, $S=S^z=0$.
We thus investigate the $(L+1)/2$ normalized states
\begin{equation}
|m\rangle = \sqrt{\frac{1}{2}} \left(
\hat{b}_{m,\uparrow}^+\hat{d}_{\downarrow}^+ -
\hat{b}_{m,\downarrow}^+\hat{d}_{\uparrow}^+
\right) |\hbox{vac}\rangle\; ,
\end{equation}
where $\hat{b}_{m,\sigma}^+$ creates an electron in the kinetic-energy
eigenstate of the half-chain with energy $\epsilon(m)$,
\begin{equation}
\langle m | \hat{T}^C |m' \rangle = \delta_{m,m'}\epsilon(m) \; .
\end{equation}
In the spin-singlet basis $|m\rangle$, 
the Kondo coupling reduces to an attractive potential scattering term,
\begin{equation}
\langle m |\hat{V}_{\rm sd} |m'\rangle
=-\frac{3J_{\rm K}}{4} \frac{2}{L+1} (-1)^{m+m'} \; . 
\end{equation}
We solve the Schr\"odinger equation
\begin{equation}
\left(\hat{T}^C +\hat{V}_{\rm sd} \right)|\psi\rangle = E |\psi\rangle
\end{equation}
with the Ansatz
\begin{equation}
|\psi\rangle=\sum_{r=1}^{(L+1)/2} (-1)^r f_r |r\rangle
\end{equation}
and find
\begin{equation}
\left(E-\epsilon(r)\right)f_r=\frac{3J_{\rm K}}{2(L+1)}
\sum_{m=1}^{(L+1)/2} f_m \; ,
\end{equation}
which has the solution
\begin{equation}
 f_r = -\frac{3J_{\rm K}}{4} \frac{\gamma(E)}{E-\epsilon(r)} \;,
\end{equation}
where $\gamma(E)$ follows from the solution of
\begin{equation}
 1 = -\frac{3J_{\rm K}}{4} \frac{2}{L+1}\sum_{r=1}^{(L+1)/2}
\frac{1}{E-\epsilon(r)} \;.
\label{eq:implicitE}
\end{equation}
This equation has a simple solution for $J_{\rm K}\gg 1$, because
we can replace the sum by $(L+1)/(2E)$ so that
$E(J_{\rm K}\gg 1)=-3J_{\rm K}/4$. For strong coupling, 
a local singlet forms between the impurity spin and the bath spin
on site $n=0$.
The analysis of the weak-coupling limit is more subtle.

\paragraph{Finite system sizes}

As long as $L$ is finite, there are no bound states, and we can set
\begin{equation}
E_0=\epsilon(1)-\frac{\delta}{L+1}\; ,
\end{equation}
where $\delta>0$ is of the order unity.
Furthermore, we use 
\begin{equation}
\epsilon(m)=\frac{2\pi}{L+1}  \left(m-\frac{(L+3)}{4}\right)
\end{equation} 
for our half-chain Hamiltonian. Then, eq.~(\ref{eq:implicitE}) becomes
\begin{equation}
\delta=\frac{3J_{\rm K}}{2} \left(
1+\frac{\delta}{2\pi} \sum_{r=1}^{(L-1)/2}
\frac{1}{r+\delta/(2\pi)}
\right) \;.
\label{eq:deltaeq}
\end{equation}
Apparently, this equation has the solution $\delta=3J_{\rm K}/2$
for small $J_{\rm K}$ so that
\begin{equation}
e_0(J_{\rm K})=E_0-\epsilon(1)=-\frac{1}{L+1}\frac{3J_{\rm K}}{2} 
+{\cal O}\left(J_{\rm K}^2\ln(L)/L
\right)
\label{eq:e0smallJK}
\end{equation}
for $J_{\rm K}\ll 1/\ln(L)$.
For finite system sizes, the energy is linear in $J_{\rm K}$ but the
prefactor is proportional to $1/L$ so that the linear term vanishes in
the thermodynamic limit. Moreover, the linear region itself
vanishes logarithmically for large system sizes.

\paragraph{Thermodynamic limit}

The sum in eq.~(\ref{eq:deltaeq}) diverges logarithmically. 
Therefore, we must use that
\begin{equation}
e_0=E_0-\epsilon(1)
\end{equation}
is finite in the thermodynamic limit. 
Then, eq.~(\ref{eq:implicitE}) becomes
\begin{equation}
 1 = \frac{3J_{\rm K}}{4} \frac{2}{L} \frac{L}{2\pi}
\int_0^{\pi} \rmd k \frac{1}{-e_0+k}
= \frac{3J_{\rm K}}{4\pi}
 \ln\left[ \frac{-e_0+\pi}{(-e_0)}\right]
\end{equation}
so that 
\begin{equation}
e_0(J_{\rm K})=-\frac{\pi}{\exp\left(4\pi/(3J_{\rm K})\right)-1} \; .
\end{equation}
Therefore, $e_0(J_{\rm K})$ is exponentially small but finite
in the thermodynamic limit. This does not come as a surprise because
attractive potentials support bound states in one-dimensional
scattering problems.
For large $J_{\rm K}$ we recover eq.~(\ref{eq:e0smallJK}).

\subsubsection{Bethe Ansatz}

We analyze the Bethe Ansatz equation for finite~$L$ because 
we can then assure that there are no bound states.
For $N^e=1$ we find from eq.~(\ref{eq:lambdadefeBA}) for $M=1$
\begin{equation}
\left[
\frac{\rmi (1-\lambda)+c/2}{\rmi(1-\lambda)-c/2}
\right]
\left[ 
\frac{\rmi (-\lambda)+c/2}{\rmi(-\lambda)-c/2}
\right]
=(-1)(-1)\; ,
\label{eq:lambdadefeBAN1}
\end{equation}
which has the solution $\lambda=1/2$. 
The energy becomes
\begin{equation}
E_0=\frac{2\pi}{L}\tilde{n}_1 -\frac{\phi}{L} -\frac{2}{L}\arctan(c)
\end{equation}
or
\begin{equation}
e_0=-\frac{1}{L} \left(\phi +2\arctan(c)\right) \; .
\end{equation}
For the two Bethe Ansatz expression we obtain the following results.

\paragraph{Andrei, Furuya, Lowenstein.}

Eq.~(\ref{eq:cvaluesBA}) together with eq.~(\ref{eq:phivaluesBA}) gives
\begin{eqnarray}
e_0^{\rm AFL}(J_{\rm K}^{\rm BA})&=&\frac{2}{L}
\arctan\left(\frac{J_{\rm K}^{\rm BA}}{4}\right)
\nonumber \\
&&-\frac{2}{L}\arctan\left(\frac{J_{\rm K}^{\rm BA}}{1-3(J_{\rm K}^{\rm BA})^2/16}\right)
\nonumber \\
&=& -\frac{1}{L}\left(\frac{3J_{\rm K}^{\rm BA}}{2}-\frac{9(J_{\rm K}^{\rm BA})^3}{32}
\right) \; ,
\end{eqnarray}
and corrections are of the order $(J_{\rm K}^{\rm BA})^5$.
Note that the solution with a finite bandwidth, eq.~(\ref{eq:e0smallJK}),
has $L$-dependent
corrections of the order $J_{\rm K}^2\ln(L)/L$ for $J_{\rm K}\ll 1/\ln(L)$.

\paragraph{Tsvelick and Wiegmann.}

Eq.~(\ref{eq:cvaluesBA}) together with eq.~(\ref{eq:phivaluesBA}) gives
\begin{equation}
  e_0^{\rm TW}(J_{\rm K}^{\rm BA})=-\frac{1}{L}\frac{3J_{\rm K}^{\rm BA}}{2}
  \label{eq:BAtwoparticleenergy}
\end{equation}
without corrections to higher orders in $J_{\rm K}^{\rm BA}$,
in agreement with eq.~(\ref{eq:e0smallJK}).

The agreement between eq.~(\ref{eq:e0smallJK})
and eq.~(\ref{eq:BAtwoparticleenergy}) can be improved when
a flexible the length scale is used in Bethe Ansatz, $L^{\rm BA}\sim L\ln(L)$,
instead of $L^{\rm BA}=L$. We shall not further elaborate this
possibility.

\section{Free energy in second-order weak-coupling perturbation theory}
\label{supp:freenergyPT}

For a better comparison, we closely follow the considerations in appendix~C of AFL.

\subsection{Formal expansion}

The unperturbed Hamiltonian is defined by
\begin{equation}
\hat{H}_0=\hat{T}+\hat{H}_{\rm mag} \; .
\end{equation}
In perturbation theory we need the Kondo term in the interaction picture,
\begin{equation}
  \hat{V}_{\rm sd}(\lambda)=e^{\lambda \hat{H}_0}\hat{V}_{\rm sd}e^{-\lambda\hat{H}_0}
  \; .
\end{equation}
We use
\begin{eqnarray}
  e^{\lambda \hat{H}_0}\hat{b}_{k,\uparrow}^+e^{-\lambda\hat{H}_0}
  &=& e^{(\epsilon(k)-B)\lambda}\hat{b}_{k,\uparrow}^+ \; , \nonumber\\
  e^{\lambda \hat{H}_0}\hat{b}_{k,\downarrow}^+e^{-\lambda\hat{H}_0}
  &=& e^{(\epsilon(k)+B)\lambda}\hat{b}_{k,\downarrow}^+  \; , \nonumber \\
e^{\lambda \hat{H}_0}\hat{d}_{\uparrow}^+e^{-\lambda\hat{H}_0}
  &=& e^{-B\lambda}\hat{d}_{\uparrow}^+  \; , \nonumber \\
  e^{\lambda \hat{H}_0}\hat{d}_{\downarrow}^+e^{-\lambda\hat{H}_0}
  &=& e^{B\lambda}\hat{d}_{\downarrow}^+  \; .
   \end{eqnarray}
Therefore,
\begin{eqnarray}
  \hat{V}_{\rm sd}(\lambda)&=&
  \frac{J_{\rm K}}{L}  \sum_{k,k'} e^{(\epsilon(k')-\epsilon(k))\lambda}
  \nonumber \\
  && \times \Bigl[
\hat{b}_{k',\uparrow}^+\hat{b}_{k,\downarrow}^{\vphantom{+}} 
\hat{d}_{\downarrow}^+\hat{d}_{\uparrow}^{\vphantom{+}} 
+ 
\hat{b}_{k',\downarrow}^+\hat{b}_{k,\uparrow}^{\vphantom{+}} 
\hat{d}_{\uparrow}^+\hat{d}_{\downarrow}^{\vphantom{+}}  \nonumber \\
&& \hphantom{\times \Bigl[}
+\frac{1}{2} (\hat{d}_{\uparrow}^+\hat{d}_{\uparrow}^{\vphantom{+}} 
- \hat{d}_{\downarrow}^+\hat{d}_{\downarrow}^{\vphantom{+}} )
(\hat{b}_{k',\uparrow}^+\hat{b}_{k,\uparrow}^{\vphantom{+}} 
-\hat{b}_{k',\downarrow}^+\hat{b}_{k,\downarrow}^{\vphantom{+}} )
\Bigr]\;,\nonumber \\
\end{eqnarray}
as in eq.~(C6) of AFL.

The partition function contains the terms
\begin{eqnarray}
  Z_0&=&\hbox{Tr}\left(e^{-\beta \hat{H}_0}\right) \; , \nonumber \\
  Z_1&=& -\hbox{Tr}\left(e^{-\beta \hat{H}_0}
  \int_0^{\beta}\rmd\lambda \hat{V}_{\rm sd}(\lambda)
  \right) \; , \nonumber \\
  Z_2&=& \hbox{Tr}\left(e^{-\beta \hat{H}_0}
  \int_0^{\beta}\rmd\lambda_1 \hat{V}_{\rm sd}(\lambda_1)
  \int_0^{\lambda_1}\rmd\lambda_2 \hat{V}_{\rm sd}(\lambda_2)
  \right) \; ,\nonumber \\
  \label{eq:defineorders}
  \end{eqnarray}
see eq.~(C5) of AFL, where $\beta=1/T$ is the inverse temperature.
The free energy ${\cal F}=-T\ln(Z)$ becomes
\begin{equation}
  {\cal F} = -T\left[
\ln(Z_0)+\frac{Z_1}{Z_0}+\frac{Z_2}{Z_0}-\frac{1}{2}\left(\frac{Z_1}{Z_0}\right)^2
\right] \;,
  \label{eq:seriesforF}
  \end{equation}
up to and including second order in $J_{\rm K}$, see eq.~(C7) of AFL.

\subsection{Terms up to first order}
\label{sec:first-orderterm}

The leading-order term in the impurity-induced contribution
to free energy is obtained from
\begin{equation}
Z_0^{\rm ii}=e^{\beta B}+e^{-\beta B}=2\cosh(\beta B) 
  \end{equation}
as
\begin{equation}
{\cal F}_0^{\rm ii}=-T\ln[2\cosh(\beta B) ] \; .
\end{equation}

{}From the definition~(\ref{eq:defineorders}) we find for the first-order term
\begin{eqnarray}
  Z_1&=&-J_{\rm K} \frac{1}{L} \sum_{k,k'}\int_0^{\beta} \rmd \lambda
  e^{\lambda(\epsilon(k')-\epsilon(k))}\nonumber \\
  && \times \hbox{Tr}
  \left(e^{-\beta \hat{H}_0}
  \frac{1}{2} (\hat{d}_{\uparrow}^+\hat{d}_{\uparrow}^{\vphantom{+}} 
- \hat{d}_{\downarrow}^+\hat{d}_{\downarrow}^{\vphantom{+}} )
(\hat{b}_{k',\uparrow}^+\hat{b}_{k,\uparrow}^{\vphantom{+}} 
-\hat{b}_{k',\downarrow}^+\hat{b}_{k,\downarrow}^{\vphantom{+}} )
\right)\nonumber \\
\label{eq:firstorderF}
  \end{eqnarray}
because the spin-flip terms do not contribute as the trace is done
over states with fixed $d$-occupation.
The impurity trace gives
\begin{equation}
 F(B/T)\equiv  \frac{Z_1^{\rm i}}{Z_0^{\rm i}}=
  \frac{\hbox{Tr}\left(e^{2 \beta B\hat{S}_z}\hat{S_z}\right)}{
    \hbox{Tr}\left(e^{2 \beta B\hat{S}_z}\right)}
  = \frac{1}{2} \tanh(\beta B) \; ,
  \end{equation}
see eq.~(C15) of AFL.
The host-electron trace is trivial because only $k=k'$ can contribute,
\begin{equation}
   \frac{Z_1^{\rm e}}{Z_0^{\rm e}}
  \equiv \frac{1}{2} \frac{N_{\uparrow}-N_{\downarrow}}{L/2}\equiv
  M\; .
\end{equation}
In the presence of a magnetic field, the host-electrons have the magnetization
\begin{eqnarray}
  M&=&\frac{B}{\pi}  \quad \hbox{for}\quad \epsilon(k)=k \; , \nonumber \\
  M&=&\frac{\arcsin(B)}{\pi}  \quad \hbox{for}\quad \epsilon(k)=\sin(k) \; .
  \end{eqnarray}
We may safely assume that, for $J_{\rm K}\ll 1$, the external field 
is weak enough
to only slightly polarize the host-electron system.
When we work with a general density of states and small fields,
we find $M=B\rho_0(0)$ for small fields, so that we obtain eq.~(C14) of AFL,
\begin{equation}
  {\cal F}_1^{\rm ii}  =  J_{\rm K} B\rho_0(0) F(B/T) \; ,
  \label{eq:Fone}
\end{equation}
where we used the fact that the integral over $\lambda$ in
eq.~(\ref{eq:firstorderF}) gives $\beta=1/T$,
and the signs in eqs.~(\ref{eq:seriesforF}) and~(\ref{eq:firstorderF})
cancel each other.

\subsection{Second-order term}

The second-order term consists of two contributions,
\begin{equation}
  {\cal F}_2^{\rm ii}={\cal F}_{2a}^{\rm ii}+{\cal F}_{2b}^{\rm ii}
  = -T \left[\frac{Z_2}{Z_0}
    -\frac{1}{2} \left(\frac{Z_1}{Z_0}\right)^2\right]\; .
  \label{eq:F14}
  \end{equation}
The second contribution is readily
calculated when we use eqs.~(\ref{eq:seriesforF})
and~(\ref{eq:Fone}),
\begin{equation}
  {\cal F}_{2b}^{\rm ii} 
  = \frac{1}{2T} \left( {\cal F}_1^{\rm ii}  \right)^2 \; .
  \label{eq:F15}
\end{equation}
In a diagrammatic formulation of perturbation theory, the second term cancels
unconnected diagrams in the first term.
We shall not use this concept here as we closely follow appendix~C of AFL.

\subsubsection{Trace over the impurity states}

The first contribution involves three terms that result from $Z_2$,
see eq.~(\ref{eq:defineorders}), because the trace is over eigenstates
of $\hat{S}_z$, namely,
\begin{eqnarray}
  \hbox{1st}&=& \hbox{Tr}\left(e^{-\beta\hat{H}_0}
\hat{b}_{k_1',\uparrow}^+\hat{b}_{k_1,\downarrow}^{\vphantom{+}} 
\hat{b}_{k_2',\downarrow}^+\hat{b}_{k_2,\uparrow}^{\vphantom{+}} 
\hat{d}_{\downarrow}^+\hat{d}_{\uparrow}^{\vphantom{+}}
\hat{d}_{\uparrow}^+\hat{d}_{\downarrow}^{\vphantom{+}}
\right) \; , \nonumber \\
\hbox{2nd}&=&
\hbox{Tr}\left(e^{-\beta\hat{H}_0}
\hat{b}_{k_1',\downarrow}^+\hat{b}_{k_1,\uparrow}^{\vphantom{+}}
\hat{b}_{k_2',\uparrow}^+\hat{b}_{k_2,\downarrow}^{\vphantom{+}}
\hat{d}_{\uparrow}^+\hat{d}_{\downarrow}^{\vphantom{+}}
\hat{d}_{\downarrow}^+\hat{d}_{\uparrow}^{\vphantom{+}}
\right) \; , \nonumber \\
\hbox{3rd}&=&
\hbox{Tr}\Bigl(e^{-\beta\hat{H}_0}
\frac{1}{4} (\hat{d}_{\uparrow}^+\hat{d}_{\uparrow}^{\vphantom{+}} 
- \hat{d}_{\downarrow}^+\hat{d}_{\downarrow}^{\vphantom{+}} )^2
\nonumber \\
&&\hphantom{}
\times(\hat{b}_{k_1',\uparrow}^+\hat{b}_{k_1,\uparrow}^{\vphantom{+}} 
-\hat{b}_{k'_1,\downarrow}^+\hat{b}_{k_1,\downarrow}^{\vphantom{+}} )
(\hat{b}_{k_2',\uparrow}^+\hat{b}_{k_2,\uparrow}^{\vphantom{+}} 
-\hat{b}_{k_2',\downarrow}^+\hat{b}_{k_2,\downarrow}^{\vphantom{+}} )
\Bigr) .\nonumber \\
  \end{eqnarray}
The resulting impurity traces over the spin terms are readily calculated,
\begin{eqnarray}
  W^+(B/T)&=&
  \frac{\hbox{Tr}\left(e^{2 \beta B\hat{S}_z}\hat{S}^-\hat{S}^+\right)}{
    \hbox{Tr}\left(e^{2 \beta B\hat{S}_z}\right)}
  = \frac{\hbox{Tr}\left(e^{2 \beta B\hat{S}_z}\hat{n}_{\downarrow}^d\right)}{
    \hbox{Tr}\left(e^{2 \beta B\hat{S}_z}\right)}\nonumber \\
    &=&\frac{e^{-\beta B}}{2\cosh(\beta B)}
    \; ,  \nonumber \\
      W^-(B/T)&=&
  \frac{\hbox{Tr}\left(e^{2 \beta B\hat{S}_z}\hat{S}^+\hat{S}^-\right)}{
    \hbox{Tr}\left(e^{2 \beta B\hat{S}_z}\right)}
  = \frac{\hbox{Tr}\left(e^{2 \beta B\hat{S}_z}\hat{n}_{\uparrow}^d\right)}{
    \hbox{Tr}\left(e^{2 \beta B\hat{S}_z}\right)}\nonumber \\
    &=&\frac{e^{\beta B}}{2\cosh(\beta B)}=W^+(-B/T)
\; ,  \nonumber \\
      W(B/T)&=&
  \frac{\hbox{Tr}\left(e^{2 \beta B\hat{S}_z}\hat{S}_z\hat{S}_z\right)}{
    \hbox{Tr}\left(e^{2 \beta B\hat{S}_z}\right)}\
  = \frac{1}{4}\frac{\hbox{Tr}\left(e^{2 \beta B\hat{S}_z}\right)}{
    \hbox{Tr}\left(e^{2 \beta B\hat{S}_z}\right)}\nonumber \\
    &=&\frac{1}{4}=W(-B/T)\; ,
\label{eq:defWs}
\end{eqnarray}
in agreement with eq.~(C18) of AFL.

\subsubsection{Trace over the host electrons}

The traces over the host electrons require
the free momentum-space occupancies
($\sigma_n=1 (-1)$ for $\sigma=\uparrow (\downarrow)$),
\begin{equation}
n_{k,\sigma}=\frac{\hbox{Tr}\left(e^{-\beta\hat{H}_0}
\hat{b}_{k,\sigma}^+\hat{b}_{k,\sigma}^{\vphantom{+}} \right)}{Z_0^e}
= \frac{1}{1+\exp\left(\beta(\epsilon(k)-\sigma_n B)\right)} \; .
\label{suppeq:freeoccupancies}
  \end{equation}
The traces over the host electrons for the spin-flip terms give
\begin{eqnarray}
  \hbox{1st}^e&=& \delta_{k_1',k_2}\delta_{k_1,k_2'}
    \hbox{Tr}\left(e^{-\beta\hat{H}_0}
    \hat{b}_{k_1',\uparrow}^+\hat{b}_{k_1',\uparrow}^{\vphantom{+}}\right)\\
    &&\hphantom{\delta_{k_1',k_2}\delta_{k_1,k_2'}}
    \times \hbox{Tr}\left(e^{-\beta\hat{H}_0}
    \hat{b}_{k_1,\downarrow}^{\vphantom{+}} \hat{b}_{k_1,\downarrow}^+
    \right) \; ,  \nonumber
\end{eqnarray}
and
\begin{eqnarray}
\hbox{2nd}^e&=&\delta_{k_1',k_2}\delta_{k_1,k_2'}
\hbox{Tr}\left(e^{-\beta\hat{H}_0}
\hat{b}_{k_1',\downarrow}^+\hat{b}_{k_1',\downarrow}^{\vphantom{+}}\right)
\\
&& \hphantom{\delta_{k_1',k_2}\delta_{k_1,k_2'}}
\times \hbox{Tr}\left(e^{-\beta\hat{H}_0}
\hat{b}_{k_1,\uparrow}^{\vphantom{+}}\hat{b}_{k_1,\uparrow}^+
\right) \; .
\nonumber
\end{eqnarray}
The traces over the host electrons for the diagonal terms give the contributions
\begin{eqnarray}
\hbox{3rd}^e&=&
-\delta_{k_1',k_1}\delta_{k_2',k_2}
\hbox{Tr}\left(e^{-\beta\hat{H}_0}
\hat{b}_{k_1,\uparrow}^+\hat{b}_{k_1,\uparrow}^{\vphantom{+}} \right)\nonumber \\
&&\hphantom{\delta_{k_1',k_2}\delta_{k_1,k_2'}}
\times \hbox{Tr}\left(e^{-\beta\hat{H}_0}
\hat{b}_{k_2,\downarrow}^+\hat{b}_{k_2,\downarrow}^{\vphantom{+}} 
\right) \nonumber \\
&&-\delta_{k_1',k_1}\delta_{k_2',k_2}
\hbox{Tr}\left(e^{-\beta\hat{H}_0}
\hat{b}_{k_1,\downarrow}^+\hat{b}_{k_1,\downarrow}^{\vphantom{+}} \right)\nonumber \\
&& \hphantom{\delta_{k_1',k_2}\delta_{k_1,k_2'}}\times 
\hbox{Tr}\left(e^{-\beta\hat{H}_0}
\hat{b}_{k_2,\uparrow}^+\hat{b}_{k_2,\uparrow}^{\vphantom{+}} 
\right) \nonumber \\
&&+\delta_{k_1',k_1}\delta_{k_2',k_2}
\hbox{Tr}\left(e^{-\beta\hat{H}_0}
\hat{b}_{k_1,\uparrow}^+\hat{b}_{k_1,\uparrow}^{\vphantom{+}} \right)\nonumber \\
&& \hphantom{\delta_{k_1',k_2}\delta_{k_1,k_2'}}\times 
\hbox{Tr}\left(e^{-\beta\hat{H}_0}
\hat{b}_{k_2,\uparrow}^+\hat{b}_{k_2,\uparrow}^{\vphantom{+}} 
\right) \nonumber \\
&&+\delta_{k_1',k_1}\delta_{k_2',k_2}
\hbox{Tr}\left(e^{-\beta\hat{H}_0}
\hat{b}_{k_1,\downarrow}^+\hat{b}_{k_1,\downarrow}^{\vphantom{+}} \right)\nonumber \\
&&\hphantom{\delta_{k_1',k_2}\delta_{k_1,k_2'}}
\times \hbox{Tr}\left(e^{-\beta\hat{H}_0}
\hat{b}_{k_2,\downarrow}^+\hat{b}_{k_2,\downarrow}^{\vphantom{+}} 
\right) \nonumber \\
&&+\delta_{k_1',k_2}\delta_{k_1,k_2'}
\hbox{Tr}\left(e^{-\beta\hat{H}_0}
\hat{b}_{k_1',\uparrow}^+\hat{b}_{k_1',\uparrow}^{\vphantom{+}} \right)\nonumber \\
&& \hphantom{\delta_{k_1',k_2}\delta_{k_1,k_2'}}\times
\hbox{Tr}\left(e^{-\beta\hat{H}_0}
\hat{b}_{k_1,\uparrow}^{\vphantom{+}} \hat{b}_{k_1,\uparrow}^+
\right) \nonumber \\
&&+\delta_{k_1',k_2}\delta_{k_1,k_2'}
\hbox{Tr}\left(e^{-\beta\hat{H}_0}
\hat{b}_{k_1',\downarrow}^+\hat{b}_{k_1',\downarrow}^{\vphantom{+}} \right)\nonumber \\
&& \hphantom{\delta_{k_1',k_2}\delta_{k_1,k_2'}}
\times \hbox{Tr}\left(e^{-\beta\hat{H}_0}
\hat{b}_{k_1,\downarrow}^{\vphantom{+}} \hat{b}_{k_1,\downarrow}^+
\right) \; .\nonumber \\
\end{eqnarray}
Using eq.~(\ref{suppeq:freeoccupancies}) we find,
in agreement with eq.~(C19) of AFL,
\begin{eqnarray}
  {\cal F}_{2a}^{\rm ii}
  &=& -\frac{J_{\rm K}^2}{2T} W(B/T)  \biggl(\frac{1}{L}\sum_{k}
  \left(n_{k,\uparrow}-n_{k,\downarrow}\right)  \biggr)^2
  \nonumber \\
&&  -T J_{\rm K}^2
  \int_0^{\beta}\rmd \lambda_1\int_0^{\lambda_1}\rmd \lambda_2 \nonumber \\
  &&\hphantom{-T J_{\rm K}^2}
  \times   \frac{1}{L^2}\sum_{k,k'} e^{(\epsilon(k')-\epsilon(k))(\lambda_1-\lambda_2)} K(k,k') \; ,\nonumber \\
\end{eqnarray}
where we defined
\begin{eqnarray}
  K(k,k')&=&  n_{k',\uparrow}(1-n_{k,\downarrow}) W^+(B/T)\nonumber \\
&&  +   n_{k',\downarrow}(1-n_{k,\uparrow}) W^-(B/T) \nonumber \\
&&  +   n_{k',\uparrow}(1-n_{k,\uparrow}) W(B/T) \nonumber \\
&&+  n_{k',\downarrow}(1-n_{k,\downarrow}) W(B/T) \;, 
   \label{eq:traceisdone}
\end{eqnarray}
where we integrated over $(\lambda_1,\lambda_2)$ for the first four terms
in 3rd$^e$ to obtain the first term in eq.~(\ref{eq:traceisdone}). It
can be written in terms of the magnetization,
\begin{eqnarray}
  \hbox{first term in eq.~(\ref{eq:traceisdone})}
  &=&-\frac{J_{\rm K}^2}{2T} W(B/T)  M^2\nonumber\\
&=&-\frac{J_{\rm K}^2}{2T} W(B/T) \bigl(B\rho_0(0)\bigr)^2 \;,\nonumber\\
  \end{eqnarray}
which coincides with the last term in eq.~(C20) of AFL.

\subsubsection{Simplified notations}

We define
\begin{eqnarray}
  I_+(\lambda)&=& \frac{2\pi}{L} \sum_k e^{\lambda\epsilon(k)}
n_{k,\uparrow}\\
  &=&  \int_{-{\cal D}}^{\cal D}\rmd \epsilon \pi\rho_0(\epsilon)
  \frac{e^{\lambda\epsilon}}{1+\exp\left(\beta(\epsilon-B)\right)} \; ,
    \nonumber
\end{eqnarray}
and
\begin{eqnarray}
  I_-(\lambda)&=& \frac{2\pi}{L} \sum_k e^{\lambda\epsilon(k)}
n_{k,\downarrow} \\
  &=&
  \int_{-{\cal D}}^{\cal D}\rmd \epsilon \pi\rho_0(\epsilon)
  \frac{e^{\lambda\epsilon}}{1+\exp\left(\beta(\epsilon+B)\right)} \; ,
\nonumber
      \end{eqnarray}
and use
\begin{eqnarray}
\frac{2\pi}{L} \sum_k e^{-\lambda\epsilon(k)}(1-n_{k,\downarrow})
&=&
I_+(\lambda)\; , \nonumber \\
\frac{2\pi}{L} \sum_k e^{-\lambda\epsilon(k)}(1-n_{k,\uparrow})
&=&
I_-(\lambda) \; ,
      \end{eqnarray}
due to particle-hole symmetry at half filling where
we used $\epsilon(-k)=-\epsilon(k)$.
Note that our definition slightly differs from the one given in eq.~(C21) of AFL.

The transformation
\begin{equation}
\lambda=\lambda_1-\lambda_2 \quad , \quad \lambda'=\lambda_1+\lambda_2
  \end{equation}
for the region $0\leq \lambda_1\leq \beta, 0\leq \lambda_2\leq \lambda_1$
leads to a Jacobi determinant of one half and an integration region
$0\leq \beta\leq \lambda, \lambda\leq \lambda'\leq 2\beta-\lambda$.
Since all quantities in eq.~(\ref{eq:traceisdone}) only depend on $\lambda$,
the integral over $\lambda'$ can be carried out to give $2(\beta-\lambda)/2=
\beta-\lambda$. Then, eq.~(\ref{eq:traceisdone}) can be cast into the
form
\begin{eqnarray}
  {\cal F}_{2a}^{\rm ii}
  &=& -\left(\frac{J_{\rm K}}{2\pi}\right)^2
  W(B/T) \frac{2 (B\pi\rho_0(0))^2}{T} \nonumber \\
  && 
  -\left(\frac{J_{\rm K}}{2\pi}\right)^2 T
  \int_0^{\beta}\rmd \lambda (\beta-\lambda) 2I_+(\lambda)I_-(\lambda)
  \nonumber \\
  && -\left(\frac{J_{\rm K}}{2\pi}\right)^2
  W^+(B/T)
T\int_0^{\beta}\rmd \lambda (\beta-\lambda) I_+(\lambda)^2
    \nonumber \\
    &&  -\left(\frac{J_{\rm K}}{2\pi}\right)^2
    W^-(B/T)
  T  \int_0^{\beta}\rmd \lambda (\beta-\lambda) I_-(\lambda)^2\; .\nonumber \\
\label{eq:F2aalmost}
  \end{eqnarray}
Note that
\begin{eqnarray}
I_{+}(\beta-\lambda)&=&e^{\beta B} I_-(\lambda) \; , \nonumber \\
I_{-}(\beta-\lambda)&=&e^{-\beta B} I_+(\lambda)\; ,
\label{eq:F28}
\end{eqnarray}
so that
\begin{eqnarray}
I_{+}(\beta-\lambda)
I_{-}(\beta-\lambda)
&=&I_{+}(\lambda)I_{-}(\lambda) \; , \nonumber \\
W^+(B/T)\left[I_{+}(\beta-\lambda)\right]^2
&=& W^-(B/T)\left[I_{-}(\lambda)\right]^2
\; , \nonumber \\
W^-(B/T)\left[I_{-}(\beta-\lambda)\right]^2
&=& W^+(B/T)\left[I_{+}(\lambda)\right]^2 \; .\nonumber \\
\label{eq:symmrelations}
\end{eqnarray}
We split the $\lambda$-integrals in eq.~(\ref{eq:F2aalmost})
according to
\begin{eqnarray}
\int_0^{\beta} \rmd \lambda (\beta-\lambda) X(\lambda) &=&
\frac{1}{2} \int_0^{\beta} \rmd \lambda (\beta-\lambda) X(\lambda)\nonumber \\
&&+ \frac{1}{2} \int_0^{\beta} \rmd \lambda \lambda X(\beta-\lambda)
\nonumber \\
\end{eqnarray}
and rewrite
\begin{eqnarray}
  {\cal F}_{2a}^{\rm ii}
  &=& -\left(\frac{J_{\rm K}}{2\pi}\right)^2
  W(B/T) \left(\frac{2 B^2}{T}
  +  \int_0^{\beta}\rmd \lambda  I_+(\lambda)I_-(\lambda)\right)
  \nonumber \\
  && -\left(\frac{J_{\rm K}}{2\pi}\right)^2
  \frac{W^+(B/T)}{2}
\int_0^{\beta}\rmd \lambda  I_+(\lambda)^2 \nonumber \\
&&      -\left(\frac{J_{\rm K}}{2\pi}\right)^2  \frac{W^-(B/T)}{2}
 \int_0^{\beta}\rmd \lambda  I_-(\lambda)^2
\label{eq:F2aalmost2}
  \end{eqnarray}
noting that $\beta T=1$. Using eq.~(\ref{eq:symmrelations}) again
this reduces to
\begin{eqnarray}
  {\cal F}_{2a}^{\rm ii}
  &=& -\left(\frac{J_{\rm K}}{2\pi}\right)^2
  W(B/T)\frac{2 \bigl(B\pi\rho_0(0)\bigr)^2}{T}\nonumber \\
  &&  -\left(\frac{J_{\rm K}}{2\pi}\right)^2W(B/T)
  \int_0^{\beta}\rmd \lambda  I_+(\lambda)I_-(\lambda)
\nonumber \\
&&-\left(\frac{J_{\rm K}}{2\pi}\right)^2  W^-(B/T)
 \int_0^{\beta}\rmd \lambda  I_-(\lambda)^2\;  ,
\label{eq:C24}
  \end{eqnarray}
in agreement with eq.~(C24) of AFL.
Following AFL further, we define
\begin{equation}
  P(x,B)=\int_{-\beta(1+B)}^{\beta(1-B)}\rmd y 
\frac{e^{xy}}{1+e^y}\pi\rho_0(y/\beta+B)
  \label{eq:defPxB}
  \end{equation}
so that
\begin{eqnarray}
  \int_0^{\beta}\rmd \lambda [I_-(\lambda)]^2
&=&   \frac{1}{\beta}\int_0^1\rmd x e^{-2\beta B x} \left[P(x,-B)\right]^2 \; , \nonumber\\
  \int_0^{\beta}\rmd \lambda  I_+(\lambda)I_-(\lambda)
   &=& \frac{1}{\beta}\int_0^1 \rmd x P(x,-B)P(x,B) \; .
\nonumber \\
\end{eqnarray}
Thus,
\begin{eqnarray}
  {\cal F}_{2a}^{\rm ii}
&  =& -\left(\frac{J_{\rm K}}{2\pi}\right)^2
W(B/T) \frac{2 \bigl(B\pi\rho_0(0)\bigr)^2}{T}\nonumber \\
&&-\left(\frac{J_{\rm K}}{2\pi}\right)^2W(B/T)
T\int_0^1 \rmd x P(x,B)P(x,-B)\nonumber \\
  &&
-\left(\frac{J_{\rm K}}{2\pi}\right)^2
W^-(B/T)T\int_0^1\rmd x e^{-2\beta B x} \left[P(x,-B)\right]^2.\nonumber \\
\label{eq:C24again}
  \end{eqnarray}
The remaining task is to find an approximation for $P(x,B)$ that works
in the limits of small fields and temperatures, $B, T\ll {\cal D}$.

\subsection{Approximate evaluation of the integrals in the wide-band limit
  for a constant density of states}

For $B,T\ll{\cal D}$ and a constant density of states we may write
\begin{eqnarray}
\frac{  P(x,B)}{ \pi \rho_0(0)}&\approx &
\int_{-\infty}^{\infty}\rmd y \frac{e^{xy}}{1+e^y}
      -    \int_{-\infty}^{-\beta(1+B)}\rmd y e^{xy}\left(1-e^y\right)\nonumber \\
    &&
-   \int_{\beta(1-B)}^{\infty}\rmd y e^{(x-1)y}\left(1-e^{-y}\right)\nonumber \\
  &=&
      \frac{\pi}{\sin(\pi x)} - \frac{e^{-\beta(1+B)x}}{x}+\frac{e^{-\beta(1+B)(1+x)}}{1+x}
      \nonumber \\
      &&
        - \frac{e^{-\beta(1-B)(1-x)}}{1-x} + \frac{e^{-\beta(1-B)(2-x)}}{2-x}
          \end{eqnarray}
with corrections of the order $\exp[-\beta(1+B)(2+x)]$ and $\exp[-(3-x)\beta(1-B)]$.
Now that $0<x<1$, we see that the terms proportional to
$\exp[-\beta(1+B)(1+x)]$ and
to $\exp[-\beta(1-B)(2-x)]$ are always exponentially small.
Thus we work with
\begin{equation}
  \frac{ P(x,B)}{\pi \rho_0(0)}\approx  
    \frac{\pi}{\sin(\pi x)} - \frac{e^{-\beta(1+B)x}}{x}
    - \frac{e^{-\beta(1-B)(1-x)}}{1-x}\; .
  \end{equation}

\subsubsection{Evaluation of the first integral}

We evaluate
\begin{eqnarray}
  A_1(B,T)&\equiv&\frac{1}{(\pi \rho_0(0))^2}
  \int_0^1 \rmd x P(x,B)P(x,-B)\nonumber \\
&=& \frac{2}{(\pi \rho_0(0))^2}\int_0^{1/2} \!\!\rmd x P(x,B)P(x,-B)\nonumber \\
\end{eqnarray}
because the integrand is symmetric under $x\leftrightarrow (1-x)$.
The terms proportional to $\exp[-(1-x)\beta]$ drop out because
$(1-x)$ is of order ${\cal D}$ and $\beta {\cal D}\gg 1$.
Thus, we are left with
\begin{eqnarray}
  A_1(B,T)  &\approx&
  2\int_0^{1/2}\rmd x
  \left[
    \frac{\pi}{\sin(\pi x)} - \frac{e^{-\beta(1+B)x}}{x}\right]\nonumber \\
  &&\hphantom{  2\int_0^{1/2}\rmd x
}
  \times
  \left[
  \frac{\pi}{\sin(\pi x)} - \frac{e^{-\beta(1-B)x}}{x}
     \right]\nonumber \\
   &\equiv& Q_1(T)+Q_2(B,T) \; .  \label{eq:defA}
\end{eqnarray}
Here, we denote
  \begin{eqnarray}
    Q_1(T) &=&
    2\int_0^{1/2}\rmd x
\left[ \left( \frac{\pi}{\sin(\pi x)}\right)^2
  +\frac{e^{-2\beta x}}{x^2}
  -\frac{2e^{-\beta x}}{x^2} \right]\nonumber \\
&\approx& \frac{4\ln(2)}{T}\; .
  \end{eqnarray}
  Moreover,
 \begin{equation}
Q_{2}(B,T) = 4 \int_0^{1/2}\rmd x
\frac{e^{-\beta x}}{x^2}\left[ 1- \frac{\pi x}{\sin(\pi x)} \cosh(\beta Bx)\right]
\; .\label{eq:defQ2}
\end{equation}
 For the evaluation of $Q_2(T,B)$ we use the series expansion
  \begin{eqnarray}
    \frac{\pi x}{\sin(\pi x)}&=& 1+2 \sum_{n=0}^{\infty}\alpha_nx^{2n+2}
    \quad \hbox{for} \quad |x|<1 \; ,\nonumber \\
\alpha_n&=&    \left(1-2^{-(2n+1)}\right)    \zeta(2n+2) \; , \nonumber \\
\zeta(n)&=& \sum_{k=1}^{\infty}\frac{1}{k^n} \nonumber \; ,\\
\zeta(n\gg 1)&\approx& 1+2^{-n}+3^{-n} +\ldots \; ,
\label{eq:seriesosineinvers}
    \end{eqnarray}
  where $\zeta(s)$ is the Riemann Zeta function.
  We have $\alpha_0=\pi^2/12\approx 0.822$ and
  $\alpha_1=7\pi^4/720\approx 0.947$.
  Note that the coefficient $\alpha_n$
  tends to unity  exponentially fast,
  $  \alpha_n\approx 1-2^{-(2n+2)}$ for $n\gtrsim 6$, with corrections of the
  order $10^{-6}$.
In eq.~(\ref{eq:defQ2}) we thus have
  \begin{eqnarray}
    Q_2(B,T)&=& Q_{2a}(B,T) +Q_{2b}(B,T) +\ldots \; , \nonumber\\
Q_{2a}(B,T) &=& 
4 \int_0^{1/2}\rmd x
\frac{e^{-\beta x}}{x^2}\left[ 1- \cosh(\beta Bx)\right]  \nonumber\\
&\approx &-\frac{4}{T}\left( B \arctanh(B) + \frac{1}{2} \ln\left(1 - B^2\right)\right)
\nonumber \\
&\approx& -\frac{2}{T} B^2
\; , \\
Q_{2b}(B,T) &=& 
-8\alpha_0 \int_0^{1/2}\rmd x
e^{-\beta x}\cosh(\beta Bx)  \nonumber \\
&=& -8\alpha_0 \frac{T}{1-B^2} 
  \end{eqnarray}
  in the limit $T/{\cal D}, B/{\cal D}\to 0$.
  It is seen that the contributions $Q_{2a,2b}(B,T)$ are small,
  of the order $(T/{\cal D})$ and $(B/{\cal D})$,
  and can thus be neglected.
  Therefore,
  \begin{equation}
    A_1(B,T)\approx Q_1(T)=\frac{4\ln(2)}{T} \; .
  \end{equation}
 For a better comparison with a direct numerical evaluation of $A_1(B,T)$,
it is helpful to keep the term proportional to $B^2/T$.
  
\subsubsection{Evaluation of the second integral}

We evaluate
\begin{eqnarray}
  A_2(B,T)&\equiv&\frac{e^{\beta B}}{(\pi \rho_0(0))^2}
  \int_0^1 \rmd x e^{-2\beta B x}\left[P(x,-B)\right]^2 \nonumber \\
&=& \frac{e^{\beta B}}{(\pi \rho_0(0))^2} \int_0^{1/2} \rmd x \Bigl[
    e^{-2\beta B x}\left[P(x,-B)\right]^2\nonumber \\
      &&\hphantom{\frac{e^{\beta B}}{(\pi \rho_0(0))^2}  \int_0^{1/2}}
      +e^{-2\beta B (1-x)}\left[P(x,B)\right]^2 \Bigr]\nonumber
  \; ,\\
  \label{eqdefA2}
\end{eqnarray}
where we included the factor $\exp(\beta B)$ from $W^-(B/T)$ to obtain
expressions that are symmetric in $B$. Moreover, we used
$P(1-x,-B)=P(x,B)$. Dropping exponentially small terms we find
\begin{eqnarray}
  A_2(B,T)    &\approx&
e^{\beta B} \int_0^{1/2} \rmd x
e^{-2\beta B x}\Bigl[\frac{\pi}{\sin(\pi x)}-\frac{e^{-\beta(1-B)x}}{x}\Bigr]^2
      \nonumber \\
      &&      +      e^{\beta B} \int_0^{1/2} \rmd x
e^{-2\beta B (1-x)}\nonumber \\
&&\hphantom{e^{\beta B} \int_0^{1/2} \rmd x}
\times \left[\frac{\pi}{\sin(\pi x)}-\frac{e^{-\beta(1+B)x}}{x}\right]^2
    \; .\nonumber \\
    \end{eqnarray}
We split the integrals as in the calculation of  $A_1(B,T)$,
\begin{eqnarray}
  A_2(B,T) &=& G_{1a}(B/T)+G_{1b}(B,T)+G_2(B,T) \; , \nonumber \\
  G_{1a}(B/T)&=& \int_0^{1/2} \rmd x
  \left[\left(\frac{\pi}{\sin(\pi x)}\right)^2-\frac{1}{x^2}\right]\nonumber \\
  &&\hphantom{ \int_0^{1/2} \rmd x}
  \times   \left(e^{-\beta B (2x-1)}+e^{\beta B (2x-1)}\right) \; ,
  \nonumber \\
  G_{1b}(B,T)&=&  \int_0^{1/2}\frac{ \rmd x}{x^2}  \Bigl[
    e^{-\beta B (2x-1)}\left(1-e^{-\beta(1-B)x}\right)^2 \nonumber \\
    &&\hphantom{\int_0^{1/2}\frac{ \rmd x}{x^2}  \Bigl[}
    +
  e^{\beta B (2x-1)}\left(1- e^{-\beta(1+B)x}\right)^2\Bigr] \nonumber, \\
  G_2(B,T)&=&
  2\int_0^{1/2} \frac{\rmd x}{x^2}
  \left[1-\frac{\pi x}{\sin(\pi x)}\right]\nonumber \\
&&
\times \left(  e^{\beta B}  e^{-\beta x(1+B)}+ e^{-\beta B}  
e^{-\beta x(1-B))}\right)\; .\nonumber \\
    \end{eqnarray}
As before, we can use the series expansion in eq.~(\ref{eq:seriesosineinvers})
to show that $G_2(B,T)={\cal O}(T/{\cal D},B/{\cal D})$ is negligible.
The first term in the expansion~(\ref{eq:seriesosineinvers})
gives
\begin{eqnarray}
  G_2^{(0)}(B,T)&=&-4\alpha_0 \int_0^{1/2} \rmd x\Bigl(
  e^{\beta B}  e^{-\beta x(1+B)}
  \nonumber \\
&& \hphantom{-4\alpha_0 \int_0^{1/2} \rmd x\Bigl(}
+ e^{-\beta B}e^{-\beta x(1-B))}\Bigr)\nonumber \\
&\approx & -8\alpha_0\cosh(\beta B)\frac{1-B\tanh(\beta B)}{(1-B^2)\beta}
\nonumber \\
&\approx & -8 \alpha_0 T \cosh(\beta B) 
\end{eqnarray}
for $B\ll 1$. The correction is of the order $T/{\cal D}$.
The terms proportional to $\alpha_n$ give corrections of the order
$T^{2n+1}$ and can safely be ignored.

The calculation of $G_{1a}(B,T)+G_{1b}(B,T)$ is done analytically using
{\sc Mathematica}~\cite{Mathematica11} to find
\begin{eqnarray}
  \frac{A_2(B,T)}{2\cosh(\beta B)}
  &\approx&
\beta \left(-2 B \arctanh(B) + \ln\left[\frac{4}{1-B^2}\right] \right)
\nonumber \\
&&-4 \frac{\alpha_0}{\beta} \frac{(1 - B\tanh(\beta B))}{1-B^2}
\nonumber \\
&&-2 \beta \tanh(\beta B)\arctanh(B) \nonumber \\
&& + \beta B \tanh(\beta B)
\Bigl(
      \ln\left[\frac{4 \pi^2}{(1-B^2) \beta^2}\right] -2\C\Bigr)
\nonumber\\
&& + \beta B\tanh(\beta B)\nonumber \\
&&\hphantom{+}
\times \bigl(\psi(\rmi \beta B/\pi) + \psi(-\rmi \beta B/\pi)+2\C\bigr) \nonumber  \; , \\
\label{eq:A2checked}
\end{eqnarray}
where $\psi(z)=\Gamma'(z)/\Gamma(z)$ is the Di-gamma function,
$\Gamma(z)$ is the gamma function,
\begin{equation}
\Gamma(z)= \int_0^{\infty}\rmd t \, t^{z-1}e^{-t} \; , 
\end{equation}
and $\C\approx 0.577216$ is Euler's constant.
In eq.~(\ref{eq:A2checked}) we kept all terms to order $T/{\cal D}$
and $B/{\cal D}$ to facilitate a numerical check of the analytic
expressions. Keeping only the leading-order 
contributions, eq.~(\ref{eq:A2checked}) reduces to
\begin{eqnarray}
  \frac{A_2(B,T)}{2\cosh(\beta B)}
  &\approx & \ln(4)\beta \label{eq:thatsA2}\\
&&-2\beta B\tanh(\beta B)\left(1+\C+\ln\left(\frac{\beta}{2\pi}\right)\right)\nonumber \\
&& + \beta B\tanh(\beta B)\nonumber \\
&&\hphantom{+ \beta B}
\times \bigl(\psi(\rmi \beta B/\pi) 
+ \psi(-\rmi \beta B/\pi)+2\C\bigr) \nonumber \; .
\end{eqnarray}

\subsubsection{Free energy to second-order in the limit of large bandwidth
for a constant density of states}
\label{subsec:freeenergyconstDOS}

Summarizing the results for a constant density of states,
the free energy to second order in the limit
$T,B \ll {\cal D} $ reads
\begin{eqnarray}
{\cal F}^{\rm ii}(B,T)&\approx & -T \ln[2\cosh(B/T) ] \label{eq:largecalDdone}
\\
&&+ \frac{J_{\rm K}(\pi\rho_0(0))}{2\pi} B\tanh(B/T) 
\nonumber \\
&& +\left(\frac{J_{\rm K}(\pi\rho_0(0))}{2\pi}\right)^2\nonumber\\
&&\times \biggl(
\frac{1}{2T} 
\bigl(B\tanh(B/T) \bigr)^2 -3\ln(2)-\frac{B^2}{2T}
\nonumber \\
&&\hphantom{\times \biggl(}
+2B\tanh(B/T)(1+\C-\ln(2\pi T)\nonumber \\
&&\hphantom{\times \biggl(} -TQ(B/T) \Bigr)\; ,
\nonumber \\
Q(b)&=&
b \tanh(b)
\bigl(\psi(\rmi b/\pi) + \psi(-\rmi b/\pi)+2\C\bigr) 
\; .\nonumber 
\end{eqnarray} 

\subsubsection{Limit of small temperatures}

Setting $g=(J_{\rm K}\pi \rho_0(0))/2$ and taking
the limit of small temperature compared to the magnetic field,
$B \gg T$, eq.~(\ref{eq:largecalDdone}) simplifies to
\begin{eqnarray}
  {\cal F}^{\rm ii}(B,0)&\approx&
  -3\left(\frac{g}{\pi}\right)^2\ln(2) -B +
  \left(\frac{g}{\pi}\right)  B \nonumber \\
&& +2\left(\frac{g}{\pi}\right)^2B\bigl(1-\ln(2)-\ln(B)\bigr) \label{eq:FreeFlargeB}
\end{eqnarray}
with corrections of the order $T^2/B$.
Eq.~(\ref{eq:FreeFlargeB}) agrees with eq.~(6.17) of AFL.

The impurity-induced magnetization becomes
\begin{equation}
  \frac{2m^{\rm ii}(B,0)}{g_e\mu_{\rm B}}
  \approx  
    1-\frac{1}{2} \left(\frac{2g}{\pi}\right)
    + \frac{1}{2} \left(\frac{2g}{\pi}\right)^2 \ln\left(\frac{2B}{{\cal D}}\right)
    \; ,
  \label{appeq:mfromPTsmall}
  \end{equation}
in agreement with eq.~(6.19) of AFL.

\subsubsection{Limit of small fields}

We grouped the expression in eq.~(\ref{eq:largecalDdone}) so that
they have a regular Taylor series around $B/T=0$.
Thus, for $B \ll T$ we find
\begin{eqnarray}
  {\cal F}^{\rm ii}(B\ll T,T)
  &\approx & 
- \Bigl(T+3\left(\frac{g}{\pi}\right)^2\Bigr) \ln(2)-\frac{B^2}{2T}
+\left(\frac{g}{\pi}\right) \frac{B^2}{T}
\nonumber \\
&&
+\left(\frac{g}{\pi}\right)^2
\frac{2 B^2}{T}
\left(\frac{3}{4} +  \C -  \ln(2\pi T)\right)\nonumber \\
&=& {\cal F}^{\rm ii}(0,T)-\frac{B^2}{2T} +\frac{g}{\pi} \frac{B^2}{T}\nonumber \\
&&
+\left(\frac{g}{\pi}\right)^2 \frac{2 B^2}{T}\ln\left(\frac{U}{2T}\right) \; , \nonumber \\
U&=& \frac{e^{3/4+\C}}{\pi} \; ,
\label{eq:Rdef}
\end{eqnarray}
where $g=(J_{\rm K}\pi \rho_0(0))/2$.
This result favorably compares with eq.~(6.17) of AFL who give
\begin{eqnarray}
U_{\rm AFL}&=&2\beta_{\rm AFL} \gamma_{\rm AFL} e^{-7/4} \; \nonumber \\
\gamma_{\rm AFL}&=&e^{\C} \; , \nonumber \\
\ln(\beta_{\rm AFL })&=& 
\int_0^1\rmd x \frac{(1-x)^2}{x}
\left[\left(\frac{\pi x}{\sin(\pi x)}\right)^2 -1\right] \nonumber \\
&=& \frac{5}{2}-\ln(2\pi) \; .
\label{eq:betaAFLdef}
\end{eqnarray}
The integral was done using {\sc Mathematica}.~\cite{Mathematica11}
Thus,
\begin{equation}
\ln(U_{\rm AFL})=\ln(2)+\frac{5}{2}-\ln(2\pi)+\C-\frac{7}{4}
=\frac{3}{4} +\C -\ln(\pi)\; ,
\end{equation}
in agreement with eq.~(\ref{eq:Rdef}).

The impurity-induced magnetization becomes
\begin{equation}
  \frac{m^{\rm ii}(B\ll T,T)}{g_e\mu_{\rm B}}
  \approx  \frac{B}{T}\frac{1}{2}
  \biggl[
    1-\left(\frac{2g}{\pi}\right)
    + \left(\frac{2g}{\pi}\right)^2 \ln\left(\frac{2T}{U{\cal D}}\right)
    \biggr]     ,
  \end{equation}
in agreement with eq.~(6.19) of AFL where, at the electrons' gyromagnetic
factor $g_{\rm e}=2$,
$m^{\rm ii}={\cal M}/(2\mu_{\rm B})$,
and we made the dependence on the bandwidth-parameter ${\cal D}\equiv 1$ explicit.

\subsection{Wide-band limit for a general density of states}

Before we compare to the Bethe-Ansatz results, we first derive the
limiting expressions for small fields and small temperatures in the wide-band limit
for the case of a general density of states. We shall see that a correction factor
enters the final equations that depends on the density of states.

We start from the general expressions~(\ref{eq:F14}), (\ref{eq:F15})
and~(\ref{eq:C24}) to calculate the second-order correction to the free energy,
\begin{eqnarray}
  {\cal F}_2^{\rm ii}(B,T)
  &=& \left(\frac{J_{\rm K}\pi \rho_0(0}{2\pi}\right)^2 f(B,T) \; , \nonumber \\
  f(B,T) &=& \frac{1}{2T} \left[B\tanh\left(\frac{B}{2T}\right)\right]^2
  -\frac{B^2}{2T} \nonumber  \; , \\
  && +{\cal A}(B,T) +{\cal B}(B,T) \label{eq:F60}\\
  {\cal A}(B,T) &=& -\frac{1}{2} \int_0^{1/(2T)}\rmd \lambda
    \frac{I_+(\lambda)I_-(\lambda)}{(\pi\rho_0(0))^2} \; , \nonumber \\
       {\cal B}(B,T) &=& -\frac{e^{B/T}}{2\cosh(B/T)} \int_0^{1/(2T)}\rmd \lambda
       \left(\frac{I_-(\lambda)}{\pi \rho_0(0)}\right)^2\nonumber \\
         && -\frac{e^{-B/T}}{2\cosh(B/T)} \int_0^{1/(2T)}\rmd \lambda
       \left(\frac{I_+(\lambda)}{\pi \rho_0(0)}\right)^2\; ,\nonumber
\end{eqnarray}
where we used eq.~(\ref{eq:F28}).
Recall that
\begin{equation}
  \frac{I_{\pm}(\lambda)}{\pi \rho_0(0)} = \int_{-1}^1 \rmd \epsilon
  \frac{\rho_0(\epsilon)}{\rho_0(0)}
  \frac{e^{\lambda\epsilon}}{1+\exp((\epsilon\mp B)/T)}
    \; .
\end{equation}

\subsubsection{Limit of small temperatures}

We are interested in the limit $T\to 0$ where we
neglect terms of the order $B^2/{\cal D}^2$.
In eq.~(\ref{eq:F60}), the first term drops out and the second term can be written as
\begin{equation}
  {\cal A}(B,0)= \frac{1}{2}
  \int_{-1}^{B} \rmd \epsilon_1 \frac{\rho_0(\epsilon_1)}{\rho_0(0)}
  \int_{-1}^{-B} \rmd \epsilon_2 \frac{\rho_0(\epsilon_2)}{\rho_0(0)}
 \frac{1}{\epsilon_1+\epsilon_2} \; .
\end{equation}
At $B=0$, the double integral converges because it is proportional to the
ground-state energy. Furthermore, it is symmetric in~$B$ so that corrections are of
the order $B^2$ for small~$B/{\cal D}$. Therefore, we can safely ignore the
contribution to ${\cal F}_2^{\rm ii}(B,0) $ from $  {\cal A}(B,0)$.

At $T=0$, the third contribution in eq.~(\ref{eq:F60}) reduces to
\begin{eqnarray}
  {\cal B}(B,0)&=& \frac{1}{2}
  \int_{-1}^{-B} \rmd \epsilon_1 \frac{\rho_0(\epsilon_1)}{\rho_0(0)}
  \int_{-1}^{-B} \rmd \epsilon_2 \frac{\rho_0(\epsilon_2)}{\rho_0(0)}
  \frac{1}{\epsilon_1+\epsilon_2} \nonumber \\
  &=& {\rm const} +\Delta_1(B) +\Delta_2(B)\nonumber \; , \\
  \Delta_1(B) &=&   - \int_{-B}^0 \rmd \epsilon_1 \frac{\rho_0(\epsilon_1)}{\rho_0(0)}
\int_{-B}^0 \rmd \epsilon_2 \frac{\rho_0(\epsilon_2)}{\rho_0(0)}
  \frac{1}{\epsilon_1+\epsilon_2}   \; , \nonumber \\
  \Delta_2(B)&=& -2   \int_{-1}^{-B} \rmd \epsilon_1 \frac{\rho_0(\epsilon_1)}{\rho_0(0)}
  \int_{-B}^0 \rmd \epsilon_2 \frac{\rho_0(\epsilon_2)}{\rho_0(0)}
  \frac{1}{\epsilon_1+\epsilon_2} \; . \nonumber \\
\end{eqnarray}
Since $B$ is small, we may approximate $\rho_0(\epsilon_{1,2})\approx \rho_0(0)$
in $\Delta_1(B)$ so that
\begin{equation}
\Delta_1(B)= 2B\ln(2) +{\cal O}(B^3) \; .
\end{equation}
The same approximation can be used for the integral over $\epsilon_2$ in $\Delta_2(B)$.
We ignore contributions of the order $B^2$ to find
\begin{eqnarray}
  \Delta_2(B) &\approx &
  2   \int_{-1}^{-B} \rmd \epsilon_1 \frac{\rho_0(\epsilon_1)}{\rho_0(0)}
  \ln\left( 1-\frac{B}{\epsilon_1}\right)  \nonumber \\
  &=&
  2   \int_{-1}^{-B} \rmd \epsilon_1 
  \ln\left( 1-\frac{B}{\epsilon_1}\right)  \nonumber \\
  && + 2   \int_{-1}^{-B} \rmd \epsilon_1 \left(\frac{\rho_0(\epsilon_1)}{\rho_0(0)}-1\right)
  \ln\left( 1-\frac{B}{\epsilon_1}\right) \; .
  \nonumber \\
\end{eqnarray}
In the second integral we can safely expand the logarithm in~$B$ for a density
of states that is regular around $\epsilon=0$. Thus, we obtain the final result
\begin{eqnarray}
  \Delta_2(B) &\approx &{\rm const}+ 2B\left[1-2\ln(2)-\ln(B)+\ln(\F)\right] \;, \nonumber \\
  \ln(\F) &=& -\int_{-1}^0 \frac{\rmd \epsilon}{\epsilon}
  \left(\frac{\rho_0(\epsilon)}{\rho_0(0)}-1\right) 
  \; . \label{eq:defF}
\end{eqnarray}
As compared to the result for the constant density of states,
we simply have to replace $B$ by $B/\F$ in the logarithmic term.

Thus, from eq.~(\ref{appeq:mfromPTsmall})
the impurity-induced ground-state magnetization for finite fields becomes
\begin{equation}
  \frac{2m^{\rm ii}(B,0)}{g_e\mu_{\rm B}}
  \approx  
    1-\frac{1}{2} \left(\frac{2g}{\pi}\right)
    + \frac{1}{2} \left(\frac{2g}{\pi}\right)^2 \ln\left(\frac{2B}{{\cal D}\F}\right)
    \; .
  \label{appeq:mfromPTsmallgeneral}
  \end{equation}

\subsubsection{Limit of small fields}

For small fields~$B$ and finite temperatures~$T$,
we expand $I_{\pm}(\lambda)$ to second order in~$B$,
\begin{eqnarray}
  I_{\pm}(\lambda) &=& I_0(\lambda) \pm B I_1(\lambda)+B^2 I_2(\lambda) \; ,\nonumber\\
  I_0(\lambda) &=& \int_{-1}^1 \rmd \epsilon
  \frac{\rho_0(\epsilon)}{\rho_0(0)}
  \frac{e^{\lambda\epsilon}}{1+\exp(\epsilon/T)}\; , \nonumber \\
  I_1(\lambda) &=&\frac{1}{4T} \int_{-1}^1 \rmd \epsilon
  \frac{\rho_0(\epsilon)}{\rho_0(0)}
  \frac{e^{\lambda\epsilon}}{[\cosh(\epsilon/(2T)]^2}\; , \nonumber \\
  I_2(\lambda) &=& \frac{1}{8T^2} \int_{-1}^1 \rmd \epsilon
  \frac{\rho_0(\epsilon)}{\rho_0(0)}
  \frac{e^{\lambda\epsilon}\tanh(\epsilon/(2T))}{[\cosh(\epsilon/(2T)]^2}
  \; . \label{appeq:Indef}
\end{eqnarray}
For a constant density of states we have 
\begin{eqnarray}
  I_0^{\rm const}(\lambda) &=& \frac{\pi T}{\sin(\pi T \lambda)}
  -\frac{e^{-1\lambda}}{\lambda}
  \; , \nonumber \\
  I_1^{\rm const}(\lambda) &=&
\frac{\pi T}{\sin(\pi T \lambda)}
  \; , \nonumber \\
  I_2^{\rm const}(\lambda) &=&
\frac{\lambda}{2} I_1(\lambda) \; ,
\end{eqnarray}
with exponentially small corrections.
These expressions can be used to check the results obtained 
in Sect.~\ref{subsec:freeenergyconstDOS}.

In eq.~(\ref{eq:F60}), the first term gives $-B^2/(2T)$, and the second term gives
\begin{eqnarray}
  {\cal A}(B,T)&=& {\rm const} \\
  && +\frac{B^2}{2} \int_0^{1/(2T)} \rmd \lambda
  \left(I_1(\lambda)^2-2I_0(\lambda)I_2(\lambda)\right) \;, \nonumber
  \label{eq:calAsmallB}
\end{eqnarray}
up to and including order~$B^2$.
The integral is finite in the limit $T\to 0$
so that this term is irrelevant for small~$B/T$.

The third term in eq.~(\ref{eq:F60}) gives
\begin{eqnarray}
  {\cal B}(B,T)&=& {\rm const}+\frac{2B^2}{T} \left[\Delta_3(T)+\Delta_4(T)\right]
  \nonumber\\
  \Delta_3(T)&=& -\frac{T}{2} \int_0^{1/(2T)} \rmd \lambda
  \left(I_1(\lambda)^2+2I_0(\lambda)I_2(\lambda)\right) \nonumber \\
  \Delta_4(T)&=& \int_0^{1/(2T)} \rmd \lambda
  I_0(\lambda)I_1(\lambda) 
\end{eqnarray}
up to order~$B^2$.
For the calculation of $\Delta_3(T)$ we use~(\ref{eq:calAsmallB})
to write for $T/{\cal D}\ll 1$
\begin{eqnarray}
\Delta_3(T)&\approx & -T \int_0^{1/(2T)} \rmd \lambda
  I_1(\lambda)^2 \nonumber \\
&\approx & -T \int_0^{1/(2T)} \rmd \lambda
  \left(I_1^{\rm const}(\lambda)\right)^2\nonumber \\
  &=&- \ln(2)
\end{eqnarray}
because, in the limit $T\to 0$, the factor $\cosh^{-2}[\epsilon/(2T)]$
in the definition of $I_1(\lambda)$ in eq.~(\ref{appeq:Indef})
restricts the integration to the region around $\epsilon\approx 0$
so that we can use the expressions for a constant density of states
to evaluate $\Delta_3(T\to 0)$.

Likewise, we have
\begin{eqnarray}
\Delta_4(T)&\approx  & \int_0^{1/(2T)} \rmd \lambda
I_0^{\rm const} (\lambda)I_1^{\rm const}(\lambda) \nonumber\\
&&  +\int_0^{1/(2T)} \rmd \lambda
\left(I_0(\lambda)-I_0^{\rm const} (\lambda)\right)I_1^{\rm const}(\lambda) \nonumber\\
&=& -\ln(T)+1+\C-\ln(\pi) \nonumber \\
&& +\int_0^{\infty} \rmd \lambda \int_{-1}^0\rmd \epsilon
\left(\frac{\rho_0(\epsilon)}{\rho_0(0)}-1\right) e^{\lambda\epsilon}\nonumber \\
& =& -\ln(T)+1+\C-\ln(\pi) \nonumber \\
&&+\ln(\F) \;, \label{eq:logTwithF}
\end{eqnarray}
where we used {\sc Mathematica}~\cite{Mathematica11} to solve the first integral;
recall that $\C\approx 0.577216$ is Euler's constant.
Moreover, we performed the limit $T\to 0$ in the second integral, namely,
\begin{equation}
  \left.I_1^{\rm const}(\lambda)\right|_{T\to 0}= 1 \quad, \quad
  \left.\frac{1}{1+\exp(\epsilon/T)}\right|_{T\to 0}=\theta_{\rm H}(-\epsilon) \; .
\end{equation}
The factor $\F$ is defined in eq.~(\ref{eq:defF}).

When we compare eq.~(\ref{eq:Rdef}) and eq.~(\ref{eq:logTwithF}) we see
that we have to replace $\ln(T)$ by $\ln(T/\F)$ to generalize the results for
the constant density of states to the case of a general density of states.
Therefore, we find for the impurity-induced magnetization 
\begin{eqnarray}
  \frac{m^{\rm ii}(B\ll T,T)}{g_e\mu_{\rm B}}
  &\approx&  \frac{B}{2T}
  \biggl[
    1-\left(\frac{2g}{\pi}\right)
    + \left(\frac{2g}{\pi}\right)^2 \ln\left(\frac{2T}{U\F{\cal D}}\right)
    \biggr] \nonumber \\
  \label{eq:maggeneralsmallB}
  \end{eqnarray}
in the limit of small fields for a general density of states.

\subsection{Comparison with Bethe Ansatz}

\subsubsection{Zero-field susceptibility at finite temperatures}

Using the result in eq.~(\ref{eq:maggeneralsmallB}), 
the zero-field impurity-induced susceptibility at finite temperatures becomes
\begin{equation}
  \frac{\chi_0^{\rm ii}(T)}{(g_e\mu_{\rm B})^2}\approx
  \frac{1}{4T}   \left[
    1-\left(\frac{2g}{\pi}\right)
    + \left(\frac{2g}{\pi}\right)^2 \ln\left(\frac{2T}{U\F{\cal D}}\right)
    \right]
  \; .
  \label{eq:chiforTsmall}
  \end{equation}
The Bethe Ansatz solution of AFL provides the following result
for $T\gg T_{\rm K}$,
\begin{eqnarray}
    \frac{\chi_0^{\rm ii}(T\gg T_{\rm K})}{(g_e\mu_{\rm B})^2}&\approx&
\frac{1}{4T}  \biggl[
    1-\left(\ln\left(\frac{T}{T_{\rm K}}\right)\right)^{-1} \nonumber \\
    && \hphantom{\frac{1}{4T}\! }
    -\frac{1}{2}\left(\ln\left(\frac{T}{T_{\rm K}}\right)\right)^{-2}
    \ln\left(\ln\left(\frac{T}{T_{\rm K}}\right)\right)
        \biggr] ,
  \nonumber \\
  \label{eq:BAforlargeT}
  \end{eqnarray}
see eq.~(5.61b) of AFL.

Note that the expansion~(\ref{eq:BAforlargeT}) is derived
for $T\gg T_{\rm K}$. However, the radius of convergence also covers the
region $T\lesssim T_{\rm K}$ so that we can use the expansion
\begin{equation}
  \frac{1}{\ln(T)+|\ln(T_{\rm K})|}=\frac{1}{|\ln(T_{\rm K})|}
  \left(1-\frac{\ln(T)}{|\ln(T_{\rm K})|}\right) +\ldots
\end{equation}
in eq.~(\ref{eq:BAforlargeT}) to find
\begin{eqnarray}
  \frac{\chi_0^{\rm ii}(T)}{(g_e\mu_{\rm B})^2}&\approx&
  \frac{1}{4T}  \biggl[
    1- \frac{1}{|\ln(T_{\rm K})|}
    + \frac{\ln(T)}{[\ln(T_{\rm K})]^2}\nonumber \\
    &&\hphantom{ \frac{1}{4T}  \biggl[}
    -\frac{1}{2} \frac{\ln(|\ln(T_{\rm K})|)}{[\ln(T_{\rm K})]^2}
        \biggr] \; .
  \label{eq:chifromlargeBA}
  \end{eqnarray}
This result can now be compared with eq.~(\ref{eq:chiforTsmall}).

To this end, we write
\begin{equation}
T_{\rm K}=t_{\rm K}T_{\rm K}^{(0)} \quad, \quad |\ln(t_{\rm K})|\ll |\ln(T_{\rm K}^{(0)})|
\; ,
\end{equation}
and find
\begin{eqnarray}
  \frac{\chi_0^{\rm ii}(T)}{(g_e\mu_{\rm B})^2}&\approx&
  \frac{1}{4T}
  \biggl[ 1- \frac{1}{|\ln(T_{\rm K}^{(0)})|} -\frac{|\ln(t_{\rm K})|}{[\ln(T_{\rm K}^{(0)})]^2}
\nonumber \\
    &&\hphantom{\frac{1}{2}  \biggl[}
    + \frac{\ln(T)}{[\ln(T_{\rm K}^{(0)})]^2}
    -\frac{1}{2} \frac{\ln(|\ln(T_{\rm K}^{(0)})|)}{[\ln(T_{\rm K})^{(0)}]^2}
        \biggr]\nonumber \\
  \label{eq:chichifromlarge}
\end{eqnarray}
so that
\begin{eqnarray}
\frac{1}{|\ln(T_{\rm K}^{(0)})|}&=&  \frac{2g}{\pi}\; ,\\
    - \ln\left(\frac{U \F{\cal D}}{2}\right) &=&
    -|\ln(t_{\rm K})| -    \frac{1}{2} \ln(|\ln(T_{\rm K}^{(0)})|)
\nonumber
\end{eqnarray}
or
\begin{eqnarray}
T_{\rm K}^{(0)}&=& \exp\left(-\frac{\pi}{2g}\right) \; , \nonumber\\
t_{\rm K} &=& {\cal D}  \frac{U\F}{2} \sqrt{\frac{2g}{\pi}}\; .\nonumber\\
T_{\rm K}&=& {\cal D} \frac{U\F}{2} \sqrt{\frac{2g}{\pi}} \exp\left(-\frac{\pi}{2g}\right)
\;.  \label{eq:linkTKandg}
\end{eqnarray}

\subsubsection{Impurity-induced magnetization at zero temperature}

The Bethe Ansatz solution of AFL provides the following result
for the impurity-induced magnetization at zero temperature for $B\gg T_{\rm H}$,
\begin{eqnarray}
  \frac{2m^{\rm ii}(B\gg T,T)}{g_e\mu_{\rm B}}
  &\approx & 
    1-\frac{1}{2} \left(\ln\left(\frac{B}{T_{\rm H}}\right)\right)^{-1}\nonumber\\
&&    -\frac{1}{4}\left(\ln\left(\frac{B}{T_{\rm H}}\right)\right)^{-2}
    \ln\left(\ln\left(\frac{B}{T_{\rm H}}\right)\right)
  \nonumber \\
  \label{appeq:BAforlargeB}
  \end{eqnarray}
with
\begin{equation}
  T_{\rm H}=
  D_{\rm AFL}\sqrt{\frac{\pi}{e}} \exp\left(-\frac{\pi}{J_{\rm K}^{\rm BA}}\right)
  \label{eq:defTH}
  \end{equation}
from eq.~(4.28) and eq.~(3.17) of AFL,
where $2D_{\rm AFL}=1$. Note that the expansion~(\ref{appeq:BAforlargeB}) is derived
for $B\gg T_{\rm H}$. However, the radius of convergence also covers the
region $B\lesssim T_{\rm H}$ so that we can use the expansion
\begin{equation}
  \frac{1}{\ln(B)+|\ln(T_{\rm H})|}=\frac{1}{|\ln(T_{\rm H})|}
  \left(1-\frac{\ln(B)}{|\ln(T_{\rm H})|}\right) +\ldots
\end{equation}
in eq.~(\ref{appeq:BAforlargeB}) to find
\begin{eqnarray}
  \frac{2 m^{\rm ii}(B,T)}{g_e\mu_{\rm B}}
  &\approx & 
      1-\frac{1}{2} \frac{1}{|\ln(T_{\rm H})|}
    +\frac{1}{2} \frac{\ln(B)}{[\ln(T_{\rm H})]^2}\nonumber \\
&&    -\frac{1}{4} \frac{\ln(|\ln(T_{\rm H})|)}{[\ln(T_{\rm H})]^2}\; .
  \label{appeq:mfromlargeBA}
  \end{eqnarray}
This result can now be compared with eq.~(\ref{appeq:mfromPTsmallgeneral}).

To this end, we write
\begin{equation}
T_{\rm H}= t_{\rm H} T_{\rm H}^{(0)} \quad, \quad
| \ln(t_{\rm H}) | \ll | \ln (T_{\rm H}^{(0)}) | \; ,
\end{equation}
and find 
\begin{eqnarray}
 \frac{2m^{\rm ii}(B\gg T,T)}{g_e\mu_{\rm B}}&\approx &
 1-\frac{1}{2} \frac{1}{|\ln(T_{\rm H}^{(0)})|}
 -\frac{1}{2} \frac{|\ln(t_{\rm H})|}{[\ln(T_{\rm H}^{(0)})]^2}
\nonumber \\
    &&    +\frac{1}{2} \frac{\ln(B)}{[\ln(T_{\rm H}^{(0)})]^2}
    -\frac{1}{4} \frac{\ln(|\ln(T_{\rm H}^{(0)})|)}{[\ln(T_{\rm H})^{(0)}]^2}\nonumber \\
  \label{appeq:mfromlarge}
\end{eqnarray}
Therefore,
\begin{eqnarray}
\frac{1}{|\ln(T_{\rm H}^{(0)})|} &=&   \frac{2g}{\pi}
  \nonumber \; , \\
\ln(2/({\cal D}\F)) &=&   -|\ln(t_{\rm H})| -    \frac{1}{2} \ln(|\ln(T_{\rm H}^{(0)})|)
\end{eqnarray}
or
\begin{eqnarray}
T_{\rm H}^{(0)}&=& \exp\left(-\frac{\pi}{2g}\right) \; , \nonumber\\
t_{\rm H} &=& \frac{{\cal D}\F}{2}\sqrt{\frac{2g}{\pi}}\; ,\nonumber\\
T_{\rm H}&=& \frac{{\cal D}\F}{2}\sqrt{\frac{2g}{\pi}} \exp\left(-\frac{\pi}{2g}\right)
\;.  \label{appeq:linkTHandg}
\end{eqnarray}
It is obvious that the prefactor does not contain the Wilson number.

\subsubsection{Wilson number}

The Wilson number is defined by ($e=\exp(1)$)
\begin{equation}
  w_{\rm AFL}= \sqrt{\frac{\pi}{e}}\frac{T_{\rm K}}{T_{\rm H}}    \; ,
  \end{equation}
see eq.~(6.14) of AFL. We use eqs.~(\ref{eq:linkTKandg}) and~(\ref{appeq:linkTHandg})
to find
\begin{equation}
  w_{\rm AFL}= \sqrt{\frac{\pi}{e}} U=\frac{e^{\C+1/4}}{\sqrt{\pi}}  \quad
    , \quad
    \frac{w_{\rm AFL}}{4\pi} \approx0.102676 \; ,
\label{eq:wAFLvalue}
      \end{equation}
in agreement with eq.~(6.23) of AFL, 
where $U$ is defined in eq.~(\ref{eq:Rdef}).
In Hewson's book,~\cite{Hewson} the Wilson number is defined as
\begin{equation}
  w=\frac{w_{\rm AFL}}{\pi} = \frac{e^{\C+1/4}}{\pi^{3/2}}
    \approx 0.410705 \; .
\label{appeq:defwHewson}
\end{equation}
Note that the Wilson number does not depend
on the factor~$\F$ in eq.~(\ref{eq:defF}) because both $T_{\rm K}$ and $T_{\rm H}$
are proportional to~$\F$. Therefore, the Wilson number is universal in the sense
that it is independent of the density of states of the host electrons.

\end{document}